\documentclass[twoside,11pt]{article}

\usepackage{blindtext}
\usepackage[abbrvbib,preprint]{jmlr2e}
\usepackage{enumerate}
\usepackage{url}

\usepackage{bm}
\usepackage{amsfonts,amscd}
\usepackage[]{units}
\usepackage{listings}
\usepackage{multicol}
\usepackage{tcolorbox}
\usepackage{physics}
\usepackage{chngcntr}
\usepackage{caption,capt-of}
\usepackage{comment}
\usepackage{booktabs}
\usepackage{array}
\usepackage{paralist}
\usepackage{verbatim}
\usepackage{subfigure}
\usepackage{multirow,makecell}
\usepackage{rotating}
\usepackage{fancyhdr}
\usepackage{hyperref}
\usepackage{xcolor}
\usepackage[ruled,lined]{algorithm2e}
\usepackage{algpseudocode}

\newcommand{\R}{\mathbb{R}}
\newcommand{\E}{\mathbb{E}}

\newcommand{\iidsim}{\overset{i.i.d}{\sim}}
\newcommand{\indsim}{\overset{ind}{\sim}}

\newcommand{\IW}{\text{IW}} 
\newcommand{\MatrixN}{\text{MN}} 
\newcommand{\Norm}{\text{N}} 
\newcommand{\Unif}{\text{Unif}} 
\newcommand{\logN}{\text{log-N}} 
\newcommand{\Beta}{\text{Beta}} 
\newcommand{\Bin}{\text{Bin}} 
\newcommand{\NB}{\text{NB}} 
\newcommand{\Cat}{\text{Cat}} 
\newcommand{\Gam}{\text{Gamma}} 
\newcommand{\DP}{\text{DP}} 
\newcommand{\Dir}{\text{Dir}} 
\newcommand{\IG}{\text{IG}} 

\newcommand{\tsum}{\text{sum}}

\newcommand{\bigO}{\mathcal{O}}

\newcommand{\xcenter}{x_{j,d}^*}
\newcommand{\tcenter}{t_{j,d}^*}
\newcommand{\plambda}{\lambda_{j,d}^*}
\newcommand{\sdcenter}{{\sigma^*_{j,d}}^2}
\newcommand{\ssdcenter}{{\sigma^*}^2}

\newcommand{\indicator}{\mathbb{I}} 
\newcommand{\vd}{\mathrm{d}}

\newcommand{\bB}{\mathbf{B}}
\newcommand{\bL}{\mathbf{L}}
\newcommand{\bV}{\mathbf{V}}
\newcommand{\bPhi}{\mathbf{\Phi}}
\newcommand{\bI}{\mathbf{I}}

\newcommand{\bS}{\mathbf{S}}
\newcommand{\bY}{\mathbf{Y}}

\newcommand{\bX}{\mathbf{X}}

\newcommand{\bZ}{\mathbf{Z}}

\newcommand{\btheta}{\bm{\theta}}

\newcommand{\bpsi}{\bm{\psi}}
\newcommand{\bmu}{\bm{\mu}}
\newcommand{\bphi}{\bm{\phi}}

\newcommand{\blambda}{\bm{\lambda}}
\newcommand{\bxi}{\bm{\xi}}
\newcommand{\bbeta}{\bm{\beta}}
\newcommand{\bsigma}{\bm{\sigma}}

\newcommand{\bSigma}{\bm{\Sigma}}

\newcommand{\bmy}{\bm{y}}
\newcommand{\bmx}{\bm{x}}

\newcommand{\bmm}{\bm{m}}
\newcommand{\bma}{\bm{a}}
\newcommand{\bmb}{\bm{b}}
\newcommand{\bmc}{\bm{c}}
\newcommand{\bmq}{\bm{q}}
\newcommand{\bmp}{\bm{p}}
\newcommand{\bmr}{\bm{r}}
\newcommand{\bmh}{\bm{h}}
\newcommand{\bmt}{\bm{t}}
\newcommand{\bmmu}{\bm{u}}

\ShortHeadings{Covariate-Dependent Hierarchical Dirichlet Processes}{Zhang, Wade and Bochkina}
\firstpageno{1}

\begin{document}

\title{Covariate-dependent hierarchical Dirichlet processes}

\author{\name Huizi Zhang \email H.Zhang-144@sms.ed.ac.uk \\
	\addr School of Mathematics and Maxwell Institute for Mathematical Sciences\\
	University of Edinburgh\\
	Edinburgh, EH9 3FD, UK 
	\AND
	\name Sara Wade \email sara.wade@ed.ac.uk \\
	\addr School of Mathematics and Maxwell Institute for Mathematical Sciences\\
	University of Edinburgh\\
	Edinburgh, EH9 3FD, UK 
	\AND
	\name Natalia Bochkina \email N.Bochkina@ed.ac.uk \\
	\addr School of Mathematics and Maxwell Institute for Mathematical Sciences\\
	University of Edinburgh\\
	Edinburgh, EH9 3FD, UK }

\maketitle

\begin{abstract}
 Bayesian hierarchical modeling is a natural framework to  effectively integrate data and borrow information across groups. 
In this paper, we address  problems related to density estimation and identifying clusters across related groups, by proposing a hierarchical Bayesian  approach  
that incorporates additional covariate information. 
To achieve flexibility,  our approach builds on ideas from Bayesian nonparametrics, combining the hierarchical Dirichlet process with dependent Dirichlet processes. The proposed model is widely applicable, accommodating multiple and mixed covariate types through appropriate kernel functions as well as different output types through suitable component-specific likelihoods.  
This extends our ability to discern the relationship between covariates and clusters, while also effectively borrowing information and quantifying differences across groups. By employing a data augmentation trick, we are able to  tackle the intractable normalized weights and construct a Markov chain Monte Carlo algorithm for posterior inference.  The proposed method is illustrated on simulated data and two real data sets on single-cell RNA sequencing (scRNA-seq) and calcium imaging. For scRNA-seq data, we show that the incorporation of cell dynamics facilitates the discovery of additional cell subgroups. On calcium imaging data, our method identifies interpretable clusters of time frames with similar neural activity, aligning with the observed behavior of the animal. 
\end{abstract}


\section{Introduction}
\label{sec:intro}
In many studies, multiple data sets are collected across groups, where each group may represent multiple experiments, geographical sites, time points, and more. To effectively integrate and borrow information across groups, the Bayesian hierarchical framework is a natural choice. This paper delves into the problems of density estimation and clustering across related groups, proposing a Bayesian  approach that extends existing approaches in the presence of  additional covariate information.

With advancements in data acquisition, the need for such tools is growing. For example, in biological studies, complex data are typically collected across multiple groups, which may correspond to repeated experiments or different treatment conditions, tissues, or time points. Moreover, additional side information is usually also available for each observation, such as patient characteristics in bulk RNA studies or cellular dynamics \citep{La_Manno2018}  in single-cell RNA sequencing (scRNA-seq). Indeed, the lack of tools for integrating and quantifying differences across multiple single-cell data sets was emphasized as one of the grand challenges in single-cell data science \citep{lahnemann2020eleven}. Another example is information retrieval scenarios dealing with raw documents from multiple corpora, with externally observed categorical information  \citep{kim2014hierarchical}. 

In such settings, we focus on two problems. On one hand, clustering is important to uncover inherent structure. For example, in scRNA-seq, clustering is used to disentangle the heterogeneous gene expression measurements across cells and discover cell subtypes with similar expression patterns. By including cellular-level covariate information, such as dynamics, we can quantify the relationship between cell subtypes and dynamics. On the other hand, covariate-dependent density estimation (also known as density regression) allows modeling  the whole density of the response, not only the mean, to change with the covariates. For example, in bulk-RNA studies, we may be interested in understanding how the distribution of expression for certain genes changes across different patients characteristics, while also borrowing information across multiple sites.  

Mixture models \citep{fruhwirth2019handbook} arise as a natural choice in this context, providing both a probabilistic framework for clustering as well as  a useful tool for density estimation due to their attractive balance between smoothness and flexibility.  
As an alternative to parametric mixture models, Bayesian nonparametric (BNP) methods \citep{Ghosal2017} are widely used to avoid pre-specifying the number of clusters, instead allowing it to grow unboundedly with the number of observations, by placing a nonparametric prior on the unknown mixing measure.  Moreover, BNP mixture models are further supported by strong theoretical properties that provide frequentist validation \citep[e.g.][]{GGR99,GV01,WG10,shen2013adaptive}. 
The Dirichlet process (DP) \citep{Ferguson1973bayesian} is unarguably the most common prior choice in BNP literature and has many desirable properties including easy elicitation of its parameters, conjugacy, large support, and posterior consistency \citep[][Chapter 4]{Ghosal2017}.

To effectively model more complex data structures, a number of extensions of the Dirichlet process have been proposed. Recent reviews and comparisons of dependent extensions to accommodate covariates are provided in \cite{quintana2022dependent} and \cite{wade2023bayesian}. Two of the most widely-used proposals include the dependent Dirichlet process (DDP) \citep{MacEachern1999DDP} and the hierarchical Dirichlet process (HDP) \citep{Teh2006}. The latter concentrates exclusively on \textit{partially exchangeable data}, when covariates represent groups and exchangleability holds within group and across group labels (e.g. for observations on patients across different hospitals, the ordering of the patients and hospitals can be shuffled, as long as the same patients belong within each hospital). Thus, the HDP  enables clustering and density estimation across related groups, and moreover allows prediction for new groups. In contrast, the DDP incorporates fixed effects of covariates for conditional density estimation and quantifying covariate effects on clustering.

\begin{figure}[tbp]
	\centering
	\includegraphics[width=0.95\textwidth]{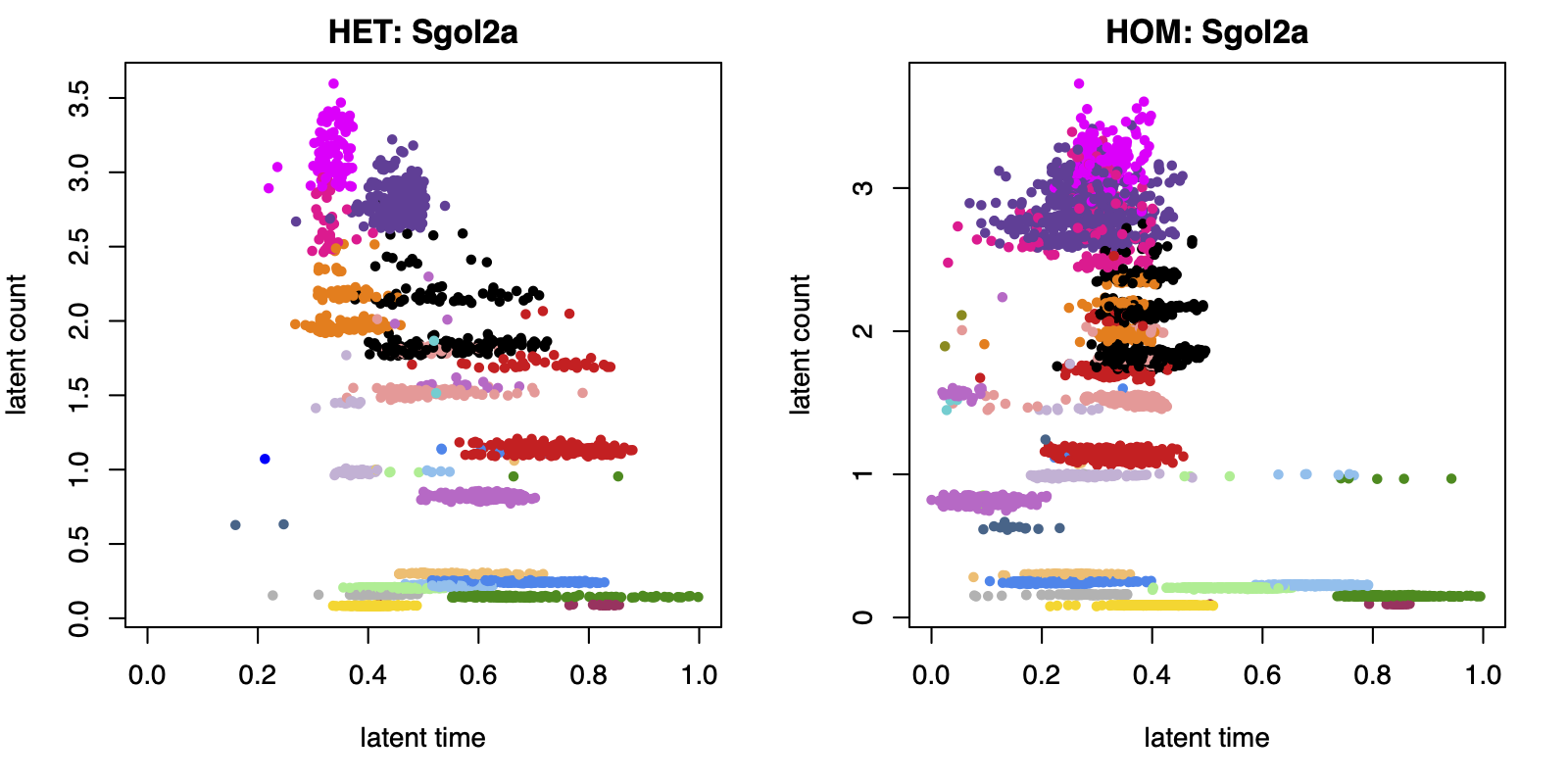}
	\caption{Latent counts versus latent time for scRNA-seq data \citep{tan2022pax6} collected under two experimental conditions (HET and HOM). Cells are colored by cluster membership from \cite{Liu2024} without information of latent time.}
	\label{fig:y0_het_vs_t}
\end{figure}

In this article, we propose a covariate-dependent hierarchical Dirichlet process (C-HDP), combining the hierarchical Dirichlet process with the dependent Dirichlet process. Our focus is the DDP that uses normalized weights and kernels to construct covariate-dependent weights \citep{foti2012slice, antoniano2014bayesian}, as they have the flexibility to recover a variety of complex data-generating scenarios and enhanced interpretability through the normalized constructions \citep{wade2023bayesian}. 
External covariates can be flexibly incorporated to facilitate clustering across groups, as well as density regression, through the use of various kernel functions. Additionally, the C-HDP can account for different response types through the choice of component-specific likelihoods. The proposed method holds utility in various settings. For instance, biological researchers may be keen to understand how cellular latent time, an indicator of cell position in the developmental path, influences the identification of cell subpopulations from different experiments (Figure \ref{fig:y0_het_vs_t}); empirically, we can appreciate a relationship between the gene expression of cells in a cluster and their latent time, with subtle differences across experiments. For efficient inference, we construct a novel Markov Chain Monte Carlo (MCMC) algorithm that employs latent variables to cope with the intractable normalized weights. We demonstrate that our model can capture the relationship between clusters and covariates, and identify meaningful clusters across groups in both simulated and real data. 

The paper is organized as follows. We commence by providing a review of the DP and its extensions DDP and HDP in Section \ref{sec:review}. Section \ref{sec:method} outlines the definition of the covariate-dependent HDP, examples of common component-specific likelihood and kernel functions. The details of posterior inference are presented in Section \ref{sec:inference}. Section \ref{sec:simlation} provides a simulation study highlighting the advantages of combining the HDP and DDP. In Section \ref{sec:case studies}, we showcase the application of C-HDP to two real-world data sets from scRNA-seq and calcium imaging, respectively. Section \ref{sec:conclusion} concludes the paper and discusses potential future directions. 

\section{Review of Dirichlet Processes and Extensions} \label{sec:review}

Our model is based on the Dirichlet process, which is reviewed in this section along with relevant extensions.

\subsection{Dirichlet Processes}
Let $P$ denote a random probability measure on the parameter space $\Theta$. The Dirichlet process \citep{Ferguson1973bayesian} is the most commonly used prior for an unknown probability measure $P$ in the Bayesian nonparametric literature, as it satisfies several desirable properties such as conjugacy, posterior consistency and large support \citep[Chapter 4]{Ghosal2017}. By definition, $P$ is said to follow a DP prior with baseline probability measure $P_0$ and concentration parameter $\alpha$, denoted by $P \sim \DP(\alpha, P_0)$, if for any finite measurable partition $\left\lbrace A_1,\ldots,A_k\right\rbrace $ of $\Theta$,
\begin{equation*}
	(P(A_1),P(A_2),\ldots,P(A_k)) \sim \Dir(\alpha P_0(A_1), \alpha P_0(A_2),\ldots, \alpha P_0(A_k)),
\end{equation*}
where $\Dir(\alpha_1, \ldots, \alpha_k)$ denotes the Dirichlet distribution with concentration parameters $\alpha_1, \ldots, \alpha_k$. 
Beyond the DP, a number of other nonparametric priors have also been proposed and studied \citep{hjort1990nonparametric, de2013gibbs, LP10}.

An important property of DP is the discrete nature of $P$ (almost surely) such that $P$ can be written as a combination of weights and point masses, $P=\sum_{j=1}^{\infty}p_j \delta_{\theta_j^*}$, where $p_j$ is the weight associated with component $j$ and $\delta_{\theta_j^*}$ is the Dirac measure at the atom $\theta_j^*$.
The atoms are independent and identically distributed from $P_0$, and independent of the weights. The weights can be defined as normalized jumps of a Gamma process $\xi(t)$ for $0\leq t \leq 1$ with shape parameter $\alpha$, defined with $\xi(0)=0$ such that the increments on disjoint intervals are independent and such that $\xi(t_2) - \xi(t_1)$ has distribution $\Gam(\alpha(t_2-t_1),1)$ for $0 \leq t_1 < t_2 \leq 1$. 
Thus, $P \sim \DP(\alpha, P_0)$ can be represented as
\begin{equation}\label{eq:dp-normgamma}
	P = \sum_{j=1}^\infty \frac{\Gamma_j}{\sum_{h=1}^\infty \Gamma_h}  \delta_{\theta_j^*}, \quad \theta_j^* \overset{i.i.d}{\sim}P_0, 
\end{equation}
where $\Gamma_j$ are the jumps of the Gamma process and $\sum_{j=1}^\infty \Gamma_j = \xi(1)-\xi(0) \sim \Gam(\alpha,1)$ is finite almost surely.
An alternative representation is given by the stick-breaking construction \cite{sethuraman1994constructive} of the DP as follows:
\begin{equation}\label{eq:stick-breaking}
	\begin{split}
		P&=\sum_{j=1}^{\infty}p_j \delta_{\theta_j^*},\\
		p_1=v_1,\quad p_j=v_j\prod_{l<j}(1-&v_l) \quad \text{for} \ j>1,\quad v_j \overset{i.i.d}{\sim}\Beta(1,\alpha),\\
	\end{split}
\end{equation}
where again $\theta_j^* \overset{i.i.d}{\sim}P_0$, and $v_j$ is independent of $\theta_j^*$. It can be shown that $\sum_{j=1}^{\infty}p_j=1$ almost surely \citep{ishwaran2001gibbs}.

If $\theta_1, \ldots, \theta_n$ represent draws from $P$, i.e. $\theta_i \iidsim P \ (i=1,\ldots,n)$, then the discreteness of $P$ implies there are ties among $\theta_1, \ldots, \theta_n$. Thus, the DP itself is typically not directly used to model the unknown data-generating distribution, but instead is convoluted with a parametric likelihood $f(y| \theta)$ on the sample space $\mathcal{Y}$, to produce a DP mixture model \citep{lo1984class}:
\begin{equation*}
	y_i | P \iidsim \int f(y_i| \theta) d P(\theta), \quad P \sim \DP(\alpha, P_0).
\end{equation*}
Hierarchically, this can be expressed as
\begin{equation*}
	y_i | \theta_i \indsim f(y_i| \theta_i), \quad \theta_i | P \iidsim P, \quad P \sim \DP(\alpha, P_0).
\end{equation*}
This model not only allows flexible density estimation, but ties among the observation-specific parameters ($\theta_1, \ldots, \theta_n$) induce a latent clustering of the data, where two points belong to the same cluster if they are generated from the same mixture component, i.e. $i$ and $i'$ are in the same cluster if $\theta_i = \theta_{i'}$. Different choices of component-specific likelihoods can be used, reflecting the shape and interpretation of a cluster for the task at hand \citep[e.g.][]{fruhwirth2010bayesian, blei2003latent, wu2022nonparametric}.
The DP does not require the specification of the number of clusters, which is
instead
data-driven and can grow with the sample size. 

\subsection{Dependent Dirichlet Processes}

When there is an exogenous covariate $x$, the random probability measure can be augmented to depend on $x$, denoted as $P_x$. A popular method is the dependent Dirichlet process (DDP) which was first introduced by \cite{MacEachern1999DDP}. The random probability measure $P_x$ is constructed following a similar stick-breaking representation to Eq.~\eqref{eq:stick-breaking}, which in full generality employs stochastic processes to model the covariate-dependent atoms $\theta_j^*(x)$ and stick-breaking proportions $v_j(x)$.
The DP can be considered as a special case when the weights and atoms are independent of the covariate $x$. In the context of mixture models, the response $y$ has the following conditional density
\begin{equation*}
	f(y|x, P_x)=\int f(y|x, \theta) d P_x(\theta)=\sum_{j=1}^{\infty}p_j(x)f(y|x, \theta_j^*(x)),
\end{equation*}
where $P_x = \sum_{j=1}^{\infty} p_j(x) \delta_{\theta_j^*(x)}$.

A common simplified DDP model is the ``single-weights'' DDP where only the atoms depend on the covariates, $P_x=\sum_{j=1}^{\infty}p_j \delta_{\theta_j^*(x)}$. It has been used in the context of ANOVA \citep{de2004anova} for categorical covariates (ANOVA-DDP). It can be generalized to linear combinations of general types of covariates, referred to as the linear DDP (LDDP) \citep{DeIorio09}.

In contrast, the ``single-atoms'' DDP assumes only covariate-dependent weights, $P_x=\sum_{j=1}^{\infty}p_j(x) \delta_{\theta_j^*}$. Constructing the covariate-dependent weights is not trivial, as one must ensure they are positive and sum to one almost surely for all $x \in \mathcal{X}$. The stick-breaking construction is the most common approach \citep[e.g.][]{Dun1, Rod}, but other methods are available including normalization \citep[e.g.][]{foti2012slice, antoniano2014bayesian}. The approach of \cite{foti2012slice}, which has been applied in \cite{rao2009spatial} for spatial applications, is based on the normalized gamma process representation of the DP and the covariate-dependent weights are constructed using bounded kernel functions on the unit interval. Instead, \cite{antoniano2014bayesian} use a parametric density function and the stick-breaking representation. These two methods allow for different types of kernel functions or density functions to induce dependence without significant modification of the sampler. For a comprehensive review of the DDP, we refer the reader to \cite{quintana2022dependent} and \cite{wade2023bayesian}.

\subsection{Hierarchical Dirichlet Processes}
The hierarchical Dirichlet process \citep{Teh2006} focuses exclusively on the partially exchangeable setting, where covariates represent groups or data sets.
For the $i$-th observation in the $d$-th data set $y_{i,d}$ $(i=1,\ldots,n_d, d=1,\ldots,D)$, and a parametric density $f(y|\theta)$ on the sample space $\mathcal{Y}$ with parameters $\theta \in \Theta$, the HDP assumes the following hierarchical structure
\begin{equation*}
	\begin{split}
		y_{i,d}|\theta_{i,d} \indsim f(y_{i,d}|\theta_{i,d}),& \quad \theta_{i,d}|P_d  \indsim P_d,\\
		P_d|\alpha,P  \indsim \DP(\alpha, P),& \quad  P | \alpha_0, P_0 \sim \DP(\alpha_0, P_0).
	\end{split}
\end{equation*}
Here, another layer of the DP prior is included to borrow strength across groups. Specifically, each group has its own mixing measure $P_d$ which are all apriori centered on the unknown global mixing measure $P$. 
The HDP can be defined through hierarchical normalized Gamma processes \citep{argiento2020hierarchical}. In particular, letting $\xi_d(t)$ denote the group-specific Gamma processes, since the base measure $P$ of each group-specific  mixing measure $P_d$ is  discrete, we have the representation \citep{kingman1975random}:
	\begin{align}\label{eq:HDP-norm-gamma}
		P_d &=\sum_{j=1}^{\infty} \frac{\xi_d(\sum_{l=1}^j p_l)- \xi_d(\sum_{l=1}^{j-1} p_l)}{\sum_{h=1}^\infty \xi_d(\sum_{l=1}^h p_l)- \xi_d(\sum_{l=1}^{h-1} p_l)} \delta_{\theta_{j}^*} =\sum_{j=1}^{\infty} \frac{\Gamma_{j,d}}{\sum_{h=1}^\infty \Gamma_{h,d}} \delta_{\theta_{j}^*}
	\end{align}
	where $\theta_{j}^* \iidsim P_0$; $(\Gamma_{j,d})$ are independent with $\Gam(\alpha p_j,1)$ distribution; and the $p_j$ are the normalized jumps of the Gamma process (Eq.~\eqref{eq:dp-normgamma}) with shape parameter $\alpha_0$.
In the following exposition, we denote a draw from the HDP prior as $P_d \sim \text{HDP}(\alpha_0, \alpha, P_0)$. This construction highlights that different measures $P_d$ share the same atoms $\theta_j^*$ but assign different weights to the atoms. In the context of clustering, this is important as it allows clusters (data points that share the same parameters) to be potentially shared across groups, but allows the weight 
(or size) of the cluster to vary across groups.

Recent research on leveraging predictors in HDP is also available. \cite{dai2014supervised} developed the supervised HDP for topic modelling that can predict continuous or categorical response associated with each document (group) using generalized linear models. The hierarchical Dirichlet scaling process \citep{kim2014hierarchical} considers documents with observed labels, and topic proportions are modelled dependent on the distance between the latent locations of the observed labels and topics. \cite{ren2008dynamic} extend HDP to dynamic HDP for time-evolving data, assuming that adjacent groups collected closer in time are more likely to share components.
\cite{ren2011logistic} incorporate spatial-temporal information using a kernel logistic regression.
\cite{diana2020hierarchical} propose the hierarchical dependent Dirichlet
process (HDDP), combining HDP and ``single-weights'' DDP. 

Lastly, while not the focus of this article, we briefly mention the nested DP \citep{rodriguez2008nested, camerlenghi2019latent} which can also be used for clustering data across groups, but uses a multi-level structure with also a latent clustering of groups. Various extensions of both the HDP and nested DP have been proposed, including the semi-HDP \citep{beraha2021semi}, the hidden HDP \citep{Lijoi2022} and the common atoms model \citep{denti2023common}.

\section{Covariate-dependent Nonparametric Models} \label{sec:method}

Real-world data sets often encompass various types of covariates for statistical modeling, in addition to collecting data across multiple groups. In order to construct a flexible BNP modeling framework for such data, we first propose a novel covariate-dependent HDP in Section \ref{sec:C-HDP-proposal}. To flexibly model the conditional density and cluster observations across groups, the C-HDP can be used as a prior for covariate- and group-dependent mixing measures in mixture models (Section \ref{sec:kernel-weights}), where we illustrate examples of kernel functions to introduce dependence on the covariate. Additionally, the C-HDP can be applied to different types of response, depending on the nature of the data and the task at hand, as detailed in Section \ref{sec:common-likelihood} where widely-used component-specific likelihoods are presented.

\subsection{Covariate-dependent Hierarchical Dirichlet Processes}  \label{sec:C-HDP-proposal}
We construct the covariate-dependent HDP that borrows ideas from the ``single-atoms'' DDP and HDP, in order to model an unknown probability measure $P_{x,d}$ that is indexed by both the covariate $x$ and group index $d$. Recalling that the HDP assumes $P_d=\sum_{j=1}^{\infty}p_{j,d} \delta_{\theta_j^*}$, with $p_{j,d}=\Gamma_{j,d}/\sum_{k=1}^\infty \Gamma_{k,d}$ defined through the normalized construction in Eq.~\eqref{eq:HDP-norm-gamma},
we propose to introduce dependence by defining the mixture weights for each group to be functions of the covariate $x$, leading to  
\begin{equation} \label{eq:P_x_d} 
	P_{x,d}=\sum_{j=1}^{\infty}p_{j,d}(x) \delta_{\theta_j^*}, \quad P_{x}=\sum_{j=1}^{\infty}p_{j}(x) \delta_{\theta_j^*},\\
\end{equation}
with $\theta_j^* \iidsim P_0$. Specifically, the covariate-dependent weights of both the group-specific and global mixing measures are defined based on a normalized construction as
\begin{equation} \label{eq:covariate-dependet-weight}
	\begin{split}
		p_{j,d}(x)=\frac{\Gamma_{j,d}K(x|\bpsi_{j,d}^*)}{\sum_{k=1}^{\infty} \Gamma_{k,d}K(x|\bpsi_{k,d}^*)}, \quad p_{j}(x)=\frac{\Gamma_{j}K(x|\bpsi_{j}^*)}{\sum_{k=1}^{\infty} \Gamma_{k}K(x|\bpsi_{k}^*)}
	\end{split}
\end{equation}
where $\Gamma_{j,d}$ are i.i.d $\Gam(\alpha p_j,1)$ as in Eq.~\eqref{eq:HDP-norm-gamma} with $p_j = \Gamma_j/\sum_{k=1}^\infty \Gamma_k$, and $K(x|\bpsi)$ is a kernel function relying on kernel parameters $\bpsi$ which may be group- and component-specific and satisfies $0\leq K(x|\bpsi)<c$ for some constant $c$ and for every $x$, at least one component $j$ satisfies $K(x|\bpsi_{j,d}^*) >0$ at the group level and $K(x|\bpsi_{j}^*) >0$ at the global level (ensuring that the normalizing constant is finite almost surely). Note that $\{\theta_j^*\}_{j=1}^\infty$, $\{\Gamma_{j,d}\}_{j=1,d=1}^{\infty,D}$ and $\{\bpsi_{j,d}^*\}_{j=1,d=1}^{\infty,D}$ are independent of each other. In addition, hierarchical priors are assigned to the kernel parameters to borrow strength across groups, such that the $\bpsi_{j,d}^*$ are independent across components ($j=1,\ldots, \infty)$ and are conditionally independent across data sets ($d=1,\ldots, D$), centered around the global kernel parameter $\bpsi_j^*$. An advantage of this hierarchical formulation is that it provides a natural, parsimonious framework to account for differences in the covariate effects on the weights across groups. Examples depend on the choice of component-specific likelihood (e.g. see Section~\ref{sec:common-likelihood}). 

\begin{figure}[tbp]
	\centering
	\includegraphics[width=0.95\textwidth]{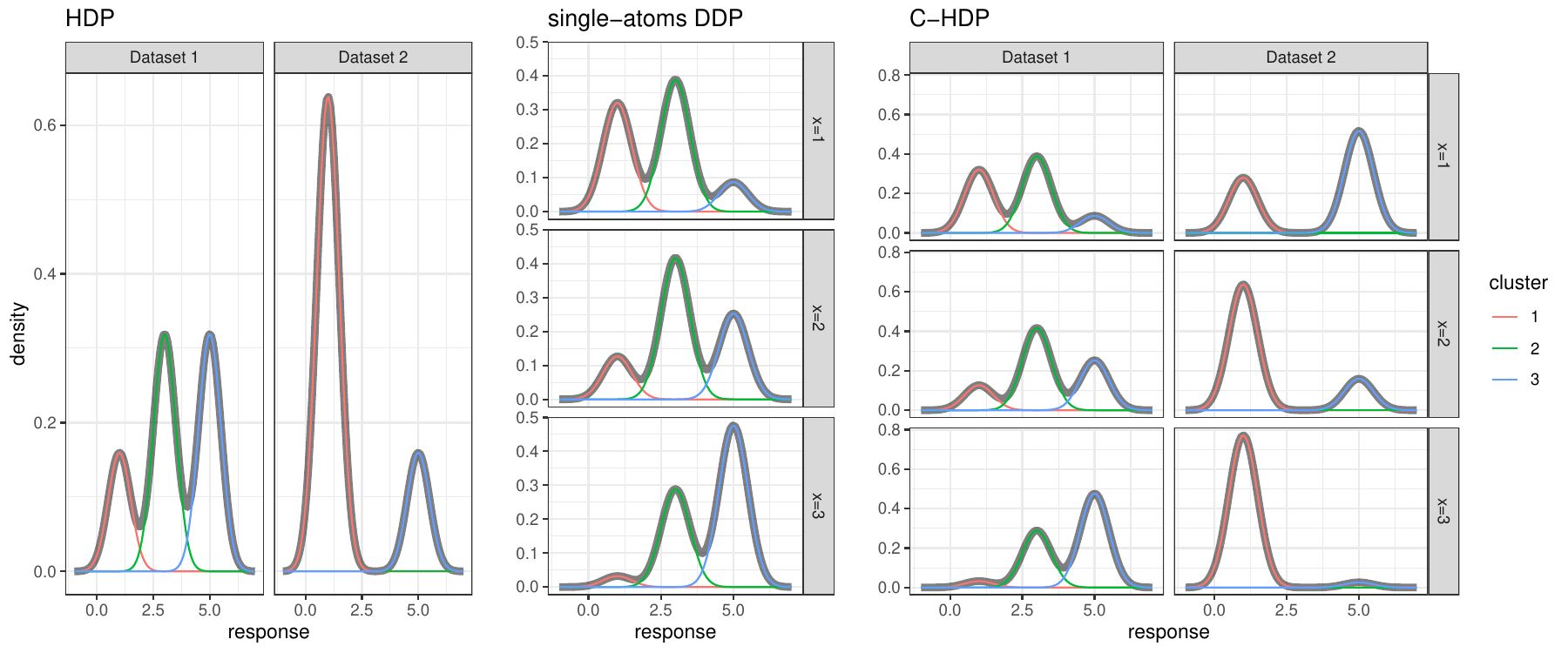}
	\caption{A demonstration of three nonparametric priors: HDP, DDP and C-HDP. The colored lines denote weighted conditional density for each cluster, with marginal density function shown in black. The HDP (left) allows for shared clusters across groups. The ``single-atoms'' DDP (middle) allows the cluster weights to vary smoothly across covariate values. The C-HDP (right) combines the HDP and DDP for shared clustering across groups, with smoothly varying covariate-dependent weights. }
	\label{fig:HDP_DDP_CHDP_weighted_density}
\end{figure}

The normalized construction of the covariate-dependent weights is motivated by \cite{foti2012slice} and 
\cite{antoniano2014bayesian}. 
Compared to alternative methods (e.g. stick-breaking), the normalized construction has the advantage of enhanced interpretability with
$p_{j,d} = \Gamma_{j,d}/ \sum_{k=1}^\infty \Gamma_{k,d}$ representing the probability that an observation in group $d$ is generated from component $j$ regardless of the covariate value, and the kernel represents how likely an observation from group $d$ that is generated from component $j$ will take the value $x$. 
Thus, this enhanced interpretability allows for more subjective or empirical specification of parameters \citep{wade2023bayesian}. On the other hand, stick-breaking constructions suffer from difficulty in selecting hyperparameters, which can significantly influence the functional shape of the weights and model performance. Figure \ref{fig:HDP_DDP_CHDP_weighted_density} illustrates how C-HDP combines the hierarchical framework of the HDP  (left) with the covariate dependence of the DDP (middle) to allow for shared clusters across groups with weights that both vary across groups and change smoothly with covariates (right).

We remark that our C-HDP prior differs from the hierarchical dependent Dirichlet process prior in \cite{diana2020hierarchical} which combines the ``single-weights'' DDP and HDP instead. In particular, in \cite{diana2020hierarchical} the covariate $x$ is introduced in the global measure $P$ instead of data-specific DPs $P_d$ as we have done, and therefore in \cite{diana2020hierarchical} the influence of the covariate is the same across data sets, whilst the effect is allowed to be different in our C-HDP model.

\subsubsection{Finite-dimensional Truncation}

For computational purposes, a finite-dimensional truncation is useful. First, a finite-dimensional truncation of the global mixing measure $P$ can be obtained from the normalized Gamma process construction. It is defined for truncation level $J$ by considering a discretization $(0, 1/J, 2/J,\ldots,1)$ of the domain of the Gamma process:
	\begin{align*}
		P^J = \sum_{j=1}^J \frac{\xi(j/J) - \xi((j-1)/J)}{\sum_{h=1}^J \xi(h/J) - \xi((h-1)/J) }  \delta_{\theta_j^*} =  \sum_{j=1}^J \frac{q_j^J}{\sum_{h=1}^J q_h^J}  \delta_{\theta_j^*},
	\end{align*}
	where by definition of the Gamma process $q_j^J \iidsim \Gam(\alpha_0/J, 1)$ and $P^J$ converges weakly to the DP \citep{kingman1975random}. Thus, a finite-dimensional truncation in the hierarchical setting, which converges weakly to the HDP \citep{Teh2006}, is similarly obtained as 
	\begin{align}\label{eq:HDP-norm-gamma-trunc}
		P_d^J &=\sum_{j=1}^{J} \frac{q^J_{j,d}}{\sum_{h=1}^\infty q^J_{h,d}} \delta_{\theta_{j}^*}. \quad q^J_{j,d} \indsim \Gam(\alpha p_j^J,1), 
	\end{align}
	where $p_j^J =q_j^J/ \sum_{h=1}^J q_h^J$ and $q_j^J \iidsim \Gam(\alpha_0/J, 1)$. 
	This allows us to construct a finite-dimensional truncation for C-HDP
\begin{equation} \label{eq:FD-CHDP}
	\begin{split}
		P_{x,d}^J&=\sum_{j=1}^{J}p_{j,d}^J(x) \delta_{\theta_j^*}, \quad P_x^J =\sum_{j=1}^{J}p_j^J(x)\delta_{\theta_j^*},
	\end{split}
\end{equation}
where
\begin{equation} \label{eq:FD-q}
	p_{j,d}^J(x) =\frac{q^J_{j,d}K(x|\bpsi_{j,d}^*)}{\sum_{k=1}^{J} q^J_{k,d}K(x|\bpsi_{k,d}^*)}, \quad p_{j}^J(x) =\frac{p^J_{j}K(x|\bpsi_{j}^*)}{\sum_{k=1}^{J} p^J_{k}K(x|\bpsi_{k}^*)},
\end{equation}
with $q^J_{j,d} \sim \Gam(\alpha p_j^J, 1)$ and $(p^J_{1},\ldots, p^J_{J}) \sim \Dir( \alpha_0/J,\ldots, \alpha_0/J)$. For notational simplicity, we will drop the superscript $J$, when the context is clear.

The weights $p_{j,d} = \Gamma_{j,d}/ \sum_{k=1}^\infty \Gamma_{k,d}$ of the C-HDP in Eq.~\eqref{eq:covariate-dependet-weight} could alternatively be constructed based on the hierarchical stick-breaking representation of the HDP \citep{Teh2006}. This leads to an equivalent nonparametric process with the same law, however, the truncated approximations differ. Our focus is the finite-dimensional approximation, which, in contrast to the stick-breaking truncation, has nice features such as exchangeability of the weights  \citep{ishwaran2002exact}; therefore, label-switching moves which are required for posterior exploration of the stochastically ordered stick-breaking weights \citep[][Section 6]{liverani2015premium} are not necessary. Moreover, as shown in \cite{catalano2024hierarchical}, the discrepancy between the DP and the stick-breaking truncation depends on the value of  $\alpha$, and for a diverging sequence of $\alpha$ it may fail to converge. Instead, the finite-dimensional approximation has polynomial convergence rate in $J$ which does not depend on $\alpha$. 

\subsection{Examples of Kernels for Dependent Weights} \label{sec:kernel-weights}

The choice of kernel has an important role in defining the dependence structure in the weights. Depending on the characteristics of the data and application,  different kernels may be appropriate.
Below we provide a few examples of the kernel functions that will be used in the paper along with a description of the type of dependence implied.

\begin{figure}[tbp]
	\centering
	\includegraphics[width=1\textwidth]{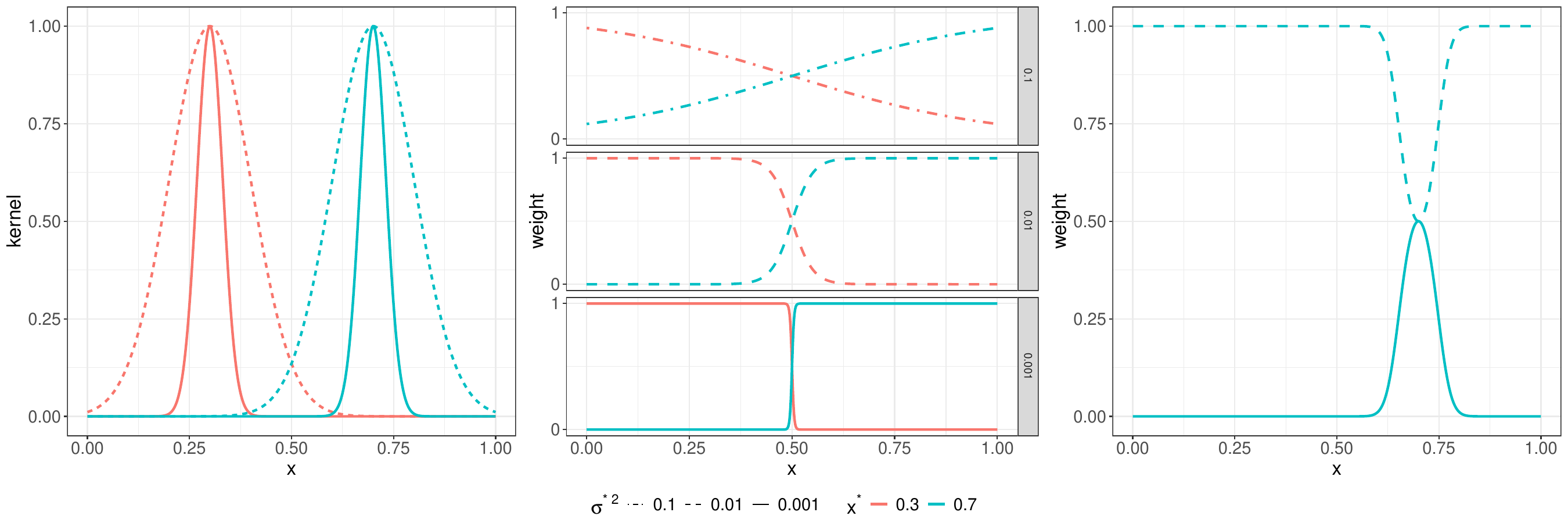}
	\caption{Left: Example of a Gaussian kernel for varying parameter values. Middle: Covariate-dependent weights for two components. In each row, both components have the same $\sdcenter$ but different $\xcenter$. When $\sdcenter$ decreases from top to bottom, the change of weights with respect to the covariate $x$ becomes more abrupt, showing a sharp transition. Right: Covariate-dependent weights for two components. Both clusters have the same $\xcenter$, but different $\sdcenter$. The weights can exhibit a bimodal behavior.}
	\label{fig:Gaussian_kernel_example}
\end{figure}

\paragraph{Gaussian Kernel.}  $$K(x|\bpsi_{j,d}^*)=\exp\left( -\frac{\left( x-\xcenter \right) ^2}{2 \sdcenter} \right),$$ where 
$\bpsi_{j,d}^*=(\xcenter, \sdcenter) \in \R \times \R^+$. The parameter $\xcenter$ describes the value in the covariate space where the $j$-th component best applies for the $d$-th group, and $\sdcenter$ affects the sharpness of  the boundary between the covariate regions associated to each component. Smaller values for $\sdcenter$ lead to more drastic change in the weights (Figure \ref{fig:Gaussian_kernel_example} middle). 
A Gaussian kernel is recommended when we believe the covariate space can be partitioned into well-behaved regions. Notice that when $\sdcenter \to \infty$ for all $j$ and $d$, the C-HDP reduces to the HDP.

\begin{figure}[tbp]
	\centering
	\includegraphics[width=1\textwidth]{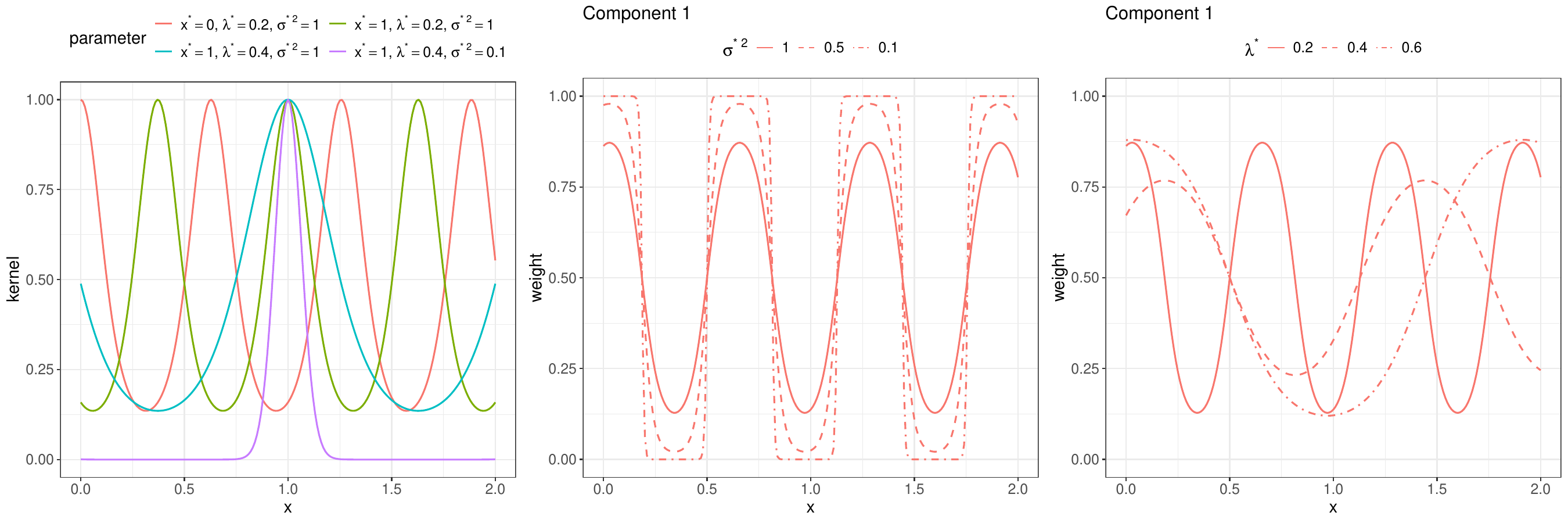}
	\caption{Left: Example of a periodic kernel for varying parameter values. Middle: Covariate-dependent weights for the first component given a total of two components (red and green on the left), for different $\sdcenter$. When $\sdcenter$ decreases, the weights show a more abrupt change. Right: Covariate-dependent weights for the first component  for different $\plambda$.}
	\label{fig:periodic_kernel_example}
\end{figure}

\paragraph{Periodic Kernel.}$$K(x|\bpsi_{j,d}^*)=\exp\left( -\frac{2}{\sdcenter}\sin^2\left(\frac{ x-\xcenter }{\plambda} \right) \right),$$ where $\bpsi_{j,d}^*=(\xcenter, \sdcenter, \plambda) \in \R \times \R^+ \times \R^+$. Note that for identifiability, $\xcenter$ needs to be restricted within one period $(\pi \plambda)$.
Figure \ref{fig:periodic_kernel_example} (left) shows the periodic kernel under different parameter values. The parameter $\xcenter$ represents the value that maximizes the kernel and changing $\xcenter$ will shift the kernel (red vs. green). The period is determined by $\plambda$ (green vs. blue), and $\sdcenter$ is related to the minimum value of the kernel and smooths the kernel (blue vs. purple). The influence of $\sdcenter$ and $\plambda$ on the covariate-dependent weight is shown in Figure \ref{fig:periodic_kernel_example} (middle and right). A periodic kernel is appropriate when there is repeated behavior over the covariate, e.g. time. Similar to the Gaussian kernel, for periodic kernels, as $\sdcenter \rightarrow \infty$, the C-HDP reduces to the HDP.

\begin{figure}[tbp]
	\centering
	\includegraphics[width=1\textwidth]{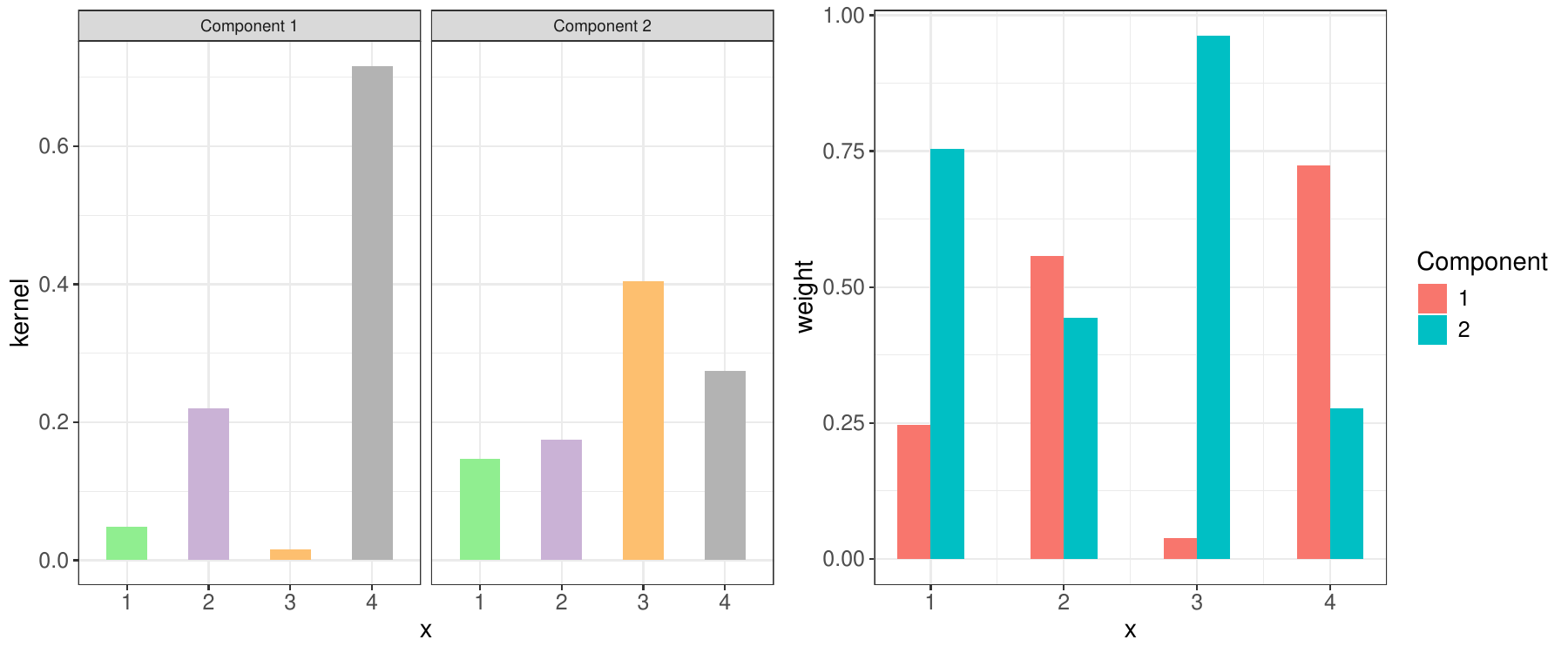}
	\caption{Left: Example of a categorical kernel for two different sets of kernel parameters (probabilities), with $L=4$. Right: Covariate-dependent weights for two components, with kernel parameters shown in the left panel.}
	\label{fig:categorical_kernel_example}
\end{figure}

\paragraph{Categorical Kernel.}$$K(x|\bpsi_{j,d}^*)=\prod_{l=1}^L(\rho_{j,d,l}^*)^{\indicator(x=l)},$$ for $x \in \{1,\ldots,L \}$ where $\indicator(\cdot)$ is the indicator function, $\bpsi_{j,d}^*=(\rho_{j,d,1}^*,\ldots,\rho_{j,d,L}^*)$ and the probabilities $\rho_{j,d,l}^*$ are positive and  sum to one. Categorical kernels are a natural choice when we have categorical covariates (see Figure \ref{fig:categorical_kernel_example} for examples).

The choice of these kernels makes the denominator in Eq.~\eqref{eq:covariate-dependet-weight} finite, ensuring $P_{x,d}$ and $P_{x}$ are valid probability measures. Following the definition in Eq.~\eqref{eq:P_x_d} and Eq.~\eqref{eq:covariate-dependet-weight}, a draw from the C-HDP prior is concisely denoted as $P_{x,d} \sim \text{C-HDP}(\alpha_0, \alpha, P_0, \bm{\Psi}^*)$, where $\bm{\Psi}^*$ denotes the collection of all kernel parameters across components and groups.

\begin{figure}[tbp]
	\centering
	\includegraphics[width=1\textwidth]{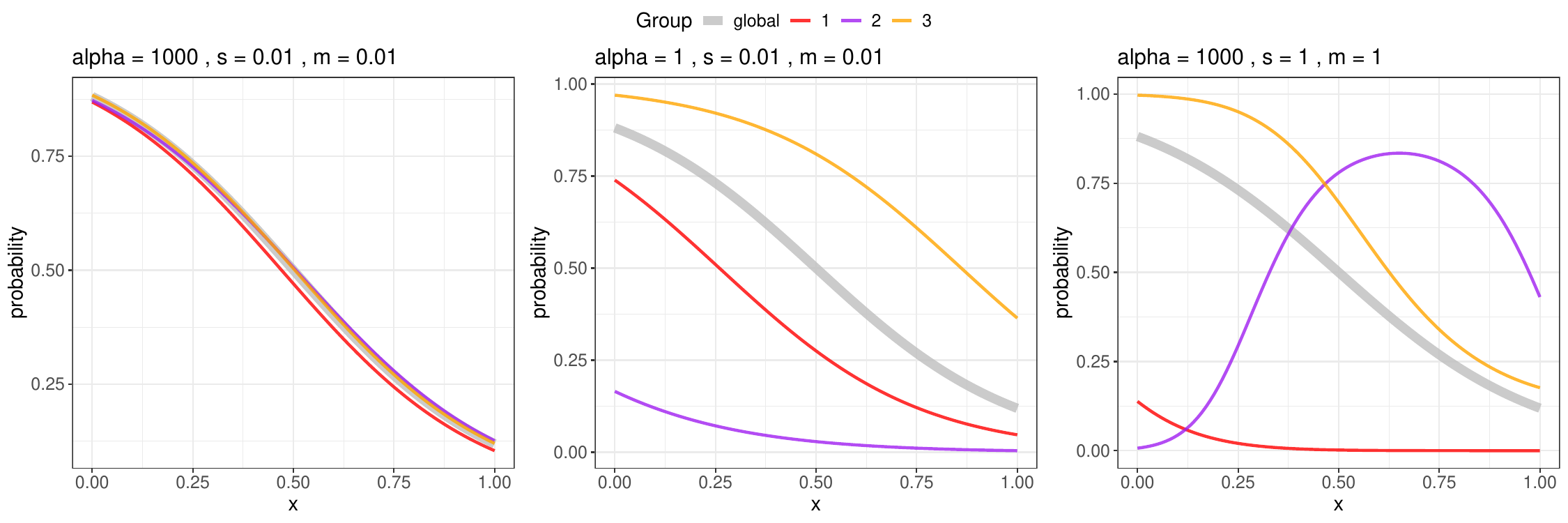}
	\caption{The effects of the concentration parameter $\alpha$ and prior variance parameters $s$ and $m$ on the covariate-dependent weights across groups. Three groups are considered here.}
	\label{fig:kernel_prior_all}
\end{figure}

In our work, hierarchical priors are assigned to kernel parameters, similar to the hierarchical prior for $q_{j,d}$, to borrow strength across groups and account for differences in the covariate effects on the weights across groups. For example, for the Gaussian kernel, the following hierarchical priors are considered in our case study later,
	\begin{equation*} \label{eq:priors-gaussian-kernel}
		\begin{split}
			\xcenter &\indsim \Norm(r_j, s^2),\quad r_j  \iidsim \Norm(\mu_r, \sigma_r^2),\quad s^2\sim \IG(\eta_1, \eta_2),\\
			\sdcenter & \indsim \logN(h_j, m^2),\quad h_j  \iidsim \Norm(\mu_h, \sigma_h^2),\quad m^2 \sim \IG(\kappa_1, \kappa_2).
		\end{split}
	\end{equation*}
	The effects of the prior parameters are demonstrated in Figure \ref{fig:kernel_prior_all}. For large values of the concentration parameter $\alpha$ and small prior variance $s^2$ and $m^2$, the differences in the relationship between the weights and covariates across groups are minimal and resemble the global relationship (left panel of Figure \ref{fig:kernel_prior_all}). On the other hand, for small $\alpha$ and/or large prior variance, the covariate-dependent weights vary across groups. In particular, for small $\alpha$, the group-level relationship still centers around the global one (middle panel of Figure \ref{fig:kernel_prior_all}), while the differences in the shape are more prominent for larger prior variance (right panel of Figure \ref{fig:kernel_prior_all}).

\subsection{Component-specific Likelihood}
\label{sec:common-likelihood}

The mixture model, under C-HDP, is given by
\begin{equation*}
	f(y| x, P_{x,d}) = \int f(y|\theta, x) d P_{x,d}(\theta),
\end{equation*}
where $P_{x,d} \sim \text{C-HDP}(\alpha_0, \alpha, P_0, \bm{\Psi}^*)$. 
This can be hierarchically written as 
	\begin{equation*}
		y_{i,d}|\theta_{i,d}, x_{i,d} \indsim f(y_{i,d}|\theta_{i,d}, x_{i,d}), \quad \theta_{i,d}|x_{i,d} = x, P_{x,d}  \indsim P_{x,d}.
	\end{equation*}
And, the latent variables $z_{i,d} \in \{1,2, \ldots \}$ can be introduced, where   $z_{i,d}=j$ if $\theta_{i,d} = \theta_j^*$.

Depending on the type of the data, different distributions can be selected for the component-specific likelihood, which should account for the characteristics and nature of the sample space $\mathcal{Y}$. For instance, a normal distribution is the most common for a continuous response, while a (skew) $t$ or skew-normal may be more appropriate when there are outliers or asymmetry \citep{fruhwirth2010bayesian, lee2014finite}. Positive responses are usually modeled from a gamma or log-normal distribution, and a beta distribution is suitable for values on the unit interval.
For a categorical response, a Bernoulli \citep{pan24} or multinomial distribution \citep{shahbaba2009nonlinear, blei2003latent} is usually selected. Poisson \citep{karlis2005mixed, krnjajic2008parametric} and negative-binomial \citep{wu2022nonparametric, Liu2024} distributions are used for count data. Below are some examples of component-specific likelihoods, with the first used in the simulation study of Section~\ref{sec:simlation} and the latter two employed in the examples of Section \ref{sec:case studies}.

\paragraph{Gaussian.} 
For a continuous response $\bm{y}_{i,d}=(y_{i,1,d},\ldots,y_{i,G,d})^T \in \mathbb{R}^G$, a normal distribution is commonly used
\begin{equation} \label{eq:likelihood-GMM}
	\bm{y}_{i,d}|z_{i,d}=j, \bmu_j^*, \bSigma_j^* \sim ~ \Norm(\bmu_j^*, \bSigma_j^*),
\end{equation}
where $\bmu_j^* \in \mathbb{R}^G$ denotes the mean and $\bSigma_j^*$ is a $G \times G$ covariance matrix, both associated with component $j$, i.e. $\theta_j^* = (\bmu_j^*, \bSigma_j^*) $.

\paragraph{Linear Regression.}
For a continuous response, where we also want to allow each component to reflect different linear relationships between the response and covariates, a linear regression model can be employed,
\begin{equation*}
	\bm{y}_{i,d}| z_{i,d}=j, \bmx_{i,d}, \bbeta_j^*, \bSigma_j^* \sim \Norm(\bmx_{i,d}^T \bbeta_j^*, \bSigma_j^*),
\end{equation*}
where $\bbeta_j^*$ is a $p$-dimensional (column) vector of coefficients for component $j$ and $\bmx_{i,d}$ is a $p$-dimensional (column) vector of covariates.

\paragraph{Vector Autoregressive (VAR).} 
An extension of the normal distribution to time-series data, based on a VAR model with lag one in the mean, is of the following form:
\begin{equation} \label{eq:likelihood-VAR}
	\bm{y}_{i,d}|\bm{y}_{i-1,d},z_{i,d}=j, \bma_j^*, \bB_j^*, \bSigma_j^* \sim ~ \Norm(\bma_j^*+\bB_j^*\bm{y}_{i-1}, \bSigma_j^*)  ,
\end{equation}
where $\bma_j^* \in \mathbb{R}^G$ denotes the intercept and $\mathbf{B}_j^*$ is a real-valued $G \times G$ matrix of the coefficients in VAR, both associated with component $j$.

\paragraph{Negative-binomial.} 
For count data, we consider a negative-binomial distribution for the within-component likelihood, with mean $\mu_{j,g}^* \in \R^+$ and dispersion $\phi_{j,g}^* \in \R^+$ in each dimension $g$ ($g=1,\ldots,G$) for component $j$:
\begin{equation*} \label{eq:likelihood-NB}
	y_{i,g,d}|z_{i,d}=j, \mu_{j,g}^*, \phi_{j,g}^* \sim \NB(\mu_{j,g}^*,\phi_{j,g}^*).
\end{equation*}

\section{Inference} \label{sec:inference}

In this section, we describe a Gibbs sampling scheme for the C-HDP mixture model, where we focus on the finite-dimensional truncation (Eq.~\eqref{eq:FD-CHDP}-\eqref{eq:FD-q}). Gibbs sampling can be applied to draw posterior samples for parameters with full conditionals of a standard form. For non-standard full conditionals, adaptive Metropolis-Hastings (AMH) is used \citep{JimEGriffin2013AiMc}. We highlight the key steps in constructing the Gibbs sampler, including an initial data augmentation trick to handle the normalizing constant of the weights.

\subsection{A Data Augmentation Trick} \label{sec:C-HDP-data-augmentation}
For mixture models, the complete data likelihood is widely employed to allow for efficient inference, which for the finite-dimensional C-HDP mixture has the form:
\begin{equation*}
  f(y_{i,d},z_{i,d}=j|\bmq_{1:J,d}, \theta_j^*, \bpsi_{1:J,d}^*, x_{i,d})=\frac{q_{j,d}K(x_{i,d}|\bpsi_{j,d}^*)}{\sum_{k=1}^{J} q_{k,d}K(x_{i,d}|\bpsi_{k,d}^*)} \times f(y_{i,d}| \theta_j^*, x_{i,d}).
 \end{equation*} 
 However, the normalizing constant in the denominator makes it difficult to obtain standard full conditional densities for $q_{j,d}$ and the kernel parameters. We propose to use a data augmentation trick, introducing a latent variable $\xi_{i,d} \in \mathbb{R}_+$, and the augmented likelihood is
\begin{align} \label{eq:full_likelihood}
    f(y_{i,d},\xi_{i,d},z_{i,d}=j|\bmq_{1:J,d}, \theta_j^*, \bpsi_{1:J,d}^*, x_{i,d})=&\exp\left( -\xi_{i,d}\sum_{j=1}^{J} q_{j,d}K(x_{i,d}|\bpsi_{j,d}^*)\right)  \times \nonumber \\  &q_{j,d}K(x_{i,d}|\bpsi_{j,d}^*) \times f(y_{i,d}| \theta_j^*, x_{i,d}).
\end{align}
 Using the fact that $\int_0^{\infty} \exp(-\xi \lambda) d\xi=\frac{1}{\lambda}$, the  complete data likelihood is restored when integrating out $\xi_{i,d}$. It is worth noticing that, unlike $z_{i,d}$, the $\xi_{i,d}$ do not have a physical interpretation. Define $N_{j,d}$ as the number of observations in component $j$ in data set $d$ and $n_d$ the size of data $d$. The augmented data likelihood yields standard full conditional distributions to enable sampling both $q_{j,d}$ and $\xi_{i,d}$ effectively:
 \begin{equation*}
	\begin{split}
		\pi(q_{j,d} | \ldots ) & \propto  \left( q_{j,d}\right) ^{N_{j,d}}\times \exp\left( -q_{j,d}\sum_{i=1}^{n_d}\xi_{i,d}K(x_{i,d}|\bpsi_{j,d}^*)\right) \times \left( q_{j,d}\right) ^{\alpha p_j -1} \exp(-q_{j,d})\\
		& \propto  \left( q_{j,d}\right) ^{N_{j,d}+\alpha p_j -1}\times \exp\left( -q_{j,d}\left[ 1+\sum_{i=1}^{n_d}\xi_{i,d}K(x_{i,d}|\bpsi_{j,d}^*)\right] \right),\\
		\Rightarrow & \quad q_{j,d} | \ldots \sim \Gam\left( N_{j,d}+\alpha p_j ,1+\sum_{i=1}^{n_d}\xi_{i,d}K(x_{i,d}|\bpsi_{j,d}^*)\right).
	\end{split}
\end{equation*}
\begin{equation*}
	\begin{split}
		\pi(\xi_{i,d} | \dots) & \propto  \exp\left( -\xi_{i,d}\sum_{j=1}^{J} q_{j,d}K(x_{i,d}|\bpsi_{j,d}^*)\right),  \\
		\Rightarrow & \quad \xi_{i,d} | \ldots \sim \Gam\left( 1 ,\sum_{j=1}^{J} q_{j,d}K(x_{i,d}|\bpsi_{j,d}^*)\right).
	\end{split}
\end{equation*}

In addition to the intricate denominator, the presence of the kernel parameters inside the exponential term in Eq.~\eqref{eq:full_likelihood} also poses challenges. For kernel parameters, we introduce another latent variable $u_{i,j,d} \in (0,1)$ to facilitate MCMC sampling. The update of $\bpsi_{j,d}^*$ and $\theta_j^*$ depends on the choice of the kernel and component-specific likelihood. If appropriate priors are chosen, it is possible to sample from the full conditionals of $\bpsi_{j,d}^*$ and $\theta_j^*$.

For example, consider the Gaussian kernel with a hierarchical normal prior $\xcenter \indsim \Norm(r_j, s^2)$ (for details see online Appendix), 
the full conditional distribution for $\xcenter$ is
\begin{equation*}
	\begin{split}
		\pi(\xcenter |  \ldots) \propto & \prod_{i: z_{i,d}=j}  K(x_{i,d}|\bpsi_{j,d}^*) \times  \prod_{i=1}^{n_d}  \exp\left( -\xi_{i,d}q_{j,d}K(x_{i,d}|\bpsi_{j,d}^*)\right) \times \Norm(\xcenter | r_j, s^2).
	\end{split}
\end{equation*}
With the introduction of $u_{i,j,d} \in (0,1)$, the above can be written as
\begin{equation*}  \label{eq:kernel_full_condition}
	\begin{split}
		\pi(\xcenter |  \ldots) \propto &  \prod_{i: z_{i,d}=j}  K(x_{i,d}|\bpsi_{j,d}^*) \times \prod_{i=1}^{n_d}  \indicator\left( u_{i,j,d}< M_{i,j,d}\right) \times \Norm(\xcenter|r_j, s^2),
	\end{split}
\end{equation*}
where $M_{i,j,d}=\exp\left( -\xi_{i,d}q_{j,d}K(x_{i,d}|\bpsi_{j,d}^*)\right)$. The full conditional of the latent variable is $u_{i,j,d}|\ldots \sim \Unif\left(0,M_{i,j,d}\right)$, and for the kernel parameter $\xcenter$ given $u_{i,j,d}$, the full conditional is a truncated normal distribution:
\begin{equation*}
	\begin{split}
		&\pi(\xcenter|\ldots) \propto \Norm(\xcenter|\hat{r}_{j,d},\hat{s}_{j,d}^2) \times \indicator(\xcenter\in A_{j,d}),
	\end{split}
\end{equation*}
where 
\begin{equation*}
	\begin{split}
		\hat{s}_{j,d}^2=\left( \frac{1}{s^2}+\frac{N_{j,d}}{\sdcenter}\right)^{-1}, \quad \hat{r}_{j,d}=\frac{r_j/s^2+\sum_{i: z_{i,d}=j}x_{i,d}/\sdcenter}{1/s^2+N_{j,d}/\sdcenter},
	\end{split}
\end{equation*}
and the truncation region is of the form
\begin{equation*}
	\begin{split}
		A_{j,d}=\bigcap_{i: -\log{u_{i,j,d}} < \xi_{i,d}q_{j,d}}  A_{i,j,d},
	\end{split}
\end{equation*}
where
\begin{equation*}
	\begin{split}
		A_{i,j,d}&=\left(-\infty, x_{i,d}- k_{i,j,d}\right) \bigcup \left( x_{i,d}+k_{i,j,d} , +\infty \right),\quad k_{i,j,d}  = \sqrt{-2\sdcenter\log\left[ -\frac{\log{u_{i,j,d}}}{\xi_{i,d}q_{j,d}}\right] }.
	\end{split}
\end{equation*}
The full derivation and details of the Gibbs sampling algorithm for all parameters under different kernel and within-component likelihood choices are presented in online Appendix.

\subsection{Clustering}

The Bayesian approach provides a collection of posterior samples of the allocation variables. To understand the posterior and uncertainty in the clustering represented by these allocation variables, we construct the posterior similarity matrix (PSM) where each entry is the posterior probability that observations $i$ and $i'$ are co-clustered, which is approximated by
$\text{PSM}_{[i,i']} \approx 1/L \sum_{l=1}^L \indicator(z_i^{(l)}=z_{i'}^{(l)})$
where $z_i^{(l)}$ is the $l$-th MCMC sample. For multiple data sets, we can compute the PSM for observations both within and across data sets to visualize the uncertainty in clustering across groups.

To summarize the posterior with a point estimate, the optimal clustering is obtained that minimizes the posterior expected variation of information (VI) \citep{wade2018bayesian}. In order to better understand the patterns within each cluster of this optimal solution, a subsequent MCMC is considered for all other parameters with the allocations fixed to the optimal one. 

For the post-processing step, to understand the uncertainty in allocations, we calculate the posterior allocation probability of each data point conditioned on all others
\begin{equation*}
	\begin{split}
		p(z_i=j | \mathcal{D}, \bm{z}_{-i}^*) = \int p(z_i = j | \mathbf{\Theta}, \mathcal{D}) p(\mathbf{\Theta} | \mathcal{D}, \bm{z}_{-i}^*) d\mathbf{\Theta}, \quad i=1,\ldots,n,
	\end{split}
\end{equation*}
where $\mathcal{D}$ denotes the observed data, $\bm{z}_{-i}^*$ denotes the optimal clustering without the $i$-th observation and $\mathbf{\Theta}$ represents all the unknown parameters. If we approximate $p(\mathbf{\Theta} | \mathcal{D}, \bm{z}_{-i}^*)$ by $p(\mathbf{\Theta} | \mathcal{D}, \bm{z}_{1:n}^*)$, which corresponds to the posterior distribution of $\mathbf{\Theta}$ from the post-processing MCMC, then the above can be approximated by the average over MCMC samples (from the post-processing step),
\begin{equation}  \label{eq:PP-approximation}
	p(z_i=j | \mathcal{D}, \bm{z}_{-i}^*) \approx \frac{1}{L} \sum_{l=1}^L p(z_i=j| \mathbf{\Theta}^{(l)}, \mathcal{D}),
\end{equation}
where $\mathbf{\Theta}^{(l)}$ denotes the $l$-th MCMC sample.

\subsection{Covariate-dependent Predictive Quantities of Interest} \label{sec:predictive mean}
In the context of density estimation, we can obtain the covariate-dependent conditional density, approximated by averaging over the MCMC samples:
\begin{equation*} 
	\begin{split}
		f(\Tilde{y}|\Tilde{x},\mathcal{D}) & = \int f(\Tilde{y}|\Tilde{x},\mathbf{\Theta}) \pi(\mathbf{\Theta}| \mathcal{D}) d\mathbf{\Theta} \approx \frac{1}{L} \sum_{l=1}^{L}f(\Tilde{y}|\Tilde{x},\mathbf{\Theta}^{(l)}),
	\end{split}
\end{equation*}
where $\Tilde{x}$ and $\Tilde{y}$ denote new data from group $d$ 
and
$f(\Tilde{y}|\Tilde{x},\mathbf{\Theta})=\sum_{j=1}^{J}p_{j,d} (\Tilde{x}) f(\Tilde{y}|\theta^*_j, \Tilde{x})$. 
Similarly, other quantities can be computed, such as the posterior predictive mean
\begin{equation*}
	\E(\Tilde{y}|\Tilde{x}, \mathcal{D})  \approx \frac{1}{L} \sum_{l=1}^{L}\E(\Tilde{y}|\Tilde{x},\mathbf{\Theta}^{(l)}).
\end{equation*}
For instance, in the case of the Gaussian within-component likelihood defined in Section \ref{sec:common-likelihood}, for an observation from group $d$, we have $\E(\Tilde{y}|\Tilde{x},\mathbf{\Theta})=\sum_{j=1}^{J}p_{j,d}(\Tilde{x}) \mu_{j}^*$. 

\section{Simulation Study} \label{sec:simlation}

We now demonstrate our method in a simulation study and compare its performance with other existing Bayesian nonparametric priors, in particular the HDP and DDP. We generate 10 replicated sets of data from a two-dimensional Gaussian mixture, consisting of $J=3$ components. The data-generating process is as follows:
	\begin{equation*}
		\begin{split}
			\bmy_{i,d} | z_{i,d}=j, \bmu^*_j, \bSigma^*_j &\indsim \Norm(\bmu^*_j, \bSigma^*_j), \quad i=1,\ldots,n_d, d=1,\ldots,D, \\
			z_{i,d} &\indsim \Cat(p_{1,d}^J(x_{i,d}),\ldots, p_{J,d}^J(x_{i,d})),
		\end{split}
	\end{equation*}
	where $D=5, n_d=300$ for all $d$. The covariate $x$ for all observations in each data set $d$ is equally spaced between 0 and 1. The parameters in each component-specific likelihood are
	\begin{equation*}
		\begin{split}
			\bmu_1^*&=(0,0), \ \bmu_2^*=(4,4), \ \bmu_2^*=(0,4),\\
			\bSigma^*_1 &= \begin{pmatrix}
				1 & 0 \\
				0 & 1
			\end{pmatrix},
			\bSigma^*_2 = \begin{pmatrix}
				1 & 0.5 \\
				0.5 & 1
			\end{pmatrix},
			\bSigma^*_3 = \begin{pmatrix}
				1 & -0.5 \\
				-0.5 & 1
			\end{pmatrix}.
		\end{split}
	\end{equation*}
	The relationship between the weights and the covariate is given by
	\begin{equation*}
		\begin{split}
			p_{j,d}^J(x)=\frac{\exp(\beta_{j,d,1}+\beta_{j,d, 2}x+\beta_{j,d, 3}x^2)}{\sum_{k=1}^J \exp(\beta_{k,d,1}+\beta_{k,d,2}x+\beta_{k,d,3}x^2)},
		\end{split}
	\end{equation*}
	where each $\bbeta_{j,d}=(\beta_{j,d,1},\beta_{j,d,2},\beta_{j,d,3})$ is generated from a diagonal Gaussian distribution, $\bbeta_{j,d} \sim \Norm(\bbeta_{j}, 5 \bI)$, with $\bbeta_{1}=(0,20,0), \bbeta_{2}=(10,-20,0), \bbeta_{3}=(0,0,40)$.

\subsection{Implementation and Results}
The Gaussian kernel defined in Section \ref{sec:common-likelihood} is used in the C-HDP and DDP to model the covariate-dependent weights. While the HDP allows for weights to be different across data sets, it does not incorporate covariate effects. Instead, for the DDP, we consider two types of models,
	with and without the group indicator as an additional covariate. In the former, the group variable is incorporated through a categorical kernel defined in Section \ref{sec:kernel-weights}. 
	For each model and replicate, the full MCMC algorithm and the post-processing step use 15000 iterations, followed by a burn-in of 12000 iterations and thinning of 3, leading to 1000 samples for posterior inference. The MCMC setup is the same across all methods and a truncation level of $J=8$ is applied.

\begin{figure}[tbp]
	\centering
	\includegraphics[width=0.32\textwidth]{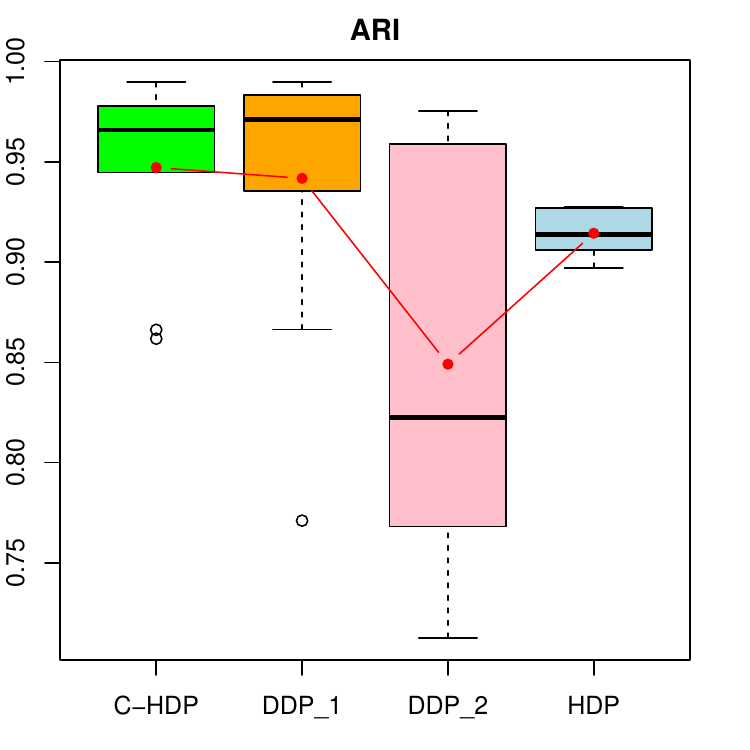}
	\includegraphics[width=0.32\textwidth]{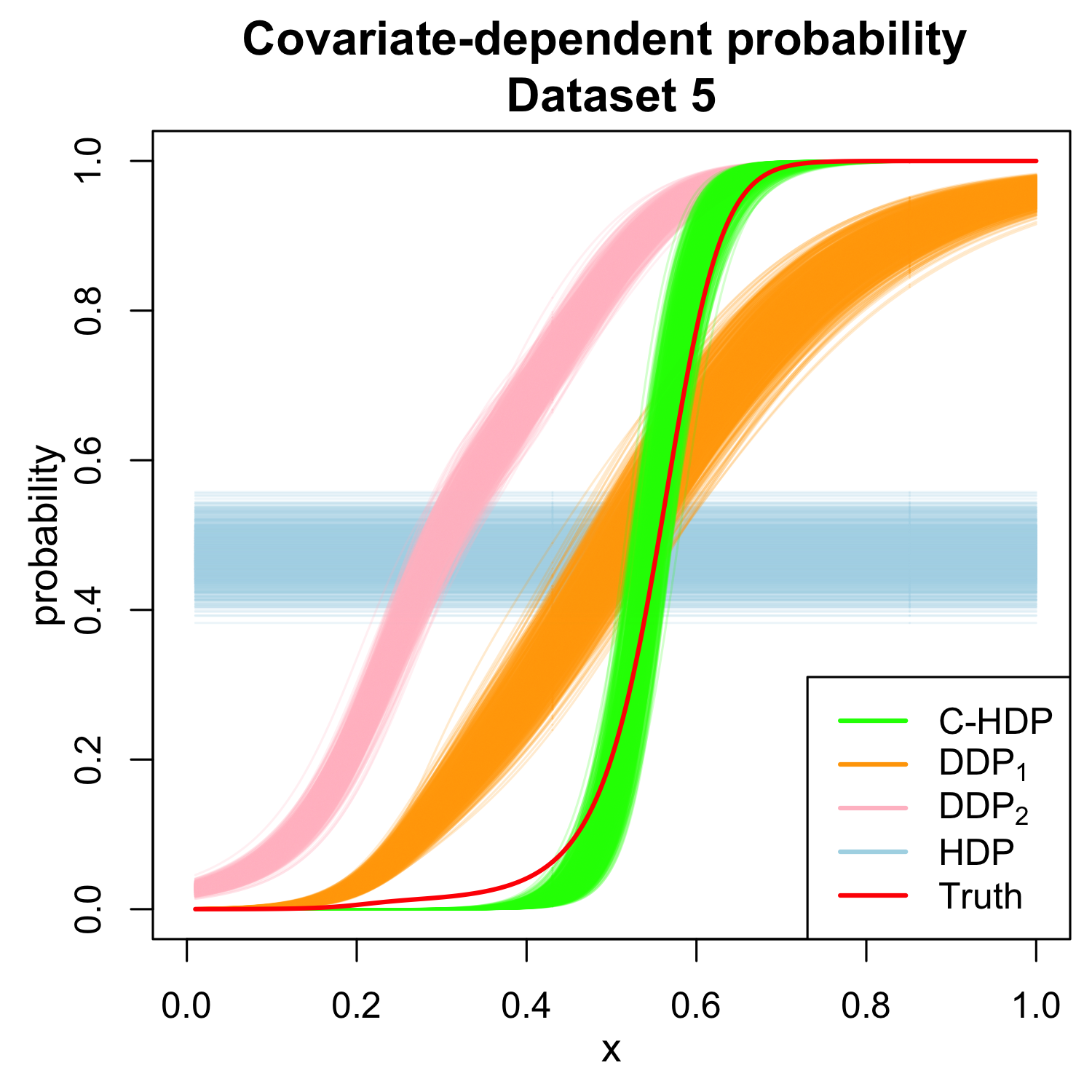}
	\includegraphics[width=0.32\textwidth]{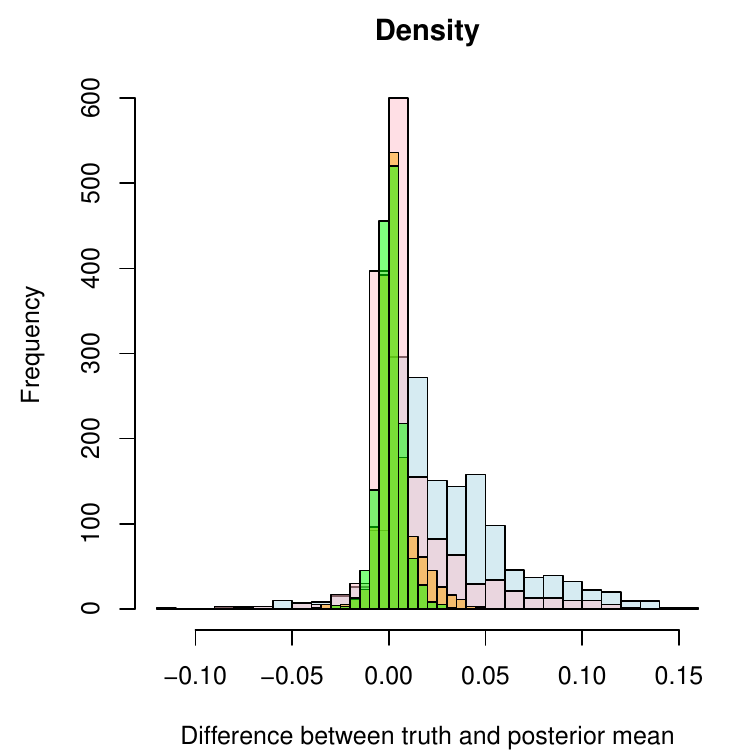}
	\caption{Left: Adjusted rand index (ARI) comparing each method to the truth across all replicates, with the red line showing the average ARI. $\text{DDP}_1$ and $\text{DDP}_2$ refer to the model with and without the group indicator, respectively. Middle: Posterior samples of covariate-dependent probabilities for all models, shown for an example cluster in data set 5 from one replicate. Right: Differences between the posterior mean of the density and the truth for all observations in one replicate, under the four methods.}
	\label{fig:gmm_simulation}
\end{figure}

	Across all replicates, the estimated clustering from our proposed C-HDP method generally shows better alignment with the truth, with a higher average adjusted rand index (ARI) in Figure \ref{fig:gmm_simulation} (left, red points). The DDP with the group indicator ($\text{DDP}_1$) also provides relatively high ARI, sometimes even higher than the C-HDP, but the difference is small and the performance of the DDP appears more unstable. Further, the ARI drops dramatically if the grouping is not considered in the DDP ($\text{DDP}_2$). As for the HDP, it is noticeably outperformed by the C-HDP and $\text{DDP}_1$, but with much smaller variability in its performance.

	Regarding the modelling of the relationship between the weights and the covariate, for an example cluster shown in Figure \ref{fig:gmm_simulation} (middle), we notice that using the Gaussian kernel for the C-HDP can estimate the relationship well, with the truth covered by the posterior samples, while in the DDP (including the groups), the true relationship is only marginally within the samples. This is worse in the simpler DDP ($\text{DDP}_2$) which does not capture the true relationship except for the upward trend. Note that since the HDP does not take into account the covariate, the probability is constant with $x$, showing much larger uncertainty. More detailed results are shown in online Appendix.

	In addition, we also compare the four methods in terms of density estimation (Figure \ref{fig:gmm_simulation} right). The estimated density from the C-HDP appears much closer to the truth with their differences concentrated around zero, followed by the DDP and HDP.

\begin{figure}[tbp]
	\centering
	\includegraphics[width=0.45\textwidth]{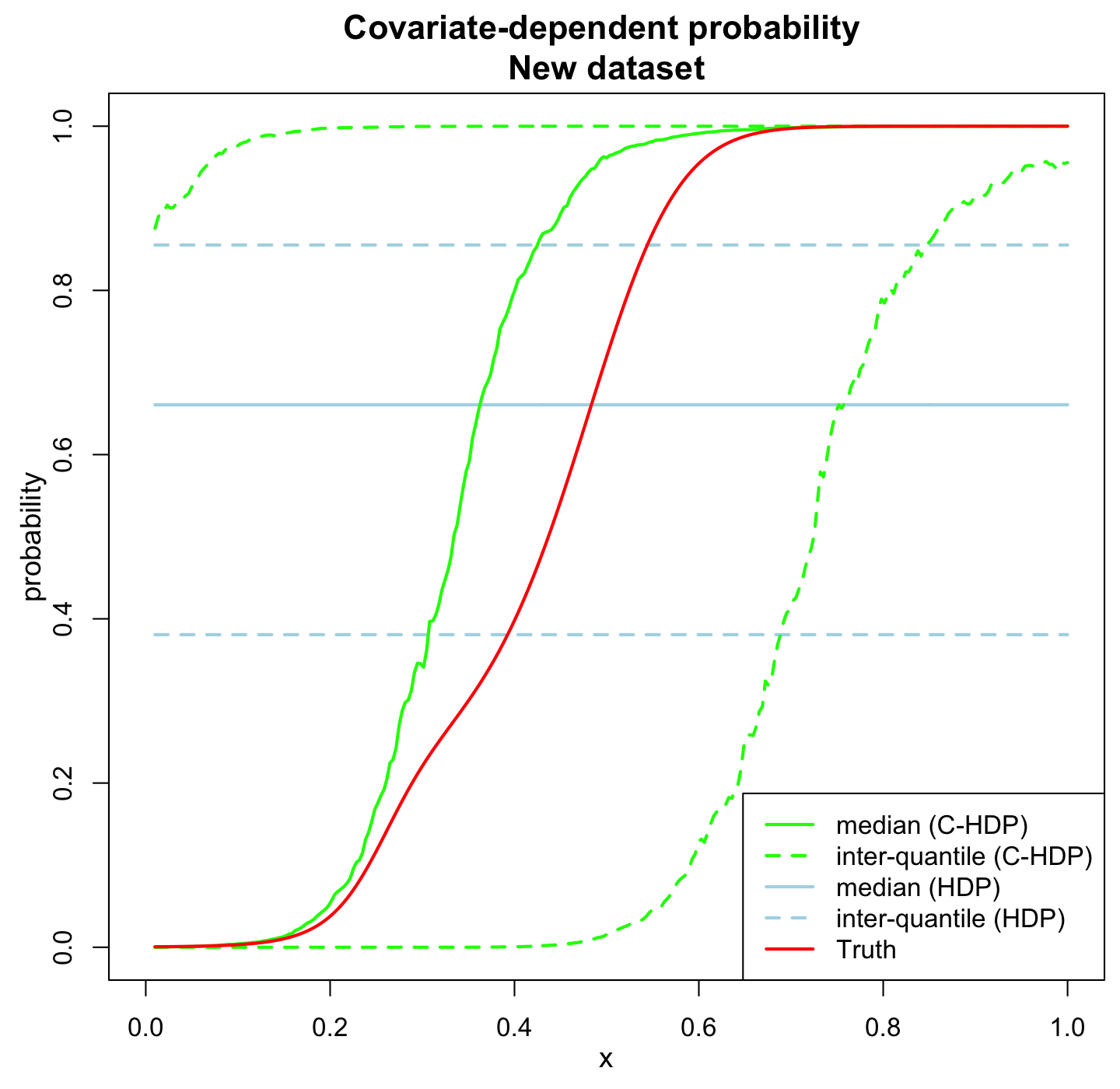}
	\includegraphics[width=0.45\textwidth]{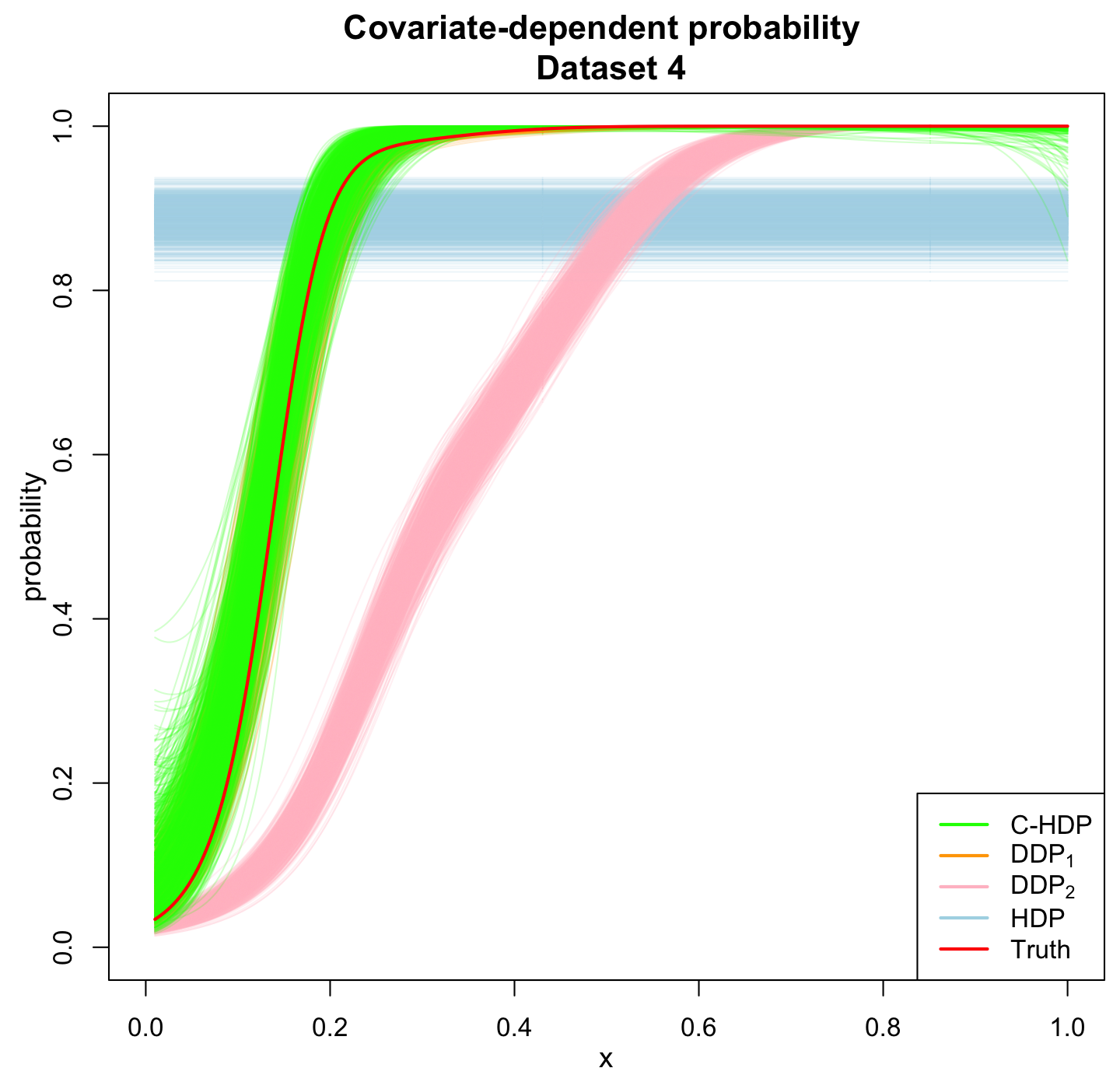}
	\caption{Left: Posterior predictive samples of covariate-dependent probabilities for the C-HDP and HDP, shown for an example cluster in a new group from one replicate. Right: Posterior samples of covariate-dependent probabilities for all models, shown for an example cluster in data set 4 from one replicate. Samples from the C-HDP and $\text{DDP}_1$ overlap.}
	\label{fig:gmm_simulation_new_group}
\end{figure}

	Lastly, one advantage of the C-HDP and HDP over the DDP is the ability to predict the (covariate-dependent) weights in new groups through posterior predictive sampling, which is not available for the DDP as the groups are treated as fixed categories. The relationship between the weights and covariate shows variability across groups (see red lines in Figure \ref{fig:gmm_simulation_new_group} and Figure \ref{fig:gmm_simulation} middle). Figure \ref{fig:gmm_simulation_new_group} (left) illustrates that the new group's weight can be reasonably estimated by the C-HDP where the posterior median (green solid line) is close to the truth and samples exhibit relatively large uncertainty (green dashed lines) compared to those for the existing groups. As for the HDP, large discrepancy from the truth is observed due to the absence of the covariate in the method.

	Overall, the simulation study on Gaussian mixtures shows that C-HDP outperforms DDP and HDP in estimating the clustering, the relationship between the weights and covariate, as well as density estimation. In particular, even though the true relationship between the weights and covariate is not correctly specified in the C-HDP model (a softmax function as opposed to a Gaussian kernel), a Gaussian kernel is still able to capture the true underlying relationship. Moreover, the hierarchical structure of the C-HDP facilitates understanding of the weights in a new group based on the posterior predictive distribution.

\section{Case Studies}
\label{sec:case studies}
We demonstrate the application of the C-HDP prior through experiments on two real data sets, using a Gaussian kernel combined with a negative-binomial component-specific likelihood and a periodic kernel combined with a VAR model, respectively.

\subsection{Application to Single-cell RNA Sequencing Data - Pax6}

The transcription factor Pax6 is believed to play an important role in the development and fates of embryonic cells. Specifically, Pax6 is critical to regulate gene expressions to avoid patterning defects in the brain and can affect signaling between neighbouring cells during development \citep{caballero2014cell}. Mutations in Pax6 can result in eye anomalies \citep{jordan1992human} and neurodevelopmental disorders, such as intellectual disability and autism spectrum disorder \citep{davis2008pax6,kikkawa2019role}. It plays a crucial role in cortical neurogenesis, controlling proliferation and differentiation of neural stem cells and progenitor cells \citep{estivill2002pax6,GOTZ19981031}.

To study and quantify the effects of Pax6 empirically  at the single-cell level, \cite{tan2022pax6} collected the scRNA-seq data at 13.5 days since conception (day E13.5) from mouse embryos 
in two groups: the mutant (HOM) group where both copies of Pax6 gene were deleted, and the control group (HET) where only one copy of Pax6 was deleted, to account for gene changes due to the process of deletion. In previous work, \cite{Liu2024} developed a model using the HDP to 
discover hidden cell subpopulations with similar gene expression and study how the proportion of cells in each cell group is affected by Pax6 activity. However, additional information is available for cells, such as the developmental trajectory, which may 
be relevant to identify clusters and quantify the influence of Pax6 across the developmental stages.  
Our aim is to extend the model of \cite{Liu2024} to incorporate this information and determine how the proportion of cells in each cell group is affected by both Pax6 activity and  cellular developmental trajectory. 
	
Each group $d\ (d=1,2)$ contains the mRNA counts $y_{c,g,d}$ for gene $g \ (g=1,\ldots,G)$ in cell $c \ (c=1,\ldots,C_d)$. Data has been preprocessed in \cite{Liu2024} using the approaches in \cite{seurat}, where low-quality cells and lowly expressed genes are removed and highly variable genes are selected. The HET and HOM data sets contain $C_1$ = 3096 and $C_2$ = 5282 cells, both with $G$ = 2529 genes. The covariate of interest is a proxy  for cell developmental trajectory, specifically, the cell-specific latent time  $t_{c,d}\in [0,1]$ \citep{Bergen2020}, which empirically shows a relation to the clustering in \cite{Liu2024} (Figure \ref{fig:y0_het_vs_t}). \cite{Bergen2020} argue that the latent time is correlated to the cellular position in the biological process, with a small value corresponding to an earlier developmental stage. 
The latent time is derived from a per-gene model based on the relative amount of unspliced mRNAs to spliced mRNAs. 
For each group, the abundances of unspliced and spliced mRNAs are obtained from the \textit{velocyto} pipeline \citep{La_Manno2018} and the latent time is computed from a generalized RNA velocity model \citep{Bergen2020}.

\subsubsection{Bayesian Model for Pax6 data}
The model for clustering the Pax6 data extends the work in \cite{Liu2024} to include latent time as a covariate.  \cite{Liu2024} employed the HDP prior for shared clustering across the HET and HOM data sets, where the clustering model is built upon the likelihood of \textit{bayNorm} \citep{Tang2020} that addresses the problem of normalization, imputation and batch effect correction in an integrated manner.

\paragraph{Likelihood.} The observed count $y_{c,g,d}$ is assumed to follow a binomial distribution given the latent true count $y_{c,g,d}^0$, with cell-specific capture efficiency $\beta_{c,d}$
\begin{equation*}
	y_{c,g,d} |  y^0_{c,g,d}, \beta_{c,d} \sim \Bin( y^0_{c,g,d}, \beta_{c,d}).
\end{equation*}
The binomial distribution accounts for the case where partial true counts are observed. The latent counts follow a negative-binomial distribution accounting for over-dispersion:
\begin{equation*}
	y^0_{c,g,d} | \mu_{c,g,d}, \phi_{c,g,d} \sim \NB( \mu_{c,g,d}, \phi_{c,g,d} ),  
\end{equation*}
with mean expression $\mu_{c,g,d}$ and dispersion $\phi_{c,g,d}$ that are both specific to each gene and cell. The latent counts can be integrated out to obtain:
\begin{equation*}
	y_{c,g,d} | \mu_{c,g,d}, \phi_{c,g,d}, \beta_{c,d} \sim \NB( \mu_{c,g,d} \beta_{c,d}, \phi_{c,g,d} ),
\end{equation*}
where it is noticed that $\mu$ and $\beta$ are not identifiable while only their product is. An informative prior for $\beta_{c,d}$ is applied to mitigate this problem \citep{Liu2024}. 

While \cite{Liu2024} assume the cell-specific mean and dispersion $$(\bm{\mu}_{c,d}, \bm{\phi}_{c,d}) |P_d \iidsim P_d,$$ 
and employ an HDP prior $P_d \sim \text{HDP}(\alpha_0, \alpha, P_0)$, we extend with a C-HDP prior 
$$P_{t_{c,d},d} \sim \text{C-HDP}(\alpha_0, \alpha, P_0, \bm{\Psi}^*),$$ with the covariate being the latent time. 
For the base measure $P_0$, a linear relationship is assumed between log mean expression and dispersion \citep{brennecke2013accounting, vallejos2015basics, eling2018correcting}. 

\paragraph{Kernel.} The latent time is included via a Gaussian kernel  with kernel parameters $\bpsi_{j,d}^*=(\tcenter, \sdcenter)$. As shown in Figure \ref{fig:Gaussian_kernel_example}, the parameter $\tcenter$ represents the value in the covariate space where the $j$-th component from the $d$-th group best applies and $\sdcenter$ controls the smoothness of the transition between components in the covariate space. The kernel parameters are data-specific, which allows for shared clusters across HET and HOM to be associated with different cellular positions and the degree of separation between clusters that can vary between two conditions.

For full details of the model and prior specifications, Gibbs sampling algorithm and implementation, see the online Appendix. Approaches for identifying marker genes that distinguish between different cell subtypes, posterior predictive checks \citep{Gelman1996posterior} that suggest no strong disagreement between the data and model, along with a simulation study, are available in online Appendix.

\subsubsection{Results on Pax6 Data}

\paragraph{Clustering.}

\begin{figure}[!tb]
	\begin{minipage}[h]{0.49\textwidth}
		\includegraphics[width=0.9\textwidth]{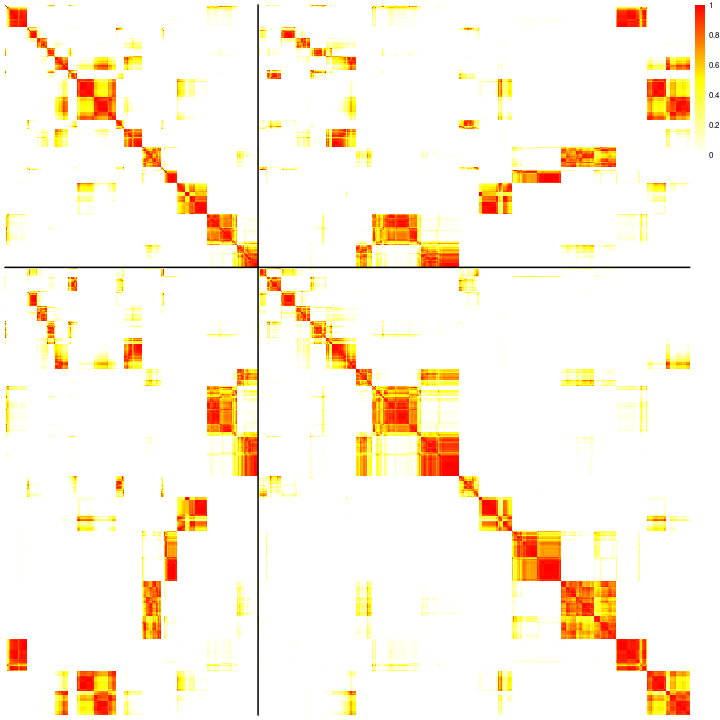}
	\end{minipage}
	\begin{minipage}[h]{0.49\textwidth}
		\includegraphics[width=0.95\textwidth]{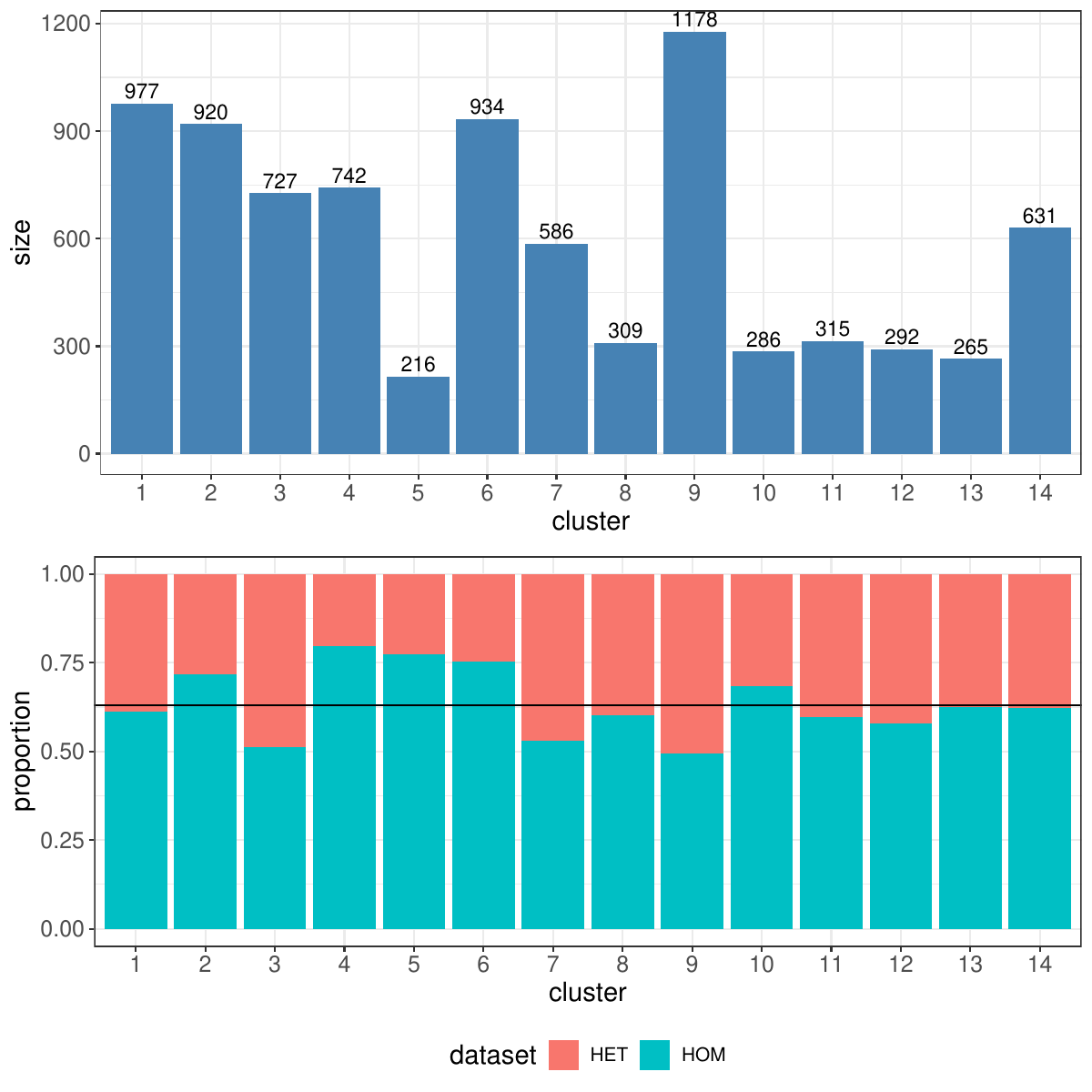}		
	\end{minipage}
	\caption{Left: Posterior similarity matrix within and between two experimental conditions. Diagonal blocks correspond to within-group PSM. Top-right: Size of each cluster found in Pax6 data. Bottom-right: Proportions of HET and HOM cells in each cluster. The black horizontal line represents the overall proportion.}
	\label{fig:psm_pax6}
\end{figure}

The Bayesian C-HDP model identifies 14 clusters using the VI criterion, whose sizes and proportion of HET and HOM cells are summarized in Figure \ref{fig:psm_pax6}. We refer to clusters as over-represented/under-represented in the mutant group if their proportion of HOM cells is greater/less than the overall proportion, and stable if the proportion is similar to the overall proportion. In this case, five clusters (2, 4, 5, 6, 10) are found to be over-represented in the mutant group, three clusters (3, 7, 9) are under-represented in the mutant group, and the remaining clusters (1, 8, 11, 12, 13, 14) show relatively stable proportions. Moreover, while we focus on the clustering estimate, the posterior similarity matrix (Figure \ref{fig:psm_pax6}) highlights some uncertainty in further splitting or merging some clusters. In addition, based on posterior allocation probability (Eq.~\eqref{eq:PP-approximation}), we observe some uncertainty in cell allocations at the boundary between clusters in the covariate space (see online Appendix).

\begin{figure}[!t]
	\centering
	\includegraphics[width=0.95\textwidth]{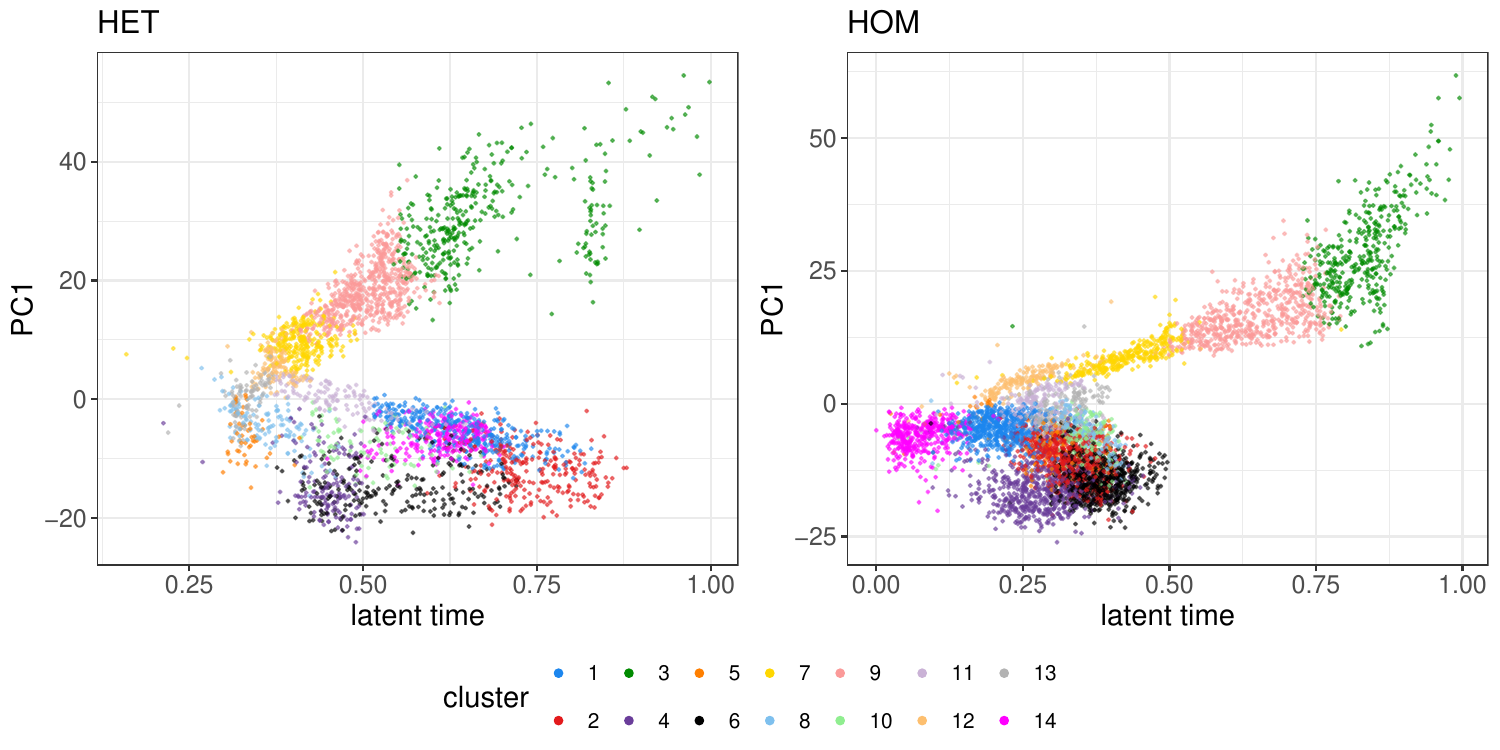}
	\caption{Plot of the first principal component (PC1) against latent time for HET (left) and HOM (right). Cells are colored by cluster membership. The three under-represented clusters 3, 7, 9 (dark green, yellow, light pink) are associated with higher latent time in the mutant group, and the over-represented clusters 2, 4, 5, 6, 10 (red, dark purple, orange, black, light green) have more moderate latent time, while the stable clusters (1, 8, 11, 12, 13, 14) are mainly associated with smaller latent time.
	}
	\label{fig:pc1_vs_t}
\end{figure}

Although all clusters are shared in both groups, suggesting that Pax6 may play a small role at this early stage in the development (day E13.5) \citep[as concluded in][]{Liu2024}, we observe some interesting patterns in the mutant group when connecting the clusters to latent time. Figure \ref{fig:pc1_vs_t} displays the first principal component computed from the observed gene expression matrix against latent time. The three under-represented clusters 3, 7, 9 (dark green, yellow, light pink) are associated with larger latent time, particularly for the mutant group, which suggests interesting implications in the role of Pax6 in cellular development, especially later in the biological process. The over-represented clusters 2, 4, 5, 6, 10 (red, dark purple, orange, black, light green) have moderate latent time in the mutant group, whereas the stable clusters have relatively earlier latent time in the mutant group.

\paragraph{Time-dependent Probabilities.}

\begin{figure}[tbp]
	\centering
	\includegraphics[width=0.95\textwidth]{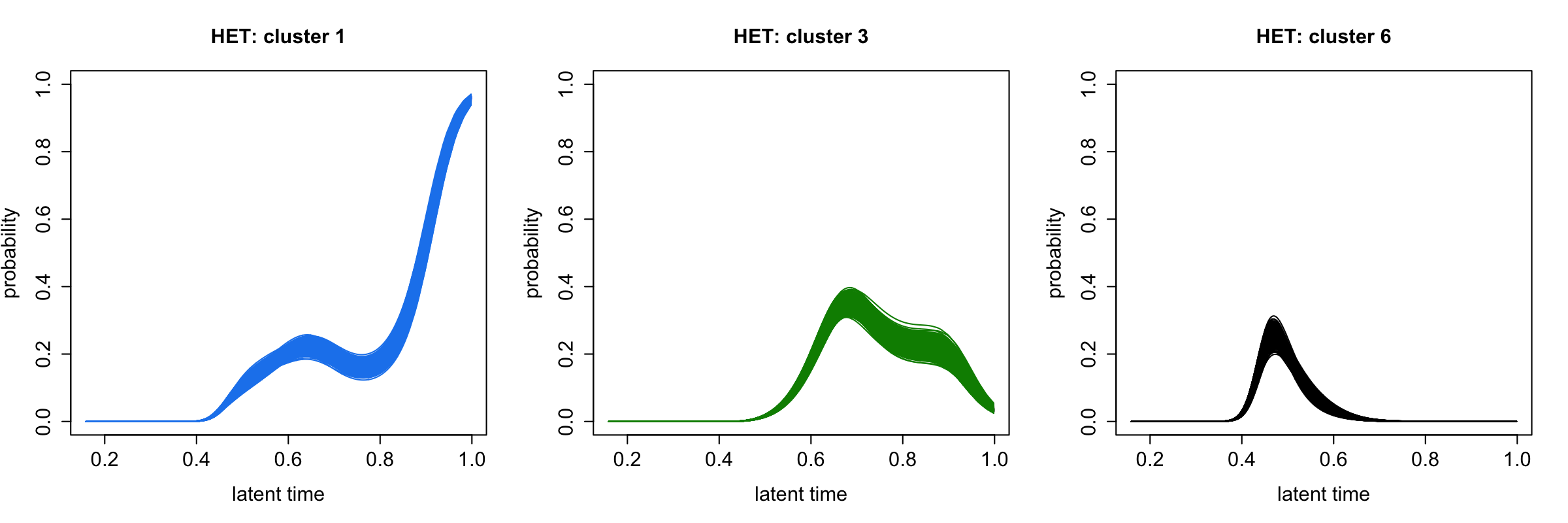}
	\includegraphics[width=0.95\textwidth]{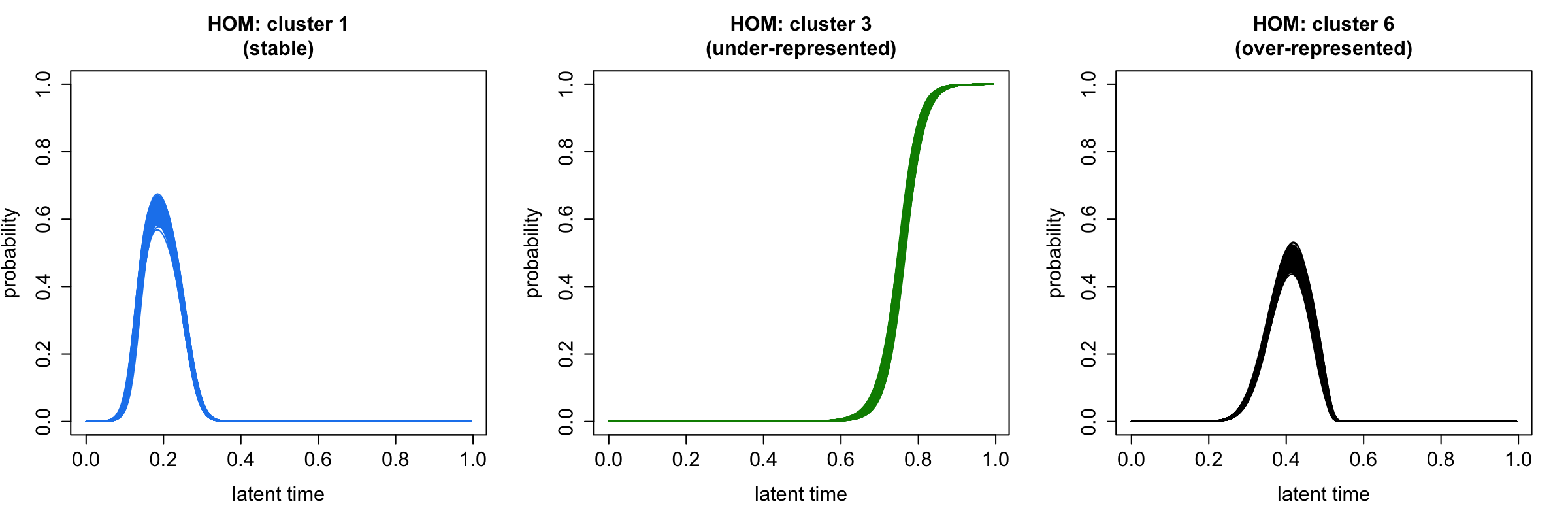}
	\caption{Time-dependent probabilities for clusters 1, 3, 6 (stable, under-represented, over-represented in HOM) in each data set for Pax6. The top row shows results for HET, with HOM shown in the bottom.}
	\label{fig:p_vs_t_subcluster}
\end{figure}

To further investigate the differences between HET and HOM, the time-dependent probabilities are visualized for stable, under-represented and over-represented clusters (1, 3, 6) in Figure \ref{fig:p_vs_t_subcluster}. The under-represented cluster is closely associated to cells with high latent time (probability close to 1) in the mutant group, while for the control group this association is not so pronounced. Specifically, although this cluster is associated with high latent time in the control group, cells with higher latent time may also belong to other clusters (probability is less than 0.5), as observed in Figure \ref{fig:pc1_vs_t} (left). The stable cluster is mainly associated to cells with small latent time in the mutant group (probability greater than 0.5), whereas it is more associated with moderate to large latent time in the control group. As for cluster 6 (over-represented in HOM), the difference between two conditions is not evident. The results for full set of clusters are shown in online Appendix.

\paragraph{Latent Counts.}

\begin{figure}[!t]
	\centering
	\includegraphics[width=0.7\textwidth]{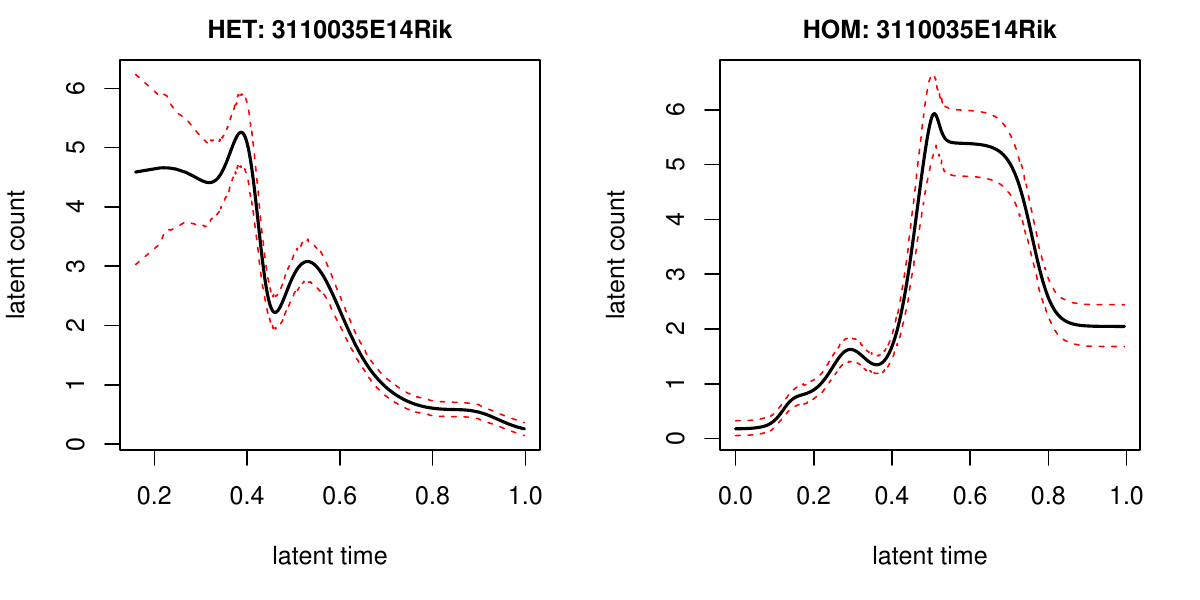}
	\includegraphics[width=0.7\textwidth]{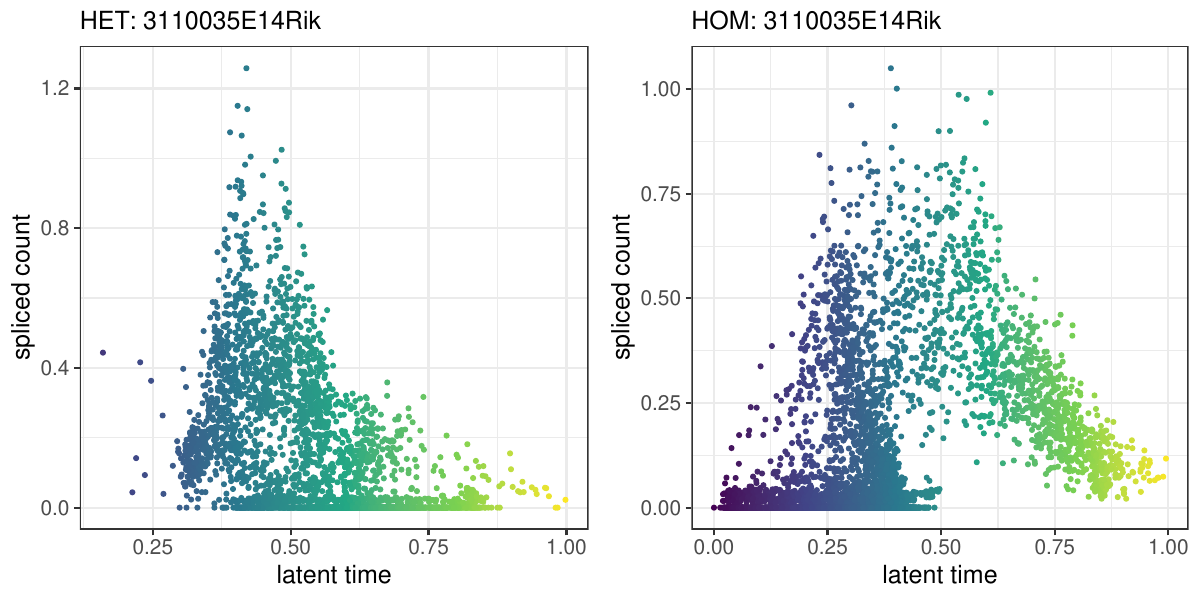}
	\caption{Top: Estimated latent counts against latent time for an example gene \textit{3110035E14Rik} identified as globally differentially expressed. The 95\% highest posterior density interval is given by dashed red lines. Bottom: Spliced mRNA counts against latent time. Cells are colored by latent time.}
	\label{fig:mean_latent_count_against_time}
\end{figure}

Beyond clustering, the C-HDP can be used for nonparametric conditional density estimation and regression (Section \ref{sec:predictive mean}). Specifically, for scRNA-seq, this is useful to understand how gene expression changes over latent time and across conditions. The expected count for a new cell $c$ from data set $d$ as a function of latent time is
\begin{equation*} \label{eq:mean of latent count}
	\E(y_{c,g,d}^0|t_{c,d}=t,\mathcal{D})= \int \sum_{j=1}^{J}p_{j,d}^J (t) \mu_{j,g}^* d \pi(\bmq^J_{1:J,d}, \bmu_{1:J,g}^*, \bpsi_{1:J,d}^* | \mathcal{D}),
\end{equation*}
which is approximated from the MCMC samples. This is illustrated in Figure \ref{fig:mean_latent_count_against_time} for an example gene \textit{3110035E14Rik} which is identified as globally differentially expressed (see online Appendix for methods for identifying marker genes). A general decreasing pattern is observed in the control group with a potentially mild increase in the beginning (small latent time), whereas in the mutant group a clear increasing trend is present followed by a decreasing trend, suggesting Pax6 may influence the expression activity of gene \textit{3110035E14Rik}. In addition, the relationship between the latent counts and latent time appears similar to that between the observed spliced mRNA counts and latent time, which implies the reliability of the nonparametric estimate.  

\subsection{Application to Calcium Imaging Data}

Most neurons interact with other neurons for communicating information rather than receiving direct inputs from the external world, which gives rise to an internal structure \citep{rubin2019revealing}. Such internal structure differs across brain regions with different computational roles, and can be used to expose unknown encoded variables, facilitating understanding of the functional roles of different brain circuits.  Therefore, understanding the internal structure has become a critical goal in neuroscience, elucidating the neural mechanisms important for behavioral variables. This can be empirically studied using the calcium imaging technique which records neural activity of behaving animals over time \citep{mukamel2009automated, rubin2015}, with a higher measurement suggesting that the neuron is firing, actively passing or receiving information. 

In this setting, clustering is useful to identify time frames when neural activity is similar. In \cite{rubin2019revealing}, it was shown that neural activity can be encoded in a lower-dimensional space and the identified clusters in the latent space correlate with behavioral measurements and positions, shedding light on the neural code associated with behavior and location. However, the authors analyzed multiple experiments individually, which makes it difficult to identify shared patterns across experiments. In addition, algorithmic clustering, e.g. k-means, is used, which is unable to account for the temporal nature of the data. To address these concerns, our aim is to use the C-HDP to integrate multiple experiments, identify shared neural activity patterns, and account for heterogeneous temporal dependence in the data (Figure \ref{fig:yt_vs_yt-1}) through a hierarchical Bayesian modelling framework.

We note that related approaches include \cite{d2023bayesian}, who propose a nested Bayesian finite mixture model with common atoms to allow shared clustering across experiments, which takes into account the time dependence. However, the method is designed for neural activity in a single neuron. In addition, \cite{denti2023common} identify clusters characterized by similar spike magnitudes and therefore temporal dependence between successive time frames is assumed homogeneous within each experiment.

We study calcium imaging data collected by the Nolan lab in the Centre for Discovery Brain Sciences at the University of Edinburgh to study neural activity in mice. Data is from the dorsal CA1 region of the hippocampus, where neurons are tuned to spatial positions of the animal \citep{o1971hippocampus}. For a single mouse, two experiments were conducted over two different days inside a linear rig. The mouse was asked to run back and forth in the rig, and each experiment records the calcium activity of hundreds of neurons at fixed time bins. Following the approach in \cite{rubin2019revealing}, the data is projected to a lower-dimensional space to explore their activity patterns. Specifically, we apply kernel PCA \citep{bishop2006} to the first experiment, which is used to transform both experiments from hundreds of neurons into a three-dimensional encoding of the neural activity of the neurons ($G=3$). Each experiment consists of $n_d=800 \ (d=1,2)$ time frames, and the covariate of interest is the observed time $t$ which is rescaled to be equally spaced on $[0,1]$.

\subsubsection{Bayesian Model for Calcium Imaging Data}
Our aim is to build on the work of \cite{rubin2019revealing} to develop a clustering model that accounts for the temporal dependence in the data through the model-based approach and integrates data across multiple experiments through the hierarchical framework of the C-HDP. For the former,
the within-component likelihood is assumed to follow  a vector autoregression with lag one (Eq.~\eqref{eq:likelihood-VAR}). Time frames with similar dependence on the previous time point and a similar covariance structure are expected to be in the same cluster. We assume the likelihood parameters for each time frame are generated from the covariate-dependent distribution $P_{t,d}$ in group $d$, which is modelled from the proposed C-HDP prior, with the covariate being the observed time:
$$
(\bL_{i,d}, \bSigma_{i,d}) | P_{t_{i,d},d} \sim P_{t_{i,d},d}, \quad P_{t_{i,d},d} \sim \text{C-HDP}(\alpha_0, \alpha, P_0, \bm{\Psi}^*),
$$
where $\bL_{i,d}=(\bma_{i,d} \quad \bB_{i,d})^T$ denote the $(G+1) \times G$ coefficient matrix for the $i$-th time frame ($i=1,\ldots,n_d$) from the $d$-th experiment, and $\bSigma_{i,d}$ is the $G \times G$ covariance matrix.

\paragraph{Base Measure.} For the base measure $P_0$ of the C-HDP, the component-specific parameters $\bL_j^*=(\bma_j^* \quad \bB_j^*)^T$ and $\bSigma_j^*$ are given conjugate priors:
\begin{equation*}
	\bL_j^* | \bSigma_j^*  \indsim \MatrixN(\bL_0, \bV_0, \bSigma_j^*), \quad \bSigma_j^*  \iidsim \IW(\omega_0, \bPhi_0),
\end{equation*}
where $\MatrixN$ and $\IW$ denote the matrix normal and inverse-Wishart distribution, $\bL_0$ is of dimension $(G+1) \times G$, $\bV_0$ is of dimension $(G+1) \times (G+1)$, $\bPhi_0$ is of dimension $G \times G$, and $\omega_0 > G-1$. Empirical estimates are used for prior specification (for details see the online Appendix).

\paragraph{Kernel.} As a cyclic pattern over time has been observed in the transformed data (see Figure \ref{fig:yt_vs_yt-1} below), we implement the periodic kernel to account for the recurring pattern, with kernel parameters $\bpsi_{j,d}^*=(\mu_{j,d}^*, \sdcenter, \plambda)$. The parameter $\mu_{j,d}^*$ represents the value that maximizes the kernel, $\plambda$ specifies the period of the kernel and $\sdcenter$ again smooths the covariate region (Figure \ref{fig:periodic_kernel_example}).

For full details of the model and prior specifications, Gibbs sampling algorithm and a simulated experiment, see the online Appendix. In addition, we provide predictions of the neural activity and covariate-dependent probabilities for future time points and assess model fit through posterior predictive checks in the online Appendix.

\subsubsection{Results on Calcium Imaging Data}

\begin{figure}[tbp]
	\begin{minipage}[h]{0.49\textwidth}
		\includegraphics[width=1\textwidth]{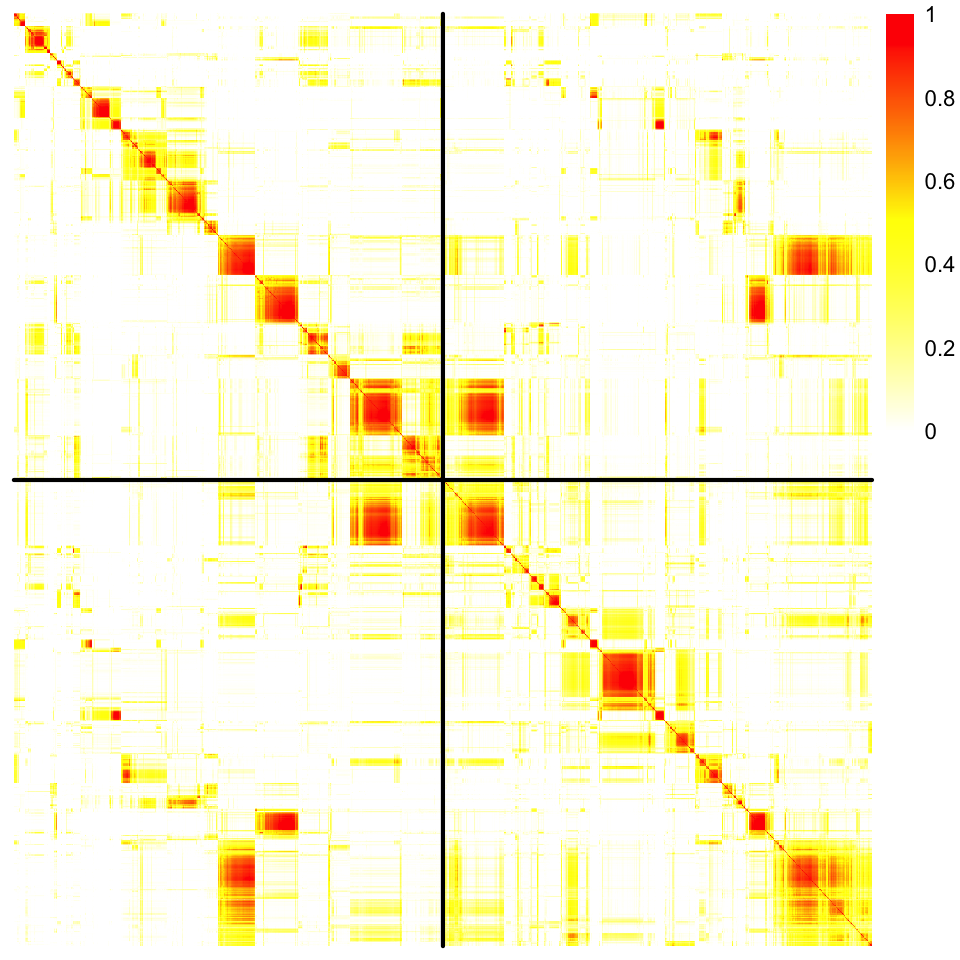}
	\end{minipage}
	\begin{minipage}[h]{0.49\textwidth}
		\includegraphics[width=0.9\textwidth]{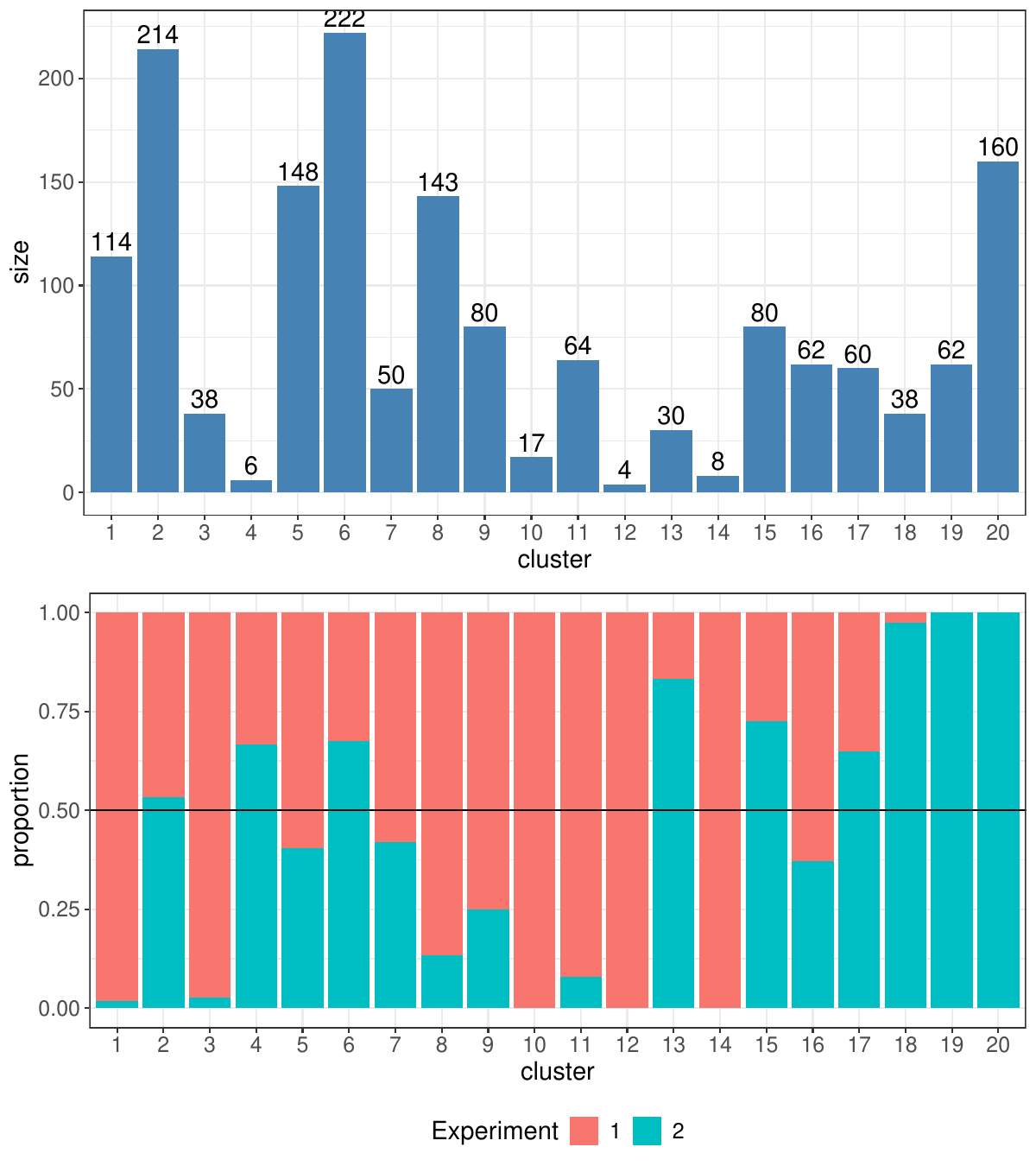}		
	\end{minipage}
	\caption{Left: Posterior similarity matrix within and between experiments. Diagonal blocks correspond to within-experiment PSM. Top-right: Size of each cluster in the calcium imaging data. Bottom-right: Proportions of time frames from the first and second experiment in each cluster. The black horizontal line shows the overall proportion.}
	\label{fig:psm_cidata}
\end{figure}

\paragraph{Clustering.}
Twenty clusters associated to different activities are identified across both experiments, with the posterior similarity matrix shown in Figure \ref{fig:psm_cidata} depicting some uncertainty in allocations. There are 15 clusters shared in both experiments, with varying proportions reflecting different amounts of time associated to patterns across the experiments.
There are also some unique clusters of neural activity as well as several small clusters with size $<20$, which may represent the noise in the data.

\begin{figure}[tbp]
	\includegraphics[width=0.92\textwidth]{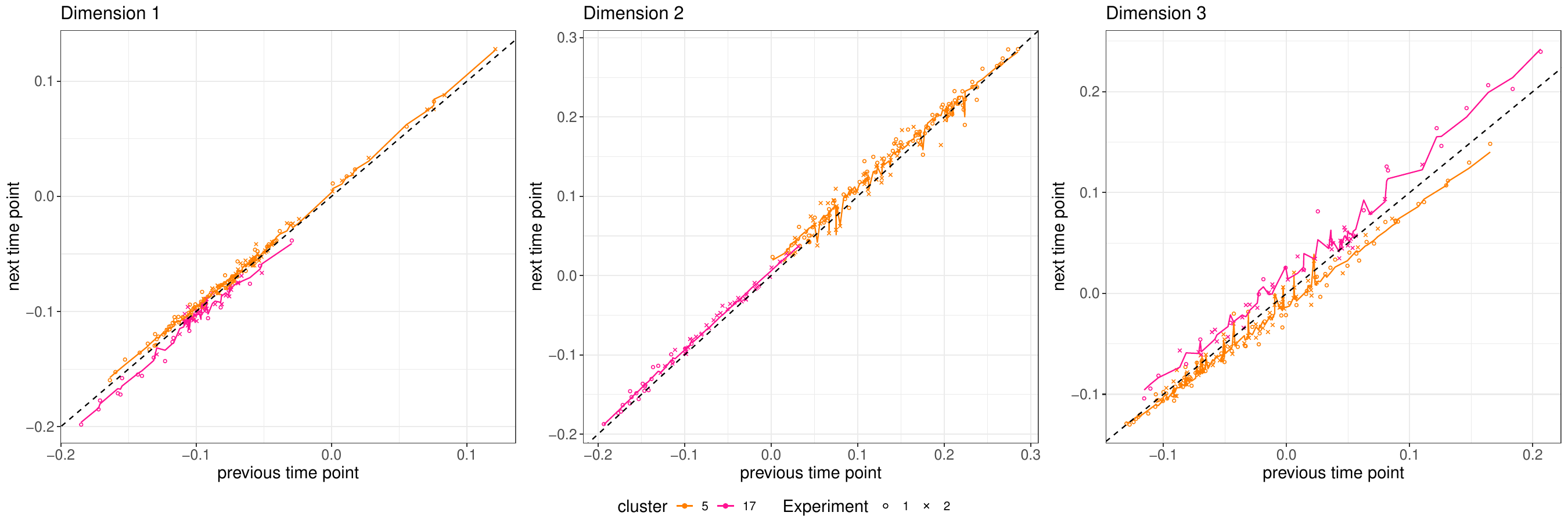}
	\includegraphics[width=0.92\textwidth]{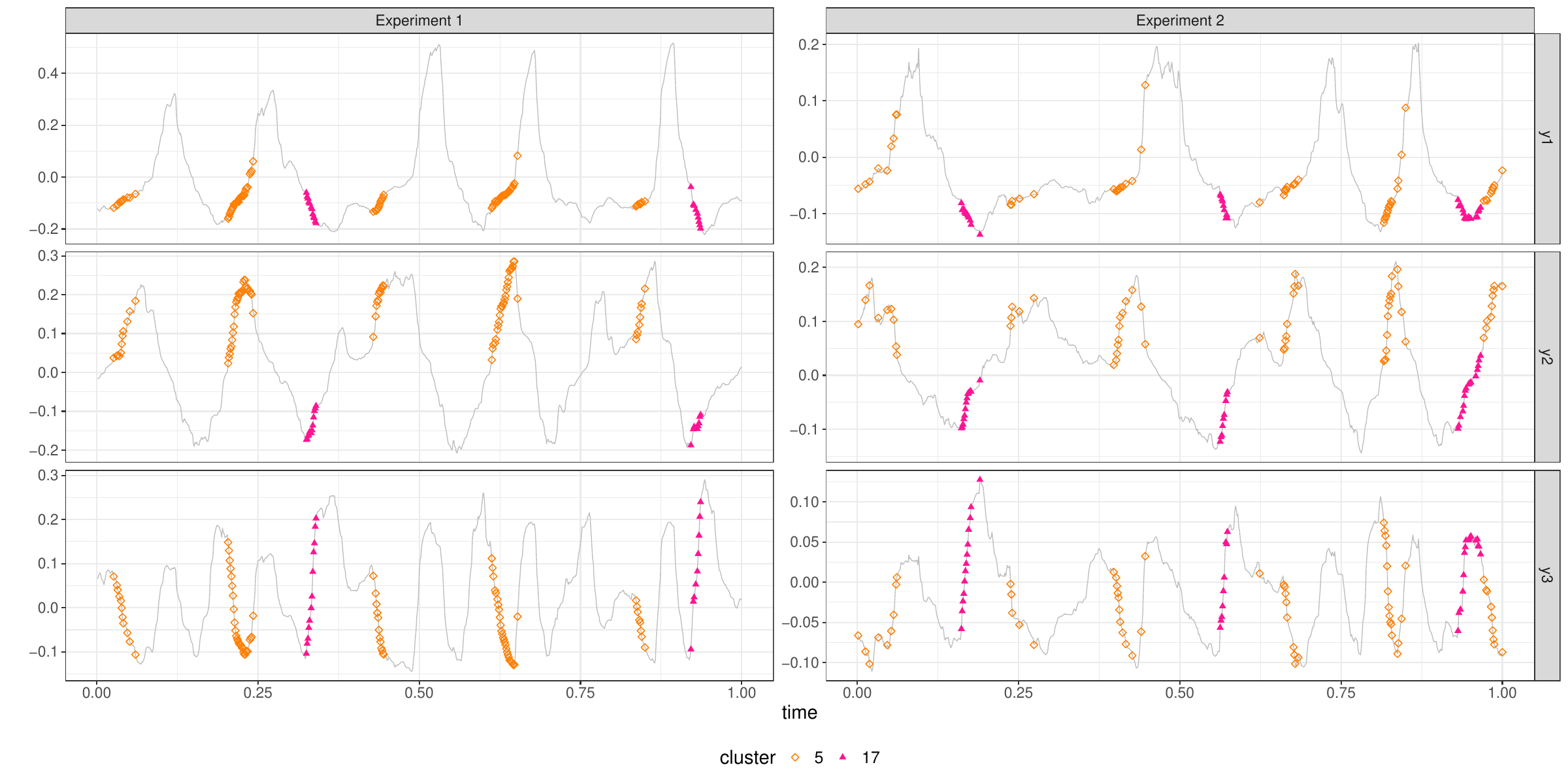}		
	\caption{Top: Plot of current time frame against the previous time frame for each dimension, for clusters 5 and 17 in both experiments. The black dashed line denotes $y=x$. The solid colored line denotes the posterior estimated relationship between successive time frames for each cluster. Bottom: Time-series plot for each dimension (row), with clusters 5 and 17 highlighted in color.}
	\label{fig:yt_vs_yt-1}
\end{figure}

To understand the neural activity patterns of the identified clusters, we plot the encoding of activity at each time frame against previous time for a specific cluster (Figure \ref{fig:yt_vs_yt-1} top). It is noticed that cluster 17 (pink) is mainly below the equivalent line $y=x$ in the first dimension, suggesting a decreasing trend in the encoded activity as time increases, whilst an increasing trend is observed in the third dimension. On the other hand, cluster 5 exhibits exactly the opposite patterns to cluster 17. Both clusters show an upward trend in the second dimension. This can be confirmed from the lower panels in Figure \ref{fig:yt_vs_yt-1} which shows a time-series plot for each reduced dimension.

Further, we compute the posterior estimated relationship between consecutive time points for each dimension based on
\begin{equation*}
	\bm{y}_i = \hat{\bma}_j^*+\hat{\bB}_j^*\bm{y}_{i-1},
\end{equation*}
where $\hat{\bma}_j^*$ and $\hat{\bB}_j^*$ denote the posterior mean of the coefficients. The estimated relationship is shown in the colored solid lines in Figure \ref{fig:yt_vs_yt-1}. It is worth noting that an almost linear relationship is observed for cluster 5 in the first dimension and cluster 17 in the second dimension, suggesting a dominant role of the past observation in the same reduced dimension. Nonlinear relationships instead indicate interactions between different lower-dimensional embeddings that represent different aspects of summaries of neuronal activities.

\begin{figure}[tbp]
	\centering
	\includegraphics[width=0.95\textwidth]{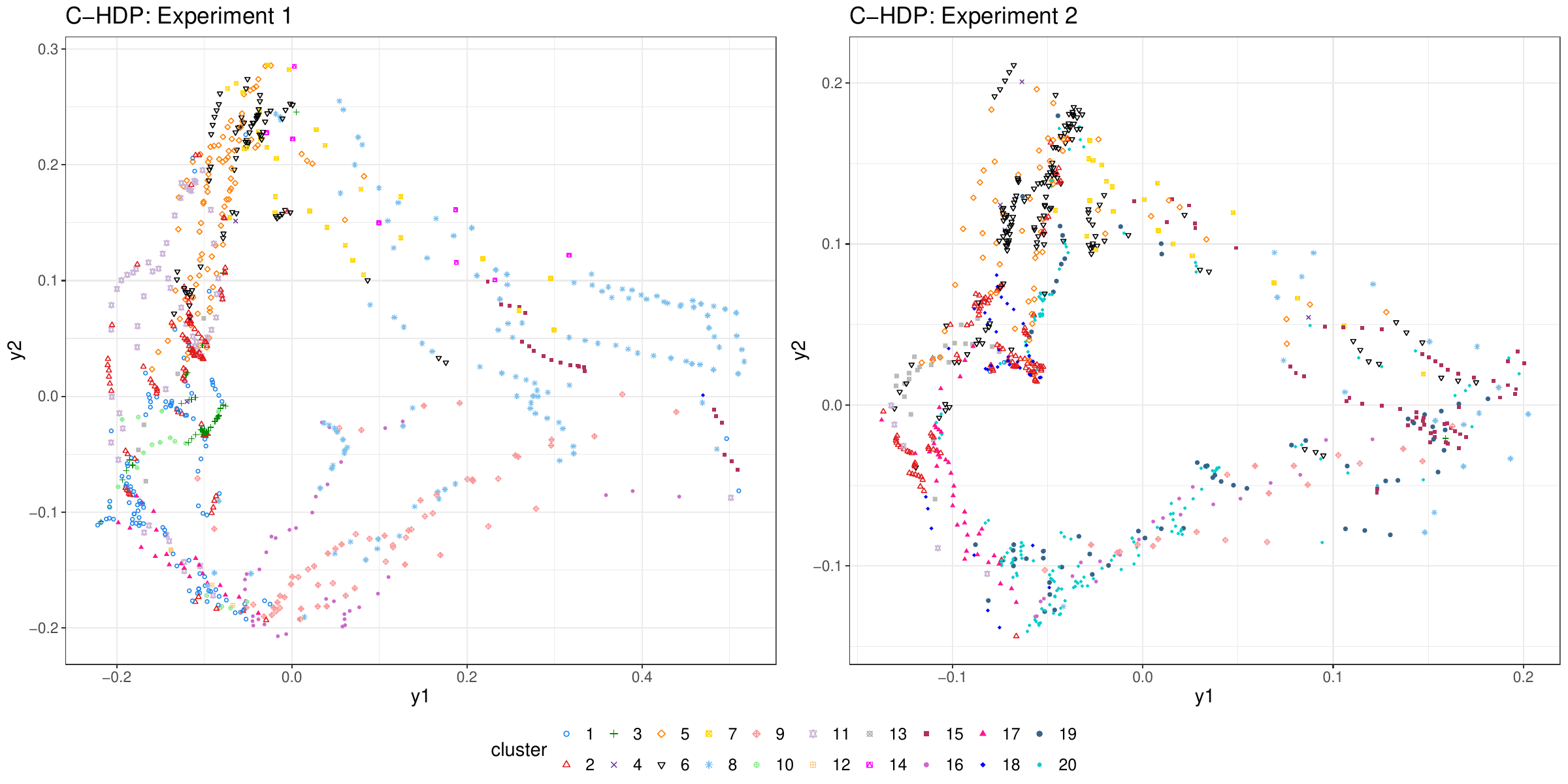}
	\includegraphics[width=0.95\textwidth]{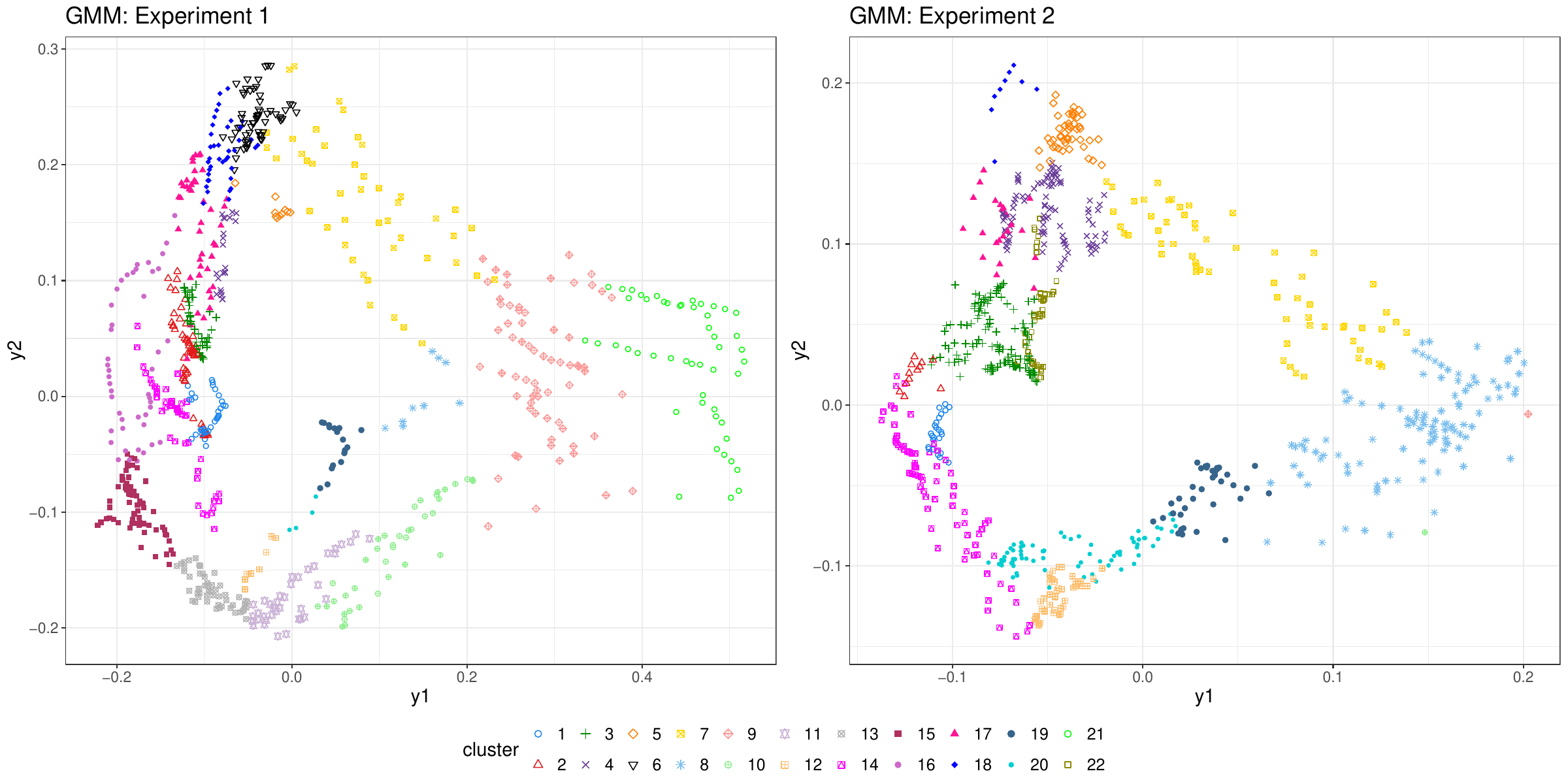}
	\caption{Pairwise scatterplots for the first two dimensions. The top panel displays results from the C-HDP, and the bottom row corresponds to a simple GMM. Columns correspond to different experiments. Time frames are colored by cluster membership.}
	\label{fig:pairplot-cidata-sub}
\end{figure}

For a comparison, we visualize the clusters obtained from fitting a Gaussian mixture model (GMM) with an unconstrained covariance matrix\footnote{GMM is implemented using R package \texttt{mclust} \citep{scrucca2016mclust}.} (Eq.~\eqref{eq:likelihood-GMM}) after pooling the data sets. The pairplots of the encoded activity (Figure \ref{fig:pairplot-cidata-sub}) displays a similar cyclic behavior in both experiments. For the C-HDP method, each shared cluster tends to accumulate at the same edge or vertex. For instance, clusters 5 and 6 (orange and black) are mainly in the top-left corner, shared in both experiments. Clusters 8 and 9 (light blue and light pink) concentrate around the upper-right and lower-right edges, respectively, and such pattern is also similar across two experiments.  However, GMM is unable to capture such characteristics. Instead, GMM simply groups time frames into blocks of similar expressions (e.g. light green and light pink clusters with $y_1>0.2$ in the first experiment), and clusters are therefore less likely to be shared at the same edge or vertex across experiments. A full set of pairwise scatterplots is shown in online Appendix.

\paragraph{Alignment with Behavioral Data.}

\begin{figure}[!tbp]
	\centering
	\includegraphics[width=1\textwidth]{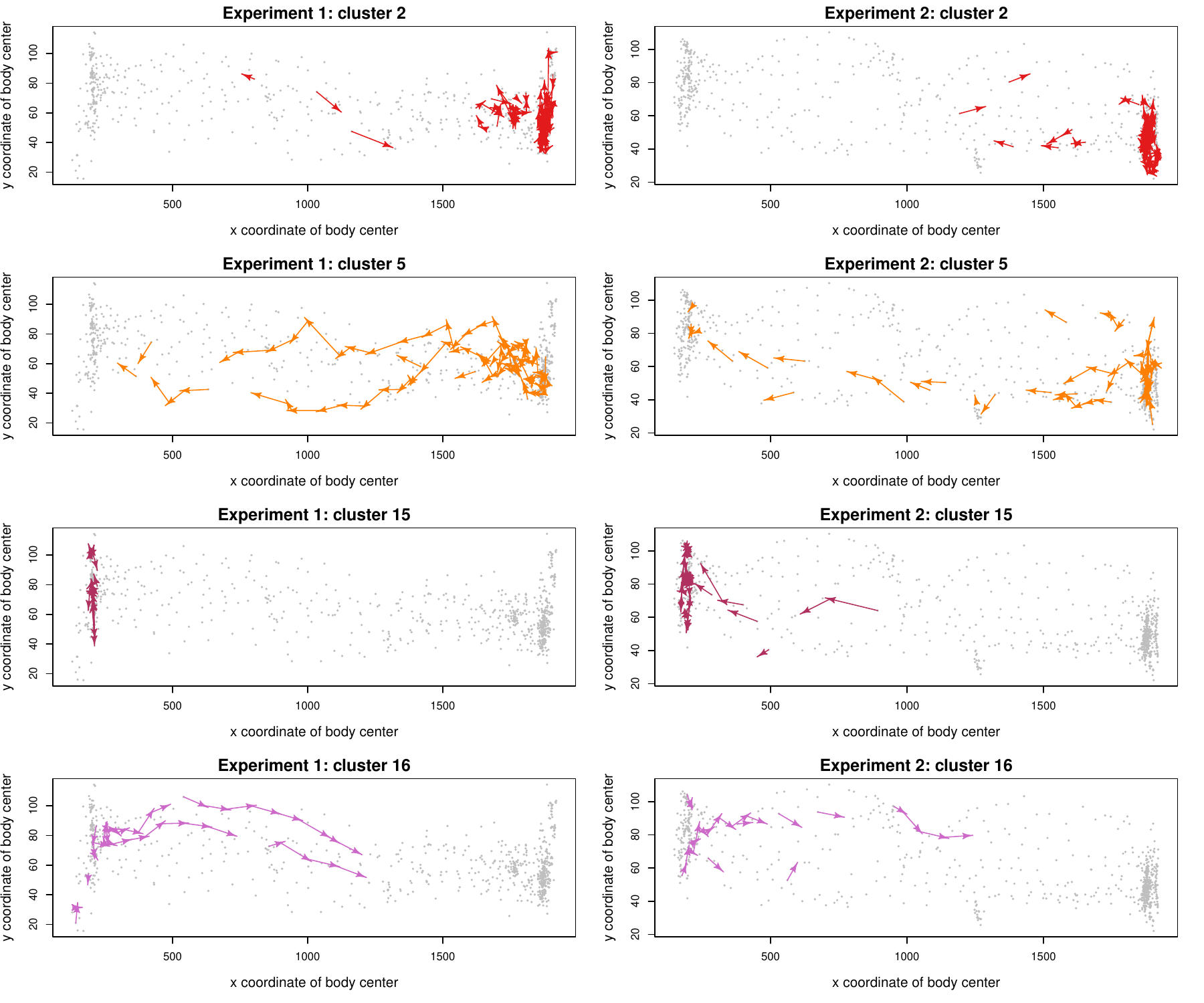}
	\caption{The body locations of the mouse. In each panel, the arrow points from the current point to the next time point, and the color indicates the cluster corresponding to the next time point (arrowhead).}
	\label{fig:body_location}
\end{figure}

To understand the relationship between clusters and externally recorded behaviors of the mouse, Figure \ref{fig:body_location} displays the spatial locations of the mouse at each time point belonging to the same cluster. Even though the spatial information is not used for modelling, clusters can still match with the spatial positions and navigation of the mouse, the known variable encoded by the hippocampus. For instance, cluster 2 is likely to correspond to the mouse scurrying around the right end of the linear rig, whereas cluster 15 may represent the mouse moving around the left end. Clusters 5 and 16 coincide with the mouse moving towards the left end and right end, respectively. In particular, cluster 5 spans almost across the whole linear track, while cluster 16 is mainly located at the left half track. This highlights that our model is able to identify meaningful neural activity patterns mapping to the behavioral variables.

\paragraph{Time-dependent Probabilities.}

\begin{figure}[tbp]
	\centering
	\includegraphics[width=0.8\textwidth]{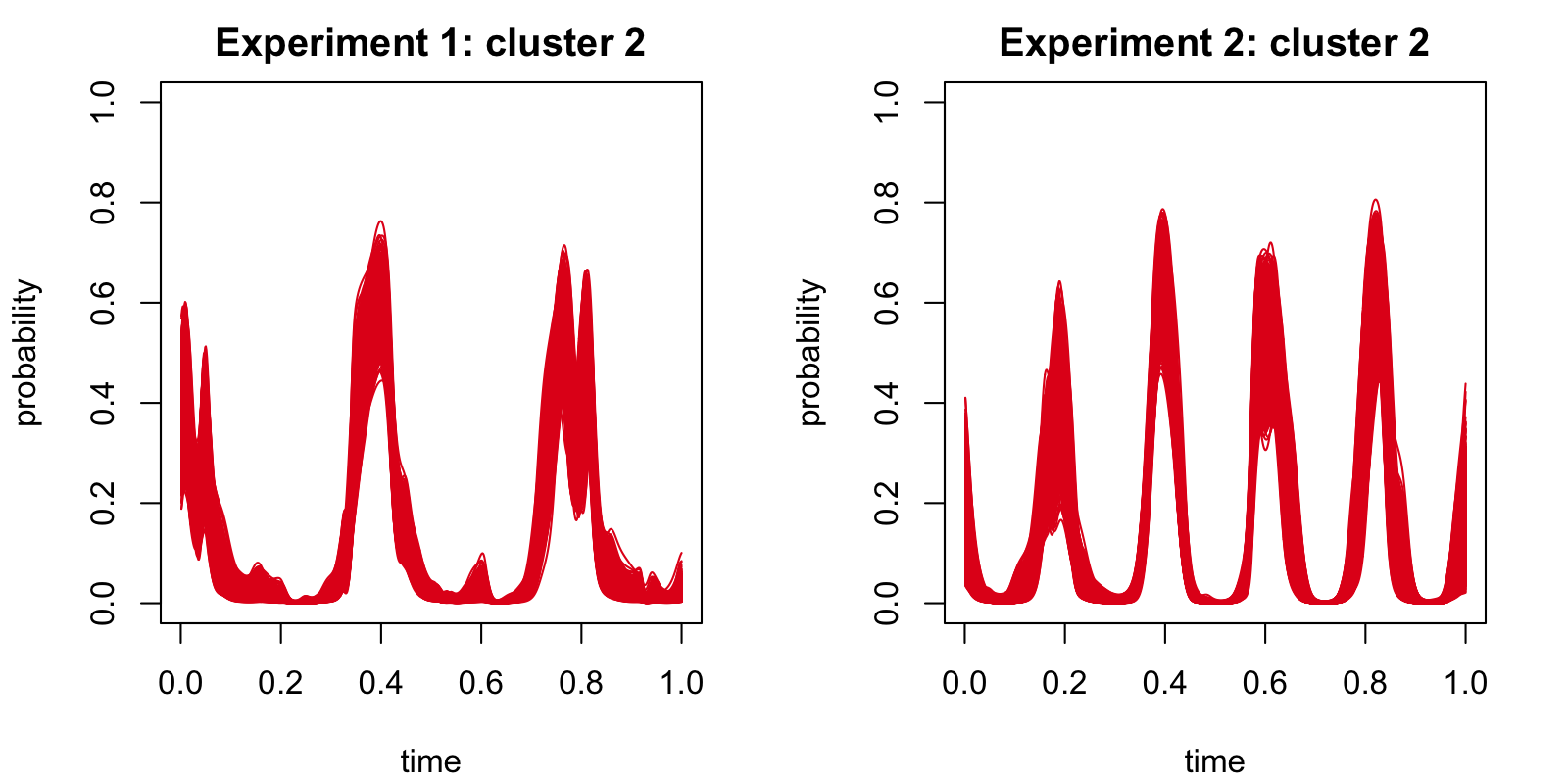}
	\caption{Time-dependent probabilities for cluster 2 for the calcium imaging data.}
	\label{fig:p_vs_t_cidata_sub}
\end{figure}

Figure \ref{fig:p_vs_t_cidata_sub} shows posterior samples of the time-dependent probabilities for cluster 2. The probabilities show an recurring pattern, and there is no substantial difference in the magnitude of uncertainty between two experiments, as they have the same data size. In addition, different experiments can have different periodicity, with cluster 2 happening more frequently in the second experiment, suggesting that the animal probably spent more time moving around the right end of the linear rig in the second experiment (Figure \ref{fig:body_location}). The posterior samples for all clusters are shown in online Appendix.

\section{Discussion}
\label{sec:conclusion}

In this paper, we have developed a covariate-dependent hierarchical Dirichlet process prior to flexibly integrate external covariates into clustering and density estimation across related groups, combining the strengths from both the HDP and DDP. An efficient MCMC sampling scheme is provided for inference based on data augmentation tricks. The dependence on the covariate is introduced by appropriate kernel functions and covariate-dependent weights are constructed based on normalization for enhanced interpretability. Our approach is fairly general in terms of the choice of the component-specific likelihood and kernel, which can be subjective and depend on the characteristics of the data.

We illustrated the utility of the C-HDP on two real data sets using two examples of kernel functions: a Gaussian kernel and a periodic kernel, and two examples of component-specific likelihood functions: negative-binomial and vector autoregression. The results demonstrate that our C-HDP prior yields meaningful clusters for both data sets. The covariate-dependent probabilities enhance our understanding of the influence of external covariates on clustering and differences between groups. For the scRNA-seq data, the identified clusters reveal separation in the lower dimensional embeddings and latent time. In particular, the under-represented clusters in the mutant group are associated with larger latent time as well as high covariate-dependent probabilities (close to 1), suggesting Pax6 plays an important role in the later stage of the biological process. On the other hand, stable clusters have relatively smaller latent time in the mutant group with moderately large probabilities (above 0.5). The C-HDP method further enables use to discern the relationship between latent counts and latent time through a nonparametric estimate, which shows a similar pattern to that of the spliced mRNA counts. For the calcium imaging data, time frames from the same cluster exhibit a homogeneous dependence structure with the previous time point, which cannot be achieved by a simple Gaussian mixture model. In addition, without using the spatial information in the model which is known to be influenced by the hippocampus where the data was collected, the clusters are still found closely aligned with the spatial locations and navigation of the animal. 

However, while we have constructed an efficient Gibbs sampling algorithm for posterior inference, the algorithm may still face the dilemma of getting trapped in the local modes and a large number of iterations is needed to reach convergence. The problem is even more severe for high-dimensional scRNA-seq data with large number of cells and genes. We have employed consensus clustering \citep{Coleman2020} to overcome these difficulties to exploit the clustering.
In the future, alternative methods will be considered, and one option is the parallelizable posterior bootstrap \citep{fong2019scalable} suitable for multimodal posteriors. Moreover, for the Pax6 application, incorporating both the clustering and estimation of latent time into a single model instead of two separate steps would better account for the uncertainty in the scRNA-seq data. As for the calcium imaging application, future work could incorporate dimension reduction within the modeling-based clustering approach \citep{chandra2023escaping}.
Finally, it is worth extending the model to encompass covariate-dependent atoms as well, enhancing its applicability to more complex data sets.

\acks{
	The authors are grateful to the Centre for Statistics for funding of the joint workshops with the Centre for Discovery Brain Sciences (CDBS) at the University of Edinburgh. The Pax6 data was provided by Prof. David Price's lab (CDBS) and the calcium imaging data was provided by Dr. Alessandro Di Filippo and Prof. Matt Nolan (CDBS).
	
	The authors declare no competing interests. 
	
	An R package to implement the proposed models is available at \url{https://github.com/huizizhang949/chdp}. Additional examples for the simulation study are available at \url{https://github.com/huizizhang949/chdp_example}.
	
}

\clearpage

\vskip 0.2in
\bibliography{export}

\begin{thebibliography}{89}
\providecommand{\natexlab}[1]{#1}
\providecommand{\url}[1]{\texttt{#1}}
\expandafter\ifx\csname urlstyle\endcsname\relax
  \providecommand{\doi}[1]{doi: #1}\else
  \providecommand{\doi}{doi: \begingroup \urlstyle{rm}\Url}\fi

\bibitem[Antoniano-Villalobos et~al.(2014)Antoniano-Villalobos, Wade, and
  Walker]{antoniano2014bayesian}
I.~Antoniano-Villalobos, S.~Wade, and S.~G. Walker.
\newblock {A Bayesian nonparametric regression model with normalized weights: A
  study of hippocampal atrophy in Alzheimer’s disease}.
\newblock \emph{Journal of the American Statistical Association}, 109\penalty0
  (506):\penalty0 477--490, 2014.

\bibitem[Argiento et~al.(2020)Argiento, Cremaschi, and
  Vannucci]{argiento2020hierarchical}
R.~Argiento, A.~Cremaschi, and M.~Vannucci.
\newblock Hierarchical normalized completely random measures to cluster grouped
  data.
\newblock \emph{Journal of the American Statistical Association}, 2020.

\bibitem[Beraha et~al.(2021)Beraha, Guglielmi, and Quintana]{beraha2021semi}
M.~Beraha, A.~Guglielmi, and F.~A. Quintana.
\newblock {The semi-hierarchical Dirichlet process and its application to
  clustering homogeneous distributions}.
\newblock \emph{Bayesian Analysis}, 16\penalty0 (4):\penalty0 1187--1219, 2021.

\bibitem[Bergen et~al.(2020)Bergen, Lange, Peidli, Wolf, and Theis]{Bergen2020}
V.~Bergen, M.~Lange, S.~Peidli, F.~A. Wolf, and F.~J. Theis.
\newblock {Generalizing RNA velocity to transient cell states through dynamical
  modeling}.
\newblock \emph{Nature Biotechnology}, 38\penalty0 (12):\penalty0 1408--1414,
  2020.

\bibitem[Bishop(2006)]{bishop2006}
C.~M. Bishop.
\newblock \emph{Pattern Recognition and Machine Learning}.
\newblock Information Science and Statistics. Springer, New York, 2006.

\bibitem[Blei et~al.(2003)Blei, Ng, and Jordan]{blei2003latent}
D.~M. Blei, A.~Y. Ng, and M.~I. Jordan.
\newblock Latent {Dirichlet} allocation.
\newblock \emph{Journal of Machine Learning Research}, 3:\penalty0 993--1022,
  2003.

\bibitem[Boon(2021)]{phdthesis_kai}
T.~K. Boon.
\newblock \emph{{Transcriptional guardian for glutamatergic fate against
  morphogens during mouse cerebral cortex development}}.
\newblock PhD thesis, University of Edinburgh, 2021.

\bibitem[Brennecke et~al.(2013)Brennecke, Anders, Kim, Ko{\l}odziejczyk, Zhang,
  Proserpio, Baying, Benes, Teichmann, Marioni,
  et~al.]{brennecke2013accounting}
P.~Brennecke, S.~Anders, J.~K. Kim, A.~A. Ko{\l}odziejczyk, X.~Zhang,
  V.~Proserpio, B.~Baying, V.~Benes, S.~A. Teichmann, J.~C. Marioni, et~al.
\newblock Accounting for technical noise in single-cell {RNA}-seq experiments.
\newblock \emph{Nature Methods}, 10\penalty0 (11):\penalty0 1093--1095, 2013.

\bibitem[Caballero et~al.(2014)Caballero, Manuel, Molinek, Quintana-Urzainqui,
  Mi, Shimogori, and Price]{caballero2014cell}
I.~M. Caballero, M.~N. Manuel, M.~Molinek, I.~Quintana-Urzainqui, D.~Mi,
  T.~Shimogori, and D.~J. Price.
\newblock Cell-autonomous repression of {S}hh by transcription factor {P}ax6
  regulates diencephalic patterning by controlling the central diencephalic
  organizer.
\newblock \emph{Cell Reports}, 8\penalty0 (5):\penalty0 1405--1418, 2014.

\bibitem[Camerlenghi et~al.(2019)Camerlenghi, Dunson, Lijoi, Pr{\"u}nster, and
  Rodr{\'\i}guez]{camerlenghi2019latent}
F.~Camerlenghi, D.~B. Dunson, A.~Lijoi, I.~Pr{\"u}nster, and A.~Rodr{\'\i}guez.
\newblock Latent nested nonparametric priors (with discussion).
\newblock \emph{Bayesian Analysis}, 14\penalty0 (4):\penalty0 1303, 2019.

\bibitem[Catalano and Lavenant(2024)]{catalano2024hierarchical}
M.~Catalano and H.~Lavenant.
\newblock Hierarchical integral probability metrics: {A} distance on random
  probability measures with low sample complexity.
\newblock In \emph{International Conference on Machine Learning}, pages
  5841--5861. PMLR, 2024.

\bibitem[Chandra et~al.(2023)Chandra, Canale, and Dunson]{chandra2023escaping}
N.~K. Chandra, A.~Canale, and D.~B. Dunson.
\newblock {Escaping the curse of dimensionality in Bayesian model-based
  clustering}.
\newblock \emph{{Journal of Machine Learning Research}}, 24\penalty0
  (144):\penalty0 1--42, 2023.

\bibitem[Coleman et~al.(2022)Coleman, Kirk, and Wallace]{Coleman2020}
S.~Coleman, P.~D. Kirk, and C.~Wallace.
\newblock {Consensus clustering for Bayesian mixture models}.
\newblock \emph{BMC Bioinformatics}, 23\penalty0 (1):\penalty0 290, 2022.

\bibitem[Dahl(2006)]{dahl2006model}
D.~B. Dahl.
\newblock {Model-based clustering for expression data via a Dirichlet process
  mixture model}.
\newblock \emph{Bayesian Inference for Gene Expression and Proteomics},
  4:\penalty0 201--218, 2006.

\bibitem[Dai and Storkey(2014)]{dai2014supervised}
A.~M. Dai and A.~J. Storkey.
\newblock {The supervised hierarchical Dirichlet process}.
\newblock \emph{IEEE Transactions on Pattern Analysis and Machine
  Intelligence}, 37\penalty0 (2):\penalty0 243--255, 2014.

\bibitem[D'Angelo et~al.(2023)D'Angelo, Canale, Yu, and
  Guindani]{d2023bayesian}
L.~D'Angelo, A.~Canale, Z.~Yu, and M.~Guindani.
\newblock Bayesian nonparametric analysis for the detection of spikes in noisy
  calcium imaging data.
\newblock \emph{Biometrics}, 79\penalty0 (2):\penalty0 1370--1382, 2023.

\bibitem[Davis et~al.(2008)Davis, Meyer, Rudd, Librant, Epping, Sheffield, and
  Wassink]{davis2008pax6}
L.~K. Davis, K.~Meyer, D.~Rudd, A.~Librant, E.~Epping, V.~Sheffield, and
  T.~Wassink.
\newblock {Pax6 3' deletion results in aniridia, autism and mental
  retardation}.
\newblock \emph{Human Genetics}, 123:\penalty0 371--378, 2008.

\bibitem[De~Blasi et~al.(2013)De~Blasi, Favaro, Lijoi, Mena, Pr{\"u}nster, and
  Ruggiero]{de2013gibbs}
P.~De~Blasi, S.~Favaro, A.~Lijoi, R.~H. Mena, I.~Pr{\"u}nster, and M.~Ruggiero.
\newblock {Are Gibbs-type priors the most natural generalization of the
  Dirichlet process?}
\newblock \emph{IEEE Transactions on Pattern Analysis and Machine
  Intelligence}, 37\penalty0 (2):\penalty0 212--229, 2013.

\bibitem[De~Iorio et~al.(2004)De~Iorio, M{\"u}ller, Rosner, and
  MacEachern]{de2004anova}
M.~De~Iorio, P.~M{\"u}ller, G.~L. Rosner, and S.~N. MacEachern.
\newblock An {ANOVA} model for dependent random measures.
\newblock \emph{Journal of the American Statistical Association}, 99\penalty0
  (465):\penalty0 205--215, 2004.

\bibitem[De~Iorio et~al.(2009)De~Iorio, Johnson, M\"uller, and
  Rosner]{DeIorio09}
M.~De~Iorio, W.~Johnson, P.~M\"uller, and G.~Rosner.
\newblock {Bayesian nonparametric non-proportional hazards survival modelling}.
\newblock \emph{Biometrics}, 65:\penalty0 762--771, 2009.

\bibitem[Denti et~al.(2023)Denti, Camerlenghi, Guindani, and
  Mira]{denti2023common}
F.~Denti, F.~Camerlenghi, M.~Guindani, and A.~Mira.
\newblock {A common atoms model for the Bayesian nonparametric analysis of
  nested data}.
\newblock \emph{Journal of the American Statistical Association}, 118\penalty0
  (541):\penalty0 405--416, 2023.

\bibitem[Diana et~al.(2020)Diana, Matechou, Griffin, and
  Johnston]{diana2020hierarchical}
A.~Diana, E.~Matechou, J.~Griffin, and A.~Johnston.
\newblock {A hierarchical dependent Dirichlet process prior for modelling bird
  migration patterns in the UK}.
\newblock \emph{The Annals of Applied Statistics}, 14\penalty0 (1):\penalty0
  473--493, 2020.

\bibitem[Dunson and Park(2008)]{Dun1}
D.~Dunson and J.~Park.
\newblock Kernel stick-breaking processes.
\newblock \emph{Biometrika}, 95:\penalty0 307--323, 2008.

\bibitem[Eling et~al.(2018)Eling, Richard, Richardson, Marioni, and
  Vallejos]{eling2018correcting}
N.~Eling, A.~C. Richard, S.~Richardson, J.~C. Marioni, and C.~A. Vallejos.
\newblock Correcting the mean-variance dependency for differential variability
  testing using single-cell {RNA} sequencing data.
\newblock \emph{Cell Systems}, 7\penalty0 (3):\penalty0 284--294, 2018.

\bibitem[Estivill-Torrus et~al.(2002)Estivill-Torrus, Pearson, van Heyningen,
  Price, and Rashbass]{estivill2002pax6}
G.~Estivill-Torrus, H.~Pearson, V.~van Heyningen, D.~J. Price, and P.~Rashbass.
\newblock {Pax6 is required to regulate the cell cycle and the rate of
  progression from symmetrical to asymmetrical division in mammalian cortical
  progenitors}.
\newblock \emph{Development}, 129\penalty0 (2):\penalty0 455--466, 01 2002.

\bibitem[Ferguson(1973)]{Ferguson1973bayesian}
T.~S. Ferguson.
\newblock A {B}ayesian analysis of some nonparametric problems.
\newblock \emph{The Annals of Statistics}, pages 209--230, 1973.

\bibitem[Fong et~al.(2019)Fong, Lyddon, and Holmes]{fong2019scalable}
E.~Fong, S.~Lyddon, and C.~Holmes.
\newblock {Scalable nonparametric sampling from multimodal posteriors with the
  posterior bootstrap}.
\newblock In \emph{Proceedings of the 36th International Conference on Machine
  Learning}, volume~97, pages 1952--1962. PMLR, 2019.

\bibitem[Foti and Williamson(2012)]{foti2012slice}
N.~Foti and S.~Williamson.
\newblock {Slice sampling normalized kernel-weighted completely random measure
  mixture models}.
\newblock In \emph{Advances in Neural Information Processing Systems},
  volume~25. Curran Associates, Inc., 2012.

\bibitem[Fraley and Raftery(2007)]{fraley2007bayesian}
C.~Fraley and A.~E. Raftery.
\newblock Bayesian regularization for normal mixture estimation and model-based
  clustering.
\newblock \emph{Journal of Classification}, 24\penalty0 (2):\penalty0 155--181,
  2007.

\bibitem[Fritsch and Ickstadt(2009)]{fritsch2009improved}
A.~Fritsch and K.~Ickstadt.
\newblock {Improved criteria for clustering based on the posterior similarity
  matrix}.
\newblock \emph{Bayesian Analysis}, 4\penalty0 (2):\penalty0 367 -- 391, 2009.
\newblock \doi{10.1214/09-BA414}.
\newblock URL \url{https://doi.org/10.1214/09-BA414}.

\bibitem[Fr{\"u}hwirth-Schnatter and Pyne(2010)]{fruhwirth2010bayesian}
S.~Fr{\"u}hwirth-Schnatter and S.~Pyne.
\newblock Bayesian inference for finite mixtures of univariate and multivariate
  skew-normal and skew-t distributions.
\newblock \emph{Biostatistics}, 11\penalty0 (2):\penalty0 317--336, 2010.

\bibitem[Fruhwirth-Schnatter et~al.(2019)Fruhwirth-Schnatter, Celeux, and
  Robert]{fruhwirth2019handbook}
S.~Fruhwirth-Schnatter, G.~Celeux, and C.~P. Robert.
\newblock \emph{Handbook of Mixture Analysis}.
\newblock CRC Press, 2019.

\bibitem[Gelman et~al.(1996)Gelman, Meng, and Stern]{Gelman1996posterior}
A.~Gelman, X.-L. Meng, and H.~Stern.
\newblock Posterior predictive assessment of model fitness via realized
  discrepancies.
\newblock \emph{Statistica Sinica}, pages 733--760, 1996.

\bibitem[Ghosal and van~der Vaart(2001)]{GV01}
S.~Ghosal and A.~van~der Vaart.
\newblock {Entropies and rates of convergence for maximum likelihood and Bayes
  estimation for mixtures of normal densities}.
\newblock \emph{The Annals of Statistics}, 29:\penalty0 1233--1263, 2001.

\bibitem[Ghosal and van~der Vaart(2017)]{Ghosal2017}
S.~Ghosal and A.~van~der Vaart.
\newblock \emph{Fundamentals of Nonparametric {B}ayesian Inference}, volume~44.
\newblock Cambridge University Press, 2017.

\bibitem[Ghosal et~al.(1999)Ghosal, Ghosh, and Ramamoorthi]{GGR99}
S.~Ghosal, J.~Ghosh, and R.~Ramamoorthi.
\newblock {Posterior consistency of Dirichlet mixtures in density estimation}.
\newblock \emph{The Annals of Statistics}, 27:\penalty0 143--158, 1999.

\bibitem[Griffin and Stephens(2013)]{JimEGriffin2013AiMc}
J.~E. Griffin and D.~A. Stephens.
\newblock {Advances in Markov chain Monte Carlo}.
\newblock In \emph{Bayesian Theory and Applications}. Oxford University Press,
  Oxford, 2013.

\bibitem[Götz et~al.(1998)Götz, Stoykova, and Gruss]{GOTZ19981031}
M.~Götz, A.~Stoykova, and P.~Gruss.
\newblock {Pax6 controls radial glia differentiation in the cerebral cortex}.
\newblock \emph{Neuron}, 21\penalty0 (5):\penalty0 1031--1044, 1998.
\newblock ISSN 0896-6273.

\bibitem[Hjort(1990)]{hjort1990nonparametric}
N.~L. Hjort.
\newblock {Nonparametric Bayes estimators based on beta processes in models for
  life history data}.
\newblock \emph{The Annals of Statistics}, pages 1259--1294, 1990.

\bibitem[Hoffman(2023)]{seurat}
P.~Hoffman.
\newblock Seurat - guided clustering tutorial, 2023.
\newblock URL \url{https://satijalab.org/seurat/articles/pbmc3k_tutorial.html}.

\bibitem[Ishwaran and James(2001)]{ishwaran2001gibbs}
H.~Ishwaran and L.~F. James.
\newblock {Gibbs sampling methods for stick-breaking priors}.
\newblock \emph{Journal of the American Statistical Association}, 96\penalty0
  (453):\penalty0 161--173, 2001.

\bibitem[Ishwaran and Zarepour(2002)]{ishwaran2002exact}
H.~Ishwaran and M.~Zarepour.
\newblock Exact and approximate sum representations for the {D}irichlet
  process.
\newblock \emph{Canadian Journal of Statistics}, 30\penalty0 (2):\penalty0
  269--283, 2002.

\bibitem[Ji and Ji(2016)]{ji2016tscan}
Z.~Ji and H.~Ji.
\newblock {TSCAN: Pseudo-time reconstruction and evaluation in single-cell
  RNA-seq analysis}.
\newblock \emph{Nucleic Acids Research}, 44\penalty0 (13):\penalty0 e117--e117,
  2016.

\bibitem[Jordan et~al.(1992)Jordan, Hanson, Zaletayev, Hodgson, Prosser,
  Seawright, Hastie, and van Heyningen]{jordan1992human}
T.~Jordan, I.~Hanson, D.~Zaletayev, S.~Hodgson, J.~Prosser, A.~Seawright,
  N.~Hastie, and V.~van Heyningen.
\newblock {The human PAX6 gene is mutated in two patients with aniridia}.
\newblock \emph{Nature Genetics}, 1\penalty0 (5):\penalty0 328--332, 1992.

\bibitem[Karlis and Xekalaki(2005)]{karlis2005mixed}
D.~Karlis and E.~Xekalaki.
\newblock {Mixed Poisson distributions}.
\newblock \emph{International Statistical Review}, 73:\penalty0 35--58, 2005.

\bibitem[Kikkawa et~al.(2019)Kikkawa, Casingal, Chun, Shinohara, Hiraoka, and
  Osumi]{kikkawa2019role}
T.~Kikkawa, C.~R. Casingal, S.~H. Chun, H.~Shinohara, K.~Hiraoka, and N.~Osumi.
\newblock {The role of Pax6 in brain development and its impact on pathogenesis
  of autism spectrum disorder}.
\newblock \emph{Brain Research}, 1705:\penalty0 95--103, 2019.

\bibitem[Kim and Oh(2014)]{kim2014hierarchical}
D.~Kim and A.~Oh.
\newblock {Hierarchical Dirichlet scaling process}.
\newblock In \emph{Proceedings of the 31st International Conference on Machine
  Learning}, pages 973--981. PMLR, 2014.

\bibitem[Kingman(1975)]{kingman1975random}
J.~F. Kingman.
\newblock Random discrete distributions.
\newblock \emph{Journal of the Royal Statistical Society: Series B
  (Methodological)}, 37\penalty0 (1):\penalty0 1--15, 1975.

\bibitem[Kiselev et~al.(2017)Kiselev, Kirschner, Schaub, Andrews, Yiu, Chandra,
  Natarajan, Reik, Barahona, Green, et~al.]{kiselev2017sc3}
V.~Y. Kiselev, K.~Kirschner, M.~T. Schaub, T.~Andrews, A.~Yiu, T.~Chandra,
  K.~N. Natarajan, W.~Reik, M.~Barahona, A.~R. Green, et~al.
\newblock {SC3: consensus clustering of single-cell RNA-seq data}.
\newblock \emph{Nature Methods}, 14\penalty0 (5):\penalty0 483--486, 2017.

\bibitem[Krnjaji{\'c} et~al.(2008)Krnjaji{\'c}, Kottas, and
  Draper]{krnjajic2008parametric}
M.~Krnjaji{\'c}, A.~Kottas, and D.~Draper.
\newblock {Parametric and nonparametric Bayesian model specification: A case
  study involving models for count data}.
\newblock \emph{Computational Statistics \& Data Analysis}, 52\penalty0
  (4):\penalty0 2110--2128, 2008.

\bibitem[{La Manno} et~al.(2018)]{La_Manno2018}
G.~{La Manno} et~al.
\newblock {RNA} velocity of single cells.
\newblock \emph{Nature}, 560:\penalty0 494--498, 2018.

\bibitem[L{\"a}hnemann et~al.(2020)L{\"a}hnemann, K{\"o}ster, Szczurek,
  McCarthy, Hicks, Robinson, Vallejos, Campbell, Beerenwinkel, Mahfouz,
  et~al.]{lahnemann2020eleven}
D.~L{\"a}hnemann, J.~K{\"o}ster, E.~Szczurek, D.~J. McCarthy, S.~C. Hicks,
  M.~D. Robinson, C.~A. Vallejos, K.~R. Campbell, N.~Beerenwinkel, A.~Mahfouz,
  et~al.
\newblock {Eleven grand challenges in single-cell data science}.
\newblock \emph{Genome Biology}, 21:\penalty0 1--35, 2020.

\bibitem[Lee and McLachlan(2014)]{lee2014finite}
S.~Lee and G.~J. McLachlan.
\newblock Finite mixtures of multivariate skew t-distributions: some recent and
  new results.
\newblock \emph{Statistics and Computing}, 24\penalty0 (2):\penalty0 181--202,
  2014.

\bibitem[Lewin et~al.(2007)Lewin, Bochkina, and Richardson]{Lewin2007fully}
A.~Lewin, N.~Bochkina, and S.~Richardson.
\newblock {Fully Bayesian mixture model for differential gene expression:
  simulations and model checks}.
\newblock \emph{Statistical Applications in Genetics and Molecular Biology},
  6\penalty0 (1), 2007.

\bibitem[Lijoi and Pr\"{u}nster(2011)]{LP10}
A.~Lijoi and I.~Pr\"{u}nster.
\newblock {Models beyond the Dirichlet process}.
\newblock In \emph{Bayesian Nonparametrics}, pages 80--136, Cambridge, UK,
  2011. Cambridge University Press.

\bibitem[Lijoi et~al.(2022)Lijoi, Prünster, and Rebaudo]{Lijoi2022}
A.~Lijoi, I.~Prünster, and G.~Rebaudo.
\newblock {Flexible clustering via hidden hierarchical Dirichlet priors}.
\newblock \emph{Scandinavian Journal of Statistics}, feb 2022.
\newblock \doi{10.1111/sjos.12578}.
\newblock URL \url{https://doi.org/10.1111%2Fsjos.12578}.

\bibitem[Lin et~al.(2017)Lin, Troup, and Ho]{lin2017cidr}
P.~Lin, M.~Troup, and J.~W. Ho.
\newblock {CIDR: Ultrafast and accurate clustering through imputation for
  single-cell RNA-seq data}.
\newblock \emph{Genome Biology}, 18:\penalty0 1--11, 2017.

\bibitem[Liu et~al.(2024)Liu, Wade, and Bochkina]{Liu2024}
J.~Liu, S.~Wade, and N.~Bochkina.
\newblock {Shared Differential Clustering across Single-cell RNA Sequencing
  Datasets with the Hierarchical Dirichlet Process}.
\newblock \emph{Econometrics and Statistics}, 2024.
\newblock ISSN 2452-3062.

\bibitem[Liverani et~al.(2015)Liverani, Hastie, Azizi, Papathomas, and
  Richardson]{liverani2015premium}
S.~Liverani, D.~I. Hastie, L.~Azizi, M.~Papathomas, and S.~Richardson.
\newblock {PReMiuM: An R package for profile regression mixture models using
  Dirichlet processes}.
\newblock \emph{Journal of Statistical Software}, 64:\penalty0 1--30, 2015.

\bibitem[Lo(1984)]{lo1984class}
A.~Y. Lo.
\newblock On a class of {B}ayesian nonparametric estimates: I. {D}ensity
  estimates.
\newblock \emph{The Annals of Statistics}, pages 351--357, 1984.

\bibitem[MacEachern(1999)]{MacEachern1999DDP}
S.~N. MacEachern.
\newblock Dependent nonparametric processes.
\newblock In \emph{ASA Proceedings of the Section on Bayesian Statistical
  Science}, pages 50--55, 1999.

\bibitem[Manuel et~al.(2022)Manuel, Tan, Kozic, Molinek, Marcos, Razak,
  Dobolyi, Dobie, Henderson, Henderson, Chan, Daw, Mason, and
  Price]{tan2022pax6}
M.~Manuel, K.~B. Tan, Z.~Kozic, M.~Molinek, T.~S. Marcos, M.~F.~A. Razak,
  D.~Dobolyi, R.~Dobie, B.~E.~P. Henderson, N.~C. Henderson, W.~K. Chan, M.~I.
  Daw, J.~O. Mason, and D.~J. Price.
\newblock {Pax6 limits the competence of developing cerebral cortical cells to
  respond to inductive intercellular signals}.
\newblock \emph{PLoS Biology}, 20\penalty0 (9):\penalty0 1--54, 09 2022.

\bibitem[Medvedovic et~al.(2004)Medvedovic, Yeung, and
  Bumgarner]{medvedovic2004bayesian}
M.~Medvedovic, K.~Y. Yeung, and R.~E. Bumgarner.
\newblock {Bayesian mixture model based clustering of replicated microarray
  data}.
\newblock \emph{Bioinformatics}, 20\penalty0 (8):\penalty0 1222--1232, 2004.

\bibitem[Mukamel et~al.(2009)Mukamel, Nimmerjahn, and
  Schnitzer]{mukamel2009automated}
E.~A. Mukamel, A.~Nimmerjahn, and M.~J. Schnitzer.
\newblock Automated analysis of cellular signals from large-scale calcium
  imaging data.
\newblock \emph{Neuron}, 63\penalty0 (6):\penalty0 747--760, 2009.

\bibitem[O'Keefe and Dostrovsky(1971)]{o1971hippocampus}
J.~O'Keefe and J.~Dostrovsky.
\newblock The hippocampus as a spatial map: preliminary evidence from unit
  activity in the freely-moving rat.
\newblock \emph{Brain Research}, 1971.

\bibitem[Pan et~al.(2024)Pan, Shen, Davis-Stober, and Hu]{pan24}
T.~Pan, W.~Shen, C.~P. Davis-Stober, and G.~Hu.
\newblock A bayesian nonparametric approach for handling item and examinee
  heterogeneity in assessment data.
\newblock \emph{British Journal of Mathematical and Statistical Psychology},
  77\penalty0 (1):\penalty0 196--211, 2024.

\bibitem[Quintana et~al.(2022)Quintana, M{\"u}ller, Jara, and
  MacEachern]{quintana2022dependent}
F.~A. Quintana, P.~M{\"u}ller, A.~Jara, and S.~N. MacEachern.
\newblock The dependent {D}irichlet process and related models.
\newblock \emph{Statistical Science}, 37\penalty0 (1):\penalty0 24--41, 2022.

\bibitem[Rao and Teh(2009)]{rao2009spatial}
V.~Rao and Y.~Teh.
\newblock {Spatial normalized gamma processes}.
\newblock In \emph{Advances in Neural Information Processing Systems},
  volume~22. Curran Associates, Inc., 2009.

\bibitem[Ren et~al.(2008)Ren, Dunson, and Carin]{ren2008dynamic}
L.~Ren, D.~B. Dunson, and L.~Carin.
\newblock {The dynamic hierarchical Dirichlet process}.
\newblock In \emph{Proceedings of the 25th International Conference on Machine
  Learning}, pages 824--831. Association for Computing Machinery, 2008.

\bibitem[Ren et~al.(2011)Ren, Du, Carin, and Dunson]{ren2011logistic}
L.~Ren, L.~Du, L.~Carin, and D.~B. Dunson.
\newblock {Logistic stick-breaking process}.
\newblock \emph{Journal of Machine Learning Research}, 12\penalty0 (1), 2011.

\bibitem[Roberts et~al.(1997)Roberts, Gelman, and Gilks]{roberts1997weak}
G.~O. Roberts, A.~Gelman, and W.~R. Gilks.
\newblock {Weak convergence and optimal ccaling of random walk Metropolis
  algorithms}.
\newblock \emph{The Annals of Applied Probability}, 7\penalty0 (1):\penalty0
  110--120, 1997.
\newblock ISSN 10505164.
\newblock URL \url{http://www.jstor.org/stable/2245134}.

\bibitem[Rodriguez and Dunson(2011)]{Rod}
A.~Rodriguez and D.~Dunson.
\newblock {Nonparametric Bayesian models through probit stick-breaking
  processes}.
\newblock \emph{Bayesian Analysis}, 6:\penalty0 145--178, 2011.

\bibitem[Rodriguez et~al.(2008)Rodriguez, Dunson, and
  Gelfand]{rodriguez2008nested}
A.~Rodriguez, D.~B. Dunson, and A.~E. Gelfand.
\newblock The nested {D}irichlet process.
\newblock \emph{Journal of the American statistical Association}, 103\penalty0
  (483):\penalty0 1131--1154, 2008.

\bibitem[Rubin et~al.(2015)Rubin, Geva, Sheintuch, and Ziv]{rubin2015}
A.~Rubin, N.~Geva, L.~Sheintuch, and Y.~Ziv.
\newblock Hippocampal ensemble dynamics timestamp events in long-term memory.
\newblock \emph{eLife}, 4:\penalty0 e12247, 2015.
\newblock ISSN 2050-084X.

\bibitem[Rubin et~al.(2019)Rubin, Sheintuch, Brande-Eilat, Pinchasof, Rechavi,
  Geva, and Ziv]{rubin2019revealing}
A.~Rubin, L.~Sheintuch, N.~Brande-Eilat, O.~Pinchasof, Y.~Rechavi, N.~Geva, and
  Y.~Ziv.
\newblock Revealing neural correlates of behavior without behavioral
  measurements.
\newblock \emph{Nature Communications}, 10\penalty0 (1):\penalty0 4745, 2019.

\bibitem[Satija et~al.(2015)Satija, Farrell, Gennert, Schier, and
  Regev]{satija2015spatial}
R.~Satija, J.~A. Farrell, D.~Gennert, A.~F. Schier, and A.~Regev.
\newblock Spatial reconstruction of single-cell gene expression data.
\newblock \emph{Nature Biotechnology}, 33\penalty0 (5):\penalty0 495--502,
  2015.

\bibitem[Scrucca et~al.(2016)Scrucca, Fop, Murphy, and
  Raftery]{scrucca2016mclust}
L.~Scrucca, M.~Fop, T.~B. Murphy, and A.~E. Raftery.
\newblock mclust 5: clustering, classification and density estimation using
  gaussian finite mixture models.
\newblock \emph{The R Journal}, 8\penalty0 (1):\penalty0 289, 2016.

\bibitem[Sethuraman(1994)]{sethuraman1994constructive}
J.~Sethuraman.
\newblock A constructive definition of {D}irichlet priors.
\newblock \emph{Statistica Sinica}, pages 639--650, 1994.

\bibitem[Shahbaba and Neal(2009)]{shahbaba2009nonlinear}
B.~Shahbaba and R.~Neal.
\newblock {Nonlinear models using Dirichlet process mixtures}.
\newblock \emph{Journal of Machine Learning Research}, 10\penalty0 (8), 2009.

\bibitem[Shen et~al.(2013)Shen, Tokdar, and Ghosal]{shen2013adaptive}
W.~Shen, S.~T. Tokdar, and S.~Ghosal.
\newblock {Adaptive Bayesian multivariate density estimation with Dirichlet
  mixtures}.
\newblock \emph{Biometrika}, 100\penalty0 (3):\penalty0 623--640, 2013.

\bibitem[Tang et~al.(2020)Tang, Bertaux, Thomas, Stefanelli, Saint, Marguerat,
  and Shahrezaei]{Tang2020}
W.~Tang, F.~Bertaux, P.~Thomas, C.~Stefanelli, M.~Saint, S.~Marguerat, and
  V.~Shahrezaei.
\newblock {bayNorm: Bayesian gene expression recovery, imputation and
  normalization for single-cell RNA-sequencing data}.
\newblock \emph{Bioinformatics}, 36\penalty0 (4):\penalty0 1174--1181, 2020.

\bibitem[Teh et~al.(2006)Teh, Jordan, Beal, and Blei]{Teh2006}
Y.~W. Teh, M.~I. Jordan, M.~J. Beal, and D.~M. Blei.
\newblock {Hierarchical Dirichlet processes}.
\newblock \emph{Journal of the American Statistical Association}, 101\penalty0
  (476):\penalty0 1566--1581, 2006.

\bibitem[Vallejos et~al.(2015)Vallejos, Marioni, and
  Richardson]{vallejos2015basics}
C.~A. Vallejos, J.~C. Marioni, and S.~Richardson.
\newblock {BASiCS}: {B}ayesian analysis of single-cell sequencing data.
\newblock \emph{PLoS Computational Biology}, 11\penalty0 (6):\penalty0
  e1004333, 2015.

\bibitem[Vallejos et~al.(2016)Vallejos, Richardson, and Marioni]{vallejos2016}
C.~A. Vallejos, S.~Richardson, and J.~C. Marioni.
\newblock Beyond comparisons of means: understanding changes in gene expression
  at the single-cell level.
\newblock \emph{Genome Biology}, 17\penalty0 (1):\penalty0 70, 2016.

\bibitem[Van~der Maaten and Hinton(2008)]{Van2008visualizing}
L.~Van~der Maaten and G.~Hinton.
\newblock {Visualizing data using t-SNE}.
\newblock \emph{Journal of Machine Learning Research}, 9\penalty0 (11), 2008.

\bibitem[Wade and Ghahramani(2018)]{wade2018bayesian}
S.~Wade and Z.~Ghahramani.
\newblock Bayesian cluster analysis: {P}oint estimation and credible balls
  (with discussion).
\newblock \emph{Bayesian Analysis}, 13\penalty0 (2):\penalty0 559--626, 2018.

\bibitem[Wade and In{\'a}cio(2025)]{wade2023bayesian}
S.~Wade and V.~In{\'a}cio.
\newblock {Bayesian dependent mixture models: A predictive comparison and
  survey}.
\newblock \emph{Statistical Science}, 40\penalty0 (1):\penalty0 81 -- 108,
  2025.

\bibitem[Wu and Luo(2022)]{wu2022nonparametric}
Q.~Wu and X.~Luo.
\newblock {Nonparametric Bayesian two-level clustering for subject-level
  single-cell expression data}.
\newblock \emph{Statistica Sinica}, 32\penalty0 (4):\penalty0 1835--1856, 2022.

\bibitem[Wu and Ghosal(2010)]{WG10}
Y.~Wu and S.~Ghosal.
\newblock {The $L_1$-consistency of Dirichlet mixtures in multivariate density
  estimation}.
\newblock \emph{Journal of Multivariate Analysis}, 101:\penalty0 2411--2419,
  2010.

\end{thebibliography}
\bibliographystyle{apalike}

\clearpage
\appendix

\section{Posterior Inference for Pax6 Data}\label{sec:full-conditionals-pax6}

We assume the following likelihood for the mRNA count $y_{c,g,d}$ for gene $g \ (g=1,\ldots,G)$ in cell $c \ (c=1,\ldots,C_d)$ from group $d \ (d=1,\ldots, D, D=2)$:
\begin{equation*}
	y_{c,g,d} | \mu_{c,g,d}, \phi_{c,g,d}, \beta_{c,d} \sim \NB( \mu_{c,g,d} \beta_{c,d}, \phi_{c,g,d} ).
\end{equation*}
Under the C-HDP mixture model, the cell-specific mean expression and dispersion parameters $\mu_{c,g,d}$ and $\phi_{c,g,d}$ are assumed
\begin{equation*} 
	\begin{split}
		(\bm{\mu}_{c,d}, \bm{\phi}_{c,d}) |P_{t_{c,d},d} &\sim P_{t_{c,d},d}, \quad 
		P_{t_{c,d},d} \sim \text{C-HDP}(\alpha_0, \alpha, P_0, \bm{\Psi}^*),
	\end{split}
\end{equation*}
where $\bm{\mu}_{c,d}= (\mu_{c,1,d},\ldots, \mu_{c,G,d})$ and $\bm{\phi}_{c,d}= (\phi_{c,1,d},\ldots, \phi_{c,G,d})$ denote the collection of parameters across all genes. 

\subsection{Prior Specification} \label{appendix:pax6-prior}
\paragraph{Base Measure} For the base measure $P_0$, we follow \cite{Liu2024} to model the relationship between $\mu_{j,g}^*$ and $\phi_{j,g}^*$ (component-specific parameters for gene $g$) as follows:
\begin{equation*}\label{eq:mu-phi-prior}
	\mu^*_{j,g} \overset{i.i.d}{\sim} \logN(0, \alpha_\mu^2),\quad \phi^*_{j,g}| \mu^*_{j,g} \indsim  \logN ( b_0 +b_1 \log(\mu_{j,g}^*), \alpha_\phi^2),
\end{equation*}
where $\logN$ denotes the log-normal distribution. The linear relationship between the logarithmic mean expression and dispersion has been observed in \cite{brennecke2013accounting}, \cite{vallejos2015basics} and \cite{eling2018correcting}. The value of $\alpha_{\mu}^2$ is set using the empirical estimates for the mean parameters from \textit{bayNorm} \citep{Tang2020} (see Section \ref{sec:baynorm_beta}). The mean-dispersion parameters $\bmb=(b_0,b_1)^T$ and $\alpha_\phi^2$ have hyper-priors as follows
\begin{equation*}
	\bmb|\alpha_\phi^2 \sim \Norm(\bmm_b, \alpha_\phi^2 \bV_b), \quad \alpha_\phi^2 \sim \IG(\nu_1, \nu_2),
\end{equation*}
where $\IG$ is the inverse-gamma distribution and by default $\bV_b=\bI$, and we use the estimated mean and dispersion parameters from bayNorm to determine $\bmm_b, \nu_1$ and $\nu_2$. 

\paragraph{Capture Efficiencies $\beta_{c,d}$}
The prior for capture efficiencies $\beta_{c,d}$ is 
\begin{equation*}
	\beta_{c,d} \indsim \Beta(a^{\beta}_{d}, b^{\beta}_{d}),
\end{equation*}
where the values of $a^{\beta}_{d}, b^{\beta}_{d}$ are based on the empirical estimates from bayNorm (see Section \ref{sec:baynorm_beta}). To avoid bimodal and exponentially decaying (increasing) shape of the Beta prior, we set $a^{\beta}_{d}, b^{\beta}_{d}>1$. For identifiability of $\beta_{c,d}$, an informative prior is used, where the mean is specified to be an estimate of global mean capture efficiency across cells 0.06 \citep{Tang2020}.

\paragraph{Kernel Parameters} \label{appendix:gaussian-kernel-prior}
For efficient MCMC sampling, a hierarchical prior for kernel parameters $\bpsi_{j,d}^*=(\tcenter, \sdcenter)$ in the Gaussian kernel is used: 
\begin{equation*}
	\begin{split}
		\tcenter &\indsim \Norm(r_j, s^2),\quad r_j  \iidsim \Norm(\mu_r, \sigma_r^2),\quad s^2\sim \IG(\eta_1, \eta_2),\\
		\sdcenter & \indsim \logN(h_j, m^2),\quad h_j  \iidsim \Norm(\mu_h, \sigma_h^2),\quad m^2 \sim \IG(\kappa_1, \kappa_2).
	\end{split}
\end{equation*}
The prior means ($r_j, h_j$) for $\tcenter$ and $\sdcenter$ are component-specific, and ($r_j, h_j$) are given Gaussian hyper-priors with global means $(\mu_r,\mu_h)$ to allow for borrowing of information across groups, which is similar to the hierarchical prior for $q_{j,d}^J$. The values of hyperparameters are set as $\mu_r=0.5, \sigma_r=0.5, \eta_1=5, \eta_2=1, \mu_h=-5, \sigma_h=0.5, \kappa_1=5, \kappa_2=1$.

\paragraph{Concentration Parameters $\alpha, \alpha_0$}
For concentration parameters, weakly informative priors are used
\begin{equation*}
	\alpha \sim \Gam(1,1), \quad \alpha_0 \sim \Gam(1,1).
\end{equation*}
If prior information on the number of clusters is available, we can use this information to set the hyperparameters. 

Using a finite-dimensional truncation at $J$, the complete model is as follows:
\begin{align*}
	y_{c,g,d} | z_{c,d}=j, \mu_{j,g}^*, \phi_{j,g}^*, \beta_{c,d} &\indsim \NB( \mu_{j,g}^* \beta_{c,d}, \phi_{j,g}^* ),\\
	z_{c,d} | p_{1,d}^J(t_{c,d}), \ldots, p_{J,d}^J(t_{c,d})&\indsim \Cat(p_{1,d}^J(t_{c,d}),\ldots, p_{J,d}^J(t_{c,d})),\\
	p_{j,d}^J(t_{c,d}) & =\frac{q_{j,d}^JK(t_{c,d}|\bpsi_{j,d}^*)}{\sum_{k=1}^{J} q_{k,d}K(t_{c,d}|\bpsi_{k,d}^*)},  \\
	q_{j,d}^J &\indsim \Gam(\alpha p_j^J, 1), \\
	p^J_{1},\ldots, p^J_{J} &\sim \Dir\left( \frac{\alpha_0}{J},\ldots, \frac{\alpha_0}{J}\right), \\
	\tcenter &\indsim \Norm(r_j, s^2), \\
	r_j & \iidsim \Norm(\mu_r, \sigma_r^2), \\
	s^2&\sim \IG(\eta_1, \eta_2), \\
	\sdcenter & \indsim \logN(h_j, m^2), \\
	h_j & \iidsim \Norm(\mu_h, \sigma_h^2), \\
	m^2 & \sim \IG(\kappa_1, \kappa_2), \\
	\mu^*_{j,g} &\iidsim \logN(0, \alpha_{\mu}^2), \\
	\phi^*_{j,g}| \mu^*_{j,g} &\indsim  \logN( b_0 +b_1 \log(\mu_{j,g}^*), \alpha_\phi^2), \\
	\beta_{c,d} &\indsim \Beta(a^{\beta}_{d}, b^{\beta}_{d}), \\
	\bmb|\alpha_\phi^2 &\sim \Norm(\bmm_b, \alpha_\phi^2 \bV_b), \\
	\alpha_\phi^2& \sim \IG(\nu_1, \nu_2), \\
	\alpha &\sim \Gam(1,1), \\
	\alpha_0 &\sim \Gam(1,1).
\end{align*}

\newpage
Define $\bZ=\left\lbrace z_{c,d}\right\rbrace_{c=1,d=1}^{C_d,D} , \bY=\left\lbrace y_{c,g,d}\right\rbrace _{c=1,g=1,d=1}^{C_d,G,D},\bmt=\left\lbrace t_{c,d}\right\rbrace_{c=1,d=1}^{C_d,D}, \bmq^J=\left\lbrace q_{j,d}^J\right\rbrace _{j=1,d=1}^{J,D}, \bmp^J=(p_1^J,\ldots,p_J^J), \bm{\mu}_j^*=(\mu_{j,1}^*,\ldots,\mu_{j,G}^*), \bm{\phi}_j^*=(\phi_{j,1}^*,\ldots,\phi_{j,G}^*), \bbeta=\left\lbrace \beta_{c,d}\right\rbrace_{c=1,d=1}^{C_d,D},\bxi=\left\lbrace \xi_{c,d}\right\rbrace_{c=1,d=1}^{C_d,D},\bmb=(b_0,b_1)^T, \bmt^*=\left\lbrace \tcenter\right\rbrace_{j=1,d=1}^{J,D},{\bsigma^*}^2=\left\lbrace\sdcenter\right\rbrace_{j=1,d=1}^{J,D} , \bmr=(r_1,\ldots,r_J), \bmh=(h_1,\ldots,h_J)$. The posterior distribution is     
\begin{equation}\label{eq:full_posterior}
	\begin{split}
		&\pi(\bZ,\bmq^J,\bmp^J, \bmu^*, \bphi^*, \bbeta, \bxi,\alpha, \alpha_0, \bmb, \alpha_\phi^2, \bmt^*,{\bsigma^*}^2, \bmr, s^2,\bmh, m^2|\bY,\bmt)\\
		\propto &\prod_{j=1}^J \prod_{(c,d): z_{c,d}=j} \prod_{g=1}^G \NB(y_{c,d,g} | \mu_{j,g}^* \beta_{c,d}, \phi_{j,g}^* )\\
		& \times \prod_{j=1}^J \prod_{d=1}^D \prod_{c: z_{c,d}=j}  K(t_{c,d}|\bpsi_{j,d}^*) \times \prod_{j=1}^J \prod_{d=1}^D \left( q_{j,d}^J\right) ^{N_{j,d}}\\
		& \times \prod_{j=1}^J \prod_{d=1}^D \prod_{c=1}^{C_d}  \exp\left( -\xi_{c,d}q_{j,d}^JK(t_{c,d}|\bpsi_{j,d}^*)\right) \\
		& \times \prod_{j=1}^J \prod_{d=1}^D \Gam(q_{j,d}^J | \alpha p_j^J,1) \times \Dir\left(\bmp^J | \frac{\alpha_0}{J},\ldots, \frac{\alpha_0}{J}\right)\\
		&\times \prod_{j=1}^J   \prod_{g=1}^G  \left[ \logN\left( \mu^*_{j,g}  | 0, \alpha_{\mu}^2\right)  \times  \logN \left(  \phi^*_{j,g} | b_0 +b_1 \log(\mu_{j,g}^*) , \alpha_\phi^2\right)\right] \\
		& \times  \prod_{d=1}^D \prod_{c=1}^{C_d} \Beta(\beta_{c,d} | a^{\beta}_{d}, b^{\beta}_{d}) \times  \Gam(\alpha | 1,1) \times \Gam(\alpha_0 | 1,1) \\
		& \times \Norm(\bmb | m_b, \alpha_\phi^2 \bV_b)\times  \IG(\alpha_\phi^2 | \nu_1, \nu_2)\\
		& \times \prod_{j=1}^J \prod_{d=1}^D \left[ \Norm(\tcenter | r_j, s^2)\times \logN(\sdcenter | h_j,m^2)\right] \times  \prod_{j=1}^J \left[ \Norm(r_j | \mu_r,\sigma_r^2)\times \Norm(h_j| \mu_h, \sigma_h^2)\right] \\
		& \times \IG(s^2 | \eta_1,\eta_2) \times \IG(m^2 | \kappa_1,\kappa_2),
	\end{split}
\end{equation}
where  
$N_{j,d}=\sum_{c=1}^{C_d} \indicator(z_{c,d}=j)$ is the number of cells in component $j$ in dataset $d$, $\indicator(\cdot)$ is the indicator function that takes the value 1 if the condition inside the bracket holds, and is 0 otherwise. Note that the first three lines in Eq. \eqref{eq:full_posterior} comes from the augmented data likelihood after introducing the latent variables $\xi_{c,d}$. The MCMC algorithm (Gibbs sampling) iteratively samples from the full conditional distributions of (blocked) parameters. For standard full conditional densities, we can draw samples directly, while adaptive Metropolis-Hastings (AMH) is used for non-standard forms. Define $\bm{C}=(C_1,\ldots,C_D)$. The time complexity for each parameter is provided below.
\begin{itemize}
	\item dataset-specific parameters for component likelihood $\bmq^J$:  $\bigO(\tsum(\bm{C})J)$.
	\item latent cell-specific parameters $\bxi$:  $\bigO(\tsum(\bm{C})J)$.
	\item latent parameters $\left\lbrace u_{c,j,d} \right\rbrace _{c=1,j=1,d=1}^{C_d,J,D}$ for efficient sampling of kernel parameters: $\bigO(\tsum(\bm{C})J)$.
	\item kernel parameters (centre) $\bmt^*$: $\bigO(\tsum(\bm{C})J)$.
	\item kernel parameters (bandwidth) ${\bsigma^*}^2$: $\bigO(\tsum(\bm{C})J)$.
	\item concentration parameter $\alpha$: $\bigO(JD)$.
	\item concentration parameter $\alpha_0$: $\bigO(J)$.
	\item allocation variables $\bZ$: $\bigO(\tsum(\bm{C})JG)$.
	\item component probabilities $\bmp^J$: $\bigO(JD)$.
	\item mean-dispersion parameters $\bmb, \alpha_\phi^2$: $\bigO(JG)$.
	\item component-specific parameters $\bmu_{1:J,1:G}^*, \bphi_{1:J,1:G}^*$: $\bigO(\tsum(\bm{C})JG)$.
	\item capture efficiencies $\bbeta$: $\bigO(\tsum(\bm{C})G)$.
	\item hyperparameters $\bmr, s^2, \bmh, m^2$ for the kernel parameters: $\bigO(JD)$.
\end{itemize}
Next we provide the details of sampling each parameter.

\subsection{Dataset-specific Parameters for Component Likelihood $q_{j,d}^J$}
For each $j$ and $d$, the full conditional distribution is
\begin{equation*}
	\begin{split}
		&\pi(q_{j,d}^J | \left\lbrace z_{c,d}\right\rbrace _{c=1}^{C_d} ,\alpha,p_j^J, \left\lbrace \xi_{c,d}\right\rbrace _{c=1}^{C_d} , \left\lbrace t_{c,d}\right\rbrace _{c=1}^{C_d} , \tcenter, \sdcenter)\\
		\propto & \left( q_{j,d}^J\right) ^{N_{j,d}}\times \exp\left( -q_{j,d}^J\sum_{c=1}^{C_d}\xi_{c,d}K(t_{c,d}|\bpsi_{j,d}^*)\right) \\
		&\times \left( q_{j,d}^J\right) ^{\alpha p_j^J -1} \exp(-q_{j,d}^J)\\
		\propto & \left( q_{j,d}^J\right) ^{N_{j,d}+\alpha p_j^J -1}\times \exp\left( -q_{j,d}^J\left[ 1+\sum_{c=1}^{C_d}\xi_{c,d}K(t_{c,d}|\bpsi_{j,d}^*)\right] \right), 
	\end{split}
\end{equation*}
i.e.
\begin{equation*}
	q_{j,d}^J | \ldots \sim \Gam\left( N_{j,d}+\alpha p_j^J ,1+\sum_{c=1}^{C_d}\xi_{c,d}K(t_{c,d}|\bpsi_{j,d}^*)\right).
\end{equation*}

\subsection{Latent Cell-specific Parameters $\xi_{c,d}$}
For each $c$ and $d$, the full conditional distribution is
\begin{equation*}
	\pi(\xi_{c,d} | \bmq^J_{1:J,d}, t_{c,d}, \bmt_{1:J,d}^*, {\bsigma_{1:J,d}^*}^2)
	\propto  \exp\left( -\xi_{c,d}\sum_{j=1}^{J} q_{j,d}^JK(t_{c,d}|\bpsi_{j,d}^*)\right),  
\end{equation*}
i.e. 
\begin{equation*}
	\xi_{c,d} | \ldots \sim \Gam\left( 1 ,\sum_{j=1}^{J} q_{j,d}^JK(t_{c,d}|\bpsi_{j,d}^*)\right).
\end{equation*}

\subsection{Kernel Parameters $\tcenter$ and $\sdcenter$} \label{sec:kernel_MCMC_pax6}
The joint full conditional distribution for $\bmt^*$ and ${\bsigma^*}^2$ is
\begin{equation*}
	\begin{split}
		&\pi(\bmt^*, {\bsigma^*}^2 |  \bZ, \bmt, \bxi, \bmq^J, \bmr, s^2, \bmh, m^2)\\
		\propto &\prod_{j=1}^J \prod_{d=1}^D \prod_{c: z_{c,d}=j}  K(t_{c,d}|\bpsi_{j,d}^*) \times \prod_{j=1}^J \prod_{d=1}^D \prod_{c=1}^{C_d}  \exp\left( -\xi_{c,d}q_{j,d}^JK(t_{c,d}|\bpsi_{j,d}^*)\right) \\
		& \times  \prod_{j=1}^J \prod_{d=1}^D \left[ \Norm(\tcenter | r_j, s^2)\times \logN(\sdcenter | h_j,m^2)\right].
	\end{split}
\end{equation*}
Due to the presence of the exponential term, it is impossible to obtain standard distributions. We introduce latent variables $\bmmu=\left\lbrace u_{c,j,d}\right\rbrace _{c=1,j=1,d=1}^{C_d,J,D} \in (0,1)$, and the joint full conditional distribution becomes 
\begin{equation} \label{eq:full_kernel}
	\begin{split}
		\pi(\bmt^*, {\bsigma^*}^2, \bmmu|\ldots)\propto &\prod_{j=1}^J \prod_{d=1}^D \prod_{c: z_{c,d}=j}  K(t_{c,d}|\bpsi_{j,d}^*) \times \prod_{j=1}^J \prod_{d=1}^D\prod_{c=1}^{C_d}  \mathbb{I}\left( u_{c,j,d}< M_{c,j,d}\right) \\
		& \times  \prod_{j=1}^J \prod_{d=1}^D \left[ \Norm(\tcenter|r_j, s^2)\times \logN(\sdcenter|h_j,m^2)\right],
	\end{split}
\end{equation}
where $M_{c,j,d}=\exp\left( -\xi_{c,d}q_{j,d}^JK(t_{c,d}|\bpsi_{j,d}^*)\right)$. Integrating out $u_{c,j,d}$ on $(0,1)$ restores the joint full conditional distribution for $\bmt^*$ and ${\bsigma^*}^2$.

\subsubsection{Latent Parameters $u_{c,j,d}$} 
For each $c,j$ and $d$, the full conditional distribution of the latent variable is
\begin{equation*}
	\pi(u_{c,j,d}|\xi_{c,d}, q_{j,d}^J, t_{c,d}, \tcenter, \sdcenter)\propto \mathbb{I}\left( u_{c,j,d}< M_{c,j,d}\right),
\end{equation*}
i.e. 
\begin{equation*}
	u_{c,j,d}|\ldots \sim \Unif\left(0,\exp\left( -\xi_{c,d}q_{j,d}^JK(t_{c,d}|\bpsi_{j,d}^*)\right) \right).
\end{equation*}

\subsubsection{Centre $\tcenter$}   \label{full_tcenter}
For each $j$ and $d$, the full conditional distribution is
\begin{equation*}
	\begin{split}
		&\pi(\tcenter|r_j, s^2, \left\lbrace z_{c,d}\right\rbrace _{c=1}^{C_d}, \left\lbrace \xi_{c,d}\right\rbrace _{c=1}^{C_d}, \left\lbrace u_{c,j,d}\right\rbrace _{c=1}^{C_d}, \left\lbrace t_{c,d}\right\rbrace _{c=1}^{C_d}, q_{j,d}^J, \sdcenter)\\
		\propto  &\prod_{c: z_{c,d}=j}  K(t_{c,d}|\bpsi_{j,d}^*) \times \Norm(\tcenter|r_j, s^2) \times \mathbb{I}(\tcenter\in A_{j,d}).
	\end{split}
\end{equation*}
Let $I_{j,d}=\left\lbrace c: z_{c,d}=j\right\rbrace $. The first two terms are proportional to 
\begin{equation*}
	\begin{split}
		& \exp\left[ -\frac{1}{2\sdcenter}\sum_{I_{j,d}}\left( \tcenter-t_{c,d}\right) ^2 \right] \times \exp\left[ -\frac{1}{2s^2}\left( \tcenter-r_j\right) ^2 \right]\\
		\propto & \exp\left[ -\frac{1}{2\sdcenter}\left(N_{j,d} {\tcenter}^2-2\tcenter \sum_{I_{j,d}}t_{c,d}\right) -\frac{1}{2s^2}\left( {\tcenter}^2-2\tcenter r_j\right) \right] \\
		\propto&  \exp\left( -\frac{1}{2\sdcenter s^2}\left[ \left( \sdcenter+N_{j,d}s^2\right) {\tcenter}^2-2\tcenter\left( r_j \sdcenter+s^2\sum_{I_{j,d}}t_{c,d}\right) \right] \right) \\
		\propto &\Norm(\tcenter|\hat{r}_{j,d},\hat{s}_{j,d}^2),
	\end{split}
\end{equation*}
where 
\begin{equation*}
	\hat{s}_{j,d}^2=\left( \frac{1}{s^2}+\frac{N_{j,d}}{\sdcenter}\right)^{-1}, \quad
	\hat{r}_{j,d}=\frac{r_j/s^2+\sum_{I_{j,d}}t_{c,d}/\sdcenter}{1/s^2+N_{j,d}/\sdcenter}.
\end{equation*}
The indicator function $\mathbb{I}(\tcenter\in A_{j,d})$ results from the one in Eq. \eqref{eq:full_kernel}, leading to a truncated normal distribution. The truncation region $A_{j,d}$ is 
\begin{equation*}
	\begin{split}
		A_{j,d}&=\bigcap_{c=1}^{C_d}A_{c,j,d}=\bigcap_{c=1}^{C_d}\left\lbrace \tcenter: u_{c,j,d} < \exp\left( -\xi_{c,d}q_{j,d}^JK(t_{c,d}|\bpsi_{j,d}^*)\right)\right\rbrace \\
		&=\bigcap_{c=1}^{C_d}\left\lbrace \tcenter: -\frac{\log{u_{c,j,d}}}{\xi_{c,d}q_{j,d}^J}> \exp\left( -\frac{1}{2 \sdcenter}\left( t_{c,d}-\tcenter \right) ^2 \right) \right\rbrace \\
		&=\bigcap_{c=1}^{C_d}\left\lbrace \tcenter: \log\left[ -\frac{\log{u_{c,j,d}}}{\xi_{c,d}q_{j,d}^J}\right] > -\frac{1}{2 \sdcenter}\left( t_{c,d}-\tcenter \right) ^2\right\rbrace. \\
	\end{split}
\end{equation*}
Since the right-hand side is always negative, if  $-\frac{\log{u_{c,j,d}}}{\xi_{c,d}q_{j,d}^J}\geq 1$, i.e. $-\log{u_{c,j,d}} \geq \xi_{c,d}q_{j,d}^J$, we will have $A_{c,j,d}=\mathbb{R}$ and hence there is no truncation. Otherwise,
\begin{equation*}
	A_{c,j,d}=\left(-\infty, t_{c,d}-\sqrt{-2\sdcenter\log\left[ -\frac{\log{u_{c,j,d}}}{\xi_{c,d}q_{j,d}^J}\right] } \right) \bigcup \left( t_{c,d}+\sqrt{-2\sdcenter\log\left[ -\frac{\log{u_{c,j,d}}}{\xi_{c,d}q_{j,d}^J}\right] } , +\infty \right).
\end{equation*}
Hence the region $A_{j,d}$ is given by 
\begin{equation*}
	A_{j,d}=\bigcap_{c: -\log{u_{c,j,d}} < \xi_{c,d}q_{j,d}^J}  A_{c,j,d}.
\end{equation*}
Note that if there is no cell in dataset $d$ that belongs to component $j$, we will sample $\tcenter$ from the prior truncated to $A_{j,d}$. Furthermore, if it satisfies that $\left\lbrace c: -\log{u_{c,j,d}} < \xi_{c,d}q_{j,d}^J\right\rbrace = \emptyset$, there is no truncation. Therefore, there are four possible cases, based on truncation or not and whether or not the component $j$ is empty in dataset $d$.

\subsubsection{Bandwidth $\sdcenter$} \label{sec:variance-kernel-pax6}
For each $j$ and $d$, the full conditional distribution is
\begin{equation*} 
	\begin{split}
		&\pi(\sdcenter|h_j, m^2,\left\lbrace z_{c,d}\right\rbrace _{c=1}^{C_d}, \left\lbrace \xi_{c,d}\right\rbrace _{c=1}^{C_d}, \left\lbrace u_{c,j,d}\right\rbrace _{c=1}^{C_d}, \left\lbrace t_{c,d}\right\rbrace _{c=1}^{C_d}, q_{j,d}^J,\tcenter)\\
		\propto  &\prod_{c: z_{c,d}=j}  K(t_{c,d}|\bpsi_{j,d}^*) \times \logN(\sdcenter|h_j,m^2) \times \mathbb{I}(\sdcenter\in B_{j,d}).
	\end{split}
\end{equation*}
The first two terms are proportional to 
\begin{equation} \label{eq:full_sd}
	\exp\left[ -\frac{1}{2\sdcenter}\sum_{I_{j,d}}\left( \tcenter-t_{c,d}\right) ^2 \right] \times \frac{1}{\sdcenter}\exp\left[ -\frac{1}{2m^2}\left( \log(\sdcenter)-h_j\right) ^2 \right],\\
\end{equation} 
which is not a standard form when component $j$ is occupied, and hence we will apply adaptive Metropolis-Hastings. The region $B_{j,d}$ is given by 
\begin{equation*}
	\begin{split}
		B_{j,d}&=\bigcap_{c=1}^{C_d}B_{c,j,d}=\bigcap_{c=1}^{C_d}\left\lbrace \sdcenter: u_{c,j,d} < \exp\left( -\xi_{c,d}q_{j,d}^JK(t_{c,d}|\bpsi_{j,d}^*)\right)\right\rbrace \\
		&=\bigcap_{c=1}^{C_d}\left\lbrace \sdcenter: -\frac{\log{u_{c,j,d}}}{\xi_{c,d}q_{j,d}^J}> \exp\left( -\frac{1}{2 \sdcenter}\left( t_{c,d}-\tcenter \right) ^2 \right) \right\rbrace \\
		&=\bigcap_{c=1}^{C_d}\left\lbrace \sdcenter: \log\left[ -\frac{\log{u_{c,j,d}}}{\xi_{c,d}q_{j,d}^J}\right] > -\frac{1}{2 \sdcenter}\left( t_{c,d}-\tcenter \right) ^2\right\rbrace. \\
	\end{split}
\end{equation*}
Similar to $\tcenter$, if  $-\frac{\log{u_{c,j,d}}}{\xi_{c,d}q_{j,d}^J}\geq 1$, there will be no truncation. Otherwise,
\begin{equation*}
	B_{c,j,d}=\left(0, -\frac{\left( t_{c,d}-\tcenter \right) ^2}{2\log\left[ -\frac{\log{u_{c,j,d}}}{\xi_{c,d}q_{j,d}^J}\right] }\right).
\end{equation*}
Hence the region $B_{j,d}$ is  
\begin{equation*}
	B_{j,d}=\bigcap_{c: -\log{u_{c,j,d}} < \xi_{c,d}q_{j,d}^J}  B_{c,j,d} = \left( 0, \sigma_{j,d}^+\right),
\end{equation*} 
where 
\begin{equation*}
	\sigma_{j,d}^+=\min_{c: -\log{u_{c,j,d}} < \xi_{c,d}q_{j,d}^J} -\frac{\left( t_{c,d}-\tcenter \right) ^2}{2\log\left[ -\frac{\log{u_{c,j,d}}}{\xi_{c,d}q_{j,d}^J}\right] }.
\end{equation*} 

\paragraph{Adaptive Metropolis-Hastings for $\sdcenter$} \label{AMH}
For notation simplicity, we will drop the subscript $j,d$ in this section.
\begin{enumerate}
	\item Apply the following transformation to $\ssdcenter$:
	\begin{equation*}
		x=g(\ssdcenter)=-\log\left( \frac{1}{\ssdcenter}-\frac{1}{\sigma^+}\right) \in \mathbb{R}.
	\end{equation*}
	The Jacobian term is 
	\begin{equation*}
		J_x=\frac{dx}{d\ssdcenter}=\frac{\sigma^+}{\ssdcenter(\sigma^+-\ssdcenter)}.
	\end{equation*}
	The inverse transformation is 
	\begin{equation*}
		\ssdcenter=\frac{1}{\exp(-x)+1/\sigma^+} \in \left( 0,\sigma^+\right).
	\end{equation*}

	\item Let $\vd$ denote the dimension of $x$ ($\vd=1$ for the case of $\sdcenter$). At the current iteration $n$, denote the sampled $\ssdcenter$ from iteration $n-1$ as ${\sigma_{old}^*}^2$. Conditional on $\sigma^+$ at the current iteration, define $x_{old}=g({\sigma_{old}^*}^2)$. 
	
	We use random walk to sample $x_{new}$. For $n \leq 100$, we draw 
	$$x_{new} \sim \Norm(x_{old}, 0.01 \times \bI_\vd).$$ 
	For $n> 100$, letting $s_\vd=2.4^2/\vd$, we propose 
	$$x_{new} \sim \Norm(x_{old}, s_\vd \times (\Sigma_{n-1}+\epsilon \bI_d)),$$
	where $\epsilon=0.01$, and $\Sigma_{n-1}$ is the sample covariance based on the past iterations, which needs to be updated at each iteration. We obtain ${\sigma_{new}^*}^2$ through the inverse transformation.
	
	\item Next, we compute the acceptance probability. Let $\pi({\sigma^*}^2)$ denote the posterior distribution, and $Q_n$ the proposal distribution at step $n$. The acceptance probability is
	\begin{equation*}
		\begin{split}
			\alpha({\sigma_{new}^*}^2,{\sigma_{old}^*}^2)&=\min \left( 1, \frac{\pi({\sigma_{new}^*}^2)Q_n({\sigma_{old}^*}^2|{\sigma_{new}^*}^2)} {\pi({\sigma_{old}^*}^2)Q_n({\sigma_{new}^*}^2|{\sigma_{old}^*}^2)}\right) \\
			&=\min\left( 1,\frac{\pi({\sigma_{new}^*}^2)\lvert J_{x_{old}} \rvert}{\pi({\sigma_{old}^*}^2)\lvert J_{x_{new}} \rvert}\right) \\
			&=\min \left(1, \exp\left[\log \pi({\sigma_{new}^*}^2) -\log \pi({\sigma_{old}^*}^2)+ \log \lvert J_{x_{old}} \rvert - \log \lvert J_{x_{new}} \rvert \right]
			\right),
		\end{split}
	\end{equation*}
	where $\pi({\sigma_{new}^*}^2)$ and $\pi({\sigma_{old}^*}^2)$ are given by Eq. (\ref{eq:full_sd}) evaluated at the new and old ${\sigma^*}^2$, and $\lvert J_x \rvert$ is provided in step 1, conditional on $\sigma^+$ in the current iteration:
	\begin{equation} \label{eq:Jacobian-ratio}
		\frac{\lvert J_{x_{old}} \rvert}{\lvert J_{x_{new}} \rvert}=\frac{{\sigma_{new}^*}^2\left( \sigma^+-{\sigma_{new}^*}^2\right) }{{\sigma_{old}^*}^2\left( \sigma^+-{\sigma_{old}^*}^2\right)}.
	\end{equation}
	
	Taking the logarithm of Eq. (\ref{eq:full_sd}) and Eq. (\ref{eq:Jacobian-ratio}) yields
	\begin{equation*}
		\begin{split}
			\log\pi\left({\sigma^*}^2\right|\ldots)&= -\frac{1}{2\ssdcenter}\sum_{I_{j,d}}\left( t^*_{j,d}-t_{c,d}\right) ^2 - \log \left( \ssdcenter\right) -\frac{1}{2m^2}\left( \log(\ssdcenter)-h_j\right) ^2+\text{const.}, \\
			\log  \left( \frac{\lvert J_{x_{old}} \rvert}{\lvert J_{x_{new}} \rvert}\right) &= \log\left( {\sigma_{new}^*}^2\right) + \log\left(\sigma^+-{\sigma_{new}^*}^2\right)  -\log\left( {\sigma_{old}^*}^2\right) - \log\left(\sigma^+-{\sigma_{old}^*}^2\right).
		\end{split}
	\end{equation*}

	\item After making the decision to accept the proposed value or not, we update the sample covariance/variance $\Sigma_{n}$. For computational purposes, \cite{Liu2024} use a recursive formulae to update $\Sigma_{n}$ sequentially at each iteration, which is described below.
	
	For $\vd=1$, the variance $\Sigma_{n}$ is computed based on two statistics: $M_2(n)$ and $\Bar{x}_{n}$, which are defined as  
	\begin{equation*}
		M_2(n) = \sum_{i=1}^n (x_i - \Bar{x}_{n})^2, \quad 
		\Bar{x}_{n} = \frac{1}{n} \sum_{i=1}^n x_i.
	\end{equation*}
	
	The following relationship is observed between $\Bar{x}_{n}$ and $\Bar{x}_{n-1}$, and between $M_2(n)$ and $M_2(n-1)$:
	\begin{align}
		\Bar{x}_n &= \left(1-\frac{1}{n} \right) \Bar{x}_{n-1} + \frac{x_n}{n}, \nonumber \\ 
		\Sigma_{n} &= \frac{1}{n-1} \sum_{i=1}^n (x_i - \Bar{x}_{n})^2 = \frac{1}{n-1}M_2(n) \nonumber \\
		&= \frac{1}{n-1} [M_2(n-1) + (x_n - \Bar{x}_{n-1})(x_n - \Bar{x}_n)].
		\label{eq:AMH for scalers}
	\end{align} 
	The proof of Eq. \eqref{eq:AMH for scalers} is as follows:
	\begin{equation*}
		\begin{split}
			(n-1)\Sigma_{n}  - (n-2)\Sigma_{n-1}  &= \sum_{i=1}^n (x_i - \Bar{x}_n)^2 - \sum_{i=1}^{n-1} (x_i - \Bar{x}_{n-1}) \\
			&= (x_n - \Bar{x}_n)^2 + \sum_{i=1}^{n-1} ((x_i - \Bar{x}_n)^2 - (x_i - \Bar{x}_{n-1})^2) \\
			&= (x_n - \Bar{x}_n)^2 + \sum_{i=1}^{n-1} (x_i - \Bar{x}_n + x_i - \Bar{x}_{n-1})(\Bar{x}_{n-1} - \Bar{x}_n) \\
			&= (x_n - \Bar{x}_n)^2 + (\Bar{x}_n - x_n)(\Bar{x}_{n-1} - \Bar{x}_n) \\
			&= (x_n - \Bar{x}_n)(x_n - \Bar{x}_n - \Bar{x}_{n-1} + \Bar{x}_n) \\
			&= (x_n - \Bar{x}_n)(x_n - \Bar{x}_{n-1}).
		\end{split}
	\end{equation*}
	Hence we first compute $\Bar{x}_n$ from $\Bar{x}_{n-1}$ and $x_n$. Then $\Bar{x}_n, \Bar{x}_{n-1},x_n$ and $M_2(n-1)$ are used for the calculation of $\Sigma_n$.

	For $\vd>1$, we will compute $\bSigma_n$ based on two statistics: $\Tilde{\bS}(n)$ and $\bmm(n)$ defined as
	\begin{equation*}
		\Tilde{\bS}(n) = 
		\begin{pmatrix}
			\sum_{i=1}^{n} (x_{i,1})^2 & \sum_{i=1}^{n} x_{i,1}x_{i,2} & \cdots & \sum_{i=1}^{n} x_{i,1}x_{i,\vd} \\
			\vdots  & \vdots  & \ddots & \vdots \\
			\sum_{i=1}^{n} x_{i,\vd} x_{i,1} & \sum_{i=1}^{n} x_{i,\vd}x_{i,2} & \cdots & \sum_{i=1}^{n} x_{i,\vd}x_{i,\vd}
		\end{pmatrix},
	\end{equation*}
	where $x_{i,l}$ is the $i$-th posterior sample of the $l$-th unknown parameter, and $\bmm(n)$ is a $\vd$-dimension row vector, with the $l$-th element $m_l(n)=\sum_{i=1}^{n} x_{i,l}/n$ ($l=1,\ldots,\vd$).
	
	The element in the sample covariance matrix after $n$ iterations is given by
	\begin{equation*}
		\begin{aligned}
			\bSigma_n(u,v) &= \frac{1}{n-1}\sum_{i=1}^n (x_{i,u}-m_u(n))(x_{i,v}-m_v(n)) \\
			&= \frac{1}{n-1} \left[ \sum_{i=1}^n x_{i,u}x_{i,v} - m_v(n)\sum_{i=1}^n x_{i,u} - m_u(n)\sum_{i=1}^n x_{i,v} + n \times m_v(n)m_u(n) \right] \\
			&= \frac{1}{n-1} \left[ \sum_{i=1}^n x_{i,u}x_{i,v} -n \times m_v(n)m_u(n) \right] \\
			&= \frac{1}{n-1} \sum_{i=1}^n x_{i,u}x_{i,v} -\frac{n}{n-1}m_v(n)m_u(n).
		\end{aligned}
	\end{equation*}
	Hence the covariance matrix could be written in the following form:
	\begin{equation*}
		\bSigma_n = \frac{1}{n-1}\Tilde{\bS}(n) - \frac{n}{n-1}\left( \bmm(n)^T \bmm(n)\right) ,
	\end{equation*}
	and we note the following relationship:
	\begin{equation*}
		\begin{split}
			\Tilde{\bS}(n) &= \Tilde{\bS}(n-1) + \bm{x}_n^T \bm{x}_n,\\
			\bmm(n) &= \left(1-\frac{1}{n}\right)\bmm(n-1) + \frac{1}{n} \bm{x}_n.
		\end{split}
	\end{equation*}
	Therefore, we first compute $\Tilde{\bS}(n)$ and $\bmm(n) $ based on $\Tilde{\bS}(n-1), \bmm(n-1) $ and the new value $\bm{x}_n$. Then the covariance matrix can be updated given $\Tilde{\bS}(n)$ and $\bmm(n) $.
\end{enumerate} 

Lastly, we note that the above AMH will be applied when component $j$ is occupied. In this case, if there is no truncation (the upper bound $\sigma_{j,d}^+=\infty$), the transformation defined in step 1 will reduce to a simple log-transformation, and the Jacobian is simply $1/\sdcenter$.  When component $j$ is empty at one iteration, we will draw a new sample from the log-normal prior (may or may not be truncated). In this case, the sample is always accepted and transformed to $x$ to update the covariance/variance.

\subsection{Concentration Parameters $\alpha$ and $\alpha_0$}
\subsubsection{Concentration Parameter $\alpha$}
The full conditional distribution of $\alpha$ is 
\begin{equation*}
	\begin{split}
		\pi(\alpha|\bmq^J,\bmp^J) & \propto  \prod_{j=1}^J \prod_{d=1}^D \Gam(q_{j,d}^J|\alpha p_j^J,1) \times \Gam(\alpha|1,1)\\
		&\propto  \prod_{j=1}^J \prod_{d=1}^D  \left[ \frac{1}{\Gamma\left(\alpha p_j^J \right) } \left( q_{j,d}^J\right) ^{\alpha p_j^J} \right]\times \exp(-\alpha).\\ 
	\end{split}
\end{equation*}
The distribution is not of a standard form and we apply the AMH scheme as described in Section \ref{AMH}. Specifically, we use the log-transformation
\begin{equation*}
	x=\log(\alpha) \in \mathbb{R}.
\end{equation*}
The Jacobian is 
\begin{equation*}
	J_x=\frac{dx}{d\alpha}=\frac{1}{\alpha}.
\end{equation*} 
and the inverse transformation is $\alpha=\exp(x)$. 

The logarithm of the full conditional density is 
\begin{equation*}
	\log \pi(\alpha|\bmq^J,\bmp^J)= -\alpha + \sum_{j=1}^J\sum_{d=1}^D  \left[ \alpha p_j^J \log(q_{j,d}^J) - \log\left( \Gamma\left( \alpha p_j^J\right) \right)  \right]+\text{const}.
\end{equation*}
Hence the acceptance probability of the new sample is
\begin{equation*}
	\begin{split}
		\alpha(\alpha_{new},\alpha_{old})&=\min \left( 1, \frac{\pi(\alpha_{new})Q_n(\alpha_{old}|\alpha_{new})} {\pi(\alpha_{old})Q_n(\alpha_{new}|\alpha_{old})}\right) \\
		&=\min \left( 1, \frac{\pi(\alpha_{new})\alpha_{new}} {\pi(\alpha_{old})\alpha_{old}}\right) \\
		&=\min \left(1, \exp \left[\log \pi(\alpha_{new})-\pi(\alpha_{old})+\log(\alpha_{new})-\log(\alpha_{old})\right]
		\right).
	\end{split}
\end{equation*}

After the decision of rejection or acceptance, we update the sample variance following step 4 of Section \ref{AMH} ($\vd=1$).

\subsubsection{Concentration Parameter $\alpha_0$}
The full conditional distribution of $\alpha$ is 
\begin{equation*}
	\begin{split}
		\pi(\alpha_0|\bmp^J) & \propto  \Gam(\alpha_0 | 1,1) \times \Dir \left(\textbf{p}^J | \frac{\alpha_0}{J}, \dots, \frac{\alpha_0}{J}\right) \\
		& \propto \exp(-\alpha_0) \times \frac{\Gamma(\alpha_0)}{[\Gamma(\frac{\alpha_0}{J})]^J} \prod_{j=1}^J {\left( p_j^J\right) }^{\frac{\alpha_0}{J}}.
	\end{split}
\end{equation*}
This distribution does not have a closed form and hence we apply AMH (Section \ref{AMH}). Same log transformation as $\alpha$ is applied to transform $\alpha_0$, and hence details are omitted here.
The logarithm of the full conditional distribution is
\begin{equation*}
	\log \pi(\alpha_0|\bmp^J)=  -\alpha_0 + \log(\Gamma(\alpha_0)) - J\log\left( \Gamma\left( \frac{\alpha_0}{J}\right) \right)  +  \frac{\alpha_0}{J}\sum_{j=1}^J \log(p_j^J)+\text{const}.
\end{equation*}

\subsection{Allocation Variables $z_{c,d}$}
The full conditional distribution is
\begin{equation*}
	\pi(z_{c,d} =j | \bmu_{1:J,1:G}^*, \bphi_{1:J,1:G}^*, \bbeta, \bY, \bmt,\bmq^J) \propto
	\prod_{g=1}^G \NB(y_{c,g,d} | \mu_{j,g}^* \beta_{c,d}, \phi_{j,g}^*) \times  q_{j,d}^JK(t_{c,d}|\bpsi_{j,d}^*).
\end{equation*}
Let $\tilde{p}_{c,d,j}$ denote the term on the right-hand side.
We have
\begin{equation*} \label{eq:full_allocation}
	\pi(z_{c,d} =j | \bmu_{1:J,1:G}^*, \bphi_{1:J,1:G}^*, \bbeta, \bY, \bmt,\bmq^J) =\frac{\tilde{K}\tilde{p}_{c,d,j}}{\tilde{K}\sum_{l=1}^{J}\tilde{p}_{c,d,l}},
\end{equation*}
where we remove the most extreme probability to avoid numerical errors,
\begin{equation*}
	\log\left( \tilde{K}\right) = -\max_{j} \log\left( \tilde{p}_{c,d,j}\right).
\end{equation*}
In all, we sample $z_{c,d}$ from $\left\lbrace 1,\ldots,J\right\rbrace $ according to $ \pi(z_{c,d} =j | \bmu_{1:J,1:G}^*, \bphi_{1:J,1:G}^*, \bbeta, \bY, \bmt,\bmq^J)$. This is repeated for every $c$ and $d$.

\subsection{Component Probabilities $p_j^J$} \label{sec:alg6_AMH}
The full conditional distribution is 
\begin{equation*}
	\begin{split}
		\pi(p_1^J,\ldots,p_J^J|\bmq^J,\alpha,\alpha_0)& \propto  \prod_{j=1}^J \prod_{d=1}^D \Gam(q_{j,d}^J|\alpha p_j^J,1)\times\Dir \left(p_1^J,\ldots,p_J^J| \frac{\alpha_0}{J}, \dots, \frac{\alpha_0}{J}\right) \\
		&\propto \prod_{j=1}^J \prod_{d=1}^D\left[ \frac{1}{\Gamma\left(\alpha p_j^J \right) } \left( q_{j,d}^J\right) ^{\alpha p_j^J} \right]\times \prod_{j=1}^J {\left( p_j^J\right) }^{\frac{\alpha_0}{J}-1},
	\end{split}
\end{equation*}
which has no closed-form, and hence AMH is applied (Section \ref{AMH}). Since $p_j^J$ sum to one, the following transformation is applied yielding $\bmx \in \mathbb{R}^{J-1}$:
\begin{equation*}
	x_j = \log \left(\frac{p_j^J}{p_J^J}\right), \quad j = 1, \dots, J-1.
\end{equation*}
The inverse transformation is given by
\begin{equation*}
	\begin{split}
		p_j^J &= \frac{\exp(x_j)}{1+\sum_{j=1}^{J-1} \exp(x_j)}, \quad j = 1, \dots, J-1,\\
		p_J^J & = 1-\sum_{j=1}^{J-1}p_j^J=\frac{1}{1+\sum_{j=1}^{J-1} \exp(x_j)}.
	\end{split}
\end{equation*}
The Jacobian matrix is
\begin{equation*}
	\begin{split}
		J_{\bmx} &= 
		\begin{pmatrix}
			\frac{dx_1}{dp_1} & \frac{dx_2}{dp_1} & \cdots & \frac{dx_{J-1}}{dp_1} \\
			\vdots  & \vdots  & \ddots & \vdots  \\
			\frac{dx_1}{dp_{J-1}} & \frac{dx_2}{dp_{J-1}} & \cdots & \frac{dx_{J-1}}{dp_{J-1}}
		\end{pmatrix} \\
		&=
		\begin{pmatrix}
			\frac{1}{p_1} + \frac{1}{p_J} & \cdots & \frac{1}{p_J} \\
			\vdots & \ddots & \vdots  \\
			\frac{1}{p_J} & \cdots & \frac{1}{p_{J-1}} + \frac{1}{p_J}
		\end{pmatrix} \\
		&=
		\begin{pmatrix}
			\frac{1}{p_J} & \dots & \frac{1}{p_J} \\
			\vdots & \ddots & \vdots \\
			\frac{1}{p_J} & \dots & \frac{1}{p_J}
		\end{pmatrix} + 
		\begin{pmatrix}
			\frac{1}{p_1} & 0 &\dots & 0 \\
			0 & \frac{1}{p_2} &\vdots & 0 \\
			\vdots & \vdots & \ddots & 0 \\
			0 & 0 & \dots & \frac{1}{p_{J-1}}
		\end{pmatrix} \\
		&= B + A.
	\end{split}
\end{equation*}
Because $\det(A + B) = \det(A) + \det(B) + Tr(A^{-1} B)\det(A)$, $\det(B) = 0$ and $\det(A) = \prod_{j=1}^{J-1} \frac{1}{p_j}$, it follows that $\det(A+B) = \prod_{j=1}^{J-1} \frac{1}{p_j} + (1-p_J) \prod_{j=1}^{J} \frac{1}{p_j} = \prod_{j=1}^J \frac{1}{p_j}$. Therefore,     
\begin{equation*}
	\log |J_{\bmx}| = \log \left[\prod_{j=1}^J \frac{1}{p_j} \right] = -\sum_{j=1}^J \log (p_j).
\end{equation*}
The log full conditional distribution is    
\begin{equation*}
	\log \pi\left(\bmp^J|\ldots\right)= \sum_{j=1}^J\sum_{d=1}^D  \left[\alpha p_j^J \log (q_{j,d}^J) - \log \Gamma(\alpha p_j^J) \right]+ \sum_{j=1}^J \left[ \left( \frac{\alpha_0}{J} - 1\right)  \log (p_j^J) \right]+\text{const}.
\end{equation*}
Combining all terms together, the acceptance probability is
\begin{equation*}
	\begin{split}
		\alpha(\bmp^J_{new}, \bmp^J_{old}) &= \min \left( 1,\frac{\pi(\bmp^J_{new})Q_n(\bmp^J_{old} | \bmp^J_{new})} {\pi(\bmp^J_{old})Q_n(\bmp^J_{new} |  \bmp^J_{old})}\right)\\
		& =  \min\left( 1,  \frac{\pi(\bmp^J_{new})|J_{\bmx_{old}}|}{\pi(\bmp^J_{old})|J_{\bmx_{new}}|}\right) \\
		&= \min \left( 1, \exp\left[\log\pi(\bmp^J_{new}) - \log \pi(\bmp^J_{old}) + \sum_{j=1}^{J} \left(\log\left( p_{j,new}^J\right) - \log\left( p_{j,old}^J\right)  \right) \right]\right).
	\end{split}
\end{equation*}

We note that the sampling of a new transformed variable $\bmx_{new}$ is slightly different from step 2 in Section \ref{AMH}. Following Algorithm 6 in \cite{JimEGriffin2013AiMc}, instead of a fixed scale parameter $s_{\vd}=2.4^2/\vd$ ($\vd=J-1$ for the case of $\bmp^J$), $s_{\vd}$ is also updated at each iteration. The idea is to adapt $s_{\vd}$ to achieve a particular average acceptance probability $\Bar{\alpha}$, e.g. 0.234, which has been shown to be optimal in the multivariate target distribution \citep{roberts1997weak}. 

We use an initial value $s_{\vd}^{(1)}=0.001$. At the current iteration $n$, let $\bmx_{new}$ denote the new sample after the decision of rejection or not. Define 
\begin{equation*}
	\omega^{(n)}=\exp\left(\log\left( s_{\vd}^{(n)}\right) + n^{-0.7}\times \left( \alpha(\bmp^J_{new}, \bmp^J_{old})-\Bar{\alpha}\right)  \right),
\end{equation*}
then 
\begin{equation*}
	s_{\vd}^{(n+1)}=\begin{cases}
		\omega^-, \quad \text{if}\ \omega^{(n)}<\omega^-,\\
		\omega^{(n)}, \ \text{if}\ \omega^{(n)}\in\left[ \omega^-,\omega^+\right],\\
		\omega^+,\quad \text{if} \ \omega^{(n)}>\omega^+,
	\end{cases}
\end{equation*}
where $\omega^-=\exp(-50)$ and $\omega^+=\exp(50)$. The update of the covariance matrix follows from step 4 (multivariate case) in Section \ref{AMH}.

\subsection{Mean-dispersion Parameters  \texorpdfstring{$\bm{b}$}{b} and $\alpha_\phi^2$}  \label{appendix:gibbs-b-alpha_phi}

The joint distribution of $\bmb=(b_0,b_1)^T$ and $\alpha_\phi^2$ is
\begin{equation} \label{eq:b_marginal}
	\begin{split}
		\pi(\bmb, \alpha_\phi^2 | \bmu_{1:J,1:G}^*, \bphi_{1:J,1:G}^*) \propto &\Norm(\bmb | \bmm_b, \alpha_\phi^2 \bI) \times \IG(\alpha_\phi^2 | \nu_1, \nu_2) \times \prod_{j=1}^J \prod_{g=1}^G \logN(\phi_{j,g}^* | b_0 + b_1 \log(\mu_{j,g}^*), \alpha_\phi^2) \nonumber \\
		\propto &\left( \alpha_\phi^2 \right)^{-\left( \nu_1+2+\frac{JG}{2}\right) } \nonumber \\
		&\times\exp \bigg( -\frac{1}{\alpha_\phi^2} \bigg[\frac{1}{2} \sum_{j=1}^J \sum_{g=1}^G \left( \log(\phi_{j,g}^*)-b_0-b_1\log(\mu_{j,g}^*)\right) ^2 + \frac{1}{2}\bmb^T \bmb\\
		& \qquad  -\bmb^T\bmm_b+\frac{1}{2}\bmm_b^T\bmm_b + \nu_2\bigg] \bigg)\nonumber. \\
	\end{split}
\end{equation}
For $\bmb$, we have
\begin{equation*}  \label{eq:b_conditional}
	\pi( \bmb | \bmu^*, \bphi^*, \alpha_\phi^2) \propto \exp \left( -\frac{1}{2\alpha_\phi^2} \left[ \sum_{j=1}^J \sum_{g=1}^G (\log(\phi_{j,g}^*)-b_0-b_1\log(\mu_{j,g}^*))^2 + \bmb^T \bmb-2\bmb^T\bmm_b \right] \right).
\end{equation*}
We have
\begin{equation*}
	\sum_{j=1}^J \sum_{g=1}^G (\log(\phi_{j,g}^*)-b_0-b_1\log(\mu_{j,g}^*))^2 = \sum_{j=1}^J (\log(\bphi_j^*) - \Tilde{\bmu}_j \bmb)^T (\log(\bphi_j^*) - \Tilde{\bmu}_j \bmb),
\end{equation*}
where
\begin{equation*}
	\log(\bphi_j^*) = 
	\begin{pmatrix}
		\log (\phi_{j,1}^*) \\
		\vdots \\
		\log (\phi_{j,G}^*)
	\end{pmatrix}, \quad
	\Tilde{\bmu}_j =
	\begin{pmatrix}
		1 & \log (\mu_{j,1}^*) \\
		\vdots & \vdots \\
		1 & \log (\mu_{j,G}^*)
	\end{pmatrix}, \quad
	\bmb =
	\begin{pmatrix}
		b_0 \\
		b_1
	\end{pmatrix}.
\end{equation*}
Therefore,
\begin{equation*}
	\begin{aligned}
		\pi( \bmb | \bmu^*, \bphi^*, \alpha_\phi^2) \propto& \exp \bigg(-\frac{1}{2 \alpha_\phi^2}
		\bigg[\bmb^T \left (\sum_{j=1}^J \Tilde{\bmu}_j^T \Tilde{\bmu}_j + \bI \right) \bmb - 2\bmb^T \left( \sum_{j=1}^J  \Tilde{\bmu}_j^T \log(\bphi_j^*) +\bmm_b\right) \\
		&\quad +  \sum_{j=1}^J  \log(\bphi_j^*)^T  \log(\bphi_j^*)  \bigg] \bigg),
	\end{aligned}
\end{equation*}
i.e. 
\begin{equation*}  \label{eq:b_standard}
	\bmb | \ldots \sim \Norm (\Tilde{\bmm}_b, \alpha_\phi^2 \Tilde{\bV}_b),
\end{equation*}
where    
\begin{equation*}
	\begin{split}
		\Tilde{\bmm}_b &= \left( \sum_{j=1}^J \Tilde{\bmu}_j^T \Tilde{\bmu}_j + \bI \right)^{-1} \left( \sum_{j=1}^J  \Tilde{\bmu}_j^T \log(\bphi_j^*)+\bmm_b \right),\\
		\Tilde{\bV}_b &= \left( \sum_{j=1}^J \Tilde{\bmu}_j^T \Tilde{\bmu}_j + 
		\bI \right)^{-1}.
	\end{split}
\end{equation*}
As for $\alpha_\phi^2$, the full condition is
\begin{equation*}
	\begin{split}
		\pi(\alpha_\phi^2 | \bmu^*, \bphi^* )= & \int \pi( \bmb, \alpha_\phi^2 | \bmu^*, \bphi^* )\, d\bmb \\
		\propto & \int \left(\frac{1}{\alpha_\phi^2}\right)^{\nu_1+1} \exp \left( -\frac{\nu_2}{\alpha_\phi^2}\right) \left( \frac{1}{\alpha_\phi^2}\right)^{JG/2} \left( \frac{1}{\alpha_\phi^2}\right) \\
		&\times \exp \bigg(-\frac{1}{2\alpha_\phi^2}\bigg[  (\bmb-\Tilde{\bmm}_b)^T \Tilde{\bV}_b^{-1} (\bmb-\Tilde{\bmm}_b) - \Tilde{\bmm}_b^T \Tilde{\bV}_b^{-1} \Tilde{\bmm}_b\\
		& \quad + \sum_{j=1}^J \log(\bphi_j^*)^T  \log(\bphi_j^*)+ \bmm_b^T\bmm_b
		\bigg]\bigg) \, d\bmb.
	\end{split}
\end{equation*}  
Conditional on $\bmu^*, \bphi^*$,
\begin{equation*}
	\int \frac{1}{\alpha_\phi^2} \exp \left( -\frac{1}{2\alpha_\phi^2} (\bmb-\Tilde{\bmm}_b)^T \Tilde{\bV}_b^{-1} (\bmb-\Tilde{\bmm}_b) \right) \, d\bmb = \text{const.},
\end{equation*}
it follows that
\begin{equation*}
	\begin{aligned}
		\pi(\alpha_\phi^2 | \bmu^*, \bphi^*)& \propto \left(\frac{1}{\alpha_\phi^2}\right)^{\nu_1+1} \exp \left( -\frac{\nu_2}{\alpha_\phi^2}\right) \left( \frac{1}{\alpha_\phi^2}\right)^{JG/2} \\
		&\quad \times \exp \left( -\frac{1}{2\alpha_\phi^2} \left[ - \Tilde{\bmm}_b^T \Tilde{\bV}_b^{-1} \Tilde{\bmm}_b + \sum_{j=1}^J \log(\bphi_j^*)^T  \log(\bphi_j^*) +\bmm_b^T\bmm_b\right] \right) \\
		&\propto \left( \frac{1}{\alpha_\phi^2} \right) ^{\nu_1+1+JG/2} \\
		&\quad  \times \exp \left( -\frac{1}{\alpha_\phi^2} \left[ \nu_2 + \frac{1}{2} \left( \sum_{j=1}^J \log(\bphi_j^*)^T  \log(\bphi_j^*) - \Tilde{\bmm}_b^T \Tilde{\bV}_b^{-1} \Tilde{\bmm}_b +\bmm_b^T\bmm_b\right) \right] \right).
	\end{aligned}
\end{equation*}
Therefore,  
\begin{equation*}
	\alpha_\phi^2 | \bmu^*, \bphi^* \sim \text{IG} (\Tilde{\nu}_1, \Tilde{\nu}_2),
	\label{eq:alpha_phi_standard}
\end{equation*}
where
\begin{equation*}
	\Tilde{\nu}_1 = \nu_1 + JG/2 ,\quad \Tilde{\nu}_2 = \nu_2 + \frac{1}{2} \left( \sum_{j=1}^J \log(\bphi_j^*)^T  \log(\bphi_j^*) - \Tilde{\bmm}_b^T \Tilde{\bV}_b^{-1} \Tilde{\bmm}_b +\bmm_b^T\bmm_b\right).
\end{equation*}

\subsection{Component-specific Parameters $\mu_{j,g}^*$ and $\phi_{j,g}^*$}

The full condition distribution is
\begin{equation*}
	\begin{split}
		\pi(\mu_{j,g}^*, \phi_{j,g}^* |\bZ,\bmb,\alpha_\phi^2,\bY,\bbeta) \propto & \logN\left( \mu^*_{j,g}  | 0, \alpha_{\mu}^2\right)  \times  \logN \left(  \phi^*_{j,g} | b_0 +b_1 \log(\mu_{j,g}^*) , \alpha_\phi^2\right)\\    
		& \times\prod_{(c,d): z_{c,d}=j}  \NB(y_{c,d,g} | \mu_{j,g}^* \beta_{c,d}, \phi_{j,g}^* )\\
		\propto& \left( \frac{1}{\mu_{j,g}^* \phi_{j,g}^*} \right) \exp \left( -\frac{1}{2 \alpha_{\mu}^2} (\log{\mu_{j,g}^*})^2 -\frac{1}{2 \alpha_{\phi}^2} (\log{\phi_{j,g}^*} - (b_0 + b_1\log\mu_{j,g}^*))^2\right) \nonumber 
		\\ & \times \prod_{(c,d):z_{c,d} =j} \binom{y_{c,g,d} + \phi_{j,g}^* - 1}{\phi_{j,g}^*-1} \left( \frac{\phi_{j,g}^*}{\mu_{j,g}^* \beta_{c,d} + \phi_{j,g}^*}\right)^{\phi_{j,g}^*} \left( \frac{\mu_{j,g}^*}{\mu_{j,g}^* \beta_{c,d} + \phi_{j,g}^*}\right)^{y_{c,g,d}}.
	\end{split}
\end{equation*}  
This is not a standard distribution and hence we will apply AMH to sample for $\left(\mu_{j,g}^*,\phi_{j,g}^*\right)$ (Section \ref{AMH}). For clarity, we will drop the subscript $j$ and $g$ here. We apply the following transformation
\begin{equation*}
	\bmx=(x_1,x_2)=\left( \log \left(\mu^*\right), \log \left(\phi^*\right)\right) \in \mathbb{R}^2,
\end{equation*}
with inverse transformation 
\begin{equation*}
	\mu^*=\exp(x_1),\quad \phi^*=\exp(x_2).
\end{equation*}
The Jacobian term is
\begin{equation*}
	J_{\bmx}= 
	\begin{pmatrix}
		\frac{dx_1}{d\mu^*} & \frac{dx_1}{d\phi^*} \\
		\frac{dx_2}{d\mu^*} & \frac{dx_2}{d\phi^*}
	\end{pmatrix} = 
	\begin{pmatrix}
		\frac{1}{\mu^*} & 0 \\
		0 & \frac{1}{\phi^*}
	\end{pmatrix},
\end{equation*}  
giving $|J_{\bmx}|=1/(\mu^*\phi^*)$.

The logarithm of the full conditional distribution is
\begin{equation*}
	\begin{split}
		\log \pi(\mu_{j,g}^*, \phi_{j,g}^* | \ldots) =&-\log (\mu_{j,g}^*\phi_{j,g}^*) -\frac{1}{2 \alpha_{\mu}^2} (\log{\mu_{j,g}^*})^2 -\frac{1}{2 \alpha_{\phi}^2} (\log{\phi_{j,g}^*} - (b_0 + b_1\log\mu_{j,g}^*))^2 \\
		&+ \sum_{(c,d):z_{c,d}=j} \log \binom{y_{c,g,d} + \phi_{j,g}^* - 1}{\phi_{j,g}^*-1} + \phi_{j,g}^* \log \left( \frac{\phi_{j,g}^*}{\mu_{j,g}^* \beta_{c,d} + \phi_{j,g}^*}\right) \\
		& + y_{c,g,d} \log \left( \frac{\mu_{j,g}^*}{\mu_{j,g}^* \beta_{c,d} + \phi_{j,g}^*}\right) + \text{const}.
	\end{split}
\end{equation*}    
Combining all terms together, the acceptance probability is
\begin{equation*}
	\begin{split}
		\alpha \left( (\mu^*,\phi^*)_{new}, (\mu^*,\phi^*)_{old}\right) = & \min \bigg( 1, \exp \bigg[  \log \pi((\mu^*,\phi^*)_{new}) - \log\pi((\mu^*,\phi^*)_{old}) \\
		& - \log(\mu_{old}^*) -\log(\phi_{old}^*) + \log(\mu_{new}^* ) +  \log(\phi_{new}^*) \bigg] \bigg).
	\end{split}
\end{equation*}    
Then the covariance matrix is updated following the multivariate case $(\vd=2)$ in step 4 of Section \ref{AMH}.

Note that due to label switching, the covariance matrix may have very large values. Therefore, to mitigate the multiplicative effect of the scale parameter $s_\vd$ on the covariance, we fix $s_{\vd}=1$ instead of $s_{\vd}=2.4^2/2$. In addition, the adaptive Metropolis-Hastings algorithm above is applied for all occupied components. For empty components, new samples are drawn from the prior directly and transformed to $\bmx$ to update the covariance matrix. 

The above step is repeated for every $j$ and $g$.  

\subsection{Capture Efficiencies $\beta_{c,d}$}
The full conditional distribution is
\begin{equation*}      \label{eq:capture_efficiencies_conditional}
	\begin{split}
		\pi( \beta_{c,d} |\left\lbrace y_{c,g,d}\right\rbrace _{g=1}^G, z_{c,d}=j, \bmu_{1:J,1:G}^*, \bphi_{1:J,1:G}^*) \propto &\Beta(\beta_{c,d} | a_d^\beta, b_d^\beta) \times \prod_{g=1}^G  \NB(y_{c,g,d} | \mu_{j,g}^* \beta_{c,d}, \phi_{j,g}^*) \nonumber \\
		\propto&(\beta_{c,d})^{a_d^\beta -1} (1-\beta_{c,d})^{b_d^\beta - 1} \\
		&\times \left[ \prod_{g=1}^G \left( \frac{1}{ \phi_{j,g}^* + \mu_{j,g}^* \beta_{c,d}} \right)^{\phi_{j,g}^* + y_{c,g,d}} (\beta_{c,d})^{y_{c,g,d}} \right]. \nonumber \\
	\end{split}
\end{equation*}
This does not have a closed form and we will apply AMH (Section \ref{AMH}) with the following variable transformation
\begin{equation*}
	x=\log\left( \frac{\beta_{c,d}}{1-\beta_{c,d}}\right) \in \mathbb{R},
\end{equation*}
with Jacobian equal to 
\begin{equation*}
	J_x=\frac{dx}{d\beta_{c,d}}=\frac{d}{d\beta_{c,d}} (\log(\beta_{c,d})-\log(1-\beta_{c,d})) = \frac{1}{\beta_{c,d}(1-\beta_{c,d})}.
\end{equation*}
The inverse transformation is given by
\begin{equation*}
	\beta_{c,d}=\frac{1}{1+\exp(-x)}.
\end{equation*}
Next, the logarithm of the full conditional distribution is
\begin{equation*}
	\begin{split}
		\log \pi(\beta_{c,d}|\ldots)=&(a_d^\beta - 1)\log (\beta_{c,d}) + (b_d^\beta - 1) \log(1-\beta_{c,d}) \\
		& - \sum_{g=1}^G \left[ (\phi_{j,g}^* + y_{c,g,d}) \log ( \phi_{j,g}^* + \mu_{j,g}^* \beta_{c,d}) - y_{c,g,d}\log(\beta_{c,d})\right]+\text{const}.
	\end{split}
\end{equation*}
Therefore, the acceptance probability is given by
\begin{equation*}
	\begin{split}
		\alpha(\beta_{new}, \beta_{old}) =& \text{min} \bigg(1, \exp \bigg[\log \pi(\beta_{new}) - \log \pi(\beta_{old})\\
		& +  \log(\beta_{new}) + \log(1-\beta_{new}) - \log(\beta_{old}) - \log(1-\beta_{old})
		\bigg] 
		\bigg),
	\end{split}
\end{equation*}
and we update the variance of the transformed variable $x$ following step 4 of Section \ref{AMH}. 

This step is repeated for every $c$ and $d$.  

\subsection{Hyperparameters $r_j, s^2, h_j$ and $m^2$}
\subsubsection{Prior Means $r_j$}
For each $j$, we have 
\begin{equation*}
	\begin{split}
		\pi(r_j|\left\lbrace \tcenter\right\rbrace _{d=1}^D,\mu_r,\sigma_r^2,s^2)&\propto \prod_{d=1}^D  \Norm(\tcenter|r_j, s^2)\times  \Norm(r_j|\mu_r,\sigma_r^2)\\
		&\propto \exp\left[ -\frac{1}{2s^2}\sum_{d=1}^{D}\left( r_j-\tcenter\right) ^2 \right] \times \exp\left[ -\frac{1}{2\sigma_r^2}\left( r_j-\mu_r\right) ^2 \right].
	\end{split}
\end{equation*}
Recall our calculation for $\tcenter$ in Section \ref{full_tcenter}. It can be noticed that the full conditional distribution for $r_j$ is a normal distribution 
\begin{equation*}
	r_j|\ldots \sim \Norm(\hat{\mu}_r,\hat{\sigma}_r^2),
\end{equation*}
where
\begin{equation*}
	\hat{\sigma}_r^2=\left( \frac{1}{\sigma_r^2}+\frac{D}{s^2}\right)^{-1},\quad
	\hat{\mu}_r=\frac{\mu_r/\sigma_r^2+\sum_{d=1}^D\tcenter/s^2}{1/\sigma_r^2+D/s^2}.
\end{equation*}

\subsubsection{Prior Variance $s^2$}
\begin{equation*}
	\begin{split}
		\pi(s^2|\left\lbrace \tcenter\right\rbrace _{j=1,d=1}^{J,D},\eta_1,\eta_2,\bmr)\propto& \prod_{j=1}^J\prod_{d=1}^D  \Norm(\tcenter|r_j, s^2)\times  \IG(s^2|\eta_1,\eta_2)\\
		\propto& (s^2)^{-\frac{JD}{2}}\exp\left[- \frac{1}{s^2}\times \frac{1}{2}\sum_{j=1}^{J}\sum_{d=1}^{D}\left( \tcenter-r_j\right) ^2 \right] \\
		&\times (s^2)^{-\eta_1-1}\exp\left[ -\frac{\eta_2}{s^2}\right] ,
	\end{split}
\end{equation*}
i.e. 
\begin{equation*}
	s^2|\ldots \sim \IG\left( \frac{JD}{2}+\eta_1,\eta_2+\frac{1}{2}\sum_{j=1}^{J}\sum_{d=1}^{D}\left( \tcenter-r_j\right) ^2\right) .
\end{equation*}

\subsubsection{Prior Means $h_j$}
For each $j$,
\begin{equation*}
	\begin{split}
		\pi(h_j|\left\lbrace \sdcenter\right\rbrace _{d=1}^D,\mu_h,\sigma_h^2,m^2)&\propto \prod_{d=1}^D  \logN(\sdcenter|h_j, m^2)\times  \Norm(h_j|\mu_h,\sigma_h^2)\\
		&\propto \exp\left[ -\frac{1}{2m^2}\sum_{d=1}^{D}\left( h_j-\log\left( \sdcenter\right) \right) ^2 \right] \times \exp\left[ -\frac{1}{2\sigma_h^2}\left( h_j-\mu_h\right) ^2 \right].
	\end{split}
\end{equation*}
Similar to $r_j$, the full conditional is a normal distribution 
\begin{equation*}
	h_j|\ldots \sim \Norm(\hat{\mu}_h,\hat{\sigma}_h^2),
\end{equation*}
where
\begin{equation*}
	\hat{\sigma}_h^2=\left( \frac{1}{\sigma_h^2}+\frac{D}{m^2}\right)^{-1},\quad
	\hat{\mu}_r=\frac{\mu_h/\sigma_h^2+\sum_{d=1}^D\log\left(\sdcenter\right)/m^2}{1/\sigma_h^2+D/m^2}.
\end{equation*}

\subsubsection{Prior Variance $m^2$}
\begin{equation*}
	\begin{split}
		\pi(m^2|\left\lbrace \sdcenter\right\rbrace _{j=1,d=1}^{J,D},\kappa_1,\kappa_2,\bmh)\propto& \prod_{j=1}^J\prod_{d=1}^D  \logN(\sdcenter|h_j, m^2)\times  \IG(m^2|\kappa_1,\kappa_2)\\
		\propto& (m^2)^{-\frac{JD}{2}}\exp\left[- \frac{1}{m^2}\times \frac{1}{2}\sum_{j=1}^{J}\sum_{d=1}^{D}\left( \log\left(\sdcenter\right)-h_j\right) ^2 \right] \\
		&\times (m^2)^{-\kappa_1-1}\exp\left[ -\frac{\kappa_2}{m^2}\right],
	\end{split}
\end{equation*}
i.e. 
\begin{equation*}
	m^2|\ldots \sim \IG\left( \frac{JD}{2}+\kappa_1,\kappa_2+\frac{1}{2}\sum_{j=1}^{J}\sum_{d=1}^{D}\left(  \log\left(\sdcenter\right)-h_j\right) ^2\right) .
\end{equation*}

\subsection{bayNorm Estimates of Capture Efficiencies, Mean and Dispersion Parameters} \label{sec:baynorm_beta}

In the bayNorm approach, capture efficiencies for cells from a dataset $d$ are estimated by
\begin{align*}
	\hat{\beta}_{c,d}^{\text{bay}} = \frac{\sum_{g=1}^G y_{c,g,d}}{\frac{1}{C_d} \sum_{c=1}^{C_d}\sum_{g=1}^G y_{c,g,d}} \times \lambda,
\end{align*}
where $\lambda$ is the an estimate of global mean capture efficiency across cells. Under the default setting of bayNorm, $\lambda = 0.06$. The estimates are used to construct empirical priors for the capture efficiencies in our C-HDP model.

As for mean and dispersion parameters, \cite{Tang2020} propose two approaches to estimate them. The first one is based on maximizing the marginal likelihood for each gene, assuming independence between cells. The log-marginal likelihood for gene $g$ is 
\begin{equation*}
	\log M_g = \sum_{d=1}^D \sum_{c=1}^{C_d} \log \NB(y_{c,g,d} | \mu_{g}\beta_{c,g}, \phi_{g}).
\end{equation*}

Alternatively, $\mu_g$ and $\phi_g$ can be estimated through the method of moments estimation (MME) by matching the empirical and theoretical moments of the normalized count $y^s_{c,g,d}=y_{c,g,d}/\beta_{c,d}$. In addition, two approaches can be combined to produce more robust and efficient estimation \citep{Tang2020}, where $\mu_g$ is obtained through MME, and $\phi_g$ from MME is adjusted by a factor obtained from fitting a linear regression between $\phi_g$ from MME and $\phi_g$ from maximizing the marginal likelihood.

\subsection{Identifying Marker Genes} \label{sec:marker-genes-method}

Following the definition in \cite{Liu2024}, we use probabilistic tools to identify differentially expressed (DE) genes and differentially dispersed (DD) genes that distinguish between clusters. The differences in each gene between clusters are measured in terms of the posterior tail probability that the absolute value of the log-fold change (LFC) of the mean expression (or dispersion) between two clusters $j$ and $j'$ is greater than the threshold $\tau_0$ and $\omega_0$, respectively. Formally, the probabilities are defined as
\begin{equation*}
	\begin{split}
		P_g (j,j', \tau_0) &= \text{Pr}\left( \left| \log \left(\frac{\mu^*_{j,g}}{\mu^*_{j',g}} \right) \right| > \tau_0 | \bZ, \bY, \tau_0 \right), \\
		L_g (j,j', \omega_0) &= \text{Pr}\left( \left| \log \left(\frac{\phi^*_{j,g}}{\phi^*_{j',g}} \right) \right| > \tau_0 | \bZ, \bY, \omega_0 \right),
	\end{split}
\end{equation*}
where $\bZ=\left\lbrace z_{c,d}\right\rbrace_{c=1,d=1}^{C_d,D} , \bY=\left\lbrace y_{c,g,d}\right\rbrace _{c=1,g=1,d=1}^{C_d,G,D}$. 

\subsubsection{Global Marker Genes}

Global DE genes are identified by considering the maximum posterior tail probabilities across all pairwise clusters,
\begin{align*}
	P_g^*(\tau_0) = \max_{(j,j')} P_g (j,j',\tau_0), 
	\quad L_g^*(\omega_0) = \max_{(j,j')} L_g (j,j',\omega_0).
\end{align*}
Then genes with $P_g^*(\tau_0)$ greater than a threshold $\alpha_M$ are identified as global DE genes, i.e. these genes have a high posterior probability that the LFC in the mean expression is greater than $\tau_0$ across at least two clusters. Similarly, those with $L_g^*(\omega_0)$ greater than a threshold $\alpha_D$ are identified as global DD genes. The thresholds are set to to control the expected false discovery rate (EFDR) to 0.05 \citep{vallejos2016}:
\begin{align*}
	\text{EFDR}_{\alpha_M} (\tau_0) &= \frac{\sum_{g=1}^G \left(1-P_g^*(\tau_0) \right) \indicator(P_g^*(\tau_0) > \alpha_M)}{\sum_{g=1}^G (1-P_g^*(\tau_0))}, \\
	\text{EFDR}_{\alpha_D} (\omega_0) &= \frac{\sum_{g=1}^G \left(1-L_g^*(\omega_0) \right) \indicator(L_g^*(\omega_0) > \alpha_D)}{\sum_{g=1}^G (1-L_g^*(\omega_0))}.
\end{align*}

\subsubsection{Local Marker Genes}
In comparison to global marker genes, local marker genes are specific to each cluster only, i.e. distinguishing the cluster from all the others. They are identified based on the minimum posterior tail probability. For cluster $j$, we compute
\begin{align*}
	P_{g,j}^*(\tau_0) = \min_{j' \neq j} P_g (j,j',\tau_0),\quad L_{g,j}^*(\omega_0) = \min_{j' \neq j} L_g (j,j',\omega_0),
\end{align*}
and genes with $P_{g,j}^*(\tau_0)$ (or $L_{g,j}^*(\omega_0)$) greater than a threshold (calibrated according to EFDR) are identified as local DE (or DD) genes. Local DE genes will have a high posterior probability that the LFC in the mean expression is greater than $\tau_0$ between cluster $j$ and any other cluster.

\subsection{Posterior Predictive Checks} \label{sec:ppc-pax6-method}

We follow the approach in \cite{Liu2024} and conduct posterior predictive checks based on the mixed predictive distribution \citep{Lewin2007fully}, given posterior samples from the post-processing step and the optimal clustering. Specifically, to generate a single replicated dataset, we draw mean expression parameters $\mu_{j,g}^*$ and capture efficiencies $\beta_{c,d}$ from the posterior samples, whereas the dispersion parameters $\phi_{j,g}^*$ are drawn from its hyper-priors given posterior samples of $\bmb, \alpha_\phi^2, \mu_{j,g}^*$.

For a single replicated dataset, we compute the following statistics for each gene: the mean of log-shifted counts across cells, i.e. $\log(y+1)$, the standard deviation of log-shifted counts, the logarithm of mean counts, and the dropout probabilities, i.e. proportion of cells with zero counts. The average in these statistics is taken across cells. We compare the relationship for two pairs of statistics between the true and replicated datasets: 1) the mean and standard deviation of log-shifted counts and 2) the logarithm of mean counts and dropout probabilities.

Furthermore, we investigate the pointwise differences between the true and replicated datasets in terms of the mean of log-shifted counts, standard deviation of log-shifted counts and dropout probabilities.

For multiple replicates, we compare the kernel density estimation of three statistics between the replicates and true data: mean and standard deviation of log shifted counts, and the dropout probabilities.

\clearpage
\section{Posterior Inference for Calcium Imaging Data} \label{sec:full-conditionals-cidata}

We use a vector autoregression model with lag one to account for the temporal nature of the data:
\begin{equation*} 
	\bm{y}_{i,d}|\bm{y}_{i-1,d},z_{i,d}=j, \bma_j^*, \bB_j^*, \bSigma_j^* \sim ~ \Norm(\bma_j^*+\bB_j^*\bm{y}_{i-1}, \bSigma_j^*)  ,
\end{equation*}
where $\bm{y}_{i,d}=(y_{i,1,d},\ldots,y_{i,G,d})^T \in \mathbb{R}^G$, $\bma_j^*$ denotes the intercept, $\mathbf{B}_j^*$ is a matrix of coefficients in VAR, and $\bSigma_j^*$ is the covariance matrix. Define $\bL_j^*=(\bma_j^* \quad \bB_j^*)^T$ and $\bmx_{i,d}=(1, y_{i-1,1,d},\ldots,y_{i-1,G,d})^T$.

\subsection{Prior Specification} \label{appendix:cidata-prior}

\paragraph{Base Measure} The component-specific parameters $\bL_j^*$ and $\bSigma_j^*$ are given conjugate priors:
\begin{equation*}
	\bL_j^* | \bSigma_j^*  \indsim \MatrixN(\bL_0, \bV_0, \bSigma_j^*), \quad \bSigma_j^*  \iidsim \IW(\omega_0, \bPhi_0).
\end{equation*}
For hyperparameters, we obtain empirical estimates for the coefficient matrix and covariance matrix, denoted by $\hat{\bL}_0$ and $\hat{\bPhi}_0$, respectively, by regressing $\bmy_{i,d}$ on $\bmy_{i-1,d}$ across all groups. Then we set $\bL_0=\hat{\bL}_0$ and $\bPhi_0=\hat{\bPhi}_0/J_0^{2/G}$ \citep{fraley2007bayesian}, where $J_0$ is a guess of the number of clusters. In addition, we use $\bV_0=100\bI$ and $\omega_0=G+2$.

\paragraph{Kernel Parameters}

For a periodic kernel with parameters $\bpsi_{j,d}^*=(\mu_{j,d}^*, \sdcenter, \plambda)$, the following hierarchical priors are used:
\begin{equation*}  \label{eq:priors-periodic-kernel}
	\begin{split}
		\mu_{j,d}^* &\indsim \Unif\left(-\frac{\pi\plambda}{2},\frac{\pi\plambda}{2}\right), \quad \plambda \indsim \logN(r_j, s^2),\quad r_j  \iidsim \Norm(\mu_r, \sigma_r^2),\quad s^2\sim \IG(\eta_1, \eta_2),\\
		\sdcenter & \indsim \IG(a_j,b_j), \quad a_j=2+\frac{h_j^2}{m^2},\quad  b_j=h_j^2+\frac{h_j^3}{m^2}, \quad
		h_j  \iidsim \logN(\mu_h, \sigma_h^2), ,\quad m^2 \sim \IG(\kappa_1, \kappa_2).
	\end{split}
\end{equation*}
Here shape $a_j$ and scale $b_j$ are modelled as functions of the mean $h_j$ and variance $m^2$ of the inverse-gamma prior. Note that $\mu_{j,d}^*$ is restricted within one period ($\pi\plambda$) for identifiability.
For calcium imaging data, we set $\mu_r=-2, \sigma_r=0.5, \eta_1=5, \eta_2=1, \mu_h=-1, \sigma_h=0.5, \kappa_1=26, \kappa_2=1$.

For concentration parameters, the priors are the same as those for Pax6 data (Section \ref{sec:full-conditionals-pax6}). With a finite-dimensional truncation at $J$, the complete model is

\begin{equation*}
	\begin{split}
		\bmy_{i,d} | z_{i,d}=j, \bmy_{i-1,d}, \bL_j^*, \bSigma_j^* &\indsim \Norm((\bL_j^*)^T \bmx_{i,d},  \bSigma_j^*), \\
		z_{i,d}| p_{1,d}^J(t_{i,d}), \ldots, p_{J,d}^J(t_{i,d})&\indsim \Cat(p_{1,d}^J(t_{i,d}),\ldots, p_{J,d}^J(t_{i,d})),\\
		p_{j,d}^J(t_{i,d}) & =\frac{q_{j,d}^JK(t_{i,d}|\bpsi_{j,d}^*)}{\sum_{k=1}^{J} q_{k,d}K(t_{i,d}|\bpsi_{k,d}^*)},  \\
		q_{j,d}^J &\indsim \Gam(\alpha p_j^J, 1), \\
		p^J_{1},\ldots, p^J_{J} &\sim \Dir\left( \frac{\alpha_0}{J},\ldots, \frac{\alpha_0}{J}\right),\\
		\mu_{j,d}^* | \plambda &\indsim \Unif\left(-\frac{\pi\plambda}{2},\frac{\pi\plambda}{2}\right), \\
		\plambda &\indsim \logN(r_j, s^2),\\
		r_j & \iidsim \Norm(\mu_r, \sigma_r^2),\\
		s^2&\sim \IG(\eta_1, \eta_2),\\
		\sdcenter & \indsim \IG(a_j, b_j),\\
		a_j=2+\frac{h_j^2}{m^2}&, \quad   b_j=h_j^2+\frac{h_j^3}{m^2}, \\
		h_j & \iidsim \logN(\mu_h, \sigma_h^2),\\
		m^2 & \sim \IG(\kappa_1, \kappa_2),\\
		\bL_j^* | \bSigma_j^*  &\indsim \MatrixN(\bL_0, \bV_0, \bSigma_j^*),\\
		\bSigma_j^*  &\sim \IW(\omega_0, \bPhi_0),\\
		\alpha &\sim \Gam(1,1),\\
		\alpha_0 &\sim \Gam(1,1).
	\end{split}
\end{equation*}

\newpage
Define $\bZ=\left\lbrace z_{i,d}\right\rbrace_{i=1,d=1}^{n_d,D} , \bY=\left\lbrace \bmy_{i,d}\right\rbrace _{i=1,d=1}^{n_d,D},\bmt=\left\lbrace t_{i,d}\right\rbrace_{i=1,d=1}^{n_d,D}, \bmq^J=\left\lbrace q_{j,d}^J\right\rbrace _{j=1,d=1}^{J,D}, \bmp^J=(p_1^J,\ldots,p_J^J), \bL^*=(\bL_1^*,\ldots,\bL_J^*), \bSigma^*=(\bSigma_1^*,\ldots,\bSigma_J^*),
{\bmu^*}=\left\lbrace\mu_{j,d}^*\right\rbrace_{j=1,d=1}^{J,D}, {\blambda^*}=\left\lbrace\lambda_{j,d}^*\right\rbrace_{j=1,d=1}^{J,D}, 
{\bsigma^*}^2=\left\lbrace\sdcenter\right\rbrace_{j=1,d=1}^{J,D} , \bmr=(r_1,\ldots,r_J), \bmh=(h_1,\ldots,h_J)$. The posterior distribution is     
\begin{equation*}
	\begin{split}
		&\pi(\bZ,\bmq^J,\bmp^J, \bL^*, \bSigma^*, \alpha, \alpha_0, \bmu^*,\blambda^*,{\bsigma^*}^2, \bmr, s^2,\bmh, m^2|\bY,\bmt)\\
		\propto &\prod_{j=1}^J \prod_{(i,d): z_{i,d}=j} \Norm(\bmy_{i,d} | (\bL_j^*)^T \bmx_{i,d}, \bSigma_j^*)\\
		& \times \prod_{j=1}^J \prod_{d=1}^D \prod_{i: z_{i,d}=j}  K(t_{i,d}|\bpsi_{j,d}^*) \times \prod_{j=1}^J \prod_{d=1}^D \left( q_{j,d}^J\right) ^{N_{j,d}}\\
		& \times \prod_{j=1}^J \prod_{d=1}^D \prod_{i=1}^{n_d}  \exp\left( -\xi_{i,d}q_{j,d}^JK(t_{i,d}|\bpsi_{j,d}^*)\right) \\
		& \times \prod_{j=1}^J \prod_{d=1}^D \Gam(q_{j,d}^J|\alpha p_j^J,1) \times \Dir\left(\bmp^J | \frac{\alpha_0}{J},\ldots, \frac{\alpha_0}{J}\right)\\
		&\times \prod_{j=1}^J  \left[ \MatrixN(\bL_j^*  | \bL_0, \bV_0, \bSigma_j^*)  \times  \IW(\bSigma_j^* | \omega_0, \bPhi_0)\right] \\
		& \times \Gam(\alpha | 1,1) \times \Gam(\alpha_0 | 1,1) \\
		& \times \prod_{j=1}^J \prod_{d=1}^D \left[ \logN(\plambda|r_j, s^2) \times \Unif\left(\mu_{j,d}^* | -\frac{\pi\plambda}{2},\frac{\pi\plambda}{2}\right ) \times \IG\left(\sdcenter | a_j, b_j\right)\right] \\
		& \times  \prod_{j=1}^J \left[ \Norm(r_j | \mu_r,\sigma_r^2)\times \logN(h_j | \mu_h, \sigma_h^2)\right] \times \IG(s^2 | \eta_1,\eta_2) \times \IG(m^2 | \kappa_1,\kappa_2),
	\end{split}
\end{equation*}
where  $N_{j,d}=\sum_{i=1}^{n_d} \indicator(z_{i,d}=j)$ is the number of observations in component $j$ in dataset $d$. The Gibbs sampling steps for updating the allocation variables $\bZ$ and parameters $\bmq^J$ are similar to those in the clustering model for Pax6 data, except that the likelihood and the kernel have changed. As for concentration parameters $\alpha, \alpha_0$ and component probabilities $\bmp^J$, the steps are the same as illustrated in Section \ref{sec:full-conditionals-pax6}. Therefore, below we only illustrate the update for component-specific parameters, kernel parameters and hyperparameters. 

\subsection{Kernel Parameters}
\subsubsection{Location $\mu_{j,d}^*$} \label{sec:mu-kernel-cidata}
For each $j$ and $d$, the full conditional distribution for $\mu_{j,d}^*$ is
\begin{equation*}
	\begin{split}
		&\pi(\mu_{j,d}^*|\left\lbrace z_{i,d}\right\rbrace _{i=1}^{n_d}, \left\lbrace \xi_{i,d}\right\rbrace _{i=1}^{n_d}, \left\lbrace t_{i,d}\right\rbrace _{i=1}^{n_d}, q_{j,d}^J, \sdcenter, \plambda)\\
		\propto  &\prod_{i: z_{i,d}=j}  K(t_{i,d}|\bpsi_{j,d}^*) \times \prod_{i=1}^{n_d}  \exp\left( -\xi_{i,d}q_{j,d}^JK(t_{i,d}|\bpsi_{j,d}^*)\right) \times \indicator\left(\mu_{j,d}* \in \left(-\frac{\pi\plambda}{2},\frac{\pi\plambda}{2}\right )\right).
	\end{split}
\end{equation*}
The distribution does not have a closed form since $\mu_{j,d}^*$ is inside $\sin(\cdot)$ function. The AMH scheme described in Algorithm 5 from \cite{JimEGriffin2013AiMc} is applied to achieve a targeted average acceptance probability for univariate parameters.

\paragraph{Adaptive Metropolis-Hastings for $\mu_{j,d}^*$} \label{AMH-univariate}
\begin{enumerate}
	\item Apply the following transformation to $\mu^*$, dropping subscript for simplicity:
	\begin{equation*}
		x=g(\mu^*)=\log\left( \frac{\mu^*-\mu^-}{\mu^+-\mu^*}\right)  \in \mathbb{R},
	\end{equation*}
	where $\mu^-=-\pi\plambda/2$ and $\mu^+=\pi\plambda/2$ are the lower bound and upper bound. The Jacobian term is 
	\begin{equation*}
		J_x = \frac{dx}{d\mu^*} = \frac{\mu^+-\mu^-}{(\mu^*-\mu^-)(\mu^+-\mu^*)}.
	\end{equation*}
	The inverse transformation is 
	\begin{equation*}
		\mu^*=\mu^++\frac{\mu^--\mu^+}{1+\exp(x)}.
	\end{equation*}
	\item At iteration $n$, letting $x_{old}=g(\mu_{old}^*)$, we propose $x_{new} \sim N(x_{old}, \zeta^{n})$ where $\zeta^{(n)}$ is the adaptive variance with an initial value $\zeta^{(1)}=0.01$, which will be updated at each iteration (see step 4 below).  Then $\mu_{new}^*$ is obtained through the inverse transformation. 
	\item The logarithm of the full conditional distribution is
	\begin{align*}
		\log \pi(\mu_{j,d}^*|\ldots) = & -\frac{2}{\sdcenter} \times \sum_{I_{j,d}}\sin^2\left(\frac{ t_{i,d}-\mu_{j,d}^* }{\plambda} \right)\\
		&- \sum_{i=1}^{n_d} \xi_{i,d}q_{j,d}^J\exp\left( -\frac{2}{\sdcenter} \times \sin^2\left(\frac{ t_{i,d}-\mu_{j,d}^* }{\plambda} \right) \right) + \text{const.},
	\end{align*}
	where $I_{j,d}=\left\lbrace i: z_{i,d}=j\right\rbrace $.
	Let $Q_n$ denote the proposal distribution at step $n$. The acceptance probability of this proposal is given by 
	\begin{equation*}
		\begin{split}
			\alpha(\mu^*_{new},\mu^*_{old})=&\min\left( 1, \frac{\pi(\mu^*_{new}) Q_n(\mu^*_{old}|\mu^*_{new})}{\pi(\mu^*_{old}) Q_n(\mu^*_{new}|\mu^*_{old})} \right) 		 \\
			=&\min\left( 1, \frac{\pi(\mu^*_{new}) (\mu^*_{new}-\mu^-)(\mu^+-\mu^*_{new})}
			{\pi(\mu^*_{old}) (\mu^*_{old}-\mu^-)(\mu^+-\mu^*_{old}) }\right)\\
			=& \min\bigg( 1, \exp \bigg[\log \pi(\mu^*_{new})-\log \pi(\mu^*_{old})+\log(\mu^*_{new}-\mu^-)+\log(\mu^+-\mu^*_{new})\\
			&-\log(\mu^*_{old}-\mu^-)-\log(\mu^+-\mu^*_{old})\bigg]\bigg).
		\end{split}
	\end{equation*}
	\item After making the decision to accept the proposed value or not, we now update the adaptive variance. Define 
	\begin{equation*}
		\omega^{(n)}=\exp\left(\log\left( \zeta^{(n)}\right) + n^{-0.7}\times \left(\alpha(\mu^*_{new},\mu^*_{old})-\Bar{\alpha}\right)  \right), 
	\end{equation*}
	where $\Bar{\alpha}$ is desired average acceptance probability (0.234 or 0.44). The updated variance is 
	\begin{equation*}
		\zeta^{(n+1)}=\begin{cases}
			\omega^-, \quad \text{if}\ \omega^{(n)}<\omega^-,\\
			\omega^{(n)}, \ \text{if}\ \omega^{(n)}\in\left[ \omega^-,\omega^+\right], \\
			\omega^+,\quad \text{if} \ \omega^{(n)}>\omega^+,
		\end{cases}
	\end{equation*}
	where $\omega^-=\exp(-50)$ and $\omega^+=\exp(50)$. 
\end{enumerate}

\subsubsection{Period $\plambda$}
As for $\plambda$, the full conditional distribution is 
\begin{equation*}
	\begin{split}
		&\pi(\plambda|r_j, s^2, \left\lbrace z_{i,d}\right\rbrace _{i=1}^{n_d}, \left\lbrace \xi_{i,d}\right\rbrace _{i=1}^{n_d}, \left\lbrace t_{i,d}\right\rbrace _{i=1}^{n_d}, q_{j,d}^J, \sdcenter, \mu_{j,d}^*)\\
		\propto  &\prod_{i: z_{i,d}=j}  K(t_{i,d}|\bpsi_{j,d}^*) \times \prod_{i=1}^{n_d}  \exp\left( -\xi_{i,d}q_{j,d}^JK(t_{i,d}|\bpsi_{j,d}^*)\right) \\
		&\times
		\logN(\plambda|r_j, s^2) \times \Unif\left(\mu_{j,d}^* | -\frac{\pi\plambda}{2},\frac{\pi\plambda}{2}\right )\\
		\propto  &\prod_{i: z_{i,d}=j} K(t_{i,d}|\bpsi_{j,d}^*) \times \prod_{i=1}^{n_d}  \exp\left( -\xi_{i,d}q_{j,d}^JK(t_{i,d}|\bpsi_{j,d}^*)\right) \\
		&\times
		\frac{1}{\plambda}\exp\left(-\frac{1}{2s^2} \left(\log(\plambda)-r_j\right)^2 \right) \times \frac{1}{\plambda} \times \indicator\left(\plambda > \frac{2|\mu_{j,d}^*|}{\pi} \right).
	\end{split}
\end{equation*}
Similar to $\mu_{j,d}^*$, the distribution is non-standard and the AMH scheme described in Section \ref{AMH-univariate} is applied. We use the following transformation to transform $\plambda$ on the real line, dropping subscript for notation simplicity:
\begin{equation*}
	x=g(\lambda^*)=\log(\lambda^*-\lambda^-) \in \mathbb{R},
\end{equation*}
where $\lambda^-=2|\mu^*|/\pi$ is the lower bound. The Jacobian term is 
\begin{equation*}
	J_x=\frac{dx}{d\lambda^*}=\frac{1}{\lambda^*-\lambda^-},
\end{equation*}
and the inverse transformation is 
\begin{equation*}
	\lambda^*=\exp(x)+\lambda^-.
\end{equation*}
The logarithm of the conditional distribution is       
\begin{equation*}
	\begin{split}
		\log \pi(\plambda|\ldots)=&  -\frac{2}{\sdcenter} \times \sum_{I_{j,d}}\sin^2\left(\frac{ t_{i,d}-\mu_{j,d}^* }{\plambda} \right) - \sum_{i=1}^{n_d} \xi_{i,d}q_{j,d}^J\exp\left[-\frac{2}{\sdcenter} \times \sin^2\left(\frac{ t_{i,d}-\mu_{j,d}^* }{\plambda} \right) \right]\\
		&-  2\log(\plambda)-\frac{1}{2s^2} \left(\log(\plambda)-r_j\right)^2 + \text{const.},
	\end{split}
\end{equation*}
where $I_{j,d}=\left\lbrace i: z_{i,d}=j\right\rbrace $.
The acceptance probability is 
\begin{equation*}
	\begin{split}
		\alpha(\lambda_{new}^*, \lambda_{old}^*)&=\min\left( 1, \frac{\pi(\lambda_{new}^*) Q_n(\lambda_{old}^*|\lambda_{new}^*)}{\pi(\lambda_{old}^*) Q_n(\lambda_{new}^*|\lambda_{old}^*)} \right) 		 \\
		&=\min\left( 1, \frac{\pi(\lambda_{new}^*) (\lambda_{new}^*-\lambda^-)}
		{\pi(\lambda_{old}^*) (\lambda_{old}^*-\lambda^-) }\right)\\
		&= \min\left( 1, \exp\left[\log \pi(\lambda_{new}^*)-\log \pi(\lambda_{old}^*)+\log(\lambda_{new}^*-\lambda^-)-\log(\lambda_{old}^*-\lambda^-)\right]\right).
	\end{split}
\end{equation*}
The adaptive variance is updated as described in Section \ref{sec:mu-kernel-cidata}.

\subsubsection{Bandwidth $\sdcenter$}
As for $\sdcenter$, the full conditional distribution is 
\begin{equation*}
	\begin{split}
		&\pi(\sdcenter|h_j, m^2, \left\lbrace z_{i,d}\right\rbrace _{i=1}^{n_d}, \left\lbrace \xi_{i,d}\right\rbrace _{i=1}^{n_d}, \left\lbrace t_{i,d}\right\rbrace _{i=1}^{n_d}, q_{j,d}^J, \mu_{j,d}^*, \plambda)\\
		\propto  &\prod_{i: z_{i,d}=j}  K(t_{i,d}|\bpsi_{j,d}^*) \times \prod_{i=1}^{n_d}  \exp\left( -\xi_{i,d}q_{j,d}^JK(t_{i,d}|\bpsi_{j,d}^*)\right) \times 
		\IG\left(\sdcenter | a_j, b_j\right),
	\end{split}
\end{equation*}
where $a_j=2+\frac{h_j^2}{m^2}, b_j=h_j^2+\frac{h_j^3}{m^2}$. Here we introduce the latent variable $u_{i,j,d}$, same as the step in Section \ref{sec:kernel_MCMC_pax6}. Then the full conditional distribution becomes
\begin{equation*}
	\begin{split}
		&\pi(\sdcenter, \left\lbrace u_{i,j,d}\right\rbrace_{i=1}^{n_d}|h_j, m^2, \left\lbrace z_{i,d}\right\rbrace _{i=1}^{n_d}, \left\lbrace \xi_{i,d}\right\rbrace _{i=1}^{n_d}, \left\lbrace t_{i,d}\right\rbrace _{i=1}^{n_d}, q_{j,d}^J, \mu_{j,d}^*, \plambda)\\
		\propto  &\prod_{i: z_{i,d}=j}  K(t_{i,d}|\bpsi_{j,d}^*) \times \prod_{i=1}^{n_d}  \mathbb{I}\left( u_{i,j,d}< M_{i,j,d}\right) \times 
		\IG\left(\sdcenter | a_j, b_j\right),
	\end{split}
\end{equation*}
where $M_{i,j,d}=\exp\left( -\xi_{i,d}q_{j,d}^JK(t_{i,d}|\bpsi_{j,d}^*)\right)$.

We sample $u_{i,j,d} \sim \Unif\left(0,\exp\left( -\xi_{i,d}q_{j,d}^JK(t_{i,d}|\bpsi_{j,d}^*)\right) \right)$. For $\sdcenter$, the conditional distribution is
\begin{equation*}
	\begin{split}
		\pi(\sdcenter|\ldots) &\propto  \prod_{i: z_{i,d}=j}  K(t_{i,d}|\bpsi_{j,d}^*) \times 
		\IG\left(\sdcenter | a_j, b_j\right) \times \indicator(\sdcenter \in E_{j,d}) \\
		&\propto \exp \left[-\frac{2}{\sdcenter} \times \sum_{I_{j,d}}\sin^2\left(\frac{ t_{i,d}-\mu_{j,d}^* }{\plambda} \right)\right] \times \left(\sdcenter\right)^{-a_j-1}\exp\left(-\frac{b_j}{\sdcenter}\right) \times \indicator(\sdcenter \in E_{j,d})  \\
		&\propto \left(\sdcenter\right)^{-a_j-1} \times \exp \left[-\frac{2}{\sdcenter} \times \sum_{I_{j,d}}\sin^2\left(\frac{ t_{i,d}-\mu_{j,d}^* }{\plambda} \right) - \frac{b_j}{\sdcenter} \right]\times \indicator(\sdcenter \in E_{j,d}).
	\end{split}
\end{equation*}
The truncation region can be derived in the same fashion as described in Section \ref{sec:variance-kernel-pax6}, and is given by
\begin{equation*}
	\begin{split}
		E_{j,d}=\left( 0, \sigma_{j,d}^+\right),  \quad \sigma_{j,d}^+=\min_{i: -\log{u_{i,j,d}} < \xi_{i,d}q_{j,d}^J} -\frac{2\sin^2\left(\frac{ t_{i,d}-\mu_{j,d}^* }{\plambda} \right)}{\log\left[ -\frac{\log{u_{i,j,d}}}{\xi_{i,d}q_{j,d}^J}\right] }.\\
	\end{split}
\end{equation*}
Thus the full conditional is a truncated inverse-gamma distribution with truncation region $E_{j,d}$
\begin{equation*}
	\sdcenter|\ldots \sim \IG\left(a_j, b_j+2\sum_{I_{j,d}}\sin^2\left(\frac{ t_{i,d}-\mu_{j,d}^* }{\plambda} \right)\right).
\end{equation*}
The latent variable is not introduced for $\mu_{j,d}^*$ and $\plambda$ as the full conditional distributions are still not of a standard form.

\subsection{Component-specific Parameters $\bL_j^*$ and $\mathbf{\Sigma}_j^*$}
Recall the density of an inverse-Wishart distribution is 
\begin{equation*}
	f(\bSigma| \omega, \bPhi) = \frac{|\bPhi|^{\omega/2}}{\frac{\omega G }{2} \Gamma_G\left(\frac{\omega}{2}\right)} |\bSigma|^{-\frac{\omega+G+1}{2}}\exp\left(-\frac{1}{2}Tr(\bPhi \bSigma^{-1})\right),
\end{equation*}
where $\bSigma$ is of dimension $G \times G$, $\Gamma_G(\cdot)$ is the multivariate gamma function, and $Tr$ denotes the trace. The density of a matrix normal distribution is
\begin{equation*}
	f(\bY| \mathbf{M}, \mathbf{U}, \bV)= \frac{\exp\left(-\frac{1}{2} Tr\left(\bV^{-1} (\bY-\mathbf{M})^T {\mathbf{U}}^{-1} (\bY-\mathbf{M})\right)\right)}{(2\pi)^{NG/2} |\bV| ^{N/2} |\mathbf{U}|^{G/2}},
\end{equation*}
where $\bY$ and $\mathbf{M}$ are of dimension $N \times G$, $\mathbf{U}$ is $N \times N$ for variance among the rows of $\bY$, $\bV$ is $G \times G$ for variance among columns, 

For empty components, new samples for $\bL_j^*,\bSigma_j^*$ are drawn from the prior directly. For occupied components, let $N_j$ denote the size of component $j$, $\bY_j$ denote the data matrix with dimension $N_j \times G$, stacking all observations belonging to component $j$, and $\bX_j$ the corresponding design matrix of dimension $N_j \times (G+1)$. The first column of $\bX_j$ is one. The likelihood for component $j$ can be expressed as
\begin{equation*}
	\bY_j \sim \MatrixN(\bX_j \bL_j^*, \bI_{N_j}, \bSigma_j^*).
\end{equation*}
The full conditional distribution for $\bL_j^*$ and $\bSigma_j^*$ is 
\begin{equation*}
	\begin{split}
		&\pi (\bL_j^*,\bSigma_j^*| \bY, \bZ, \bL_0, \bV_0, \bPhi_0, \omega_0)\\
		\propto & \MatrixN(\bY_j | \bX_j L_j^*, \bI_{N_j}, \bSigma_j^*) \times \MatrixN(\bL_j^*  | \bL_0, \bV_0, \bSigma_j^*)  \times  \IW(\bSigma_j^* | \omega_0, \bPhi_0) \\
		\propto & |\bSigma_j^*|^{-N_j/2} \exp \left(-\frac{1}{2} Tr\left({\bSigma_j^*}^{-1} (\bY_j-\bX_j \bL_j^*)^T (\bY_j-\bX_j \bL_j^*)\right)\right)\\
		&\times |\bSigma_j^*|^{-(G+1)/2} \exp \left(-\frac{1}{2} Tr\left({\bSigma_j^*}^{-1} (\bL_j^*-\bL_0)^T \bV_0^{-1} (\bL_j^*-\bL_0)\right)\right)\\
		&\times |\bSigma_j^*|^{-(\omega_0 + G +1 )/2} \exp \left\lbrace -\frac{1}{2} Tr\left( \bPhi_0  {\bSigma_j^*}^{-1} \right) \right\rbrace.
	\end{split}
\end{equation*}
Consider the first two exponential terms. It can be shown that 
\begin{equation*}
	\begin{split}
		A &= (\bY_j-\bX_j \bL_j^*)^T (\bY_j-\bX_j \bL_j^*) + (\bL_j^*-\bL_0)^T \bV_0^{-1} (\bL_j^*-\bL_0) \\
		& = \bY_j^T \bY_j + \bL_0^T \bV_0^{-1}\bL_0 -2\bY_j^T\bX_j\bL_j^* + (\bL_j^*)^T \bX_j^T \bX_j \bL_j^* + (\bL_j^*)^T\bV_0^{-1}\bL_j^*-2(\bL_j^*)^T\bV_0^{-1}\bL_0\\
		& = \bY_j^T \bY_j + \bL_0^T \bV_0^{-1}\bL_0 + (\bL_j^*)^T(\bX_j^T \bX_j +\bV_0^{-1}) \bL_j^* - 2(\bL_j^*)^T (\bX_j^T \bY_j+ \bV_0^{-1}\bL_0).
	\end{split}
\end{equation*}
Let $\bV_n=(\bX_j^T \bX_j +\bV_0^{-1})^{-1}$. Completing the square for $\bL_j^*$ yields
\begin{equation*}
	\begin{split}
		A & = \bY_j^T \bY_j + \bL_0^T \bV_0^{-1}\bL_0 + (\bL_j^*)^T \bV_n^{-1} \bL_j^* -2 (\bL_j^*)^T \bV_n^{-1} \underbrace{\bV_n (\bX_j^T \bY_j+ \bV_0^{-1}\bL_0)}_{\bL_n}\\
		&= \bY_j^T \bY_j + \bL_0^T \bV_0^{-1}\bL_0 + (\bL_j^* - \bL_n)^T \bV_n^{-1}(\bL_j^*-\bL_n) - \bL_n^T\bV_n^{-1}\bL_n.
	\end{split}
\end{equation*}
Since $Tr(A+B)=Tr(A)+Tr(B)$ and $Tr(AB)=Tr(BA)$, the joint full conditional distribution can be written as
\begin{equation*}
	\begin{split}
		\pi (\bL_j^*,\bSigma_j^*| \ldots) \propto & |\bSigma_j^*|^{-(G+1)/2} \exp \left(-\frac{1}{2} Tr\left({\bSigma_j^*}^{-1}  (\bL_j^* - \bL_n)^T \bV_n^{-1}(\bL_j^*-\bL_n) \right) \right)\\
		\times & |\bSigma_j^*|^{-(N_j+\omega_0+G+1)/2} \exp \left(-\frac{1}{2} Tr\left({\bSigma_j^*}^{-1}  \left(\bPhi_0+ \bY_j^T \bY_j + \bL_0^T \bV_0^{-1}\bL_0 - \bL_n^T\bV_n^{-1}\bL_n
		\right) \right)\right),
	\end{split}
\end{equation*}
which corresponds to the following full conditional distributions:
\begin{equation*}
	\bL_j^*|\bSigma_j^*,\ldots \sim \MatrixN(\bL_n, \bV_n, \bSigma_j^*) , \quad \bSigma_j^*|\ldots \sim \IW(\omega_n, \bPhi_n),
\end{equation*}
where 
\begin{equation*}
	\begin{split}
		&\bL_n = \bV_n (\bX_j^T \bY_j+ \bV_0^{-1}\bL_0),  \quad \bV_n=(\bX_j^T \bX_j +\bV_0^{-1})^{-1},\\
		&\bPhi_n = \bPhi_0+ \bY_j^T \bY_j + \bL_0^T \bV_0^{-1}\bL_0 - \bL_n^T\bV_n^{-1}\bL_n, \quad \omega_n = N_j+\omega_0.
	\end{split}
\end{equation*}

\subsection{Hyperparameters $r_j, s^2, h_j$ and $m^2$}
\subsubsection{Prior Means $r_j$}
For each $j$, we have 
\begin{equation*}
	\begin{split}
		\pi(r_j|\left\lbrace \plambda\right\rbrace _{d=1}^D,\mu_r,\sigma_r^2,s^2)&\propto \prod_{d=1}^D  \logN(\plambda|r_j, s^2)\times  \Norm(r_j|\mu_r,\sigma_r^2)\\
		&\propto \exp\left[ -\frac{1}{2s^2}\sum_{d=1}^{D}\left( r_j-\log(\plambda)\right) ^2 \right] \times \exp\left[ -\frac{1}{2\sigma_r^2}\left( r_j-\mu_r\right) ^2 \right].
	\end{split}
\end{equation*}
The full conditional distribution for $r_j$ is a normal distribution 
\begin{equation*}
	r_j|\ldots \sim \Norm(\hat{\mu}_r,\hat{\sigma}_r^2),
\end{equation*}
where
\begin{equation*}
	\hat{\sigma}_r^2=\left( \frac{1}{\sigma_r^2}+\frac{D}{s^2}\right)^{-1},\quad
	\hat{\mu}_r=\frac{\mu_r/\sigma_r^2+\sum_{d=1}^D\log(\plambda)/s^2}{1/\sigma_r^2+D/s^2}.
\end{equation*}

\subsubsection{Prior Variance $s^2$}
\begin{equation*}
	\begin{split}
		\pi(s^2|\left\lbrace \plambda\right\rbrace _{j=1,d=1}^{J,D},\eta_1,\eta_2,\bmr)\propto& \prod_{j=1}^J\prod_{d=1}^D  \logN(\plambda|r_j, s^2)\times  \IG(s^2|\eta_1,\eta_2)\\
		\propto& (s^2)^{-\frac{JD}{2}}\exp\left[- \frac{1}{s^2}\times \frac{1}{2}\sum_{j=1}^{J}\sum_{d=1}^{D}\left( \log(\plambda)-r_j\right) ^2 \right] \\
		&\times (s^2)^{-\eta_1-1}\exp\left[ -\frac{\eta_2}{s^2}\right] ,
	\end{split}
\end{equation*}
i.e. 
\begin{equation*}
	s^2|\ldots \sim \IG\left( \frac{JD}{2}+\eta_1,\eta_2+\frac{1}{2}\sum_{j=1}^{J}\sum_{d=1}^{D}\left( \log(\plambda)-r_j\right) ^2\right) .
\end{equation*}

\subsubsection{Prior Means $h_j$}
For each $j$,
\begin{equation*}
	\begin{split}
		\pi(h_j|\left\lbrace \sdcenter\right\rbrace _{d=1}^D,\mu_h,\sigma_h^2,m^2) \propto &\prod_{d=1}^D  \IG\left(\sdcenter | a_j, b_j\right)\times  \logN(h_j|\mu_h,\sigma_h^2),\\
		\propto & \prod_{d=1}^D \frac{b_j^{a_j}}{\Gamma(a_j)}(\sdcenter)^{-a_j-1}\exp(-\frac{b_j}{\sdcenter})\\
		& \times  \frac{1}{h_j} \exp\left(-\frac{1}{2\sigma_h^2}(\log(h_j) -\mu_h)^2 \right),
	\end{split}
\end{equation*}
where $a_j=2+\frac{h_j^2}{m^2}, b_j=h_j^2+\frac{h_j^3}{m^2}$. The distribution has no closed form and hence the AMH scheme described in Section \ref{AMH-univariate} is applied. The log transformation is applied with $x=g(h_j)=\log(h_j) \in \mathbb{R}$. The Jacobian term is $J_x=dx/dh_j=1/h_j$,
and the inverse transformation is $h_j=\exp(x)$.

The logarithm of the full conditional density is 
\begin{equation*}
	\begin{split}
		\log \pi(h_j|\ldots) =& D a_j \log(b_j) - D\log(\Gamma(a_j))-(a_j+1) \sum_{d=1}^D \log(\sdcenter) \\
		& -b_j \sum_{d=1}^D \frac{1}{\sdcenter}-\log(h_j) - \frac{1}{2\sigma_h^2} (\log(h_j) -\mu_h)^2+\text{const}.
	\end{split}
\end{equation*}
Hence the acceptance probability of the new sample is
\begin{equation*}
	\begin{split}
		\alpha(h_{new},h_{old})&=\min \left( 1, \frac{\pi(h_{new})Q_n(h_{old}|h_{new})} {\pi(h_{old})Q_n(h_{new}|h_{old})}\right) \\
		&=\min \left( 1, \frac{\pi(h_{new})h_{new}} {\pi(h_{old})h_{old}}\right) \\
		&=\min \left(1, \exp \left[\log \pi(h_{new})-\log\pi(h_{old})+\log(h_{new})-\log(h_{old})\right]
		\right).
	\end{split}
\end{equation*}

\subsubsection{Prior Variance $m^2$}
The full conditional distribution is 
\begin{equation*}
	\begin{split}
		\pi(m^2|\left\lbrace \sdcenter\right\rbrace _{j=1,d=1}^{J,D},\kappa_1,\kappa_2,\bmh) \propto &\prod_{j=1}^J \prod_{d=1}^D  \IG\left(\sdcenter | a_j, b_j\right)\times  \IG(m^2 | \kappa_1,\kappa_2),\\
		\propto & \prod_{j=1}^J \prod_{d=1}^D \frac{b_j^{a_j}}{\Gamma(a_j)}(\sdcenter)^{-a_j-1}\exp(-\frac{b_j}{\sdcenter})\\
		& \times  (m^2)^{-\kappa_1-1}\exp\left(-\frac{\kappa_2}{m^2}\right),
	\end{split}
\end{equation*}
which is not of a standard form and we apply adaptive Metropolis-Hastings (Section \ref{AMH-univariate}). Similar to $h_j$, a log transformation is applied. The logarithm of the full conditional density is 
\begin{equation*}
	\begin{split}
		\log \pi(m^2|\ldots) = & \sum_{j=1}^J \left(D a_j \log(b_j) - D\log(\Gamma(a_j))-(a_j+1) \sum_{d=1}^D \log(\sdcenter) -b_j \sum_{d=1}^D \frac{1}{\sdcenter}\right)\\
		&-(\kappa_1+1)\log(m^2)-\frac{\kappa_2}{m^2} + \text{const}.
	\end{split}
\end{equation*}
Hence the acceptance probability is
\begin{equation*}
	\begin{split}
		\alpha(m^2_{new},m^2_{old})&=\min \left( 1, \frac{\pi(m^2_{new})Q_n(m^2_{old}|m^2_{new})} {\pi(m^2_{old})Q_n(m^2_{new}|m^2_{old})}\right) \\
		&=\min \left( 1, \frac{\pi(m^2_{new})m^2_{new}} {\pi(m^2_{old})m^2_{old}}\right) \\
		&=\min \left(1, \exp \left[\log \pi(m^2_{new})-\log \pi(m^2_{old})+\log(m^2_{new})-\log(m^2_{old})\right]
		\right).
	\end{split}
\end{equation*}

\clearpage

\section{Clustering}
In this section, we summarize methods to obtain an optimal clustering from posterior samples, compare different clusterings, and introduce consensus clustering for better exploration of the posterior space of the clustering. We define posterior allocation probability which will be used to measure the uncertainty in allocations.

\subsection{Variation of Information (VI)} \label{sec:VI}
Let $\bmc$ and $\hat{\bmc}$ denote the true clustering and an estimate of the clustering, each consisting of $k$ and $k'$ clusters. Define $C_i$ ($i=1,\ldots,k$) to be the set of observation indices for cluster $i$ under $\bmc$, and $\hat{C}_j$ ($j=1,\ldots, k'$) is for cluster $j$ under $\hat{\bmc}$. The number of data points shared between $C_i$ and $\hat{C}_j$ is $n_{ij}=|C_i \cap \hat{C}_j|$, and the size of each cluster is $n_{i+}=\sum_{j=1}^{k'}n_{ij}$ under $\bmc$, and $n_{{+j}}=\sum_{i=1}^{k}n_{ij}$ under $\hat{\bmc}$.

The entropy $H(\bmc)$ of a clustering represents the uncertainty in assigning an unknown observation to clusters under $\bmc$, and the mutual information $I(\bmc,\hat{\bmc})$ between two clusterings measures the reduction in the uncertainty of the allocation of a data point in $\bmc$ if we know its allocation in $\hat{\bmc}$. They are defined as  
\begin{equation*}
	\begin{split}
		H(\bmc)&=-\sum_{i=1}^{k}\frac{n_{i+}}{n}\log\frac{n_{i+}}{n},\\
		I(\bmc,\hat{\bmc})&=\sum_{i=1}^{k}\sum_{j=1}^{k'}\frac{n_{ij}}{n}\log\frac{n_{ij}n}{n_{i+}n_{+j}},
	\end{split}
\end{equation*}
where $n$ is the total number of data points. Variation of information is defined as 
\begin{equation*}
	\begin{split}
		\text{VI}(\bmc,\hat{\bmc})=H(\bmc)+H(\hat{\bmc})-2I(\bmc,\hat{\bmc}).
	\end{split}
\end{equation*}

The optimal clustering $\bmc^*$ is 
\begin{equation*}
	\bmc^*=\arg \min_{\hat{\bmc}} \mathbb{E}\left[ \text{VI}(\bmc,\hat{\bmc})|\mathcal{D}\right],
\end{equation*}
where $\mathcal{D}$ is the data.

\subsection{Adjusted Rand Index (ARI)} \label{sec:ARI}
Following the notation in the last section, ARI between two clusterings $\bmc$ and $\hat{\bmc}$ is 
\begin{equation*}
	\text{ARI} = \frac{\sum_{ij} \binom{n_{ij}}{2} - \left[\sum_i \binom{n_{i+}}{2} \sum_j \binom{n_{+j}}{2}\right]/\binom{n}{2}}{\frac{1}{2} \left[\sum_i \binom{n_{i+}}{2} + \sum_j \binom{n_{+j}}{2}\right] - \left[\sum_i \binom{n_{i+}}{2} \sum_j \binom{n_{+j}}{2}\right]/\frac{n}{2}}.
\end{equation*}
Values closer to 1 indicate better agreement between $\bmc$ and $\hat{\bmc}$. Negative ARI is possible indicating less agreement by chance, but is of less interest. The ARI is a common metric used to compare the estimated clustering with the truth \citep{dahl2006model,fritsch2009improved,medvedovic2004bayesian}.

\subsection{Consensus Clustering} \label{sec:consensus-clustering}
In practice, it is common to run multiple MCMC chains with a large number of iterations to account for sensitivity to different initial values and to ensure convergence. Nevertheless, for high-dimensional data, chains can easily get trapped into local posterior modes even after sufficiently long time. To overcome such problems and reduce computational costs, \cite{Coleman2020} develop a general method to exploit the posterior distribution of data partitions through an ensemble of Bayesian clustering results. The method does not require the chain to reach convergence and hence is expected to relieve computational burden. In addition, it can be readily integrated into existing Bayesian clustering frameworks without requiring substantial redevelopment of the original method.

The core idea behind consensus clustering is to run a large number of chains, denoted as chain width $W$, each for a small number of iterations, denoted as chain depth $D$. Then the $D$-th sample in each chain is combined to produce a posterior similarity matrix (PSM), based on which the optimal clustering can be obtained, e.g. by minimizing VI.

\cite{Coleman2020} propose a heuristic way to choose appropriate values for $W$ and $D$. The rational is that increasing $W$ and $D$ may improve the performance substantially in the beginning, but the improvement will gradually diminish with further increases. This is similar to PCA where more variance will always be captured for more principal components, but the gain in variance will be smaller and smaller, and eventually we will have few returns. 

To choose $D$ and $W$, one begins with candidate sets $D'=(d_1,\ldots,d_I)$ and $W'=(w_1,\ldots,w_J)$ arranged in increasing order. For a given number of chains $w_j$, the PSM is computed based on the samples at the $d_{i}$-th iteration across $w_j$ chains, and compared with the PSM obtained from the $d_{(i-1)}$-th iteration across $w_j$ chains. The mean absolute difference (MAD) between the two matrices is a measurement of how stable the clustering is. Plotting these values as a function of $D$, it is expected to see an elbow-shaped curve, and a suitable $D$ can be selected at which the curve plateaus. Similarly, to choose $W$, we can fix $D$ and compute MAD between $w_{(j-1)}$ and $w_j$.

\clearpage
\section{Simulation Study} \label{appendix:C-HDP-simualtion-result}

\begin{figure}[tbp]
	\centering
	\includegraphics[width=0.9\textwidth]{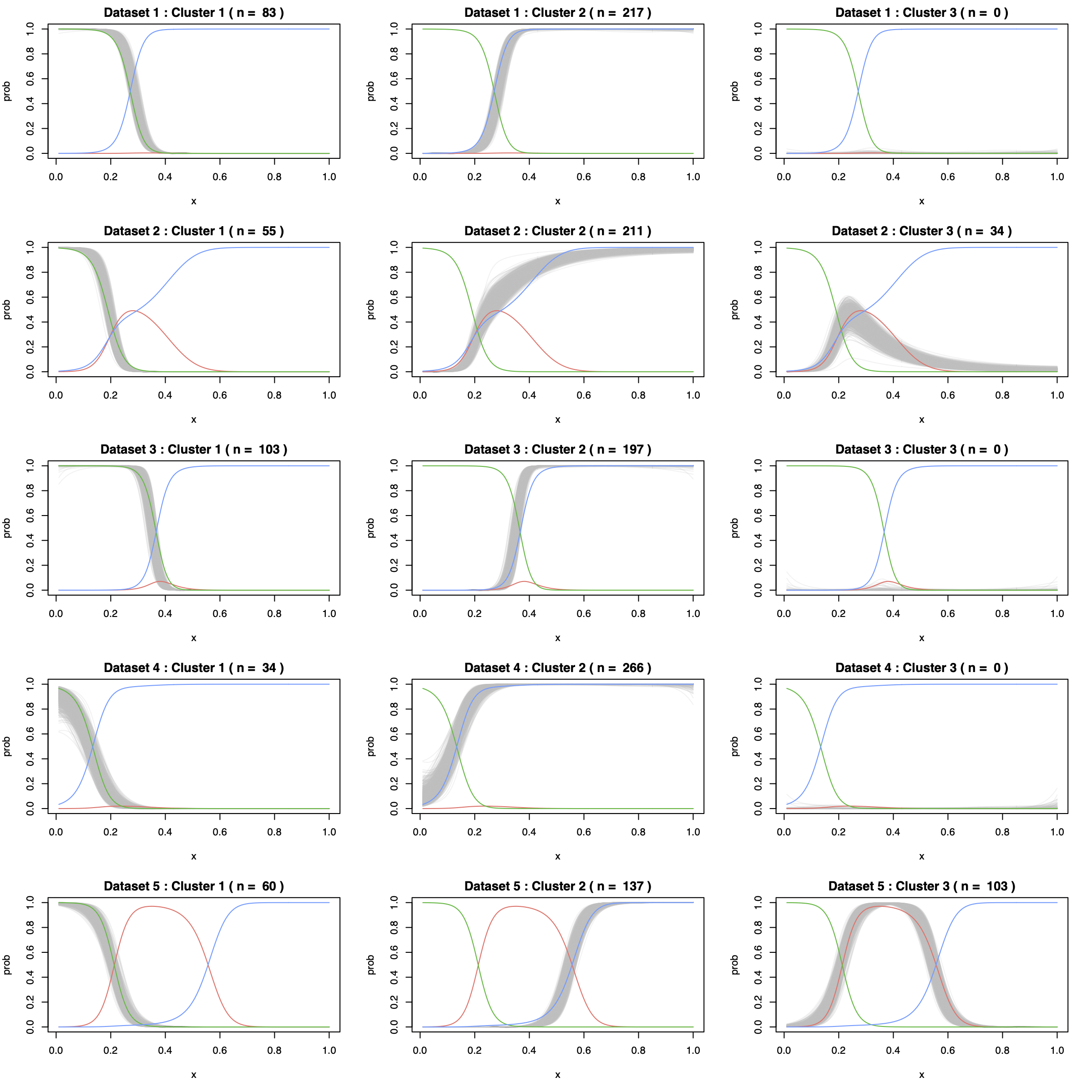}
	\caption{C-HDP: Posterior samples of the covariate-dependent probabilities (grey) for each cluster in each dataset in a selected replicate. In each panel, the true relationships for three clusters are shown in colored lines. The size of each estimated cluster is indicated in the title.}
	\label{fig:10_px_chdp}
\end{figure}

\begin{figure}[tbp]
	\centering
	\includegraphics[width=0.9\textwidth]{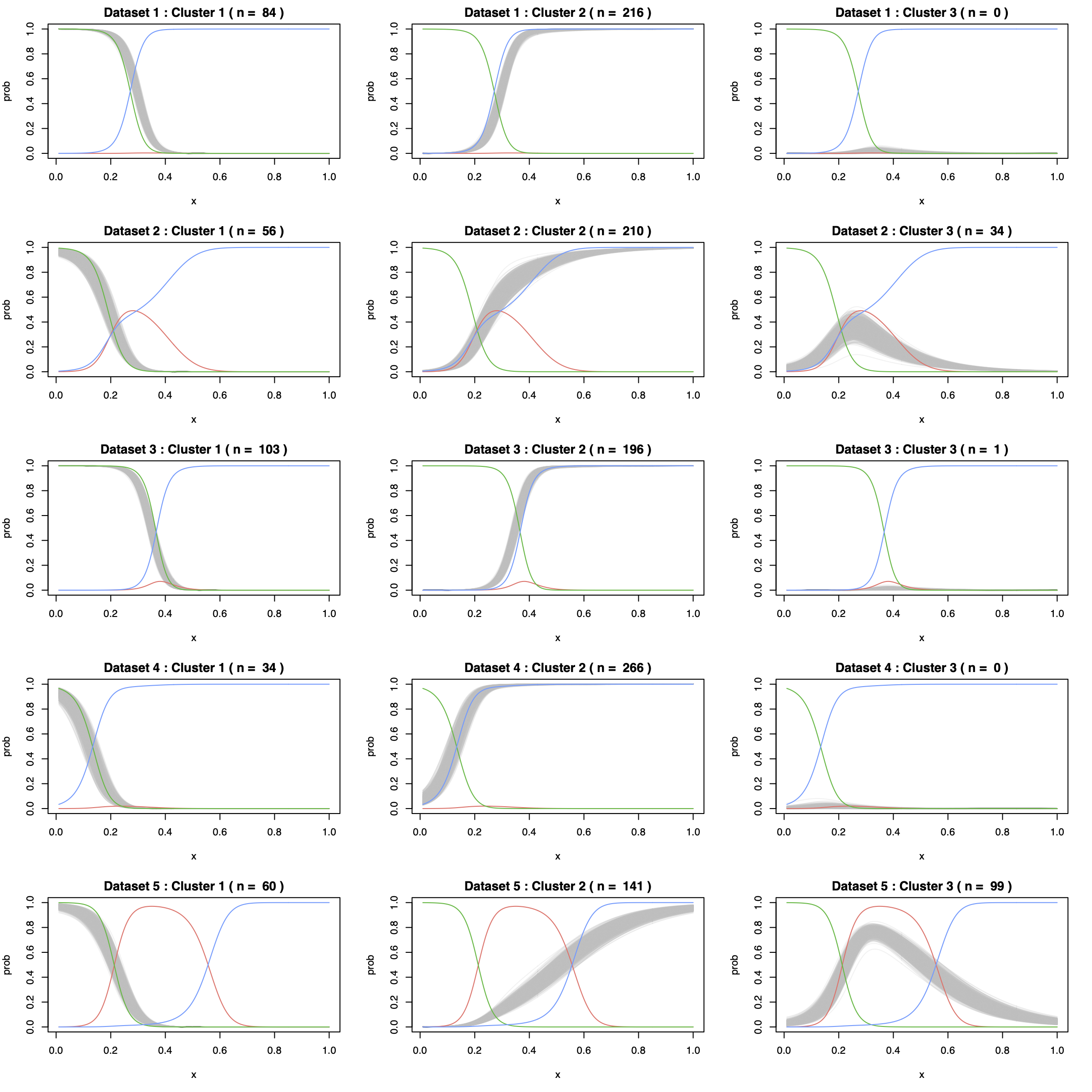}
	\caption{DDP (with group): Posterior samples of the covariate-dependent probabilities (grey) for each cluster in each dataset in a selected replicate. In each panel, the true relationships for three clusters are shown in colored lines. The size of each estimated cluster is indicated in the title.}
	\label{fig:10_px_ddp_cat}
\end{figure}

\begin{figure}[tbp]
	\centering
	\includegraphics[width=0.9\textwidth]{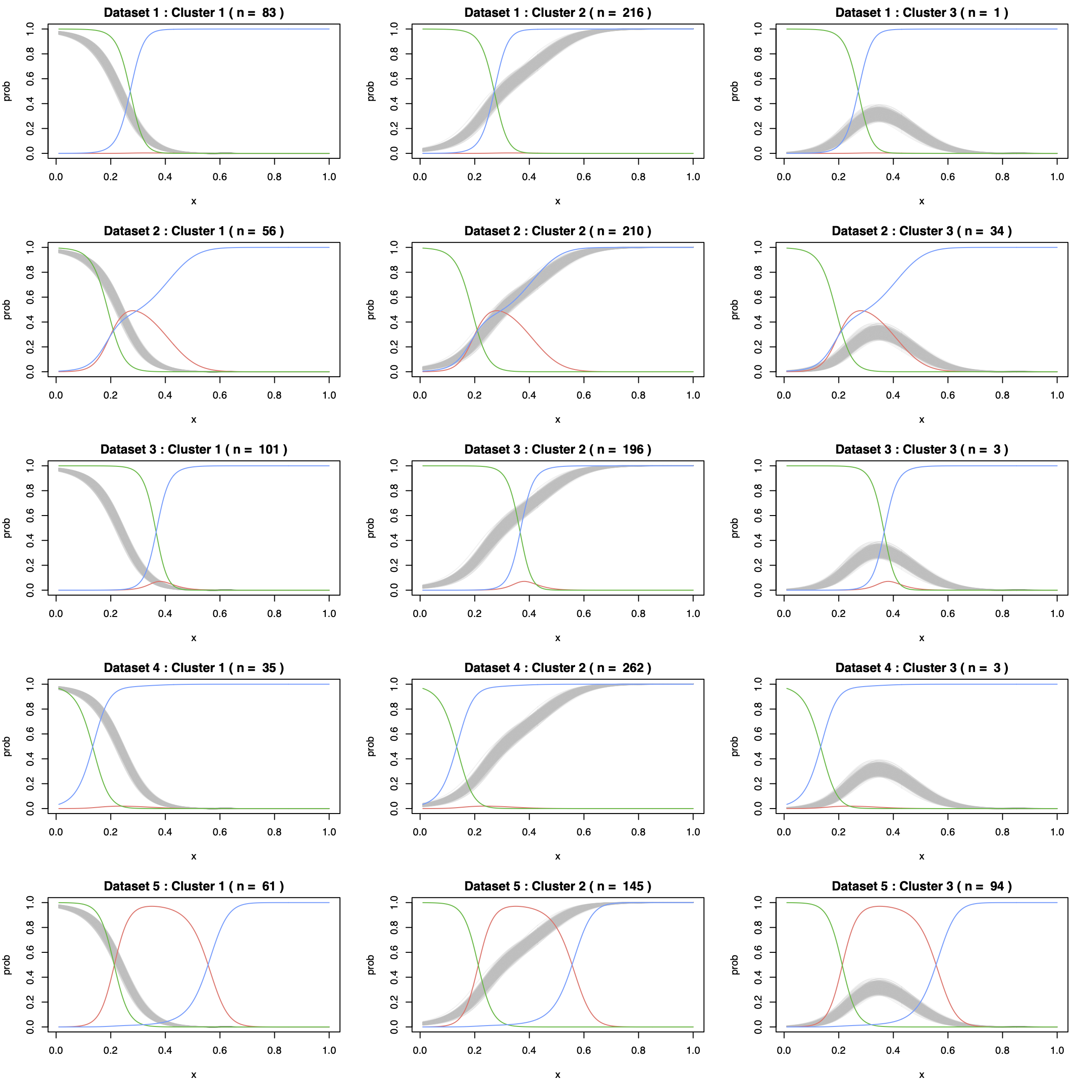}
	\caption{DDP (without group): Posterior samples of the covariate-dependent probabilities (grey) for each cluster in each dataset in a selected replicate. In each panel, the true relationships for three clusters are shown in colored lines. The size of each estimated cluster is indicated in the title.}
	\label{fig:10_px_ddp}
\end{figure}

\begin{figure}[tbp]
	\centering
	\includegraphics[width=0.9\textwidth]{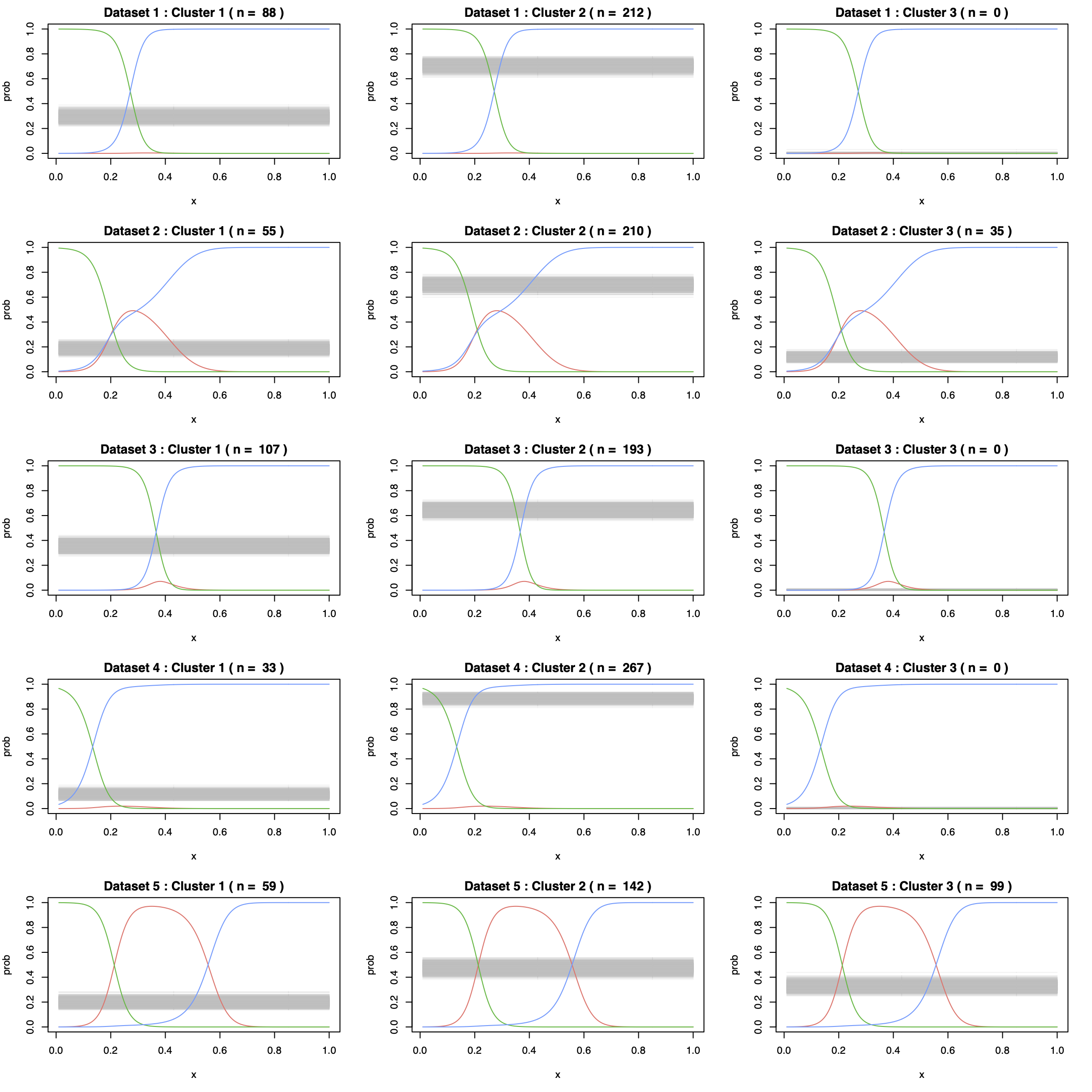}
	\caption{HDP: Posterior samples of the probabilities (grey) for each cluster in each dataset in a selected replicate. In each panel, the true relationships for three clusters are shown in colored lines. The size of each estimated cluster is indicated in the title.}
	\label{fig:10_px_hdp}
\end{figure}

In the first part of this section, we provide additional results for the simulation study on the Gaussian mixtures discussed in the manuscript. Recall that the true relationship between the weights and the covariate follows a softmax function, while a Gaussian kernel is applied in the C-HDP and DDP. Nevertheless, Figure \ref{fig:10_px_chdp} shows that the covariate-dependent probabilities can be accurately estimated in the C-HDP, with the truth mostly covered by the samples. On the other hand, although the group indicator is included in the DDP as an additional categorical covariate, the posterior samples may still fail to cover the true relationship, e.g. in clusters 2 and 3 of Dataset 5 (Figure \ref{fig:10_px_ddp_cat}). When it comes to the DDP without using the group indicator (Figure \ref{fig:10_px_ddp}), the estimated relationship is the same across groups, which appears as an average across datasets and exhibits much smaller uncertainty in the posterior samples. Finally, as the HDP does not account for the covariate, the probabilities are only constant with $x$, showing large uncertainty (Figure \ref{fig:10_px_hdp}).

\begin{figure}[tbp]
	\centering
	\includegraphics[width=1\textwidth]{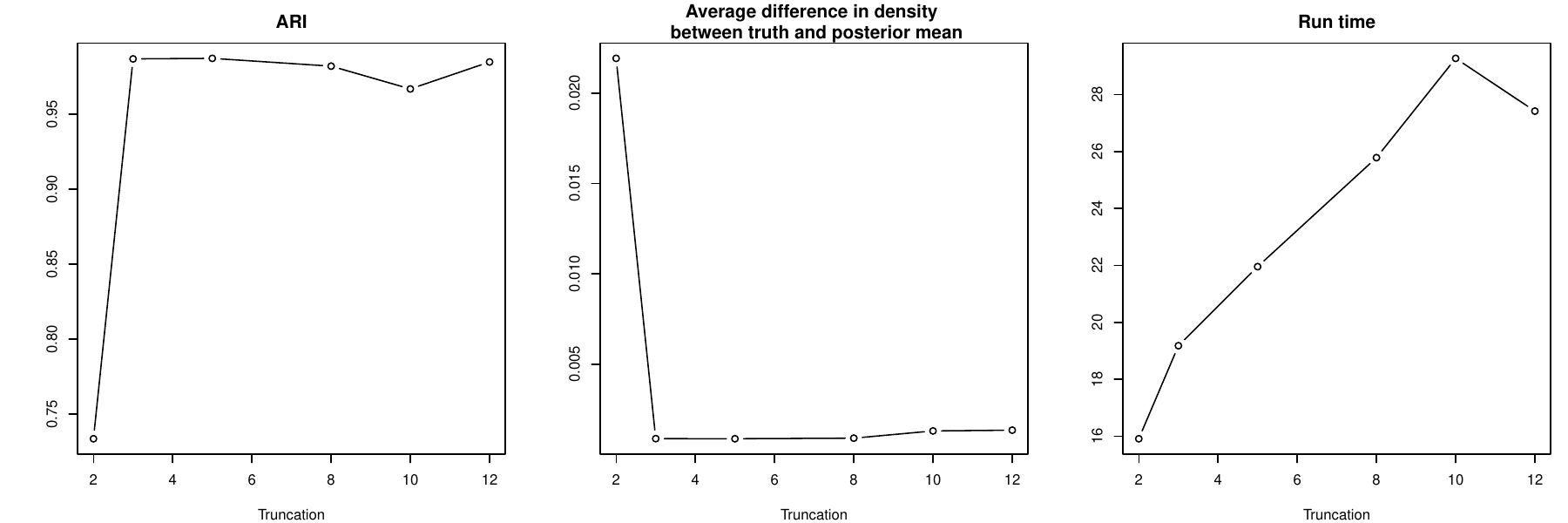}
	\caption{For a dataset simulated from Gaussian mixtures with three clusters, adjusted rand index (ARI) comparing the truth to the estimated clustering (left), average difference in the density between the truth and posterior mean across observations (middle), and computational run time in minutes versus the truncation level used in the C-HDP (right).}
	\label{fig:convergence_plot}
\end{figure}

	Further, to better understand how the quality of the approximation depends on the truncation level $J$, we compare the performance of the C-HDP across different truncation levels. The results demonstrate that inference is stable (Figure \ref{fig:convergence_plot} left for clustering and middle for density estimation) as long as we use a large enough $J$, which in this case is three as the true number of components is three. The computational run time is approximately linear with $J$.

Next, we conduct more simulation studies to assess the covariate-dependent HDP model. Two settings are considered for two different types of kernels: a Gaussian kernel and a periodic kernel, and two types of likelihoods: negative-binomial and vector autoregression. Posterior inference is performed based on the MCMC algorithms detailed in Section \ref{sec:full-conditionals-pax6} and Section \ref{sec:full-conditionals-cidata}. 

We are interested in the posterior inference of the clustering, the covariate-dependent probabilities, and component-specific parameters. For the clustering, we compare the optimal clustering with the true clustering based on adjusted Rand index (ARI) and variation of information (VI; normalized on $[0,1]$). A large value for ARI and a small value for VI suggest good performance. For covariate-dependent probabilities, we check if the true relationship is covered by the posterior samples. Credible intervals are computed to verify if parameters can be correctly estimated.

\subsection{Simulation Setting 1: A Gaussian Kernel} \label{appendix:simulation-pax6}
Below we consider three simulation scenarios for the single-cell clustering model. The first simulated dataset is generated from the proposed model to test the ability to recover true clustering and parameters, especially the time-dependent probabilities. For the second simulated data, we investigate the robustness of the model under misspecification of the relationship between the probability and covariate. Simulation 3 is used to investigate if marker genes can be correctly identified based on the probabilistic approach described in in Section \ref{sec:marker-genes-method}.

\subsubsection{Simulation 1 and 2}

In Simulation 1, the relationship between the probability and covariate is based on the proposed Gaussian kernel, while in Simulation 2, the relationship follows from a logistic regression. In each simulation, two synthetic datasets have the same number of cells $C_1=C_2=100$ and genes $G=10$, and two clusters are generated. To avoid excessive zero counts in the data, we generate capture efficiencies with a mean of 0.6 rather than 0.06 suggested in bayNorm \citep{Tang2020}. The empirical estimates for capture efficiencies in bayNorm are proportional to cell specific global scaling factors with a default global mean of 0.06 (Section \ref{sec:baynorm_beta}). However, due to the identifiability problem of $\mu_{j,g}^*$ and $\beta_{c,d}$, an informative prior for capture efficiencies is used by setting the global mean capture efficiency in bayNorm to 0.6. The data-generating process is detailed below.

\paragraph{Data-generating Process}

In simulation 1 and 2, the data is generated from the following:
\begin{equation*}
	\begin{split}
		y_{c,g,d}|y_{c,g,d}^0, \beta_{c,d}&\indsim \Bin( y^0_{c,g,d}, \beta_{c,d}), \\
		y_{c,g,d}^0|z_{c,d} = j,  \mu_{j,g}^*, \phi_{j,g}^*  &\indsim \NB(\mu_{j,g}^*, \phi_{j,g}^*),\\
		z_{c,d}| p_{1,d}^J(t_{c,d}), \ldots, p_{J,d}^J(t_{c,d})&\indsim \Cat(p_{1,d}^J(t_{c,d}),\ldots, p_{J,d}^J(t_{c,d})),\\
		t_{c,d} &\iidsim \Unif(0,1).
	\end{split}
\end{equation*}
The component-specific parameters and capture efficiencies are simulated from
\begin{equation*}
	\begin{split}
		\mu_{j,g}^*|j=1 \iidsim \logN(1,\alpha_{\mu}^2),&\quad \mu_{j,g}^*|j=2 \iidsim \logN(3,\alpha_{\mu}^2), \\
		\phi_{j,g}^* | \mu_{j,g}^* &\indsim  \logN(b_0 + b_1 \log(\mu_{j,g}^*), \alpha_\phi^2), \\
		\beta_{c,d} &\indsim \Beta(a_d^\beta,b_d^\beta).\\
	\end{split}
\end{equation*}
where $b_0=0.25, b_1=0.5, \alpha_{\mu}^2=\alpha_{\phi}^2=0.1, a_d^\beta=3$ and $ b_d^\beta=2$.

For simulation 1, the time-dependent probabilities are based on the proposed Gaussian kernel with the following parameters:
\begin{equation*}
	\begin{split}
		\text{Dataset 1}: \quad (t_{1,1}^*,t_{2,1}^*)&=(0.4,0.9),\quad (\sigma_{1,1}^*,\sigma_{2,1}^*)=(0.08,0.15),\quad (q_{1,1},q_{2,1})=(0.5,0.5),\\
		\text{Dataset 2}: \quad (t_{1,2}^*,t_{2,2}^*)&=(0.8,0.3),\quad (\sigma_{1,2}^*,\sigma_{2,2}^*)=(0.1,0.1),\quad \quad (q_{1,2},q_{2,2})=(0.3,0.7).
	\end{split}
\end{equation*}

For simulation 2, the time-dependent probabilities $p_{j,d}^J(t)$ are based on a logistic regression where the probability of belonging to the first cluster $(j=1)$ in each dataset is given by
\begin{equation*}
	\begin{split}
		p_{1,1}^J(t) = \frac{1}{1+\exp(-4+20t^2)}, \quad    p_{1,2}^J(t) = \frac{1}{1+\exp(4-10t)},
	\end{split}
\end{equation*}
and $p_{2,d}^J(t)=1-p_{1,d}^J(t)$.

\begin{figure}[tbp]
	\centering
	\includegraphics[width=0.7\textwidth]{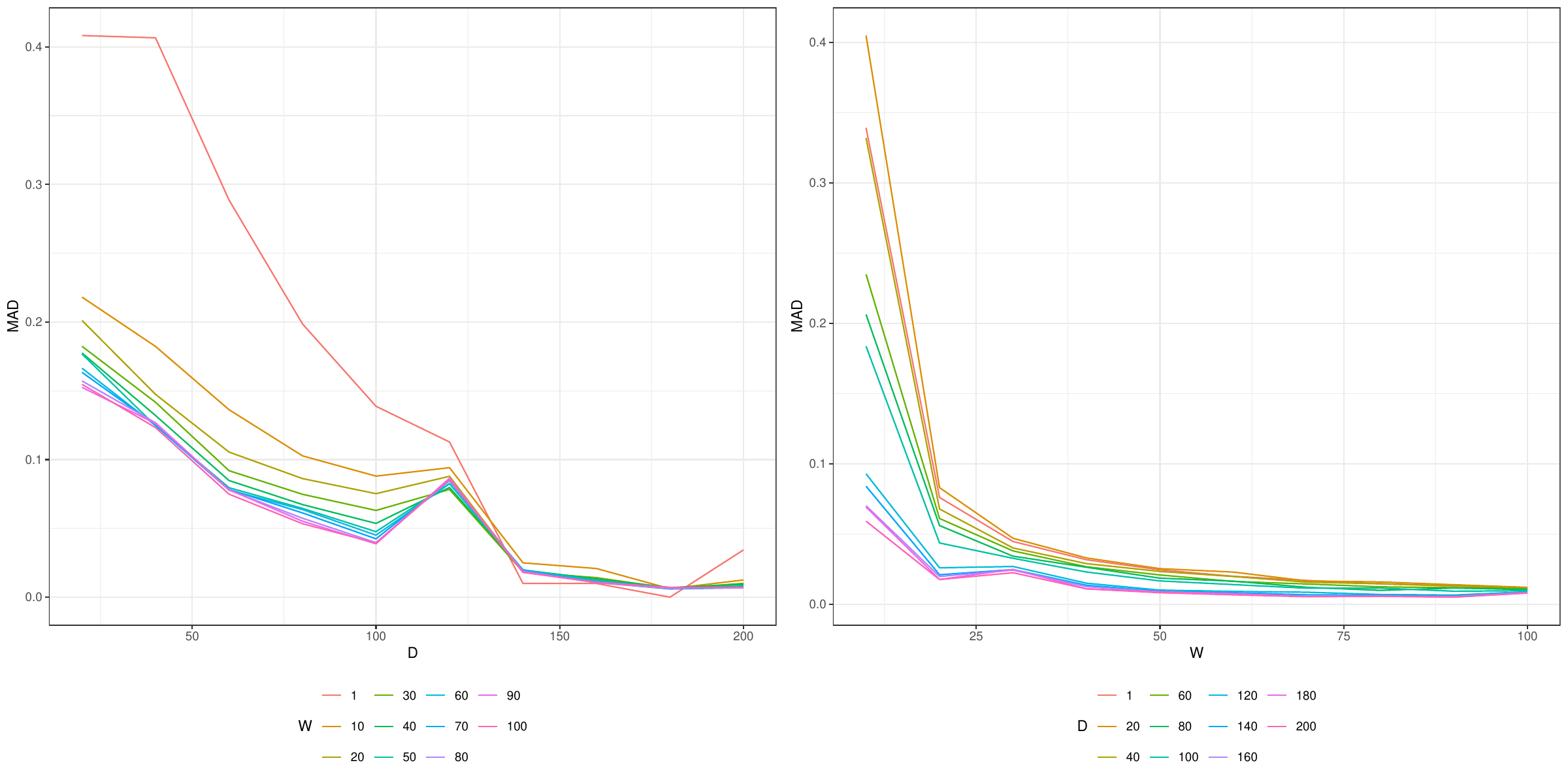}
	\includegraphics[width=0.7\textwidth]{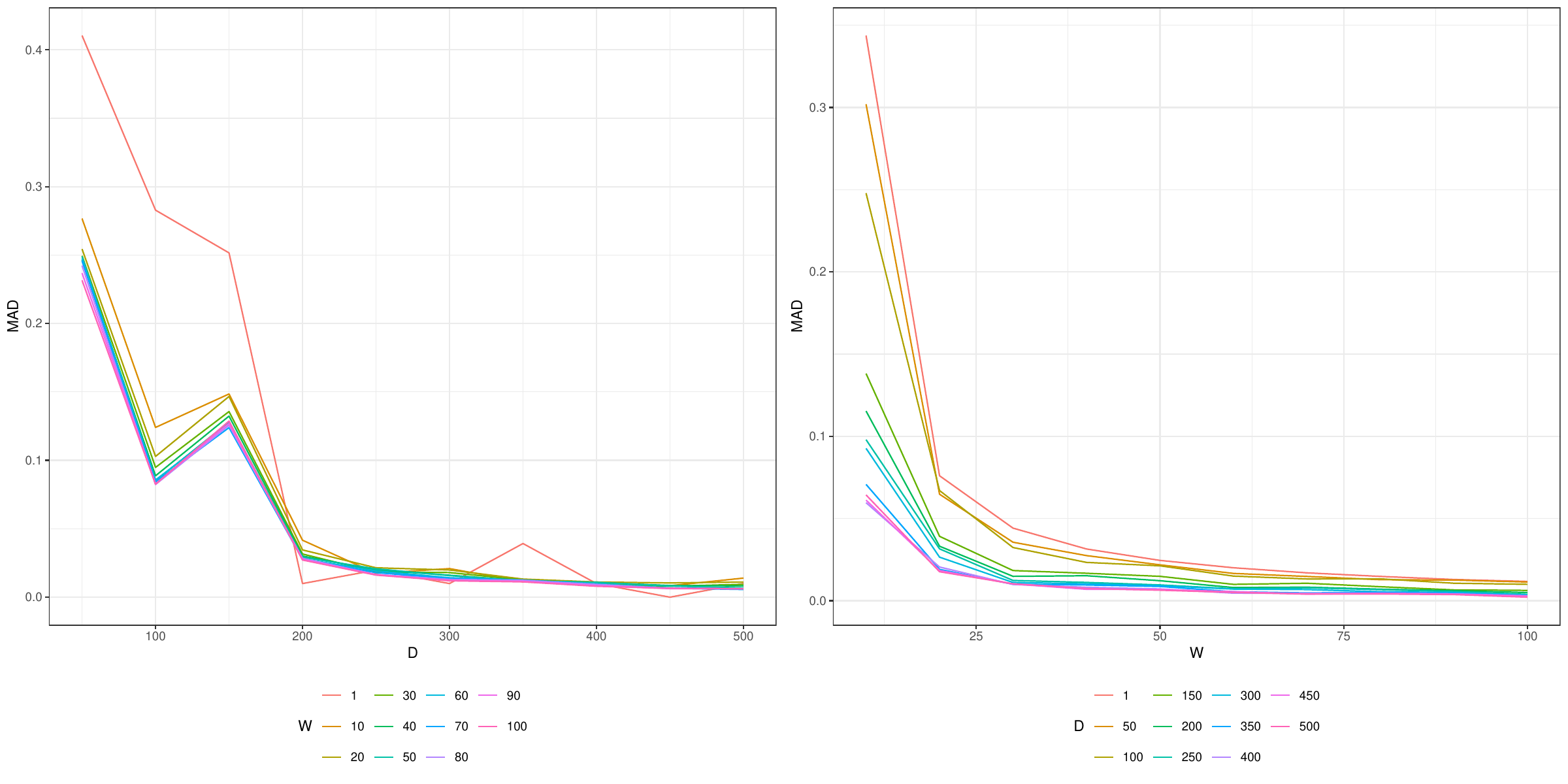}
	\caption{Choice of $W$ and $D$ in consensus clustering for Simulation 1 (top) and Simulation 2 (bottom).}
	\label{fig:sim1_6_1_2_MAD}
\end{figure}

\begin{figure}[tbp]
	\centering
	\includegraphics[width=0.8\textwidth]{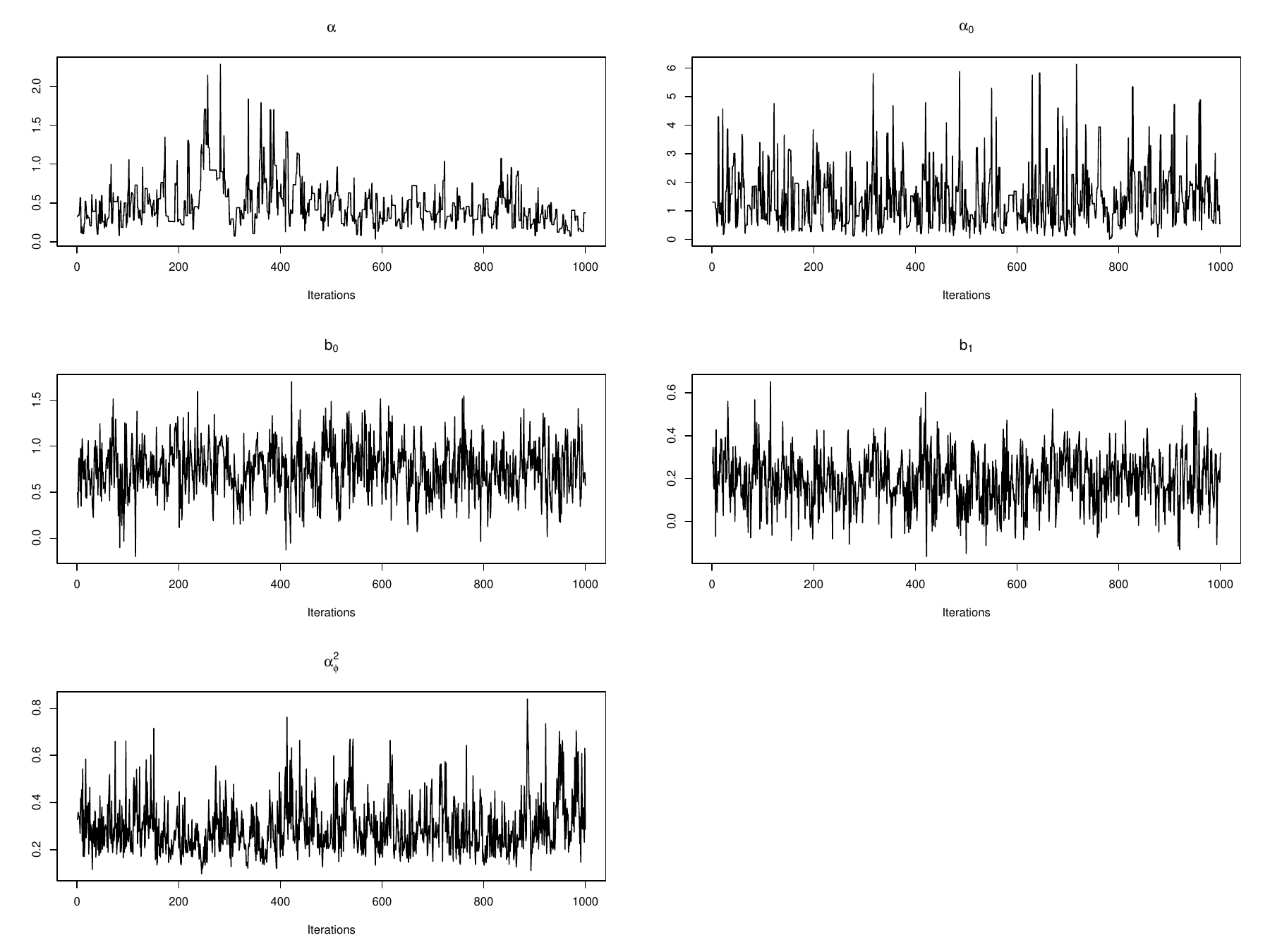}
	\includegraphics[width=0.8\textwidth]{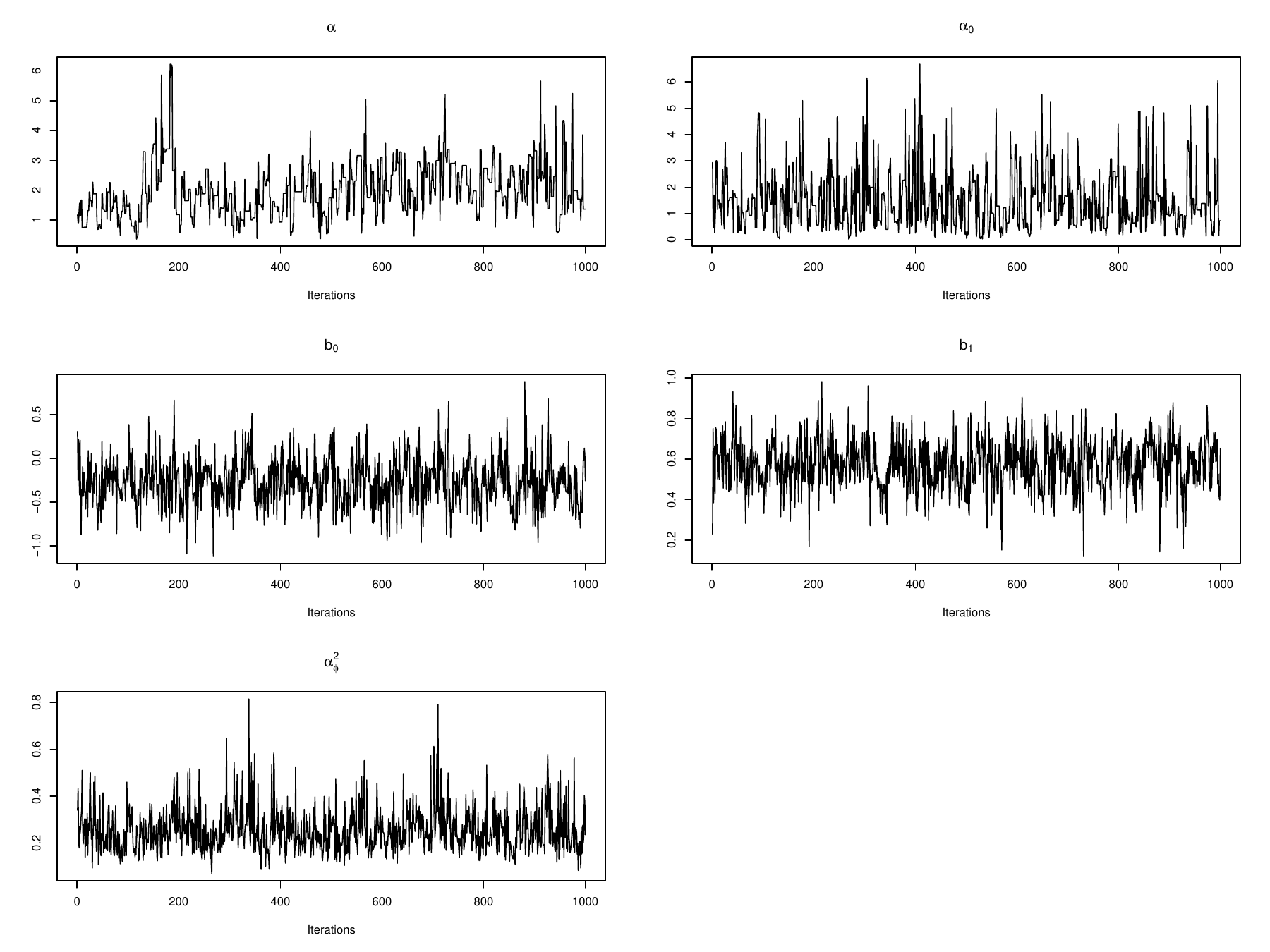}
	\caption{Traceplots for concentration parameters and regression parameters, for Simulation 1 (top 3 rows) and 2 (bottom 3 rows).}
	\label{fig:sim1_6_1_2_trace}
\end{figure}

To fit the proposed model, we use a truncation level at $J=4$ in both settings. Consensus clustering is performed with 100 parallel chains for both settings, and 200 iterations and 500 iterations, respectively (see Figure \ref{fig:sim1_6_1_2_MAD} for the choice of tuning parameters). The MCMC setup for the post-processing step with a fixed clustering is the same in both simulation scenarios. One chain of length 10000 is run, and the first 8000 iterations are thrown away, followed by a thinning of 2. Traceplots for the post processing step is shown in Figure \ref{fig:sim1_6_1_2_trace}, suggesting convergence. 

\paragraph{Results}

\begin{figure}[tbp]
	\centering
	\begin{minipage}[h]{0.4\textwidth}
		\includegraphics[width=0.95\textwidth]{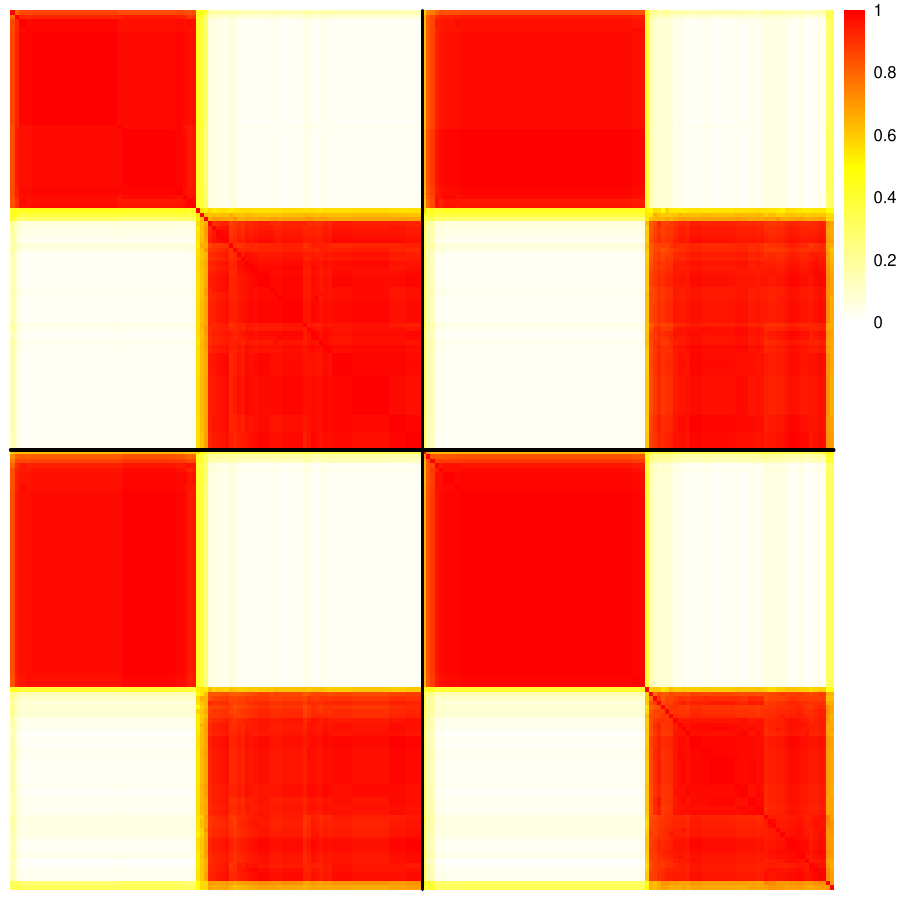}
	\end{minipage}
	\begin{minipage}[h]{0.4\textwidth}
		\includegraphics[width=0.95\textwidth]{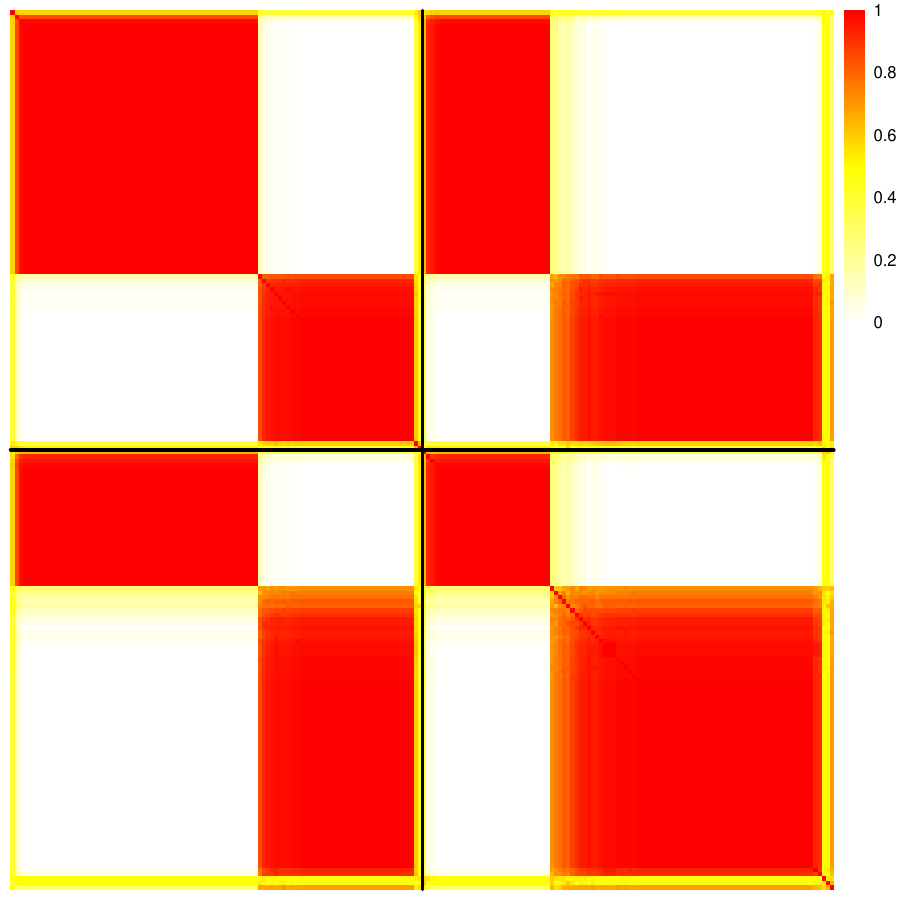}	
	\end{minipage}
	\caption{Posterior similarity matrix for Simulation 1 (left) and Simulation 2 (right). Diagonal blocks correspond to within-group PSM. The black solid line separates two datasets.}
	\label{fig:sim1_6_1_2_psm}
\end{figure}

The posterior similarity matrix from consensus clustering is shown in Figure \ref{fig:sim1_6_1_2_psm}, suggesting some uncertainty in cell allocations. 
Further, we compare our method with four popular methods for clustering scRNA-seq data: 1) Seurat \citep{satija2015spatial}, 2) CIDR \citep{lin2017cidr}, 3) TSCAN \citep{ji2016tscan}, and 4) SC3 \citep{kiselev2017sc3}. Table \ref{tab:sim1_6_1_2_comparison} shows that our C-HDP method performs the best in both simulation settings, with ARI close to 1 and VI close to 0, followed by SC3, TSCAN, Seruat and CIDR.

\begin{table}[h]
	\centering
	\caption{Simulation 1 and 2: Comparison of different estimated clusterings from the proposed C-HDP model and other four competing methods, based on ARI and VI. The best result is highlighted in bold.}
	\scalebox{1}{
		\begin{tabular}{llllll}
			& C-HDP & Seurat & CIDR & TSCAN & SC3 \\ 
			\hline
			Simulation 1 - ARI &\textbf{0.9602}& 0.5908& 0.0712& 0.8456& 0.9020    \\               
			Simulation 2 - ARI &\textbf{0.9020}& 0.4874& 0.0826& 0.7911& 0.8272    \\       
			Simulation 1 - VI &\textbf{0.0185}&  0.1344 & 0.3419& 0.0532& 0.0377    \\       
			Simulation 2 - VI &\textbf{0.0423}& 0.1589 & 0.4080& 0.0664& 0.0578    \\       
			\hline
		\end{tabular}
	}
	\label{tab:sim1_6_1_2_comparison}
	\bigskip
\end{table}

\begin{figure}[tbp]
	\includegraphics[width=0.95\textwidth]{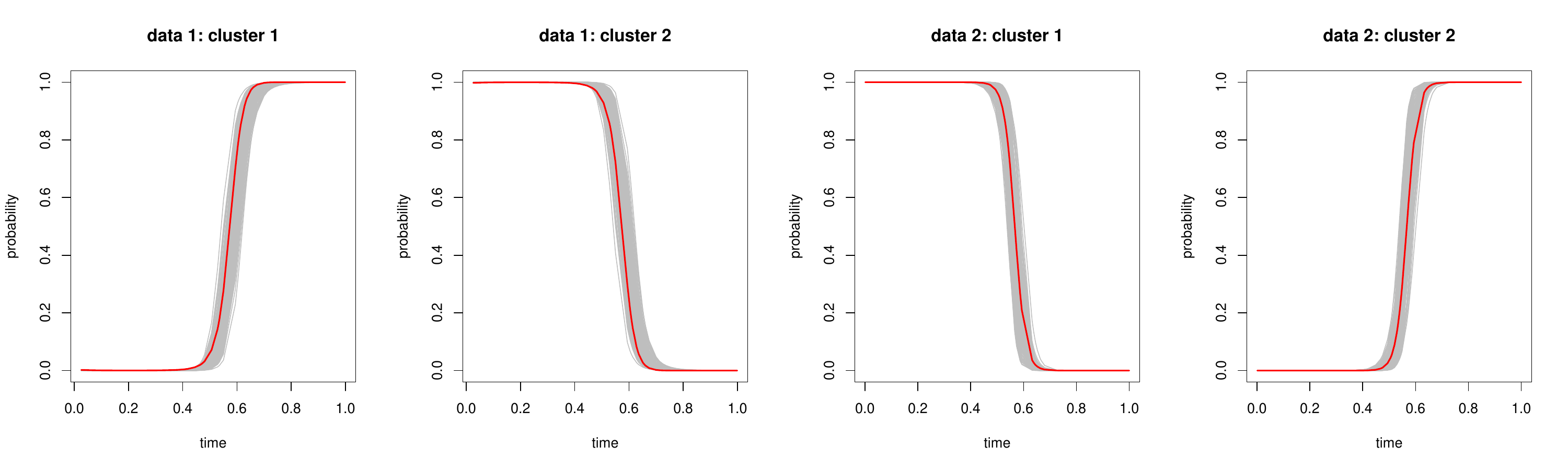}
	\includegraphics[width=0.95\textwidth]{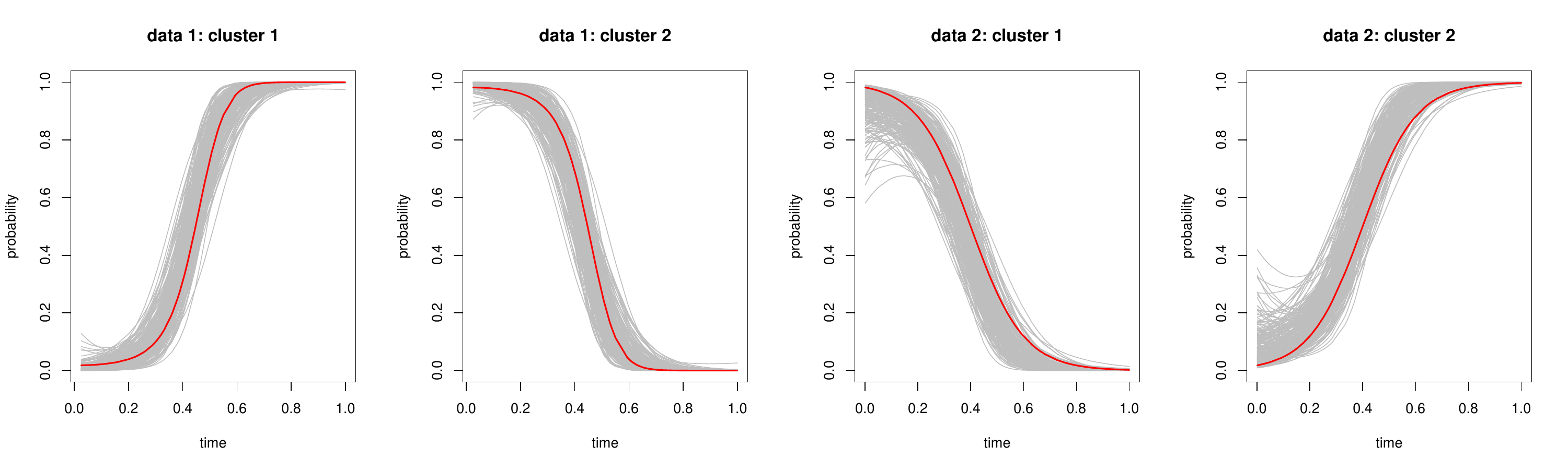}	
	\caption{Posterior samples for time-dependent probabilities for Simulation 1 (top) and Simulation 2 (bottom). The red solid line denotes the truth.
	}
	\label{fig:sim1_6_1_2_p_vs_t}
\end{figure}

\begin{figure}[tbp]
	\centering
	\includegraphics[width=0.9\textwidth]{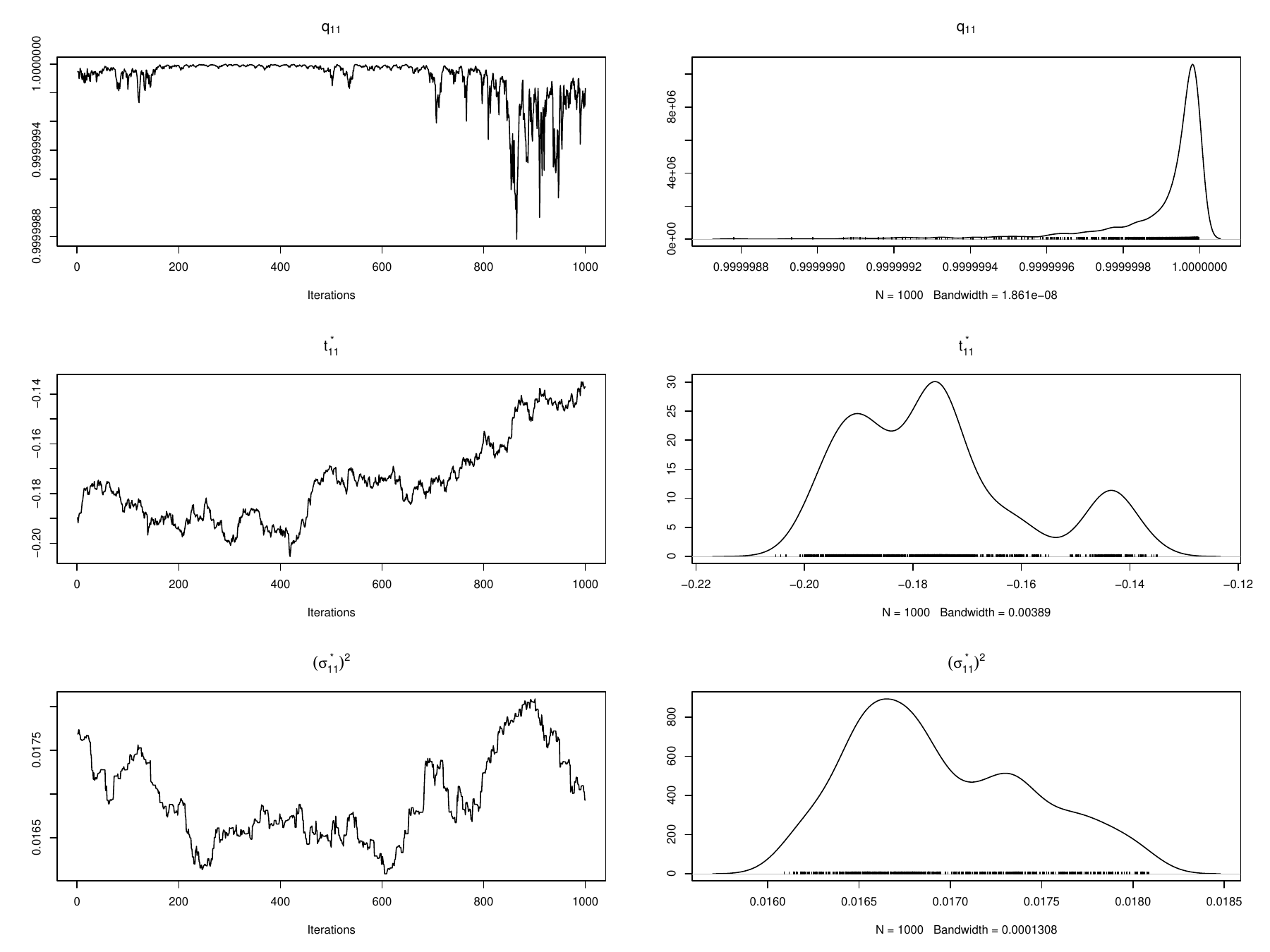}
	\caption{Traceplots and density plots for normalized $q_{1,1}$ and kernel parameters $t_{1,1}^*, {\sigma_{1,1}^*}^2$ for Simulation 1. True values are not covered.}
	\label{fig:sim1_6_1_kernel}
\end{figure}

As for time-dependent probabilities, Figure \ref{fig:sim1_6_1_2_p_vs_t} shows that the true relationship is covered by the posterior samples in both simulation scenarios, implying the robustness of the proposed kernel-based constructions for covariate-dependent probabilities. The posterior variability is comparatively smaller in the first simulation setting, which is probably because the true probabilities change more abruptly with latent time in Simulation 1 (probability close to 0 or 1), showing more decisive cluster classifications. In addition, it is worth noticing that the individual kernel parameters $\tcenter, \sdcenter$ and $q_{j,d}$ may have identifiability issue (Figure \ref{fig:sim1_6_1_kernel}), where true values are not covered by the posterior samples, but the relationship between the probability and weight can still be accurately recovered. 

\begin{figure}[tbp]
	\centering
	\subfigure[Simulation 1]{\includegraphics[width=0.95\textwidth]{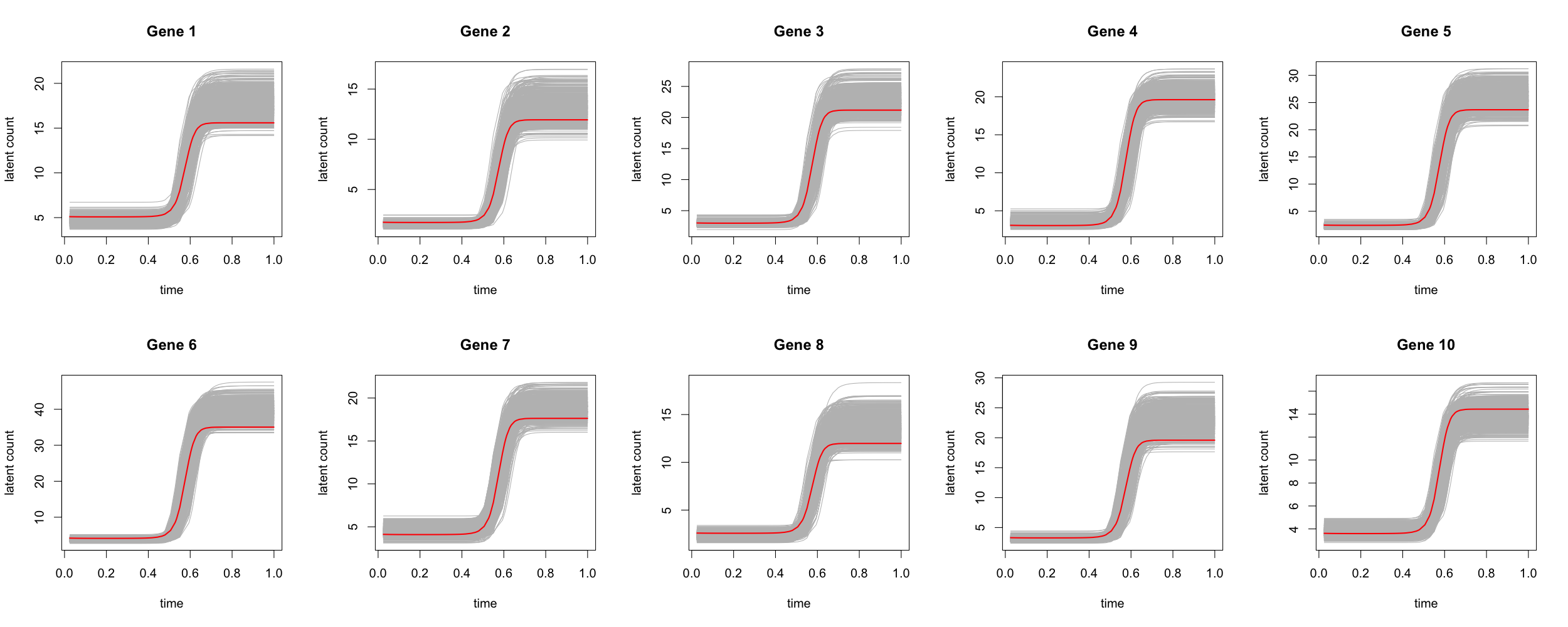}}
	\subfigure[Simulation 2]{\includegraphics[width=0.95\textwidth]{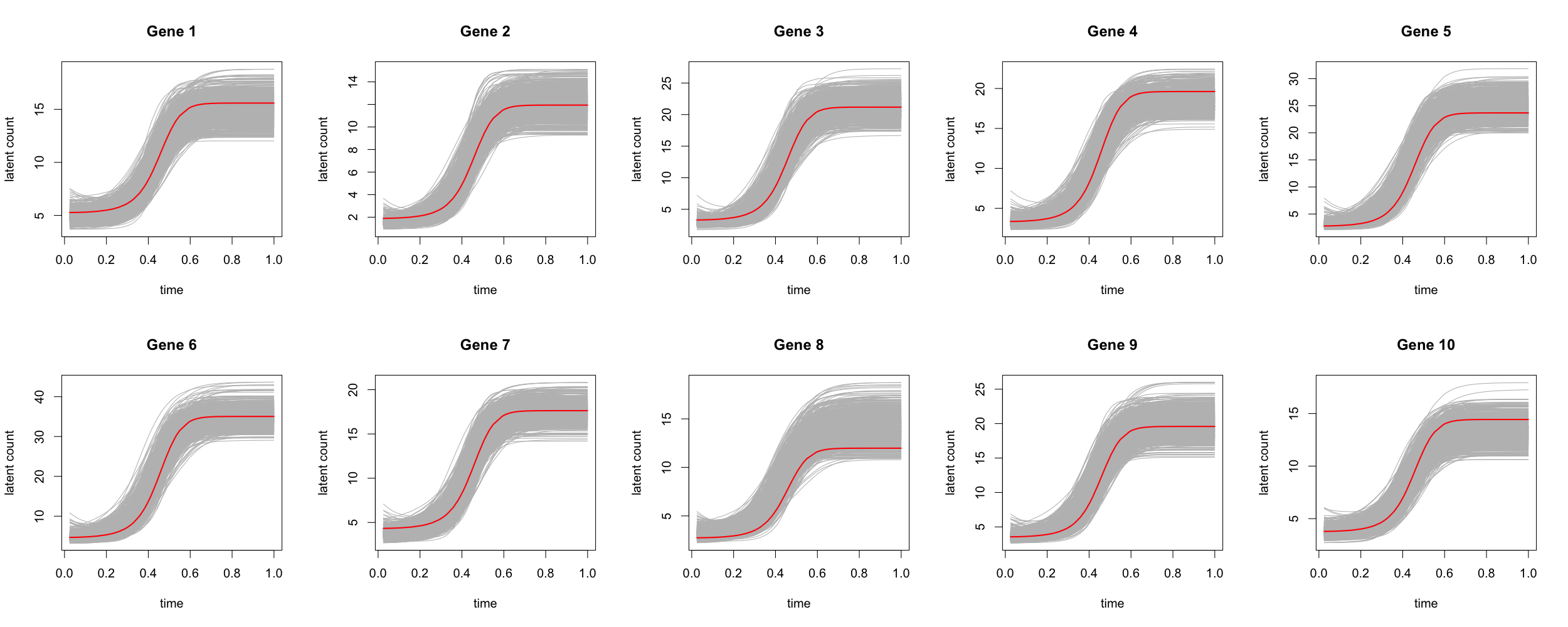}}
	\caption{Estimated latent counts against time for the first dataset in Simulation 1 (top 2 rows) and Simulation 2 (bottom 2 rows). The red solid line denotes the truth.}
	\label{fig:sim1_6_1_2_mean_latent_count_against_time}
\end{figure}

Next, we compute the expected count for a new cell $c$ from dataset $d$ as a function of latent time:
\begin{equation*}
	\E(y_{c,g,d}^0|t_{c,d}=t,\bX, \bY)= \int \sum_{j=1}^{J}p_{j,d}^J (t) \mu_{j,g}^* d \pi(\bmq_{1:J,d}, \bmu_{1:J,g}^*, \bpsi_{1:J,d}^* | \bX, \bY),
\end{equation*}
which is approximated from the MCMC samples. Figure \ref{fig:sim1_6_1_2_mean_latent_count_against_time} shows that the truth is well covered by the posterior samples. The large uncertainty in time-dependent probabilities for Simulation 2 is propagated into the expected count.

\begin{figure}[tbp]
	\centering
	\subfigure[Simulation 1]{\includegraphics[width=0.66\textwidth]{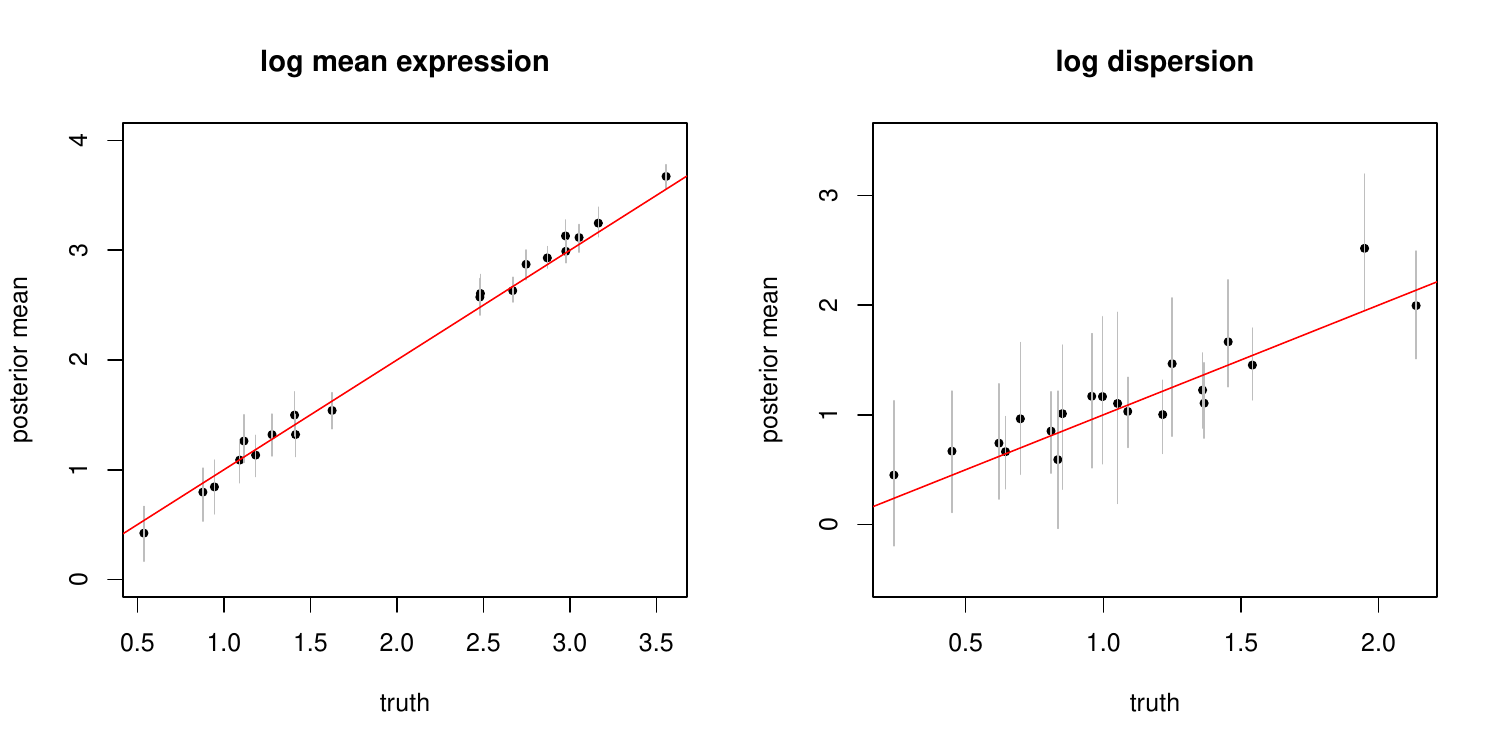}
		\includegraphics[width=0.33\textwidth]{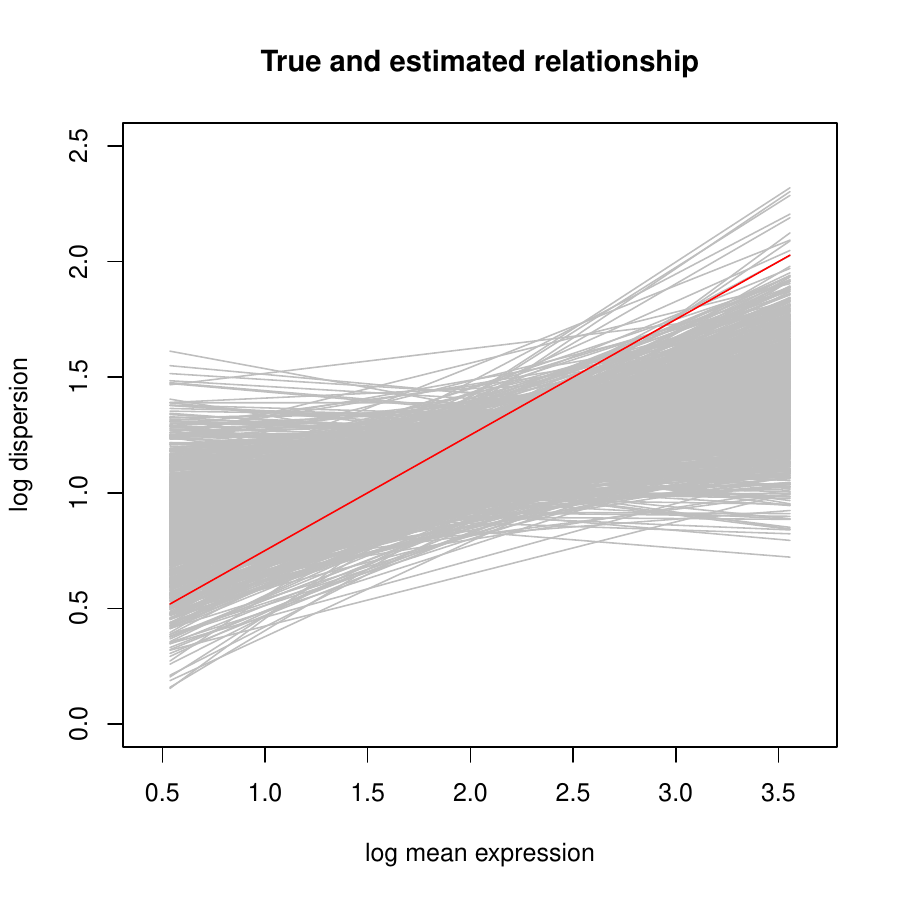}}
	\subfigure[Simulation 2]{\includegraphics[width=0.66\textwidth]{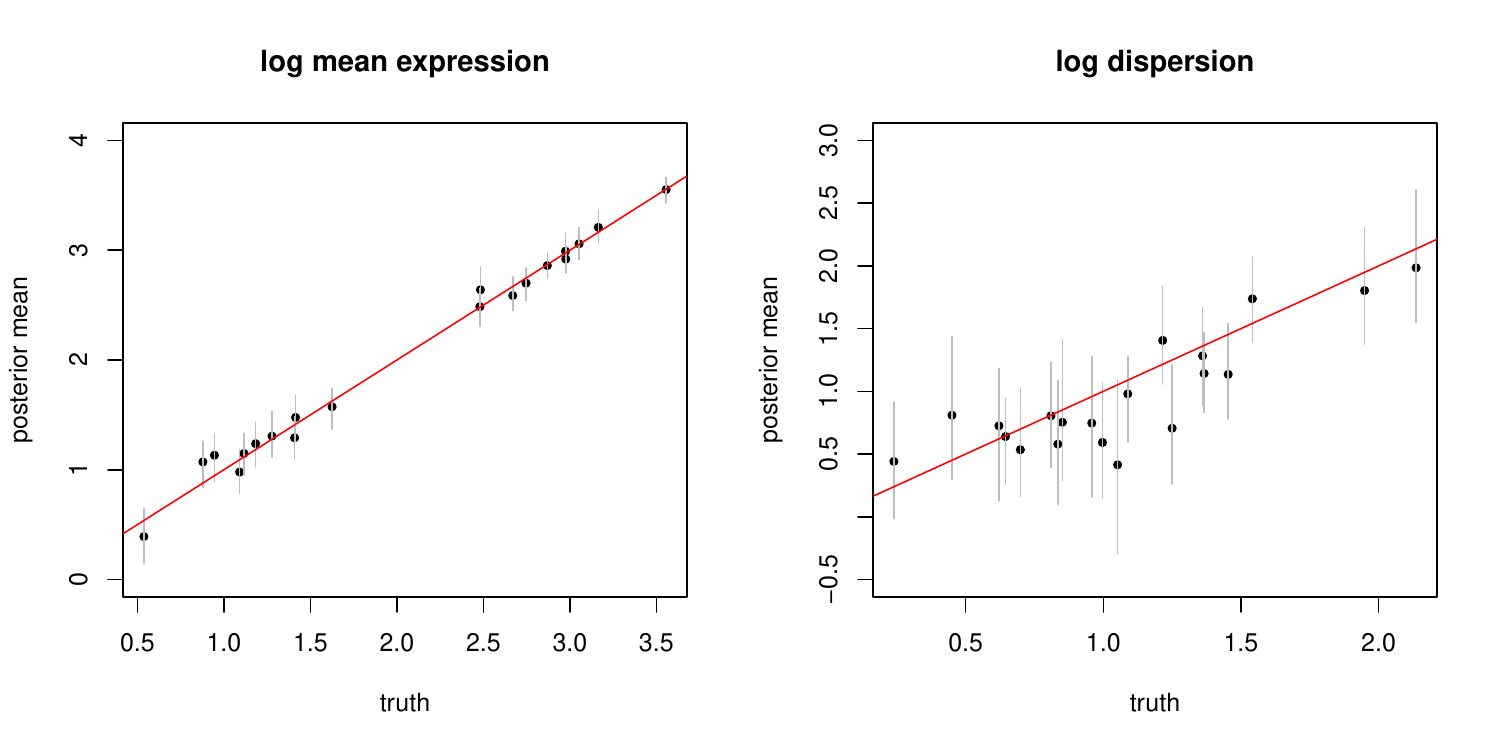}
		\includegraphics[width=0.33\textwidth]{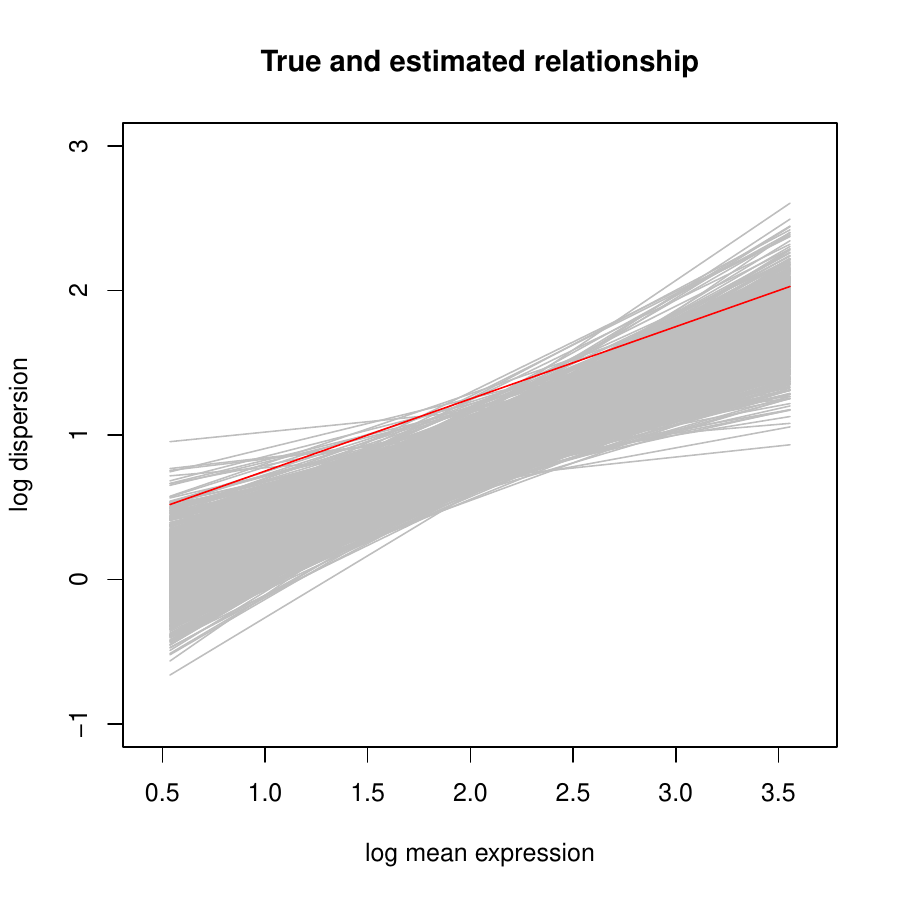}}
	\caption{Left: Posterior mean of log mean expression against truth, with 95\% HPD CIs shown in grey. Middle: Posterior mean of log dispersion against truth, with 95\% HPD CIs shown in grey. The red line denotes $y=x$. Right: Posterior samples for mean-dispersion relationships. The red line denotes truth.}
	\label{fig:sim1_6_1_2_muphi}
\end{figure}

\begin{figure}[tbp]
	\centering
	\subfigure[Simulation 1]{\includegraphics[width=0.45\textwidth]{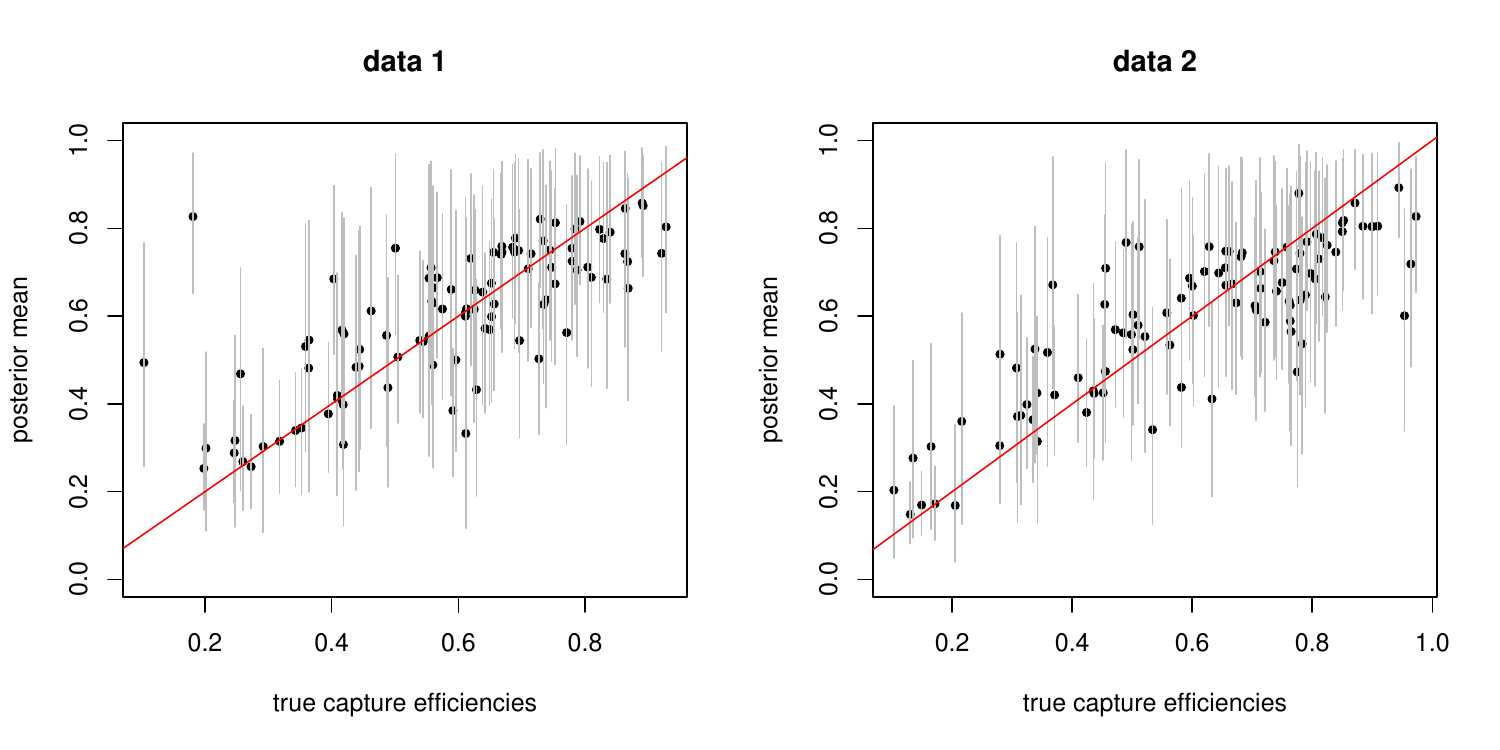}}
	\subfigure[Simulation 2]{\includegraphics[width=0.45\textwidth]{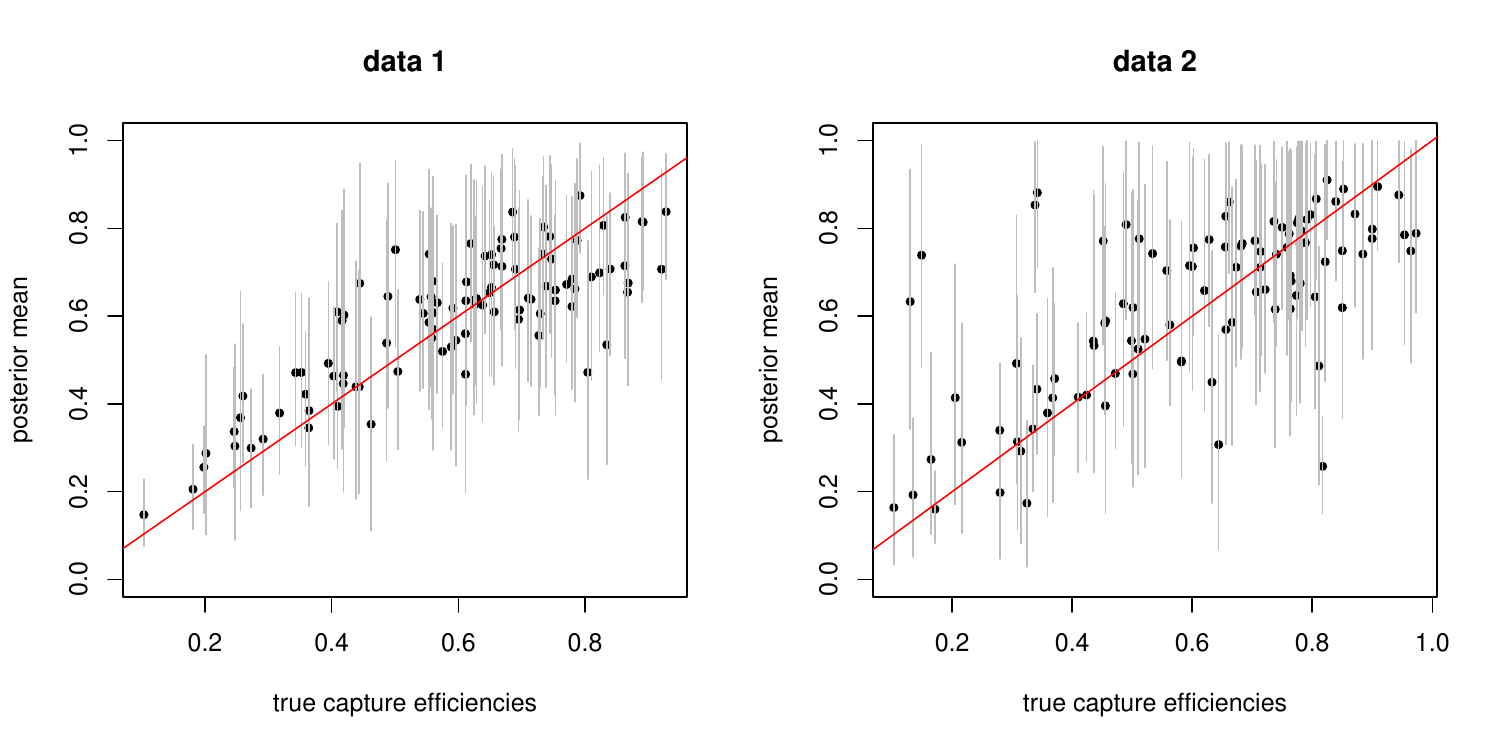}}
	\caption{Posterior mean of cell capture efficiencies against truth, with 95\% HPD CIs shown in grey. The red line denotes $y=x$.}
	\label{fig:sim1_6_1_2_beta}
\end{figure}

As for component-specific parameters, Figure \ref{fig:sim1_6_1_2_muphi} demonstrates that mean expressions, dispersion parameters and their relationship can be accurately inferred, with larger uncertainty in dispersion than mean expression. Cellular capture efficiencies are generally estimated well and showcase high uncertainty, probably due to the identifiability issue (Figure \ref{fig:sim1_6_1_2_beta}).

\begin{figure}[tbp]
	\centering
	\includegraphics[width=0.7\textwidth]{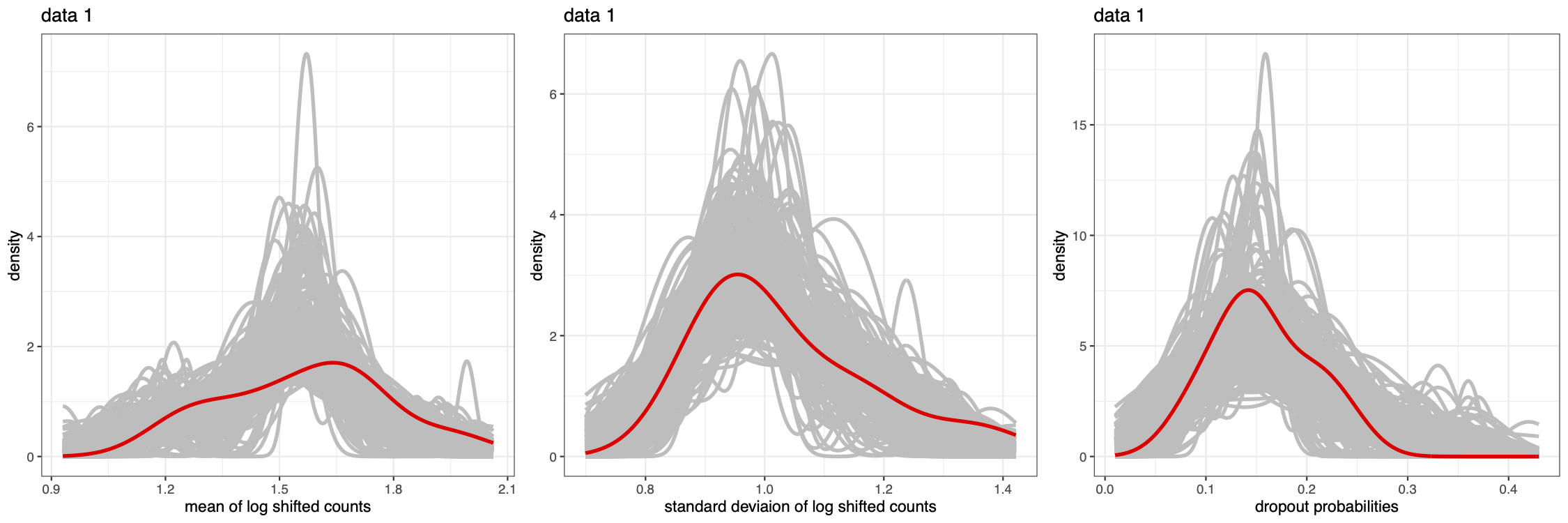}
	\includegraphics[width=0.7\textwidth]{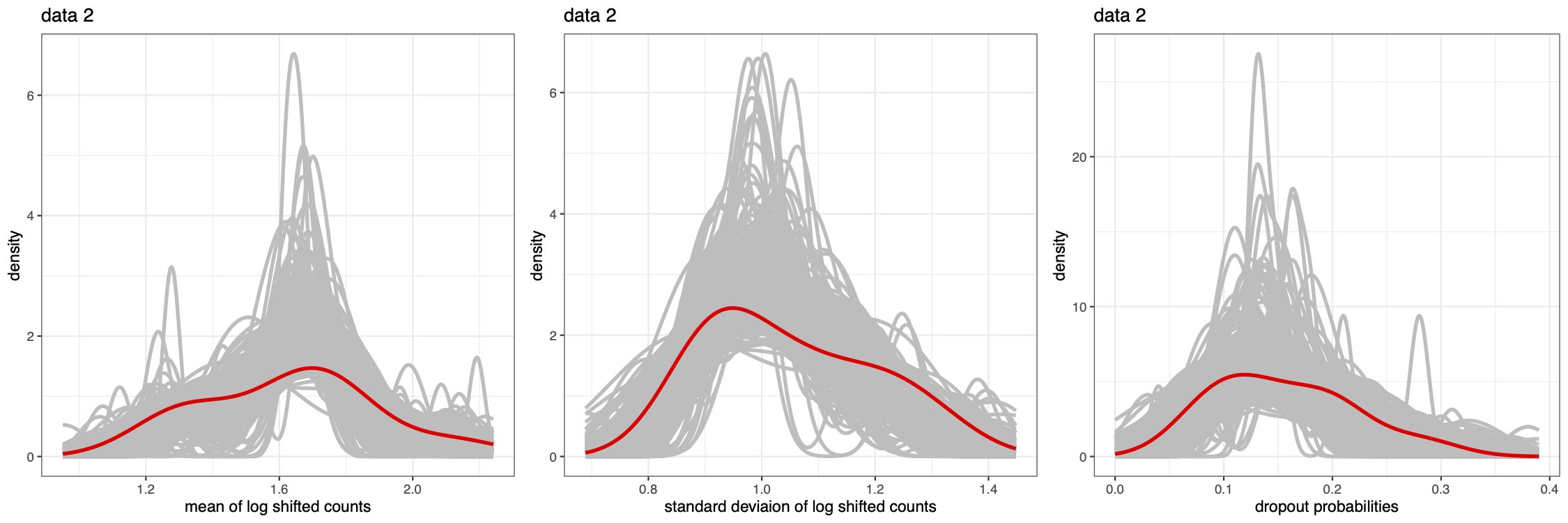}
	\caption{Posterior predictive checks for Simulation 1. 
		Each panel shows the kernel density estimation of one statistic, with replicated and true datasets in grey and red, respectively. Left to right: mean of log shifted counts, standard deviation of log shifted counts and dropout probabilities.}
	\label{fig:sim1_6_1_ppc}
\end{figure}

\begin{figure}[tbp]
	\centering
	\includegraphics[width=0.7\textwidth]{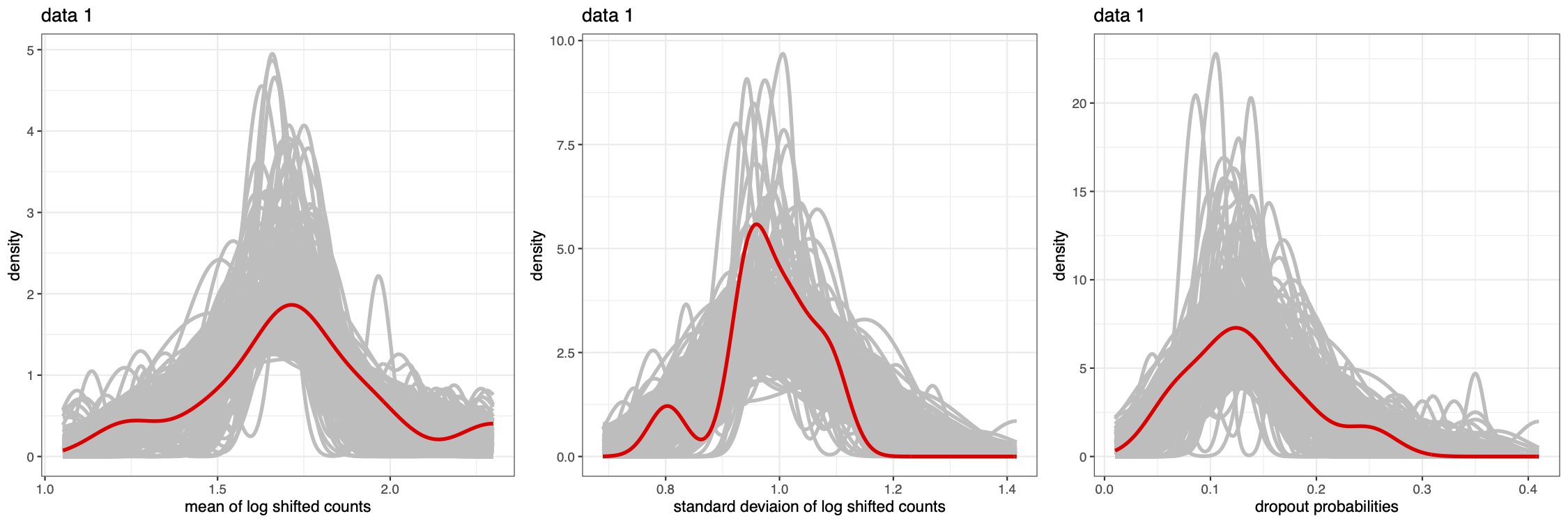}
	\includegraphics[width=0.7\textwidth]{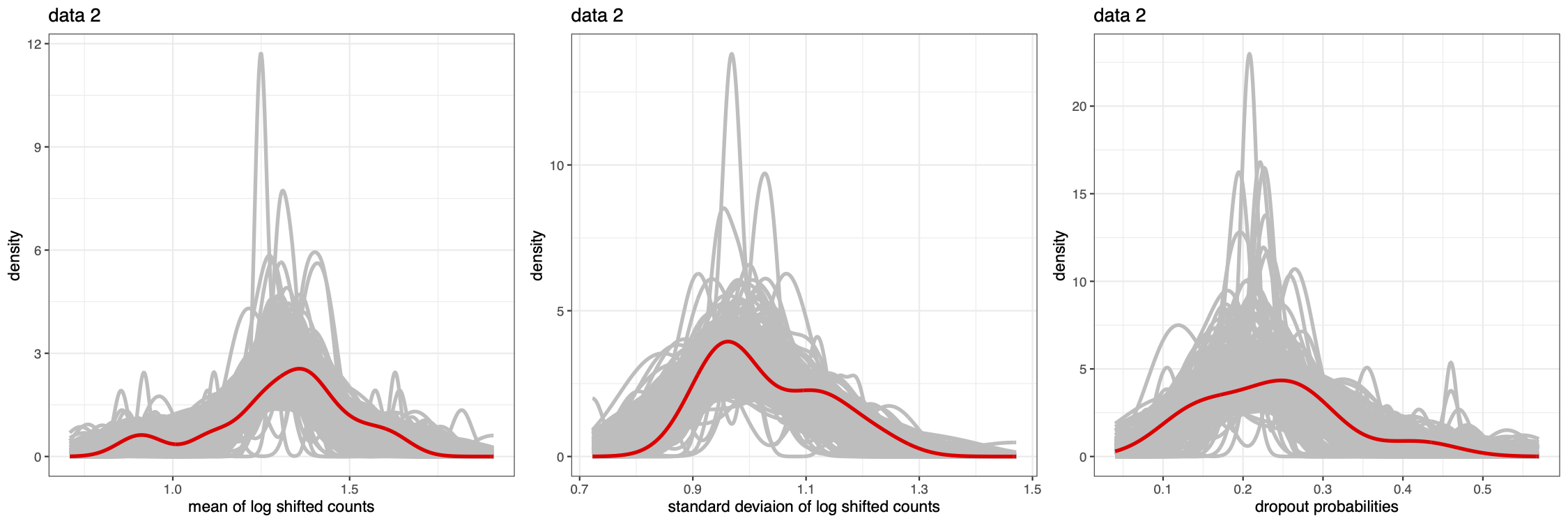}
	\caption{Posterior predictive checks for Simulation 2. 
		Each panel shows the kernel density estimation of one statistic, with replicated and true datasets in grey and red, respectively. Left to right: mean of log shifted counts, standard deviation of log shifted counts and dropout probabilities.}
	\label{fig:sim1_6_2_ppc}
\end{figure}

Results for posterior predictive checks are shown in Figure \ref{fig:sim1_6_1_ppc} and Figure \ref{fig:sim1_6_2_ppc}. Following the approach described in Section \ref{sec:ppc-pax6-method}, we compare the kernel density estimation (KDE) of three statistics, and the KDEs of the observed data is contained within the replicated data, indicating a reasonable fit of the model.

\newpage
\subsubsection{Simulation 3}
In Simulation 3, we generate two datasets comprising $C_1=200$, and $C_2=300$ cells, respectively, with $G=100$ genes. A total of 3 clusters are generated. In addition, we assume the first 70 genes are both global DE and DD genes and are simulated from
\begin{equation*}
	\begin{split}
		\mu_{j,g}^* \indsim \logN(\theta_j,0.1), \quad
		\phi_{j,g}^* | \mu_{j,g}^* &\indsim  \logN(b_0 + b_1 \log(\mu_{j,g}^*), 0.1), \\
	\end{split}
\end{equation*}
where $\btheta_{1:3}=(-2,2,4),b_0=-1, b_1=1$. For non-DE and non-DD marker genes, we set $\mu_{j,g}^*=\mu_g^*$ with $\mu_g^* \iidsim \logN(3.5,0.5)$, and $\phi_{j,g}^*=\phi_g^*$ with $\phi_g^* \indsim \logN(b_0 + b_1 \log(\mu_{g}^*), 0.1)$. The time-dependent probabilities are based on the following parameters:
\begin{equation*}
	\begin{split}
		\text{Dataset 1}: \quad t_{1:3,1}^*&=(0.4,0.9,0.1),\quad \sigma_{1:3,1}^*=(0.08,0.15,0.1),\quad \bmq_{1:3,1}=(0.5,0.5,0),\\
		\text{Dataset 2}: \quad t_{1:3,2}^*&=(0.8,0.5,0.3),\quad \sigma_{1:3,2}^*=(0.1,0.05,0.1),\quad \quad \bmq_{1:3,2}=(0.7,0,0.3).
	\end{split}
\end{equation*}
Cell capture efficiencies and latent time are simulated from the same distributions as Simulation 1 and 2. 

For model implementation, a truncation level of $J=6$ is applied. Consensus clustering is implemented with 100 chains and 100 iterations. The post-processing step has the same MCMC setup as Simulation 1 and 2.

\begin{figure}[bp]
	\centering
	\includegraphics[width=0.4\textwidth]{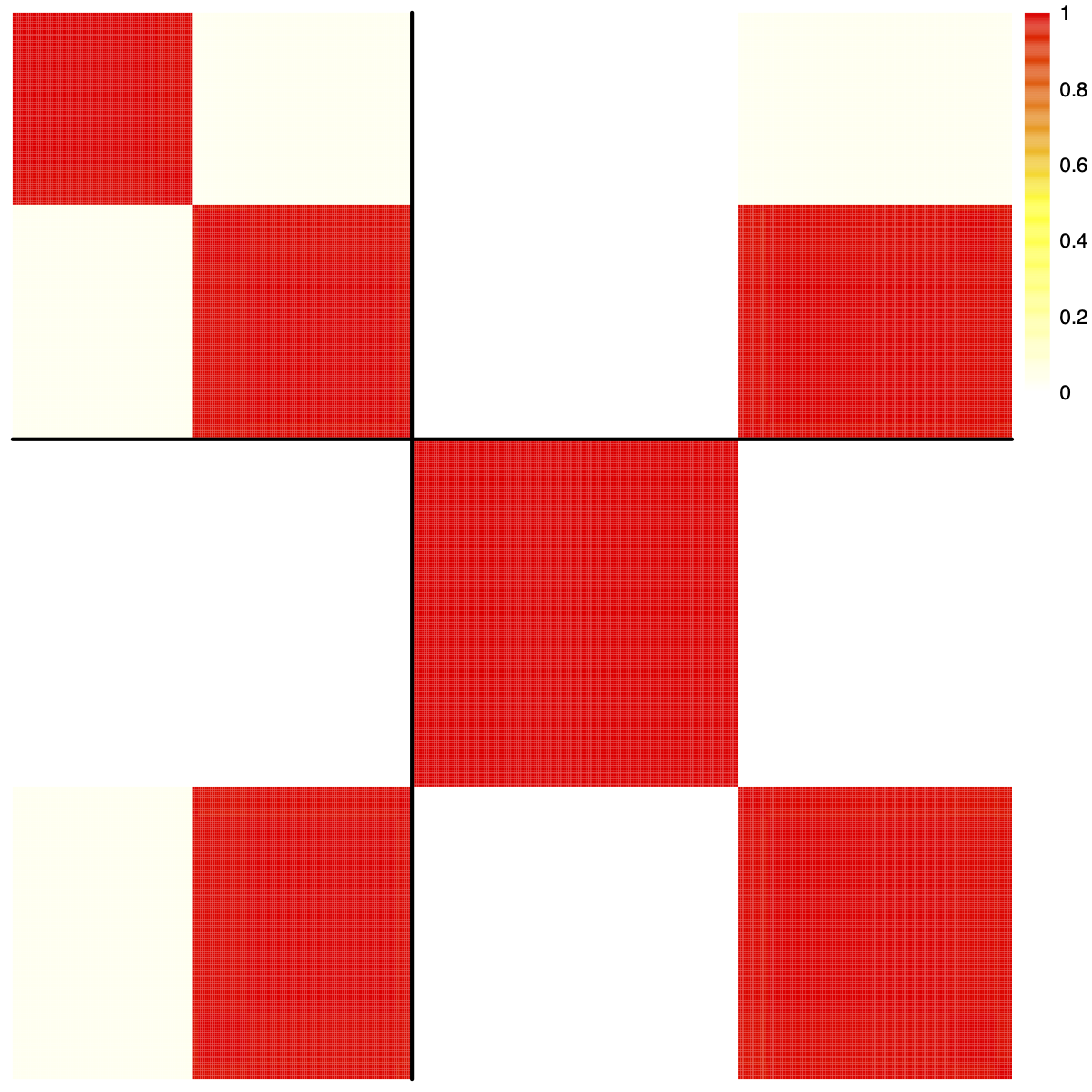}
	\caption{Posterior similarity matrix for Simulation 3. Diagonal blocks correspond to within-group PSM.}
	\label{fig:sim1_10_psm}
\end{figure}

\begin{figure}[h]
	\includegraphics[width=0.95\textwidth]{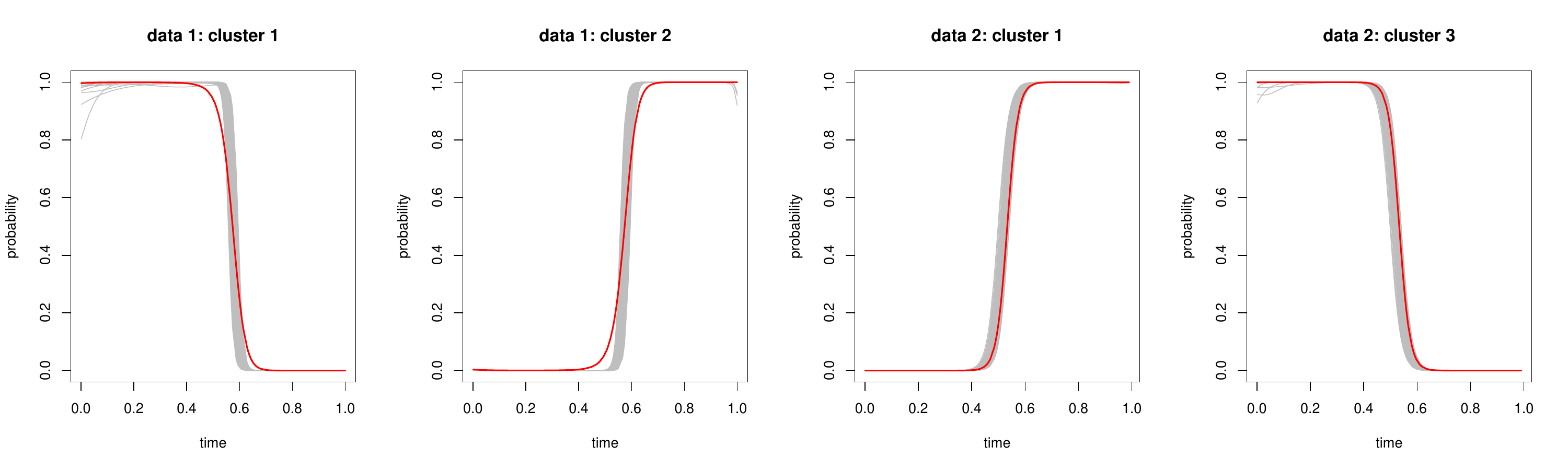}
	\includegraphics[width=0.95\textwidth]{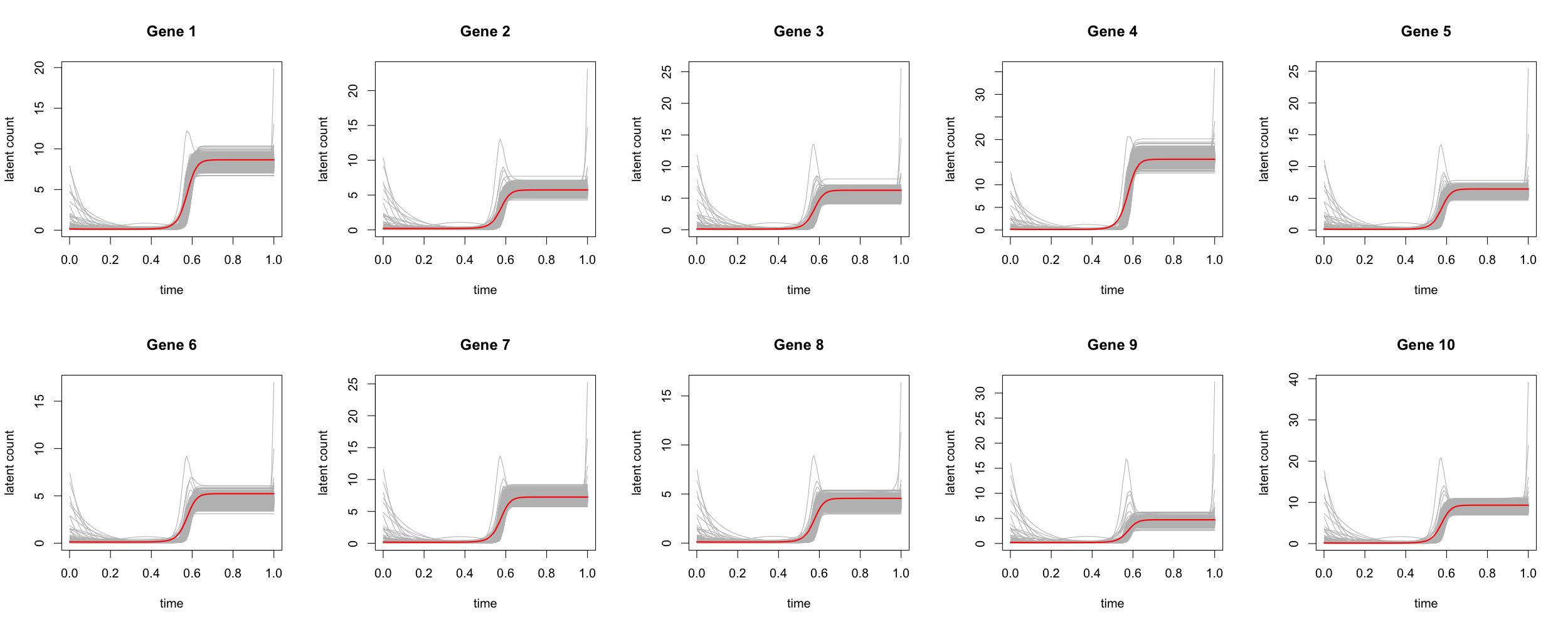}	
	\caption{Top: Posterior samples for time-dependent probabilities for Simulation 3. Only non-empty components are shown. 
		Bottom: Estimated latent counts against latent time for the first 10 genes in the first dataset. The red solid line denotes the truth.}
	\label{fig:sim1_10_p_vs_t_and_mean_latent_count}
\end{figure}

\paragraph{Results}
The optimal clustering from consensus clustering matches the truth exactly with ARI of 1, with posterior similarity matrix shown in Figure \ref{fig:sim1_10_psm} exhibiting small uncertainty in cell allocations. Figure \ref{fig:sim1_10_p_vs_t_and_mean_latent_count} shows that the time-dependent probabilities and expected counts are correctly inferred with slightly larger uncertainty around $t=0, 0.5, 1$.

\begin{figure}
	\centering
	\includegraphics[width=0.8\textwidth]{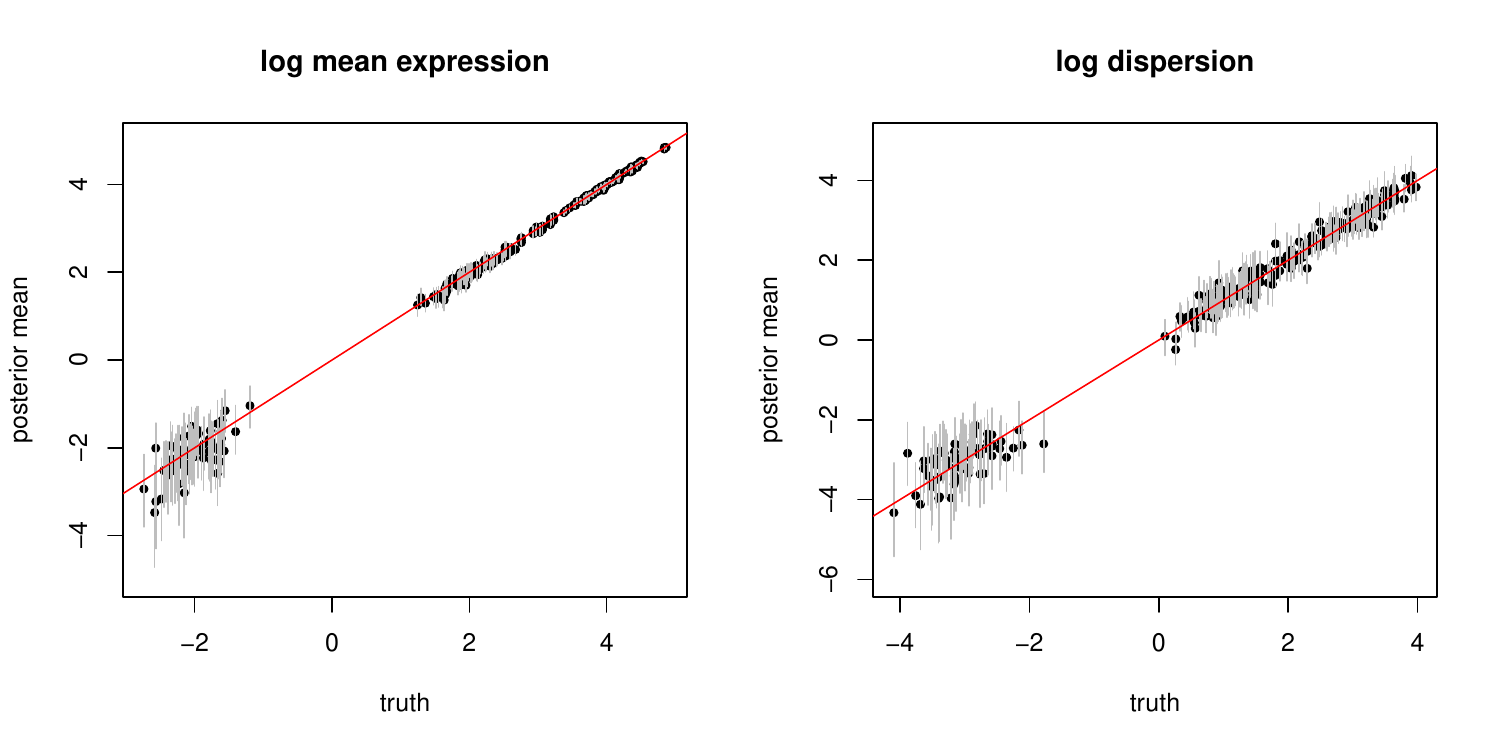}
	\caption{Left: Posterior mean of log mean expression against truth, with 95\% HPD CIs shown in grey. Right: Posterior mean of log dispersion against truth, with 95\% HPD CIs shown in grey. The red line denotes $y=x$.}
	\label{fig:sim1_10_muphi}
\end{figure}

\begin{figure}[h]
	\centering
	\includegraphics[width=0.8\textwidth]{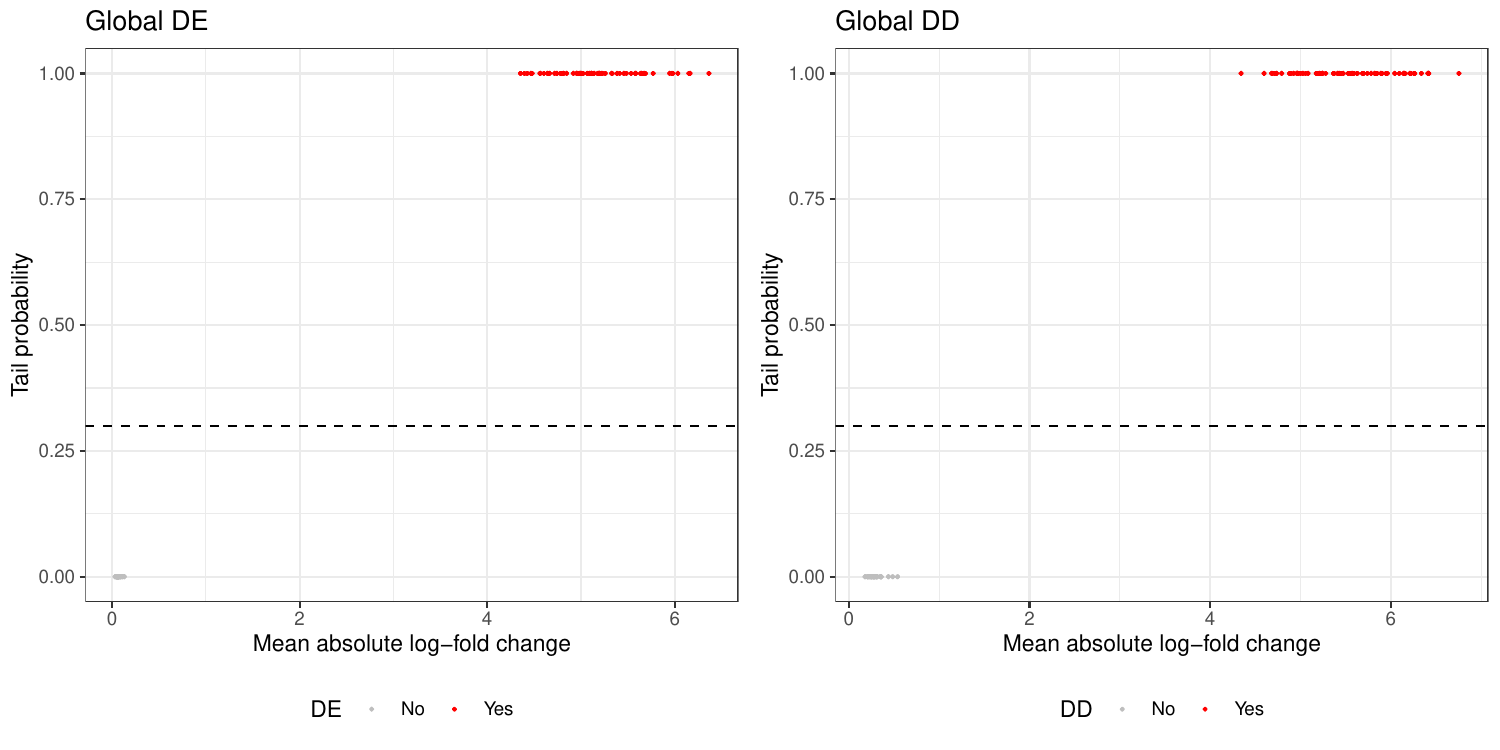}
	\caption{Posterior tail probabilities against mean absolute LFCs for global DE (left) and global DD genes (right). The black dashed line indicates the threshold to determine global genes.}
	\label{fig:sim1_10_global_LFC}
\end{figure}

Figure \ref{fig:sim1_10_muphi} illustrates that the estimated mean expression and dispersion are close to truth, with higher uncertainty for the cluster associated with small mean expressions and dispersion. Using the tools proposed in Section \ref{sec:marker-genes-method}, we apply thresholds $\tau_0=\omega_0=2.5$ to identify marker genes. In this case, all 70 global DE and DD genes are correctly uncovered, with the relationship between tail probabilities and mean absolute log-fold changes (LFCs) shown in Figure \ref{fig:sim1_10_global_LFC}. 

\clearpage
\subsection{Simulation Setting 2: A Periodic Kernel} \label{appendix:simulation-cidata}

For a VAR model with a periodic kernel, two datasets consisting of $J=3$ clusters are generated from the proposed model in Section \ref{sec:full-conditionals-cidata}, each consisting of $n_1=n_2=151$ observations with dimension $G=2$. The time associated with each dataset is equally spaced on $[0,1]$.  The data is generated from
\begin{equation*}
	\begin{split}
		\bmy_{i,d} | \bmy_{i-1,d}, z_{i,d}=j, \bL_j^*, \bSigma_j^* &\indsim \Norm((\bL_j^*)^T \bmx_{i,d},  \bSigma_j^*), \\
		z_{i,d}| p_{1,d}^J(t_{i,d}), \ldots, p_{J,d}^J(t_{i,d})&\indsim \Cat(p_{1,d}^J(t_{i,d}),\ldots, p_{J,d}^J(t_{i,d})),\\
	\end{split}
\end{equation*}
where $\bmx_{i,d}=(1, y_{i-1,1,d},y_{i-1,2,d})^T$. The component-specific parameters are given by
\begin{equation*}
	\begin{split}
		\bL_1^* = 
		\begin{pmatrix}
			0 & 0.9 & -0.1 \\
			0 & 0.1 & 0.8
		\end{pmatrix}^T, &\quad \bL_2^* =
		\begin{pmatrix}
			1 & 0.5 & 0.1& \\
			1 & -0.1 & 0.5
		\end{pmatrix}^T, \quad \bL_3^* =
		\begin{pmatrix}
			-1 & 0.9 & -0.2& \\
			-1 & 0 & 0.9
		\end{pmatrix}^T, \\
		\bSigma_1^*=
		\begin{pmatrix}
			0.001 & 0.002\\
			0.002 & 0.004
		\end{pmatrix}, &\quad  \bSigma_2^*=
		\begin{pmatrix}
			0.005 & 0\\
			0 & 0.005 
		\end{pmatrix}, \quad  \bSigma_3^*=
		\begin{pmatrix}
			0.05 & -0.02\\
			0.02 & 0.05 
		\end{pmatrix}.
	\end{split}
\end{equation*}
where $\bL_j^*= (\bma_j^* \quad \bB_j^*)^T$. For time-dependent probabilities, we set $\bmq_{1:3,1}=(0.5,0.5,0)$ and $\bmq_{1:3,2}=(0.6,0,0.4)$, leading to one shared cluster across groups and one unique cluster in each group. The kernel parameters are 
\begin{equation*}
	\begin{split}
		\text{Dataset 1}: \quad \mu_{1:3,1}^*&=(0,0.2,-0.1),\quad \lambda_{1:3,1}^*=\left(\frac{0.3}{\pi},\frac{0.6}{\pi},\frac{0.2}{\pi}\right),\quad {\sigma_{1:3,1}^*}^2=(0.2,0.1,0.05),\\
		\text{Dataset 2}: \quad \mu_{1:3,2}^*&=(0.1,-0.1,0),\quad \lambda_{1:3,2}^*=\left(\frac{0.6}{\pi},\frac{0.4}{\pi},\frac{0.3}{\pi}\right),\quad {\sigma_{1:3,2}^*}^2=(0.05,0.2,0.1).
	\end{split}
\end{equation*}

For each dataset, the first observation is not clustered as its previous time point is not observed. To infer the clustering, we run the MCMC algorithm described in Section \ref{sec:full-conditionals-cidata} with $J=6$ for $5000$ iterations, followed by a burnin of $3000$. For the post-processing step with fixed clustering, we run one chain with 16000 iterations, and apply a burnin of 12000, followed by a thinning of 2, giving 2000 MCMC samples. 

\subsubsection{Results}
\begin{figure}[bp]
	\centering
	\includegraphics[width=0.3\textwidth]{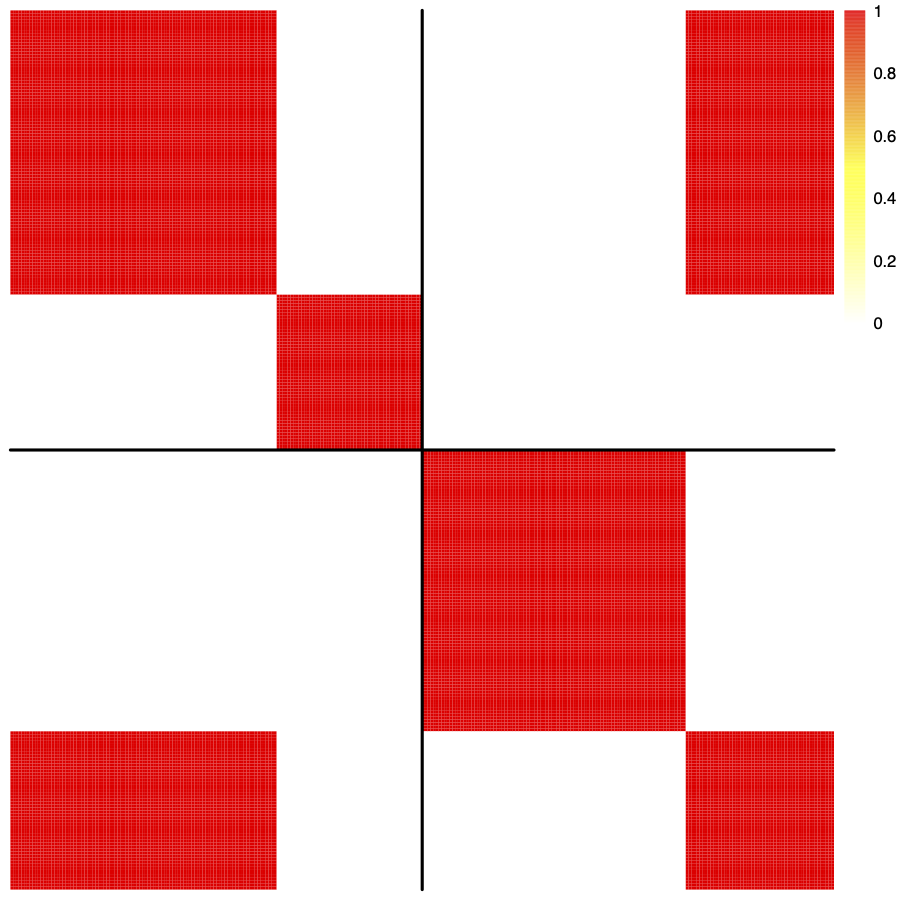}
	\includegraphics[width=0.6\textwidth]{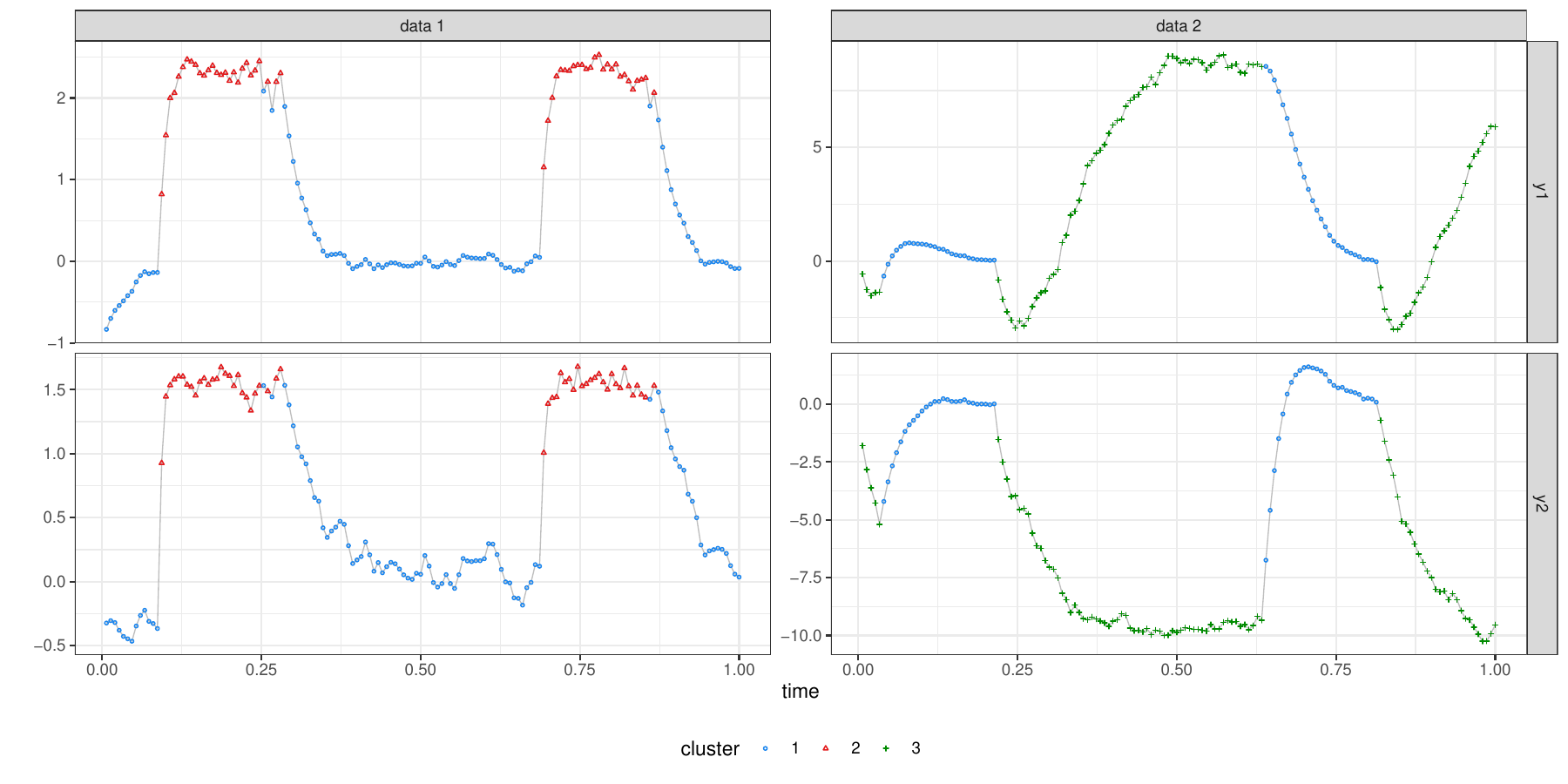}
	\caption{Left: Posterior similarity matrix. Diagonal blocks correspond to within-group PSM. Right: Plot of each dimension against time $t$ for each dataset, with observations colored by cluster labels.}
	\label{fig:Sim_5_pktest41_psm_and_y_vs_t}
\end{figure}

Based on the VI criterion, the C-HDP model correctly finds the true clustering ($\text{ARI}=1$), with the posterior similarity matrix shown in Figure \ref{fig:Sim_5_pktest41_psm_and_y_vs_t}, demonstrating small uncertainty. 

\begin{figure}[tbp]
	\centering
	\includegraphics[width=0.8\textwidth]{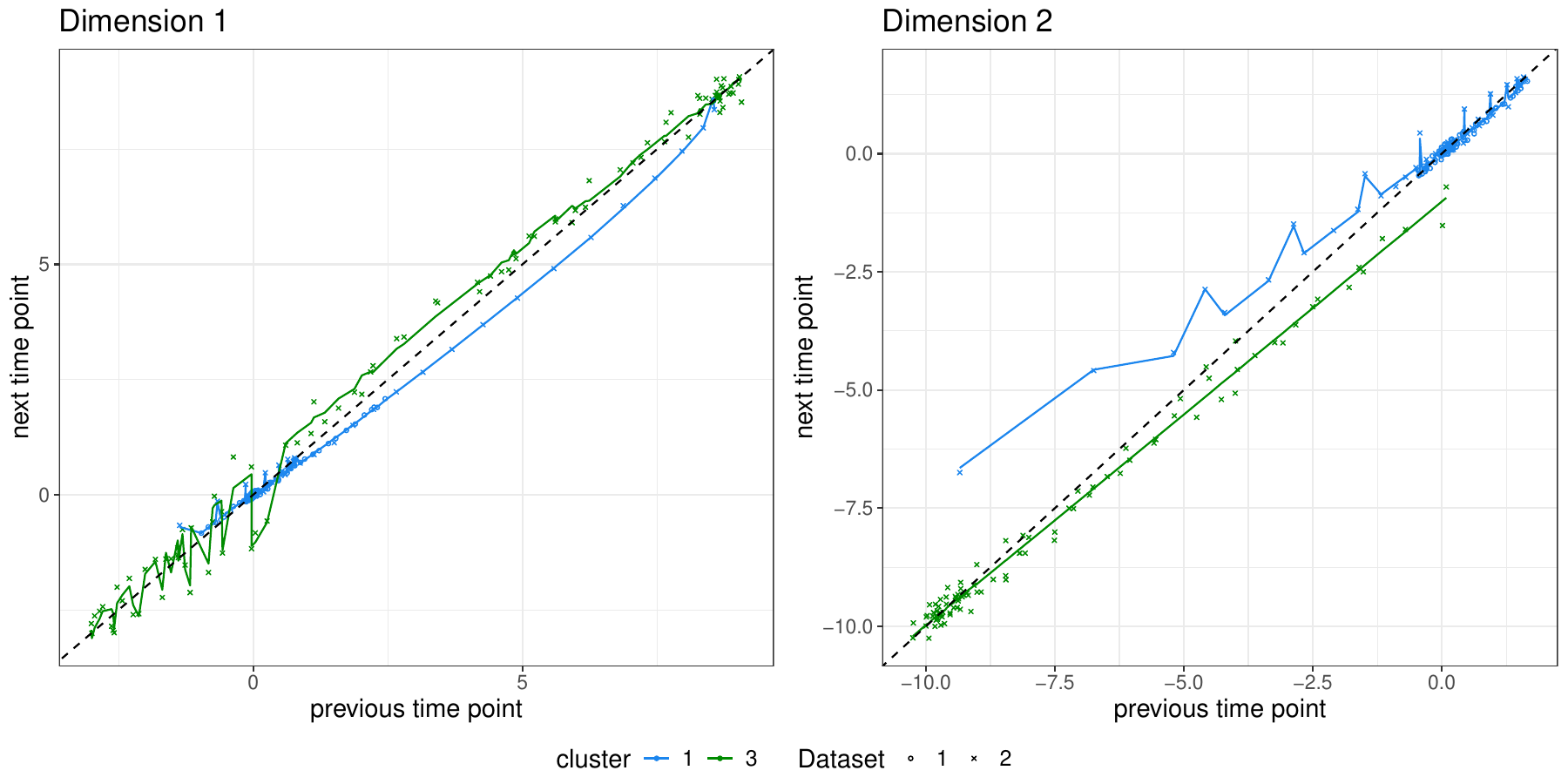}
	\caption{Plot of $y_{i,g}$ against $y_{i-1,g}$ for each dimension $g$, for clusters 1 and 3 in both datasets. The black dashed line denotes $y=x$. The solid colored line denotes the posterior estimated relationship between successive time frames for each cluster. }
	\label{fig:Sim_5_pktest41_psm_and_yt_vs_yt-1_cl1_3}
\end{figure}

In addition, to interpret the clusters, we show the time-series plot for each dimension (Figure \ref{fig:Sim_5_pktest41_psm_and_y_vs_t} right), the identified clusters appear to share homogeneous dependence on the previous time. For instance, cluster 3 mainly shows an increasing trend in the first dimension, which can be further confirmed in Figure \ref{fig:Sim_5_pktest41_psm_and_yt_vs_yt-1_cl1_3} where cluster 3 is mostly above the equivalent line $y=x$ in dimension 1. On the contrary, cluster 1 generally decreases in the first dimension. In addition, by computing the posterior estimated relationship between consecutive time points:
\begin{equation*}
	\bm{y}_i = \hat{\bma}_j^*+\hat{\bB}_j^*\bm{y}_{i-1},
\end{equation*}
where $\hat{\bma}_j^*$ and $\hat{\bB}_j^*$ denote the posterior mean of the coefficients, we notice that the estimated relationship for cluster 3 in the second dimension is almost linear, as shown in the green solid line in Figure \ref{fig:Sim_5_pktest41_psm_and_yt_vs_yt-1_cl1_3}. This implies a dominant influence from the past observation in the same reduced dimension, which corresponds to the zero coefficient in $\bL_3^*$ representing the impact of the first dimension on the second dimension.

\begin{figure}
	\centering
	\includegraphics[width=0.8\textwidth]{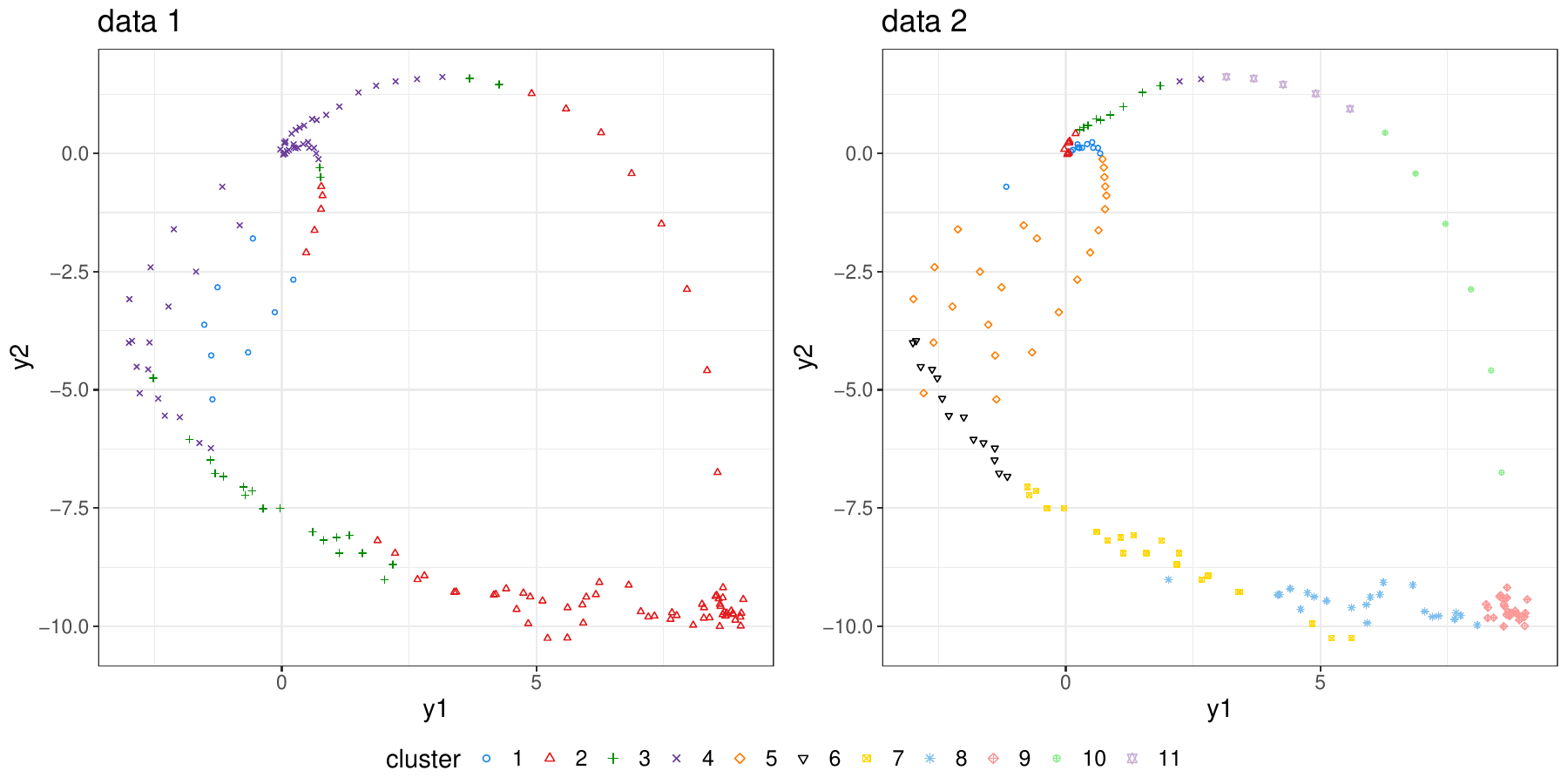}
	\caption{Pairwise scatterplots for two datasets, with observations colored by cluster membership from a simple Gaussian mixture model.}
	\label{fig:Sim_5_pairplot_mclust}
\end{figure}

For a comparison, we also fit a simple Gaussian mixture model (GMM) with an unconstrained covariance matrix\footnote{GMM is implemented using R package \texttt{mclust} \citep{scrucca2016mclust}.}. However, GMM finds 11 clusters between 2 to 15 based on BIC, yielding a poor ARI of 0.3507 with the true clustering. Figure \ref{fig:Sim_5_pairplot_mclust} shows that GMM groups observations into small blocks with similar observed values, rather than considering their relative dependence.

\begin{figure}[tbp]
	\centering
	\includegraphics[width=0.9\textwidth]{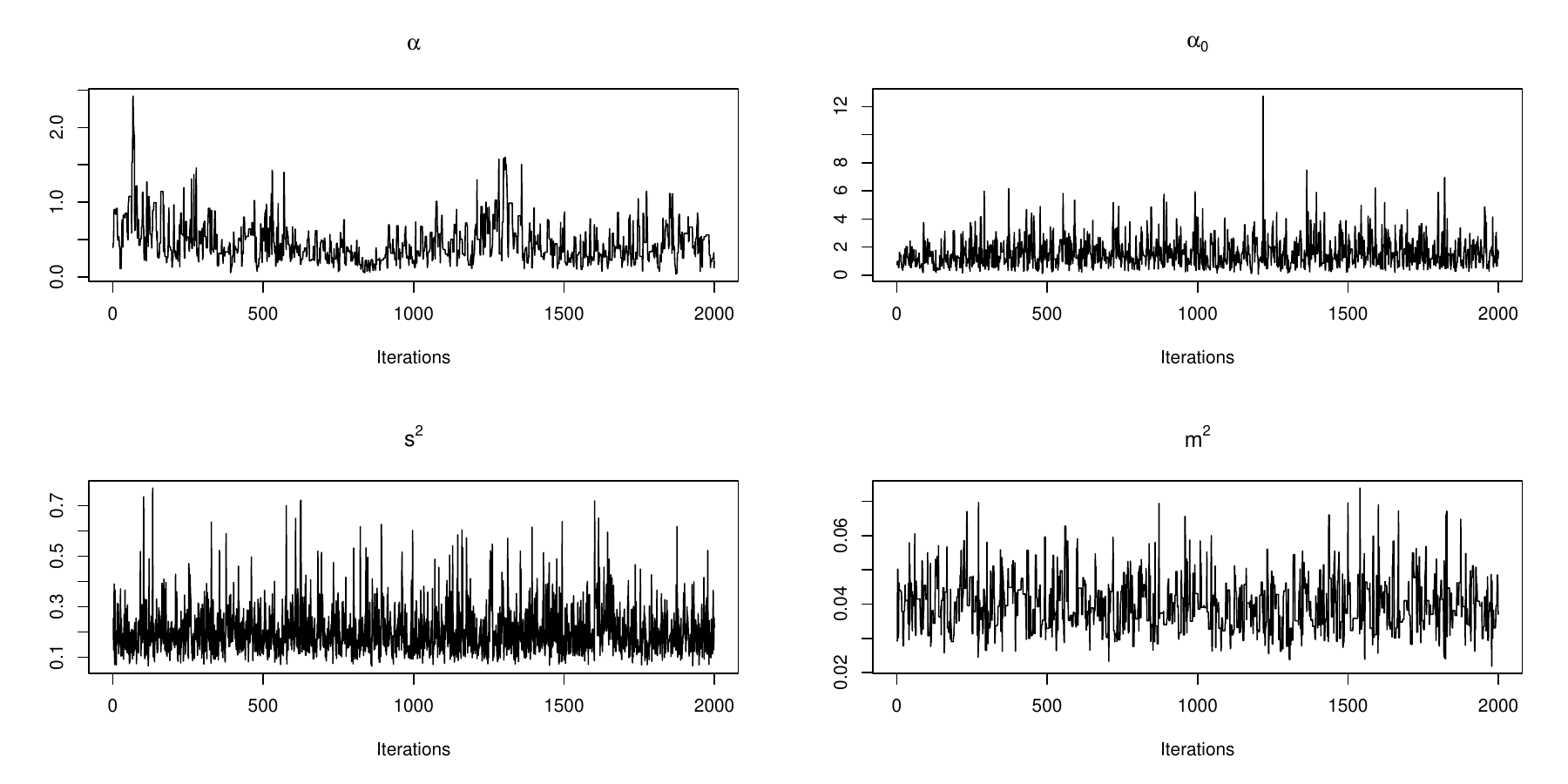}
	\caption{Traceplots for concentration parameters and hyperparameters for Simulation 3.}
	\label{fig:Sim_5_alphas_s2_m2_trace_post}
\end{figure}

\begin{figure}[tbp]
	\centering
	\includegraphics[width=0.9\textwidth]{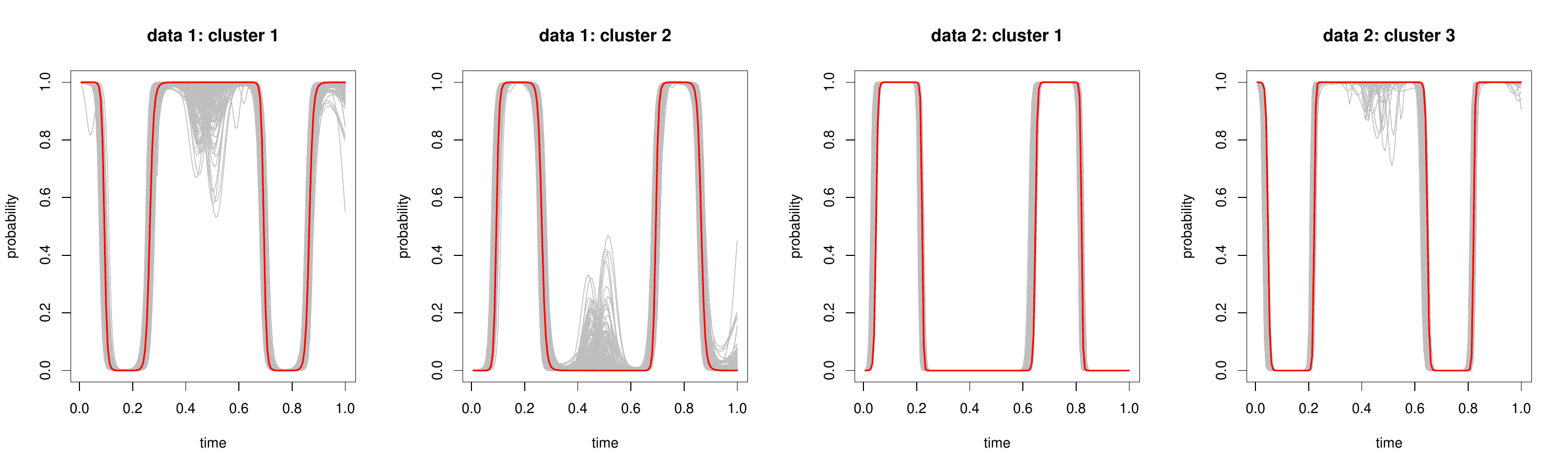}
	\caption{Posterior samples for time-dependent probabilities for Simulation 3. 
		The red solid line denotes the truth.}
	\label{fig:Sim_5_p_vs_t_thin5}
\end{figure}

Traceplots from the post-processing step are shown in Figure \ref{fig:Sim_5_alphas_s2_m2_trace_post}, suggesting convergence. The true time-dependent probabilities are contained within the MCMC samples (Figure \ref{fig:Sim_5_p_vs_t_thin5}), with large uncertainty at time point around 0.5. In addition, similar to the Gaussian kernel, individual kernel parameters and $q_{j,d}^J$ still cannot be contained in the MCMC samples, suggesting weak identifiability.

\begin{figure}[tbp]
	\centering
	\includegraphics[width=0.8\textwidth]{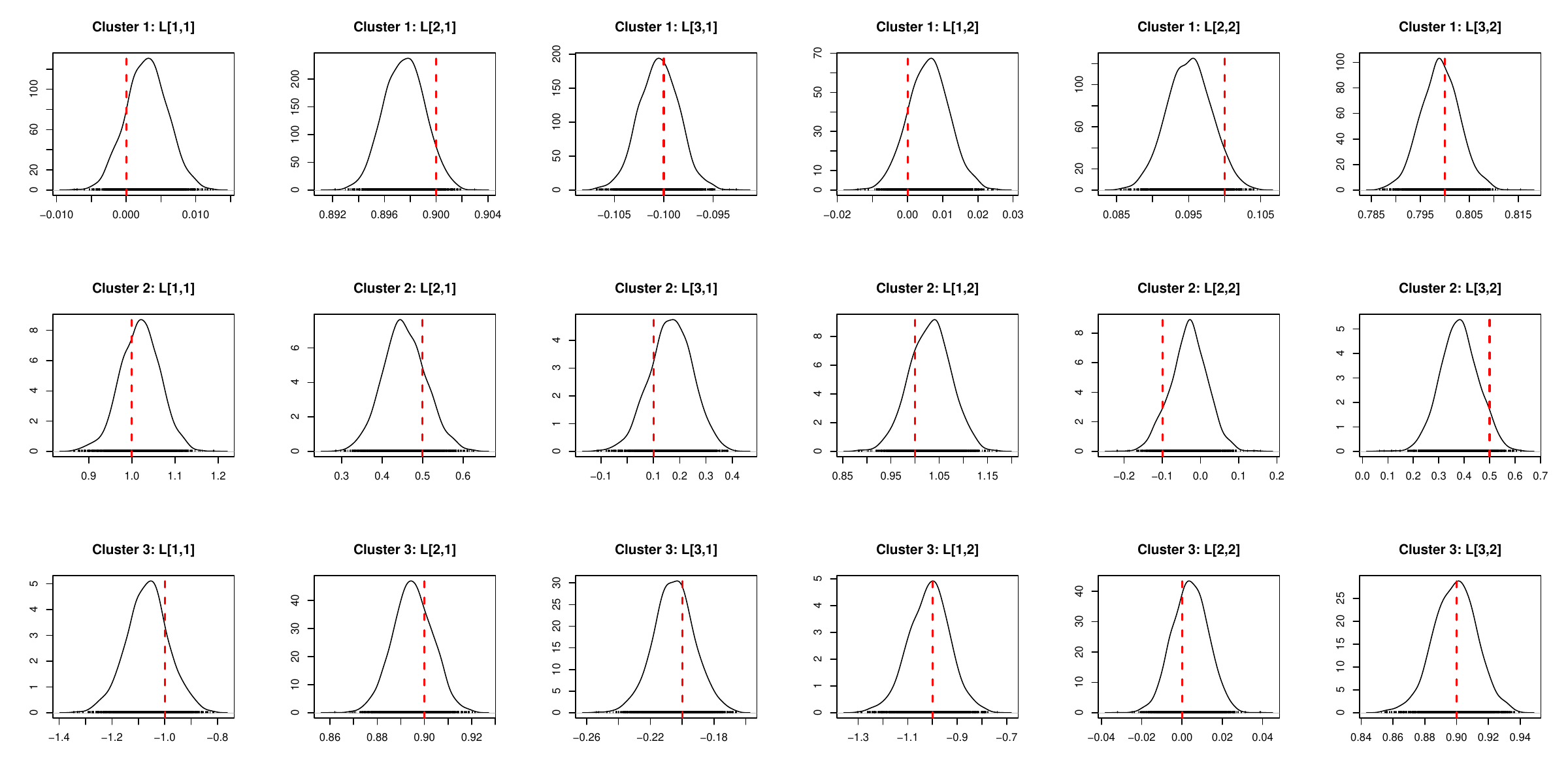}
	\caption{Density plots for each element in the coefficient matrix $\bL_j^*$. The red dashed line denotes the truth.}
	\label{fig:Sim_5_L_dens}
\end{figure}

\begin{figure}[tbp]
	\centering
	\includegraphics[width=0.8\textwidth]{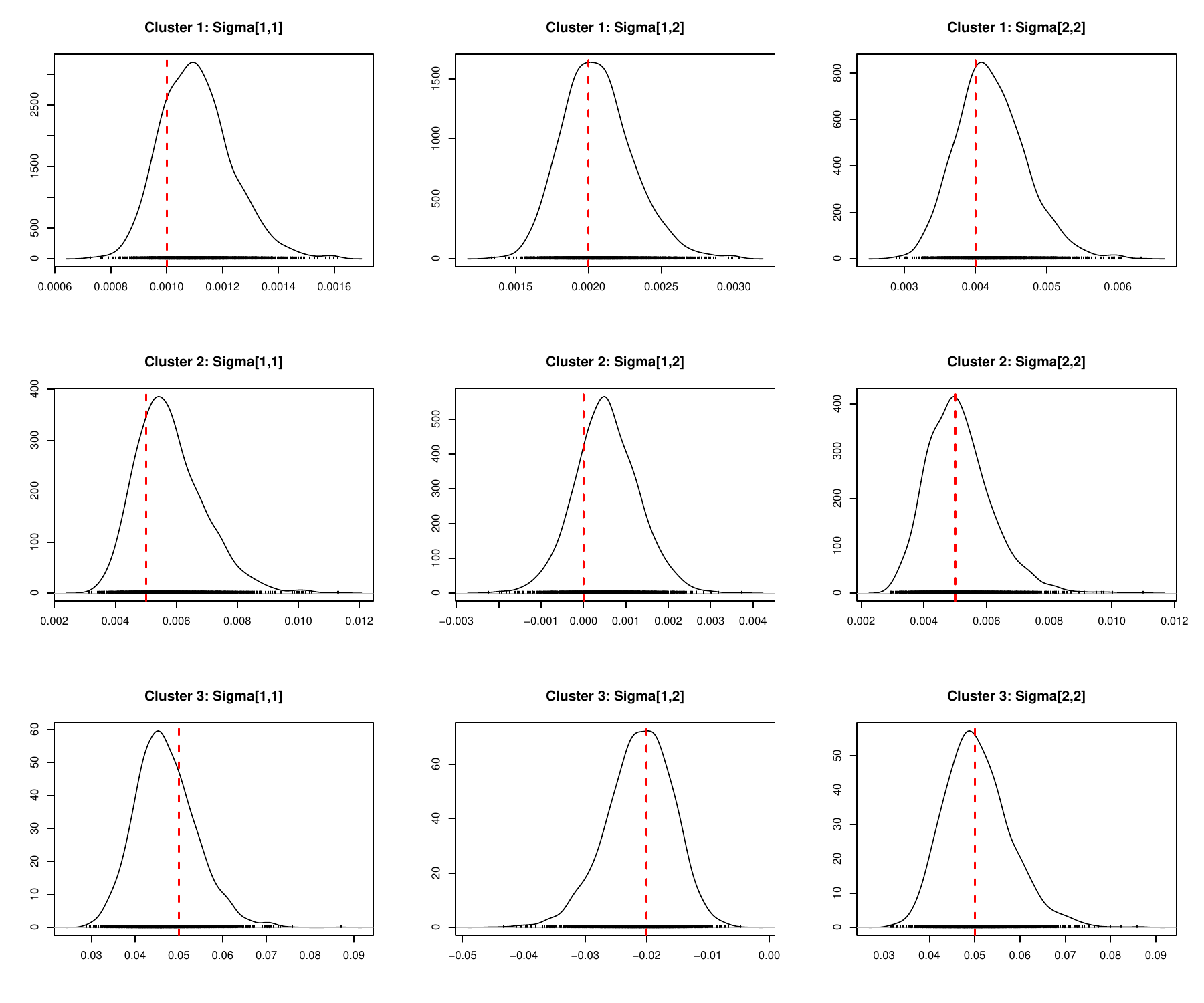}
	\caption{Density plots for each element from the upper triangular part of the covariance matrix $\bSigma_j^*$. The red dashed line denotes the truth.}
	\label{fig:Sim_5_Sigma_dens}
\end{figure}

For component-specific parameters, each element in the coefficient matrix $\bL_j^*$ and covariance matrix $\bSigma_j^*$ is investigated and all values are well estimated falling within the 99\% HPD CIs (Figure \ref{fig:Sim_5_L_dens} and Figure \ref{fig:Sim_5_Sigma_dens}).

For posterior predictive checks, one replicate for observation $i$ in dataset $d$ is generated from 
\begin{equation*}
	\bmy_{i,d}^{rep,(l)} \sim \Norm\left(\left({\bL_{j}^*}^{(l)}\right)^T \bmx_{i,d},{\bSigma_{j}^{*}}^{(l)}\right),
\end{equation*}
where $\bmx_{i,d}=(1, y_{i-1,1,d},y_{i-1,2,d})^T$, and ${\bL_{j}^*}^{(l)}$ and ${\bSigma_{j}^{*}}^{(l)}$ denote the $l$-th posterior MCMC draw. 

\begin{figure}[tbp]
	\begin{minipage}[h]{0.5\textwidth}
		\includegraphics[width=0.95\textwidth]{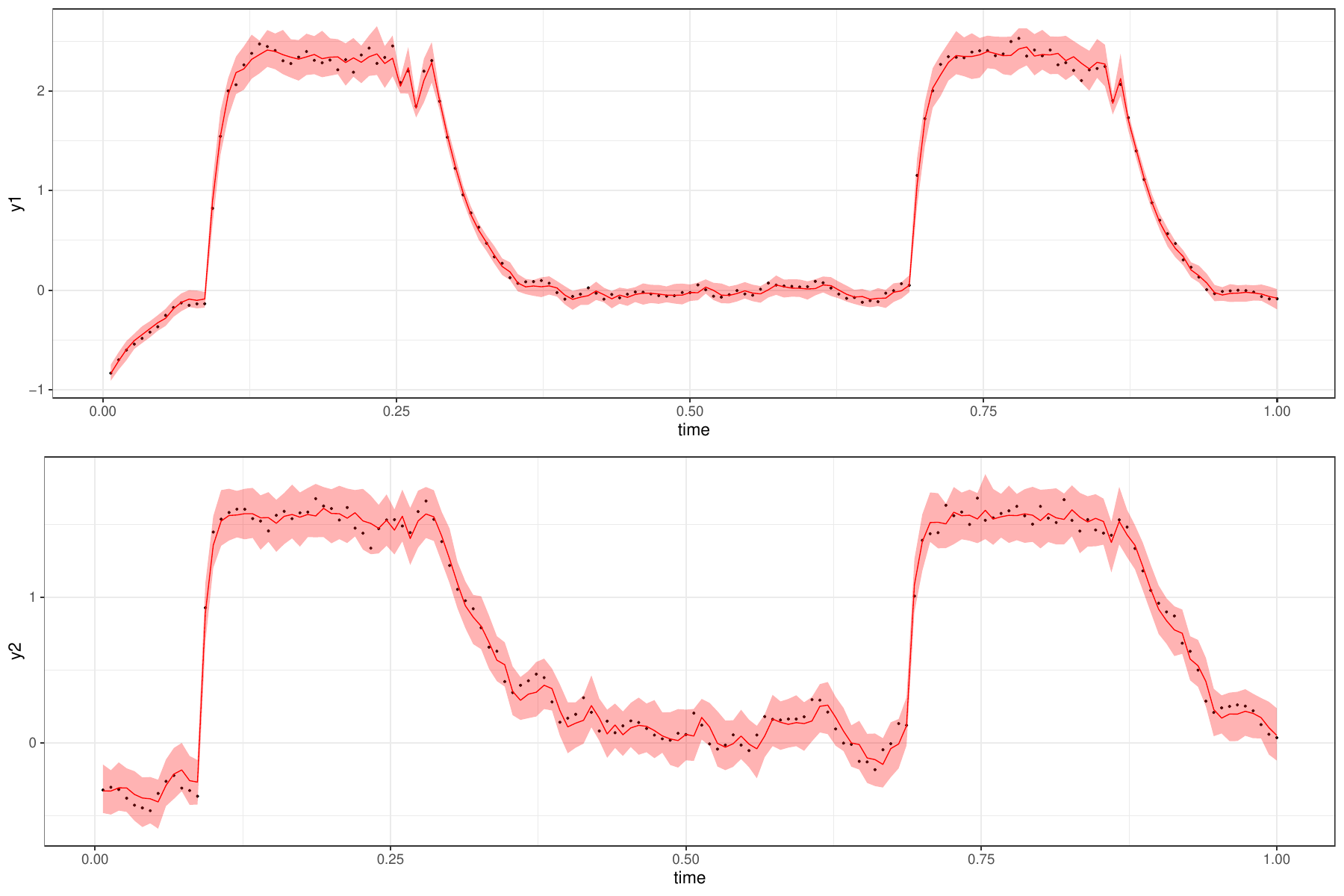}
	\end{minipage}
	\begin{minipage}[h]{0.5\textwidth}
		\includegraphics[width=0.95\textwidth]{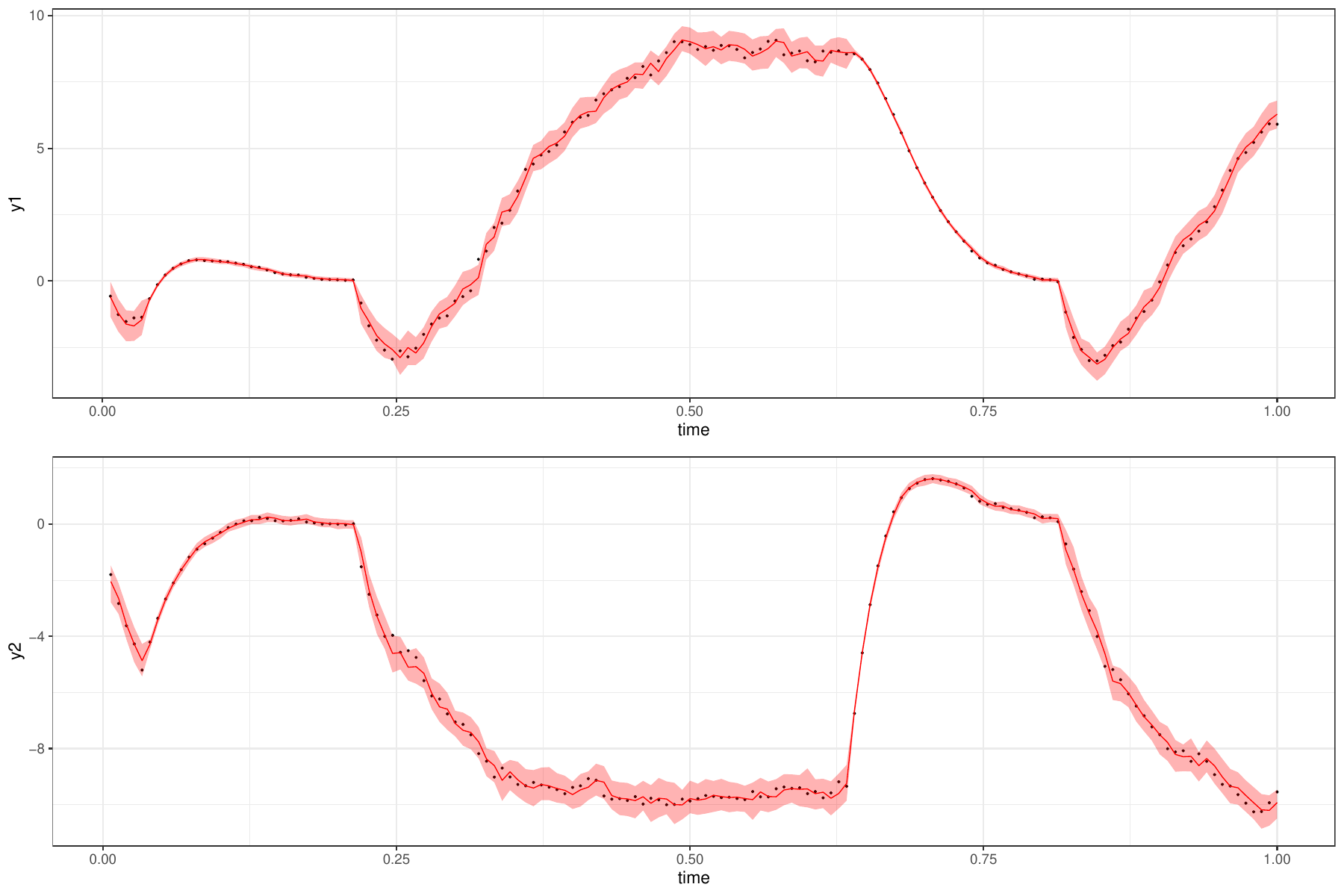}	
	\end{minipage}
	\caption{Posterior predictive checks based on 200 replicates. Red lines denote the posterior mean of the replicates, with 99\% HPD CIs shown in the red area. Black points denote the observed data. Rows correspond to dimensions. Left: Dataset 1. Right: Dataset 2.}
	\label{fig:Sim5_ppc}
\end{figure}

Conditional on the optimal clustering and observed data, we generate 200 replicated datasets based on samples from the post-processing MCMC. Figure \ref{fig:Sim5_ppc} shows that the replicated data closely resembles the actual data, thus supporting the model fit.

\clearpage
\section{Additional Results for Pax6 Data} \label{appendix:additional-results-pax6}

\begin{figure}[hbp]
\centering
\includegraphics[width=0.9\textwidth]{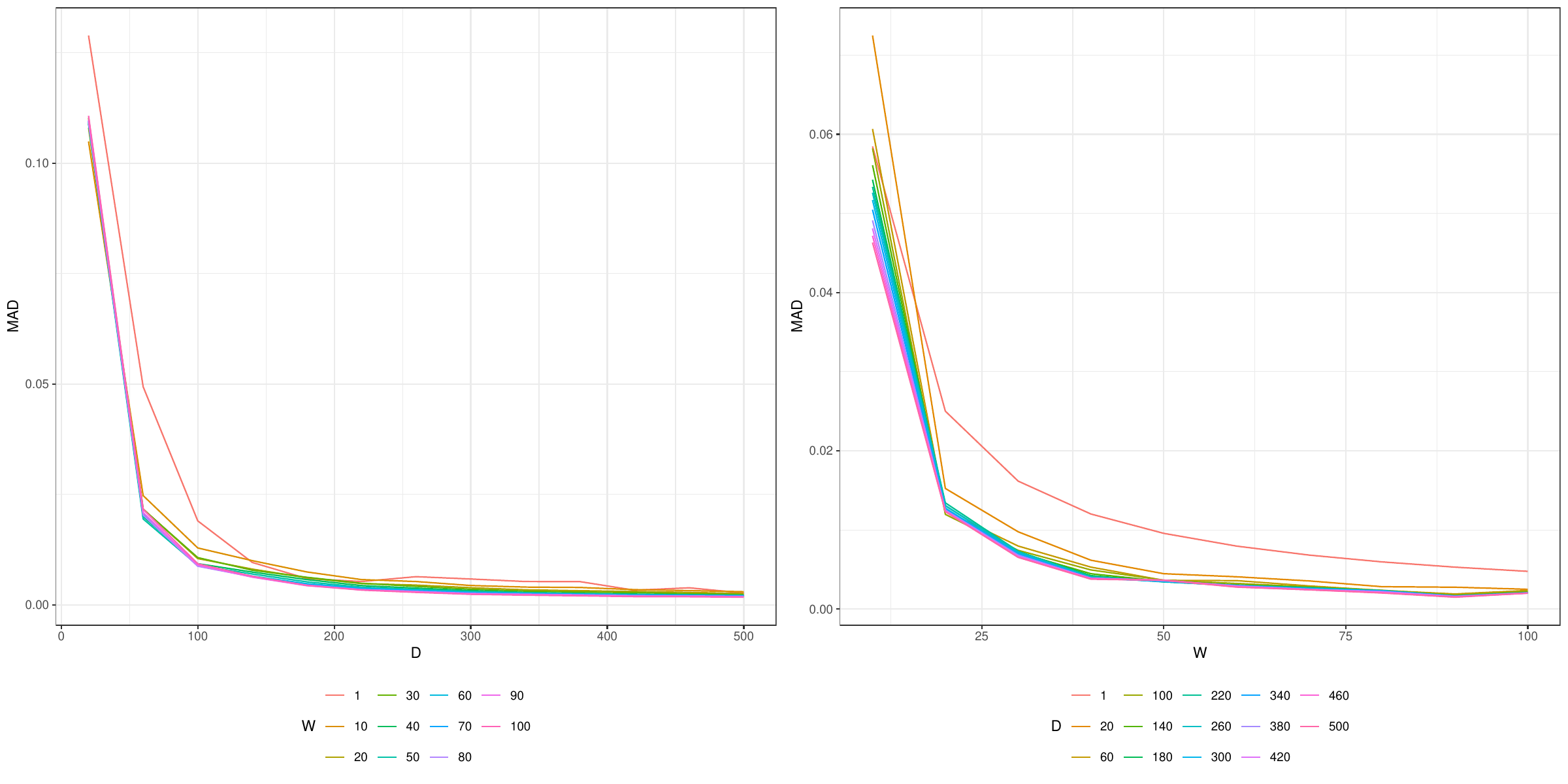}
\caption{Choice of $W$ and $D$ in consensus clustering for Pax6.}
\label{fig:Plot_MAD_W100_D500}
\end{figure}

\begin{figure}[htbp]
\centering
\includegraphics[width=0.8\textwidth]{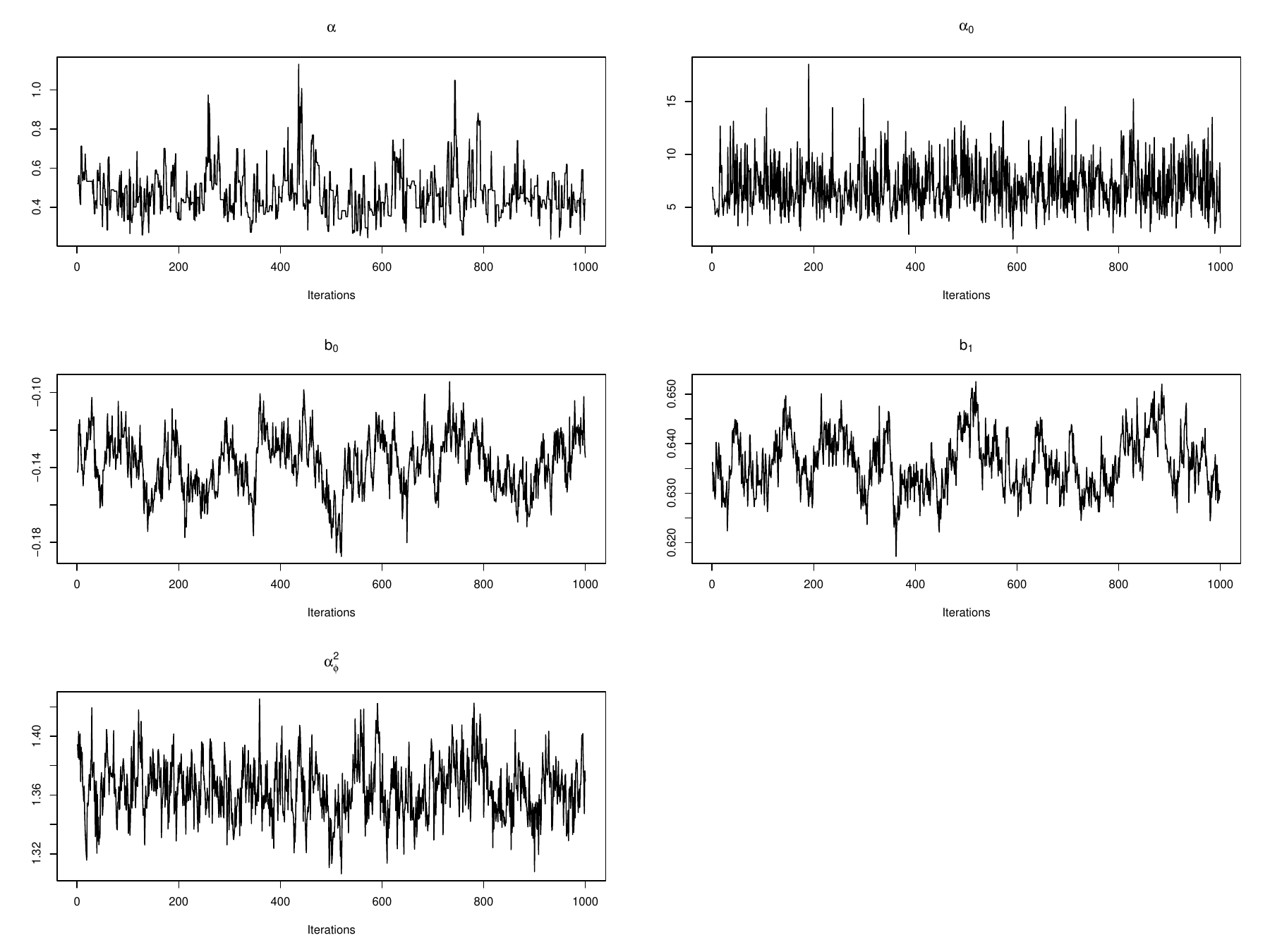}
\caption{Traceplots for concentration parameters and regression parameters for Pax6. }
\label{fig:trace_pax6}
\end{figure}

In this section, we present additional findings for clustering the Pax6 data using our C-HDP model. To infer the clustering, we perform consensus clustering \citep{Coleman2020} (see Section \ref{sec:consensus-clustering} for a review), which runs large numbers of chains (100) with a small number of iterations (500) to better exploit the posterior distribution of the clustering. Figure \ref{fig:Plot_MAD_W100_D500} shows the decision for choosing tuning parameters in consensus clustering. The truncation level is $J=30$. For the post-processing step with a fixed optimal clustering, the total number of MCMC iterations is 48000, followed by a burn-in of 43000 iterations and a thinning of 5, leading to 1000 samples in total. Traceplots for the post-processing step are shown in Figure \ref{fig:trace_pax6}, suggesting convergence.

\subsection{General Results} \label{appendix:additional-clustering-comparison-pax6}

\begin{figure}[htb]
\centering
\includegraphics[width=0.95\textwidth]{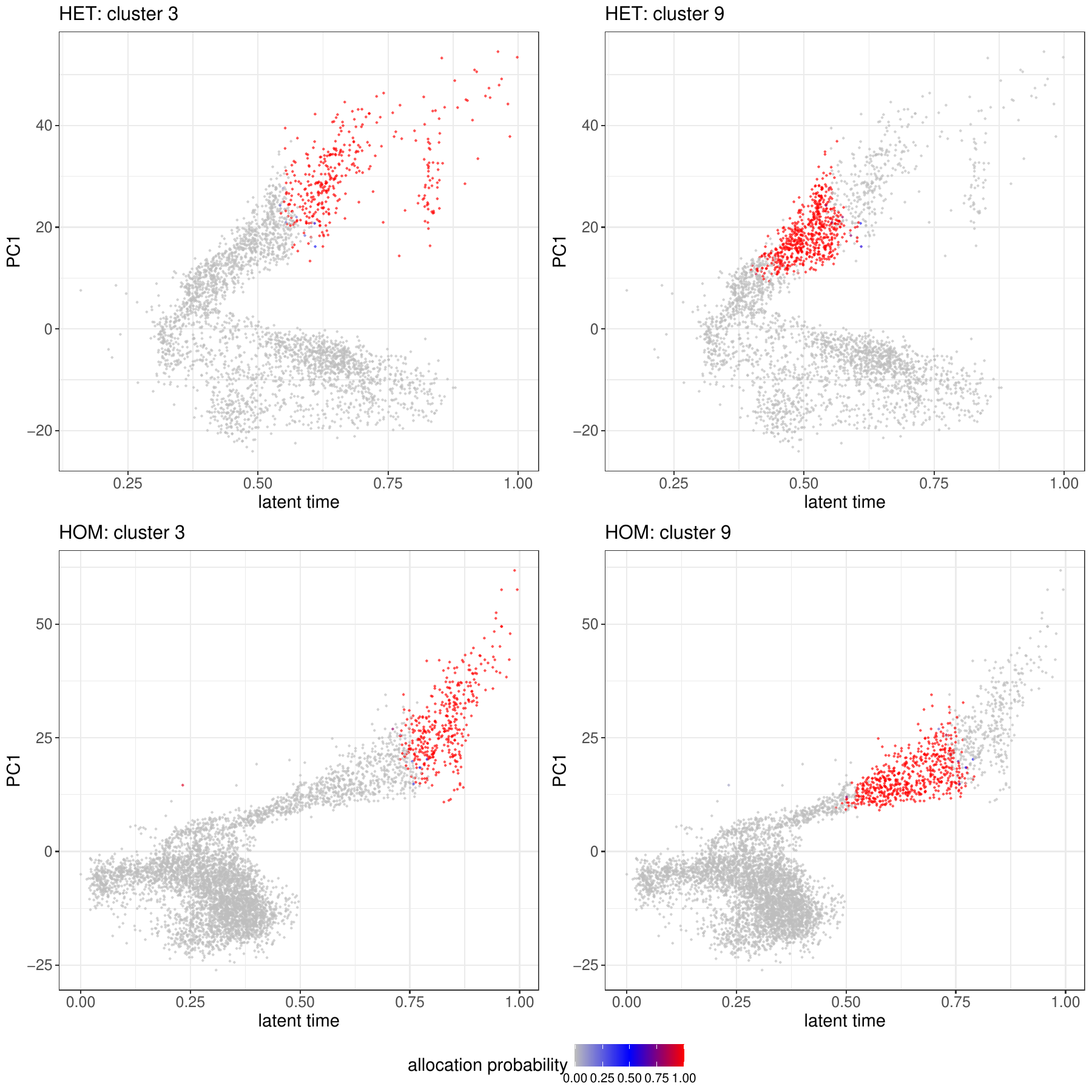}	
\caption{
	Plot of the first principal component against latent time for two example clusters in two experimental conditions. In each panel, cells are colored by the posterior allocation probability of belonging to the specific cluster indicated in the title.
}
\label{fig:pc1_vs_t_pp}
\end{figure}

By examining the posterior allocation probability for each cell,
Figure \ref{fig:pc1_vs_t_pp} suggests there is some uncertainty in cell allocations at the boundary between clusters 3 and 9, with moderate allocation probabilities (between 0.25 and 0.75 (dark blue)).

\begin{figure}[htb]
\centering
\includegraphics[width=0.5\textwidth]{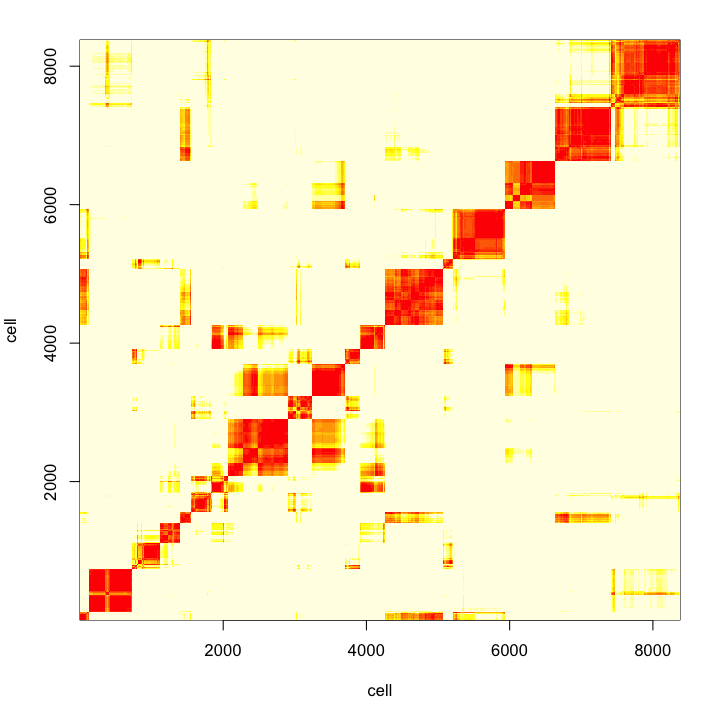}
\caption{Posterior similarity matrix for all cells without distinguishing between experimental conditions.}
\label{fig:psm_pax6_all}
\end{figure}

The posterior similarity matrix for both experimental conditions considered together is shown in Figure \ref{fig:psm_pax6_all}, where we observe some uncertainty in merging or splitting clusters.

\begin{figure}[htb]
\centering
\includegraphics[width=0.6\textwidth]{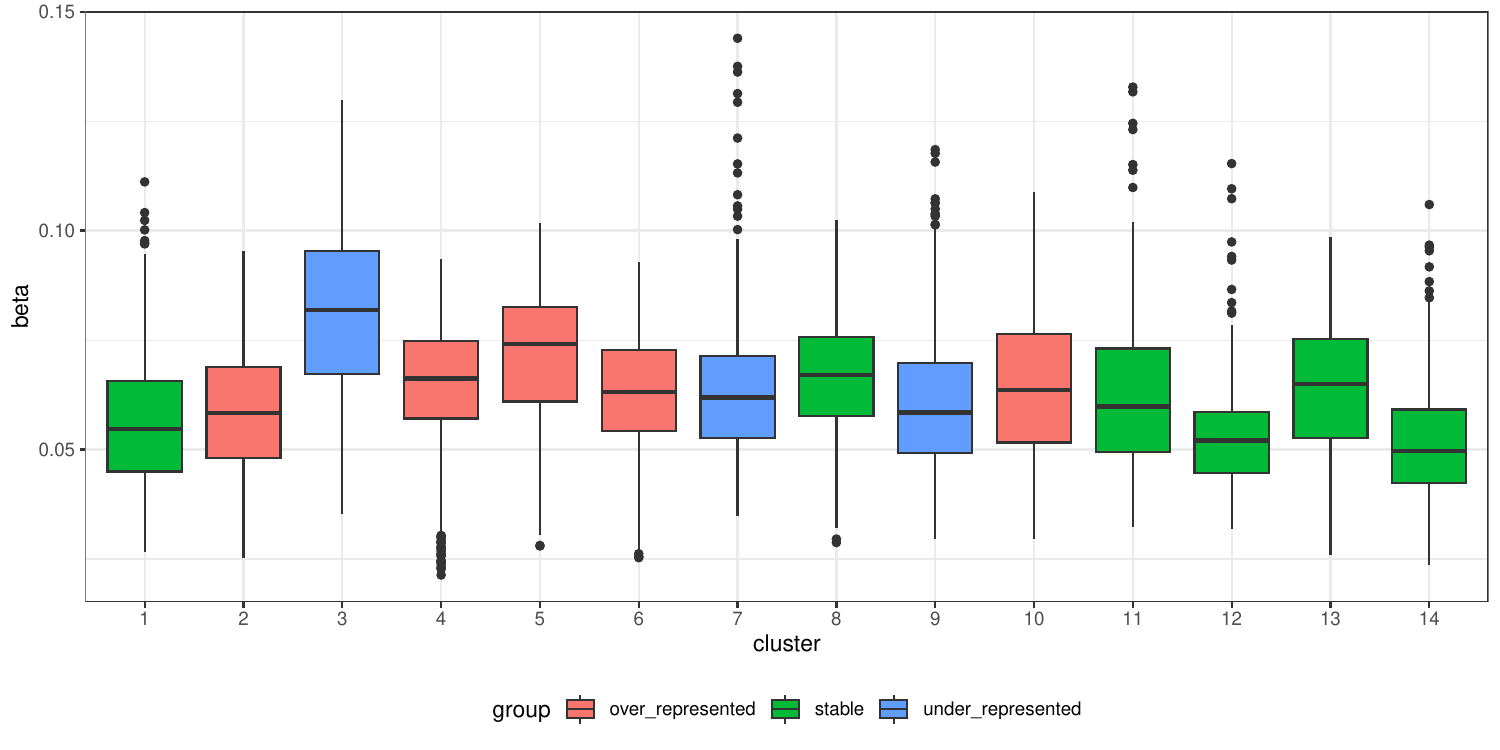}
\caption{Boxplots of posterior mean of capture efficiencies for each cluster. Clusters are colored by classification into over/under-represented and stable clusters in HOM.}
\label{fig:boxplot-beta-pax6}
\end{figure}

Figure \ref{fig:boxplot-beta-pax6} shows the boxplots for estimated capture efficiencies for each cluster. Except for cluster 3 showing comparatively larger capture efficiencies, the differences between other clusters are not evident.

\begin{figure}[htb]
\centering
\includegraphics[width=0.95\textwidth]{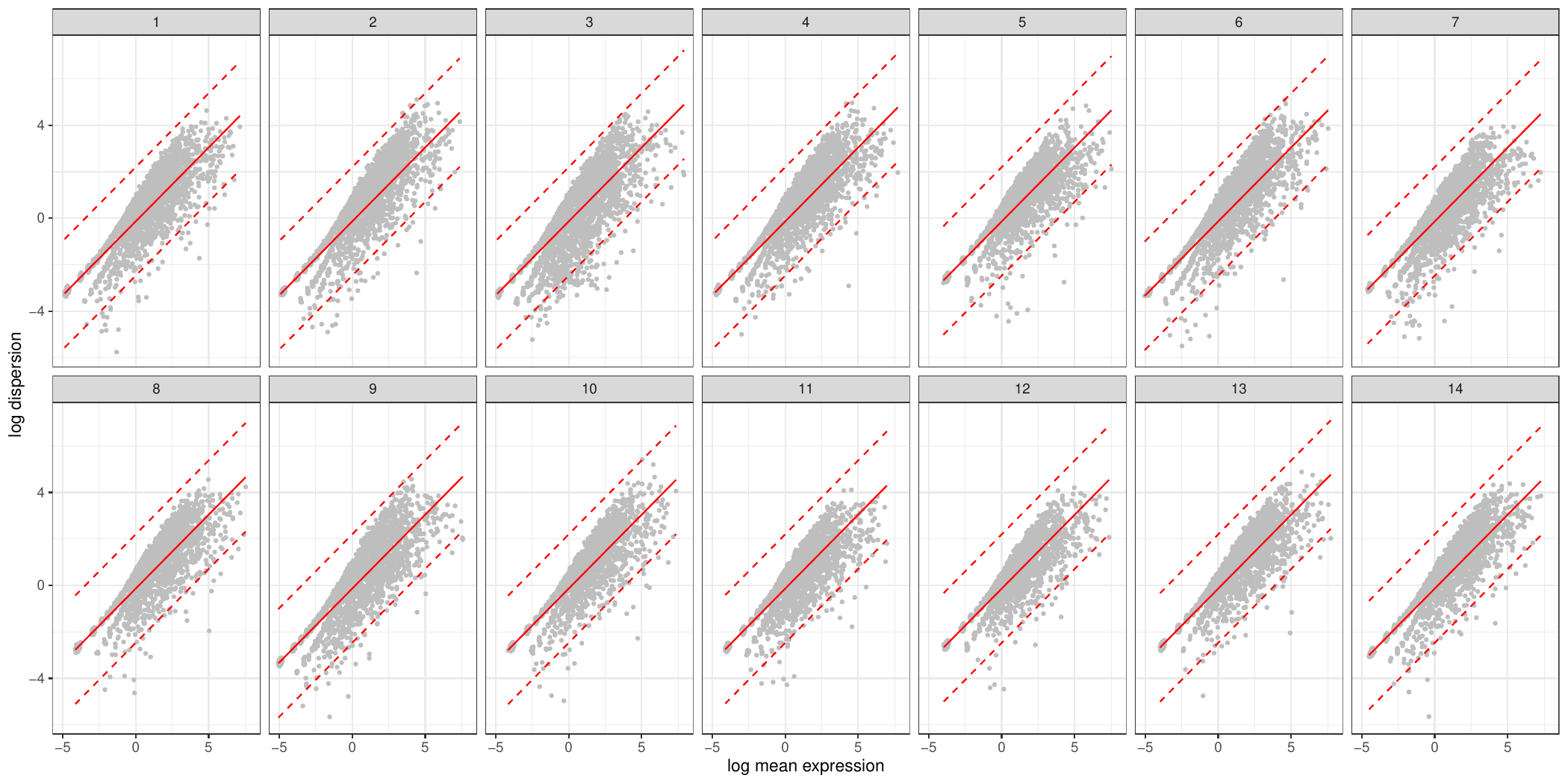}
\caption{Posterior mean of log dispersion against log mean expression for each cluster. The red solid line represents the estimated mean-dispersion relationship, with dashed lines denoting 95\% quantiles.}
\label{fig:mean_disperion_est_relation_by_cluster}
\end{figure}

As for component-specific parameters, Figure \ref{fig:mean_disperion_est_relation_by_cluster} shows that the fitted relationship between the mean expression and dispersion on the log scale is similar across clusters. The estimated relationship is based on the posterior mean of the regression parameter $\bmb$, and the 95\% CIs are computed from the posterior mean of the variance parameter $\alpha_\phi^2$.

\begin{figure}[htb]
\centering
\includegraphics[width=0.95\textwidth]{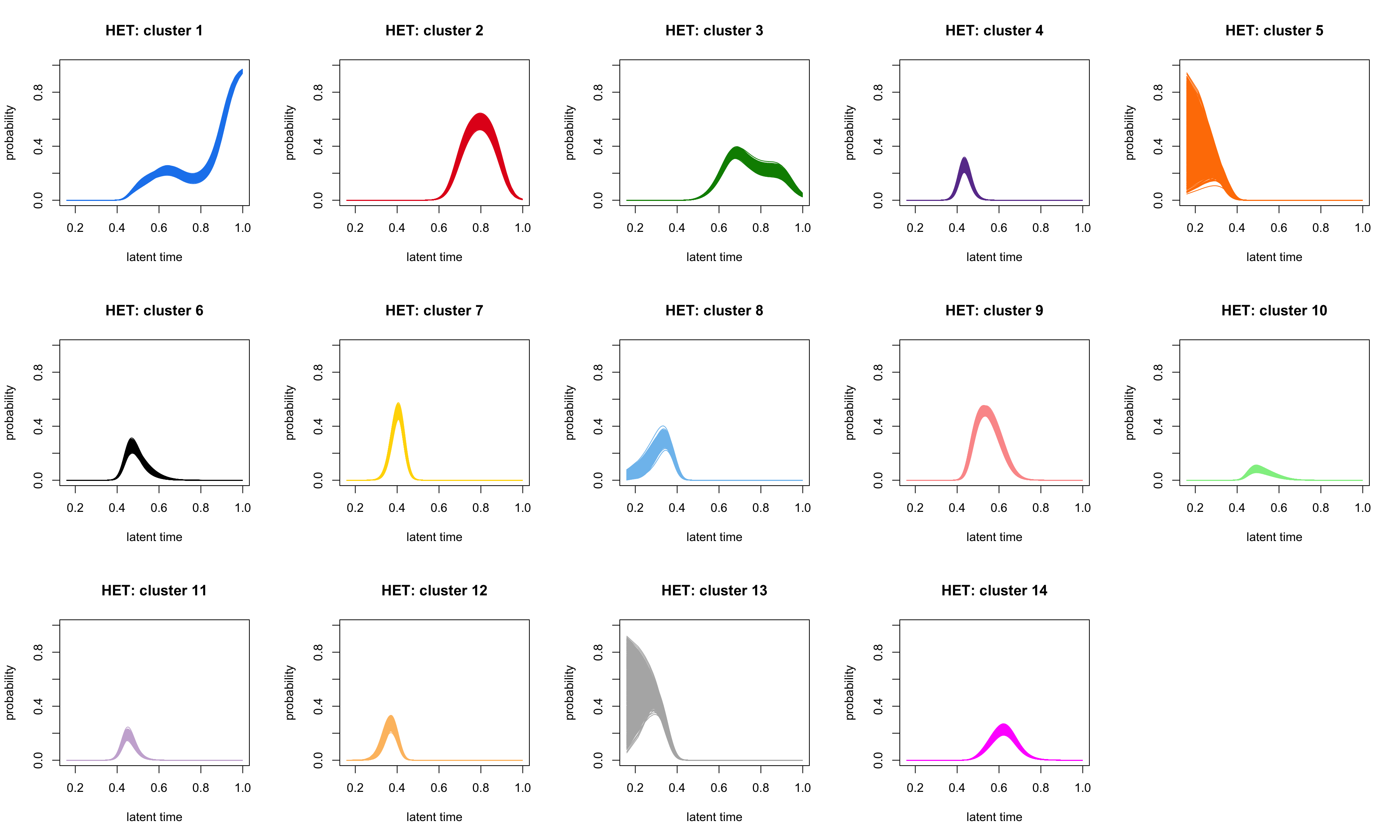}
\includegraphics[width=0.95\textwidth]{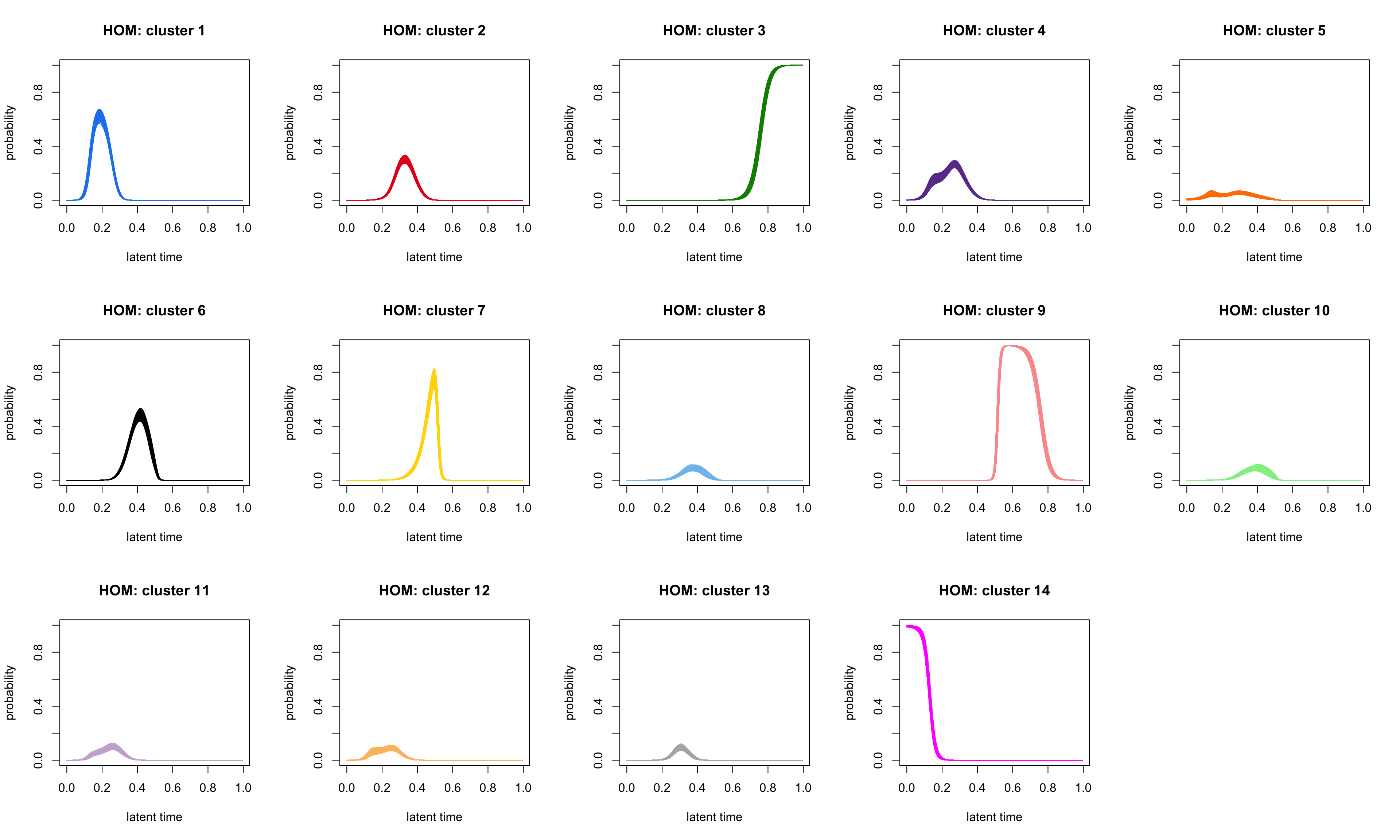}
\caption{Time-dependent probabilities for each cluster in each group for Pax6. Top 3 rows show the results for HET, with bottom 3 rows for HOM. Clusters 3, 7, 9 are under-represented in HOM.}
\label{fig:p_vs_t}
\end{figure}

Figure \ref{fig:p_vs_t} shows the time-dependent probabilities for all 14 clusters from two groups. All the three under-represented clusters in HOM (3, 7, 9) are associated with larger latent time and high probabilities that are close to 1, while for the control group the probabilities are relatively lower ($<0.6$). 

\clearpage

\subsection{Marker Genes} \label{appendix:pax6-marker-genes}

Following the marker gene detection method proposed in Section \ref{sec:marker-genes-method}, we identify global and local marker genes in Pax6. 

\subsubsection{Global Marker Genes}

\begin{figure}[!htb]
\begin{minipage}[h]{0.5\textwidth}
	\includegraphics[width=0.9\textwidth]{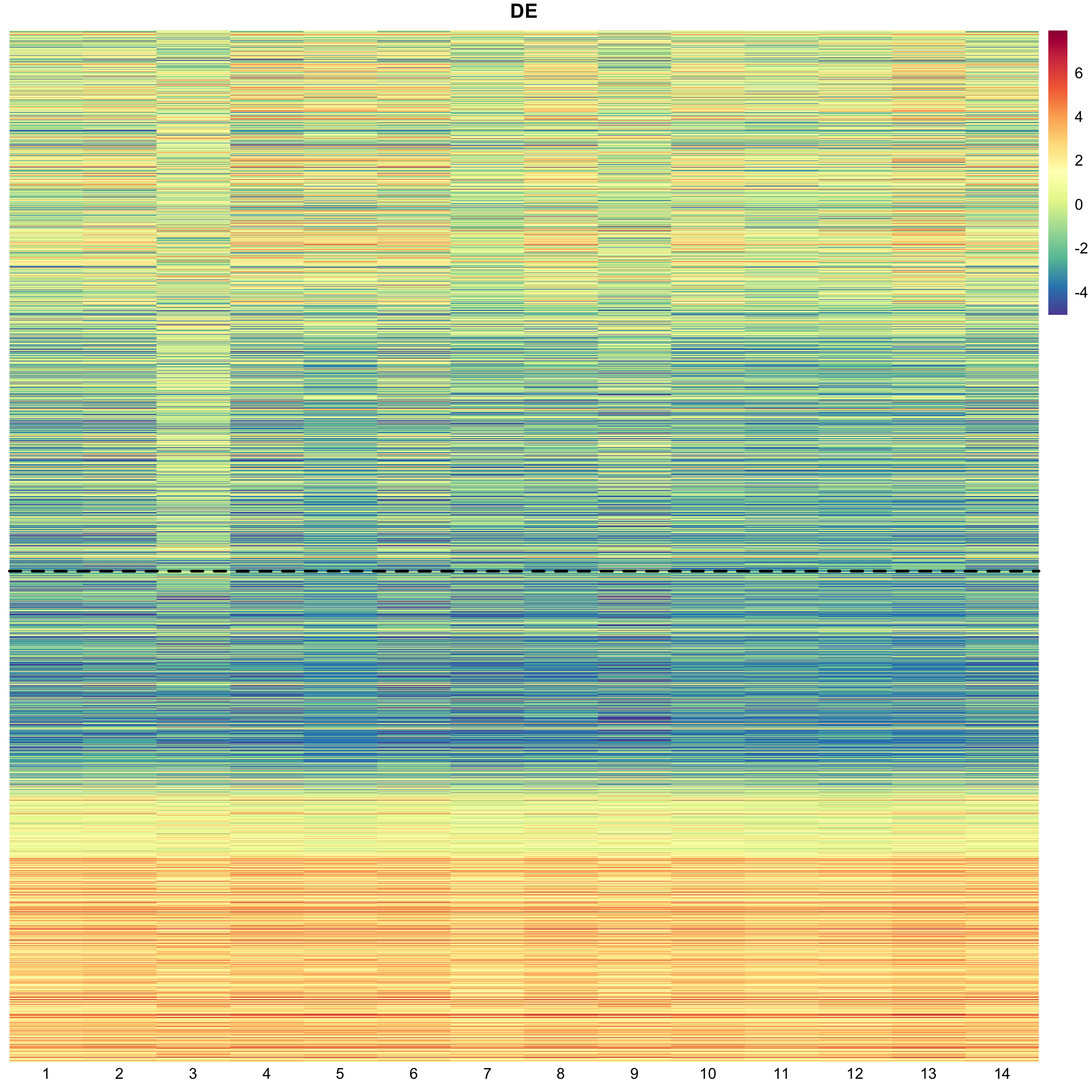}
\end{minipage}
\begin{minipage}[h]{0.5\textwidth}
	\includegraphics[width=0.9\textwidth]{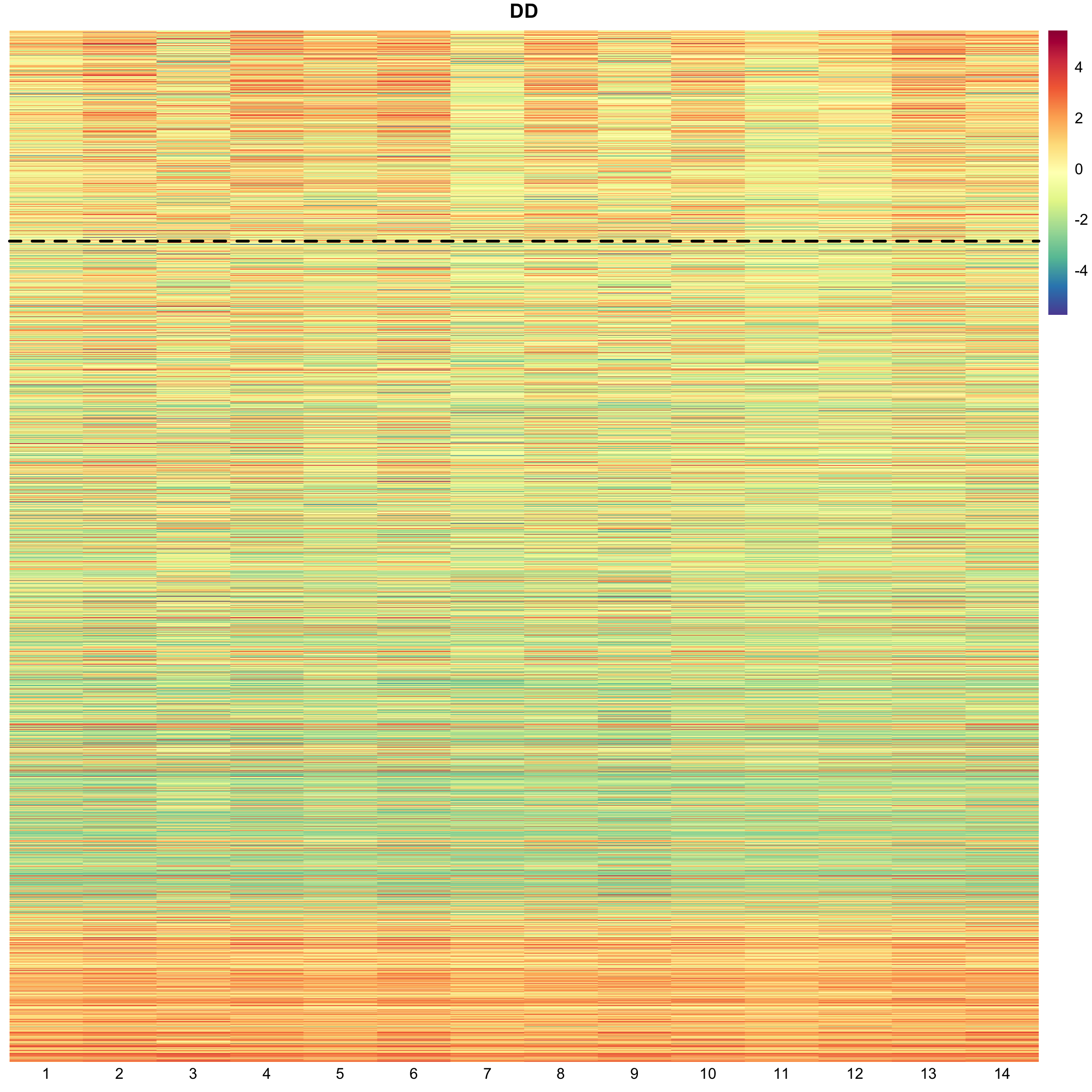}		
\end{minipage}
\caption{Global marker genes: Heatmaps of posterior mean of the logarithm of mean parameters $\mu_{j,g}^*$ (left) and dispersion parameters $\phi_{j,g}^*$ (right). Rows correspond to genes and columns correspond to clusters. Genes are ordered by decreasing maximum posterior tail probabilities $P_g^*$ and $L_g^*$ from top to bottom. Genes above the black dashed lines are global marker genes. }
\label{fig:global_genes_heatmap_by_cluster}
\end{figure}

\begin{figure}[!htb]
\begin{minipage}[h]{0.5\textwidth}
	\includegraphics[width=0.9\textwidth]{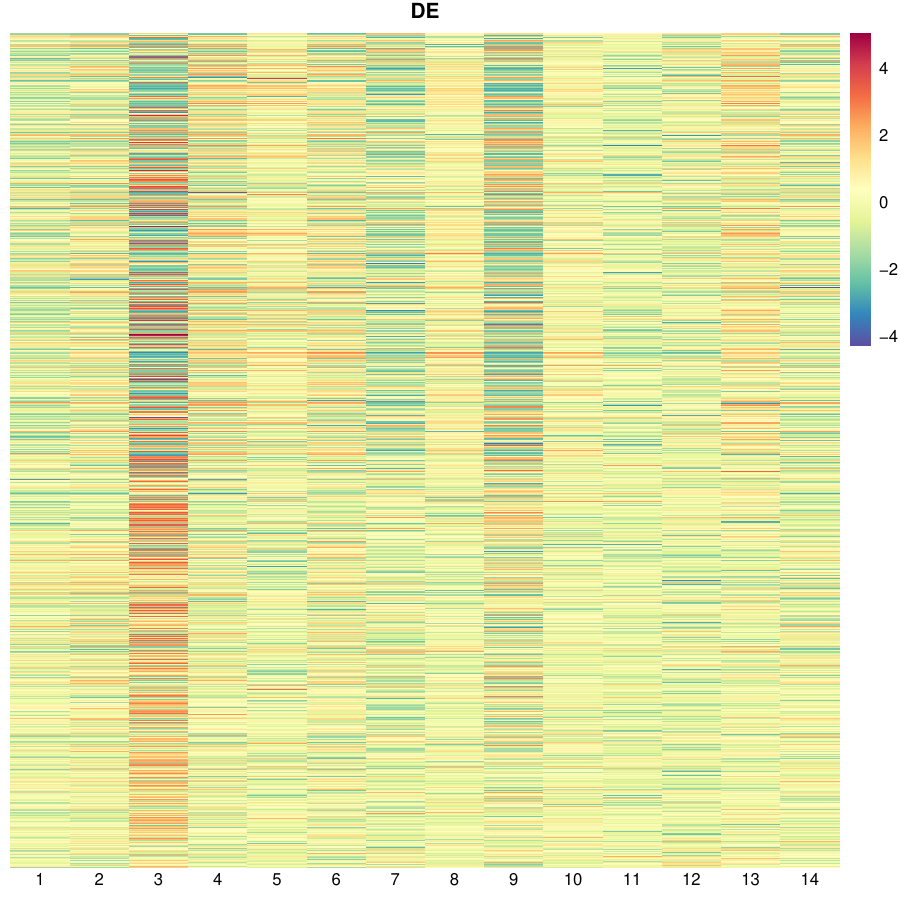}
\end{minipage}
\begin{minipage}[h]{0.5\textwidth}
	\includegraphics[width=0.9\textwidth]{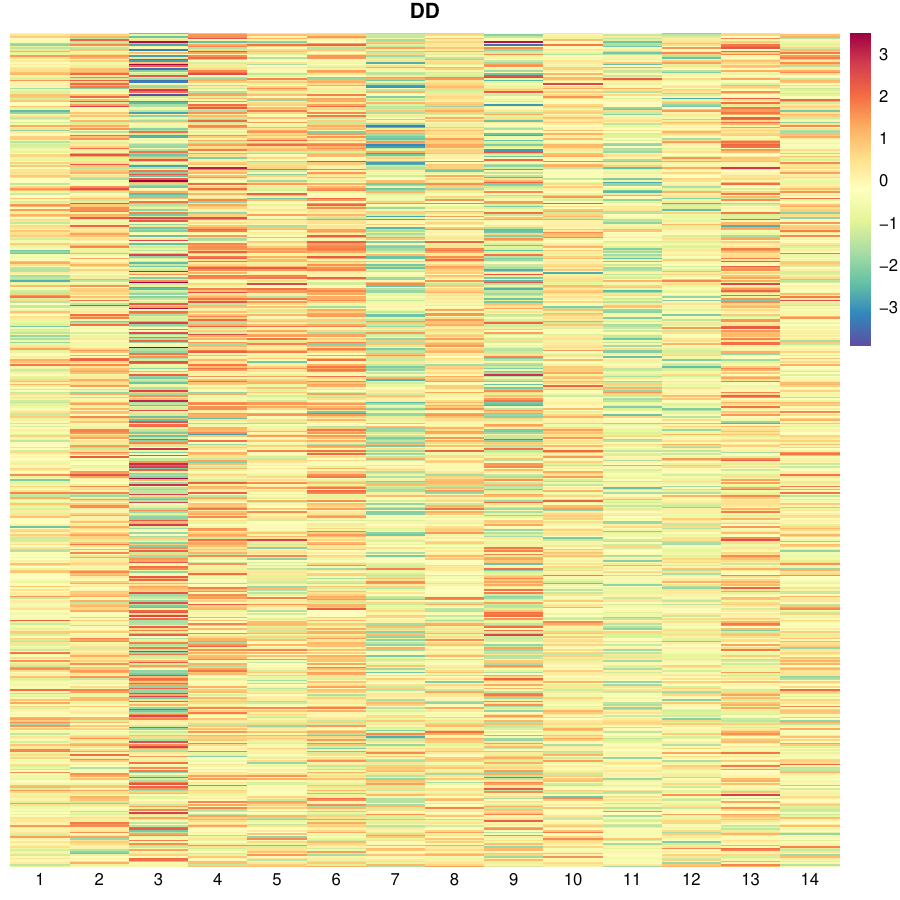}		
\end{minipage}
\caption{Global marker genes: Heatmaps of relative log mean parameters $\mu_{j,g}^*$ (left) and dispersion parameters $\phi_{j,g}^*$ (right). Rows correspond to genes and columns correspond to clusters. Only global marker genes are plotted and they are ordered by decreasing maximum posterior tail probabilities $P_g^*$ and $L_g^*$ from top to bottom.}
\label{fig:global_genes_relative}
\end{figure}

The thresholds for global DE and for DD genes are both set to 2.5. There are 52.39\% and 20.40\% of the total genes classified as global DE and global DD genes. The identified global DE genes match the findings in \cite{phdthesis_kai}. Figure \ref{fig:global_genes_heatmap_by_cluster} displays heatmaps of the estimated mean and dispersion parameters for all genes in all 14 clusters. Cluster 3 exhibits distinct mean expression patterns compared to the other clusters, with most global DE genes showing higher expression levels. As for global DD genes, clusters 7, 11 and 12 tend to have smaller dispersion levels. To better visualize the differences across clusters, we compute the relative values of mean and dispersion parameters, defined as the difference relative to the average values across clusters. Figure \ref{fig:global_genes_relative} suggests that in addition to cluster 3, clusters 4, 6, 13 also seem to have mean expressions moderately higher than the average, whereas clusters 7 and 9 have below-average mean expressions. As for dispersion, clusters 2, 4, 5, 6, 8, 10, 13, 14 tend to have higher dispersion parameters, which are either over-represented or stable clusters in HOM.

Additionally, a list of 70 ``important'' genes is provided by our collaborators from the Centre for Discovery Brain Sciences, University of Edinburgh, which is shown in Table \ref{tab:important genes}. Among these 70 important genes, 51 are classified in global DE genes and 27 are classified as global DD genes, which are shown in Table \ref{tab:DE important genes} and Table \ref{tab:DD important genes}.

\begin{table}[!htb]
\centering
\caption{List of important genes.}
\label{tab:important genes}
\scalebox{0.8}{\begin{tabular}{lllllllllll}
		\hline
		Dlx6os1 & Dlx2 & Sp9 & Nrxn3 & Dlx1 & Ccnd2 & Arx & Dlx5 & Top2a & Rrm2 & Pclaf \\ 
		Ube2c & Pfn2 & Hmgb2 & Cdca7 & Gm13889 & Sox2 & Etv1 & Cenpf & Gm26917 & Slain1 & Sp8 \\ 
		Gad2 & Hmgn2 & Cenpe & Smc2 & Insm1 & Nusap1 & Tpx2 & Neurod6 & Neurod2 & Sox5 & Cntn2 \\ 
		Mef2c & Mapt & Tbr1 & Nrp1 & Wnt7b & Nfix & Id2 & Neurod1 & Fezf2 & Nrxn1 & Satb2 \\ 
		Neurog2 & Fgfr1 & Crabp1 & Lhx2 & Zic1 & Mfap4 & Nrp2 & Zbtb20 & Ccnd2 & Nhlh1 & Plcb1 \\ 
		Nhlh2 & Lhx9 & Lmo4 & Prdm13 & Emx2 & Cited2 & Insm1 & Pou3f2 & Robo2 & Pou3f1 & Ptn \\ 
		Cux2 & Wnt7b & Pou3f3 & Cux1 & & & &  & &  &  \\    \hline
\end{tabular}}
\end{table}

\begin{table}[!htb]
\centering
\caption{List of important genes that are classified as globally differentially expressed (DE).}
\label{tab:DE important genes}
\scalebox{0.8}{\begin{tabular}{lllllllllll}
		\hline
		Pou3f3 & Satb2 & Nrp2 & Cntn2 & Lhx9 & Nhlh1 & Cenpf & Tbr1 & Dlx1 & Dlx2 & Cdca7 \\ 
		Sp9 & Neurod1 & Gm13889 & Nusap1 & Plcb1 & Insm1 & Tpx2 & Ube2c & Arx & Sox2 & Nhlh2 \\ 
		Neurog2 & Cenpe & Lmo4 & Prdm13 & Pou3f2 & Smc2 & Pou3f1 & Dlx6os1 & Dlx5 & Ptn & Neurod6 \\ 
		Ccnd2 & Sox5 & Hmgb2 & Nrp1 & Crabp1 & Pclaf & Mfap4 & Neurod2 & Top2a & Mapt & Mef2c \\ 
		Rrm2 & Etv1 & Nrxn3 & Sp8 & Wnt7b & Robo2 & Nrxn1 & &  &  &  \\ 		
		\hline
\end{tabular}}
\end{table}

\begin{table}[!htb]
\centering
\caption{List of important genes that are classified as globally differentially dispersed (DD).}
\label{tab:DD important genes}
\scalebox{0.8}{\begin{tabular}{lllllllllll}
		\hline
		Lhx9 & Nhlh1 & Cenpf & Dlx2 & Cdca7 & Neurod1 & Gm13889 & Nusap1 & Insm1 & Cenpe & Smc2 \\ 
		Dlx5 & Neurod6 & Ccnd2 & Hmgb2 & Nrp1 & Pclaf & Mfap4 & Neurod2 & Top2a & Etv1 & Wnt7b \\ 
		Zbtb20 & Robo2 & Gm26917 & Nrxn1 & Emx2 &  &  &  & &  &  \\ 		
		\hline
\end{tabular}}
\end{table}

\subsubsection{Local Marker Genes}

\begin{figure}[!bp]
\centering
\includegraphics[width=0.8\textwidth]{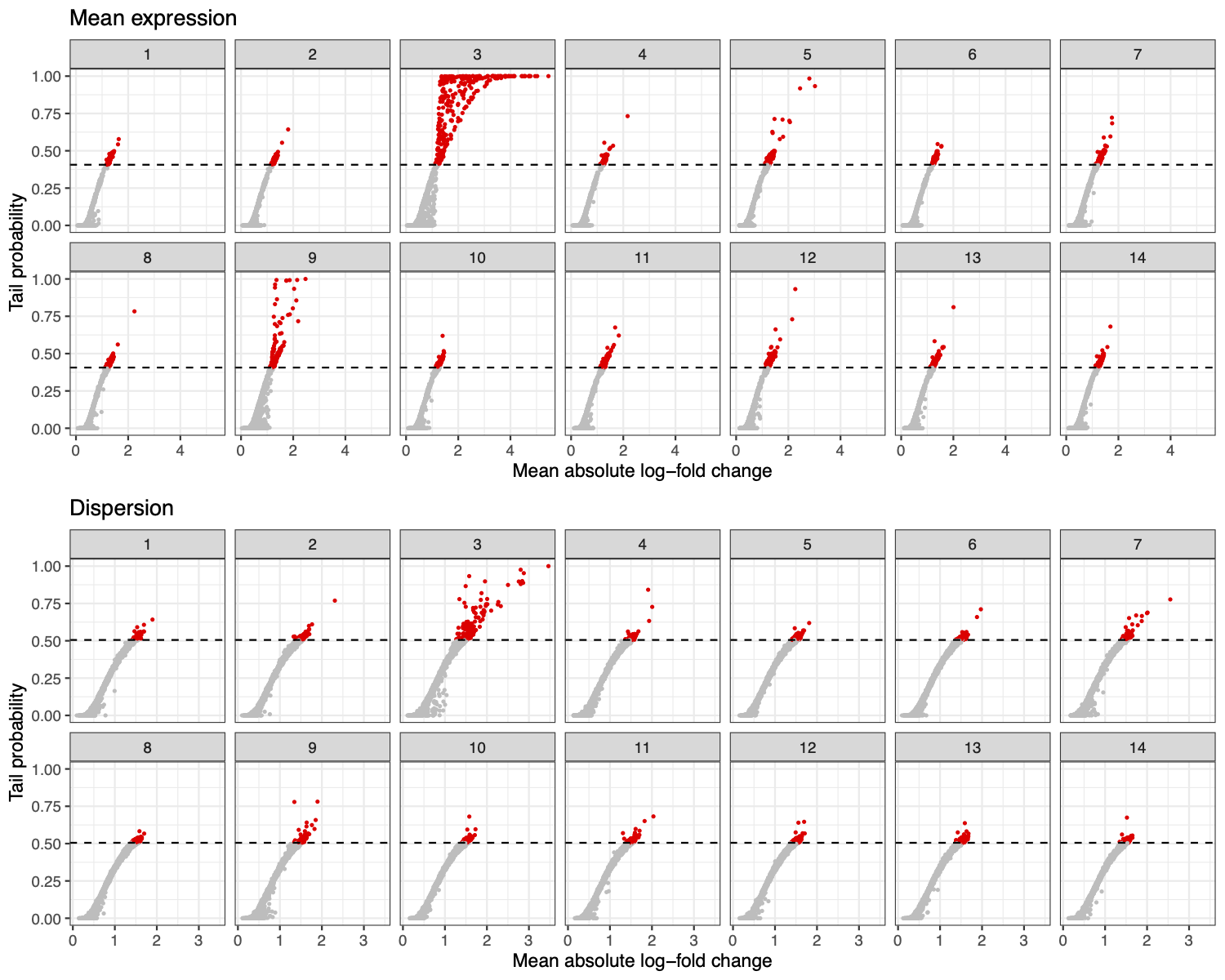}
\includegraphics[width=0.8\textwidth]{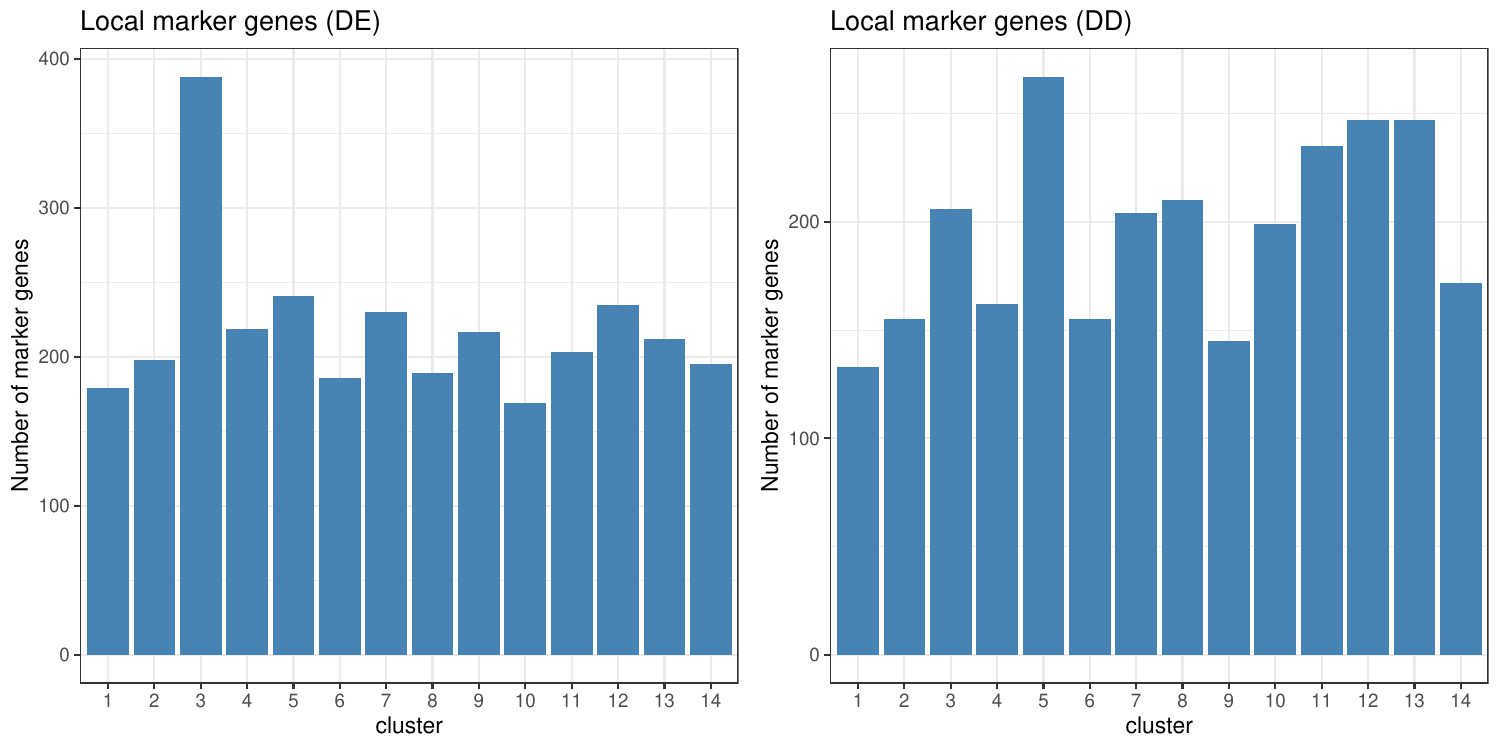}
\caption{Local marker genes: Top 2 panels show the minimum tail probabilities $P_{g,j}^*$ and $L_{g,j}^*$ against mean absolute LFCs for each cluster (Row 1: mean expression, Row 2: dispersion). The black dashed line indicates the threshold to determine local genes. The barplots on the bottom summarize the number of local DE (left) and DD (right) genes in each cluster.}
\label{fig:local_genes_summary}
\end{figure}

In terms of local marker genes, the thresholds for local DE and DD genes are set to 1.2. Figure \ref{fig:local_genes_summary} shows the minimum posterior tail probabilities against the mean absolute LFCs for each cluster, along with the number of local genes in each cluster. Cluster 3 has the largest number of local DE genes, whereas the numbers of local DD genes are more evenly spread across 14 clusters.

\begin{figure}[!htb]
\centering
\includegraphics[width=1\textwidth]{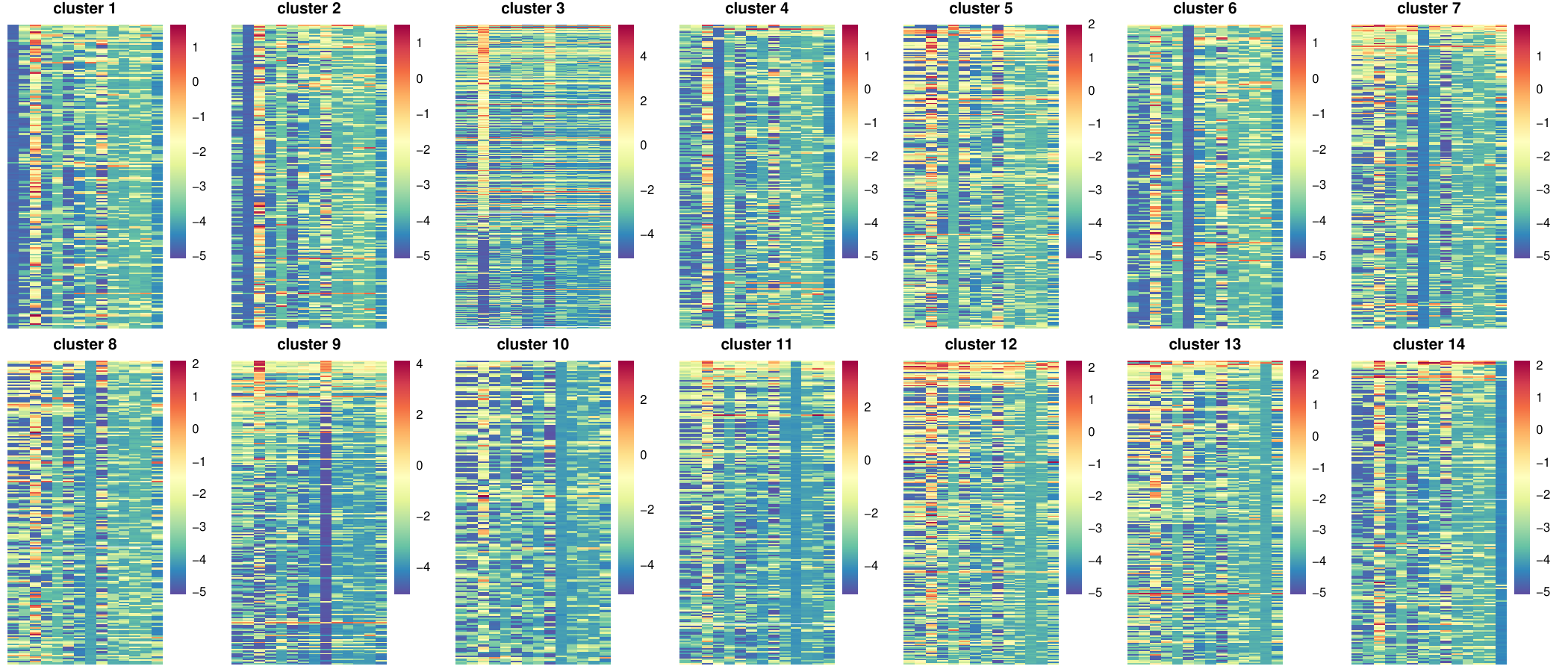}
\includegraphics[width=1\textwidth]{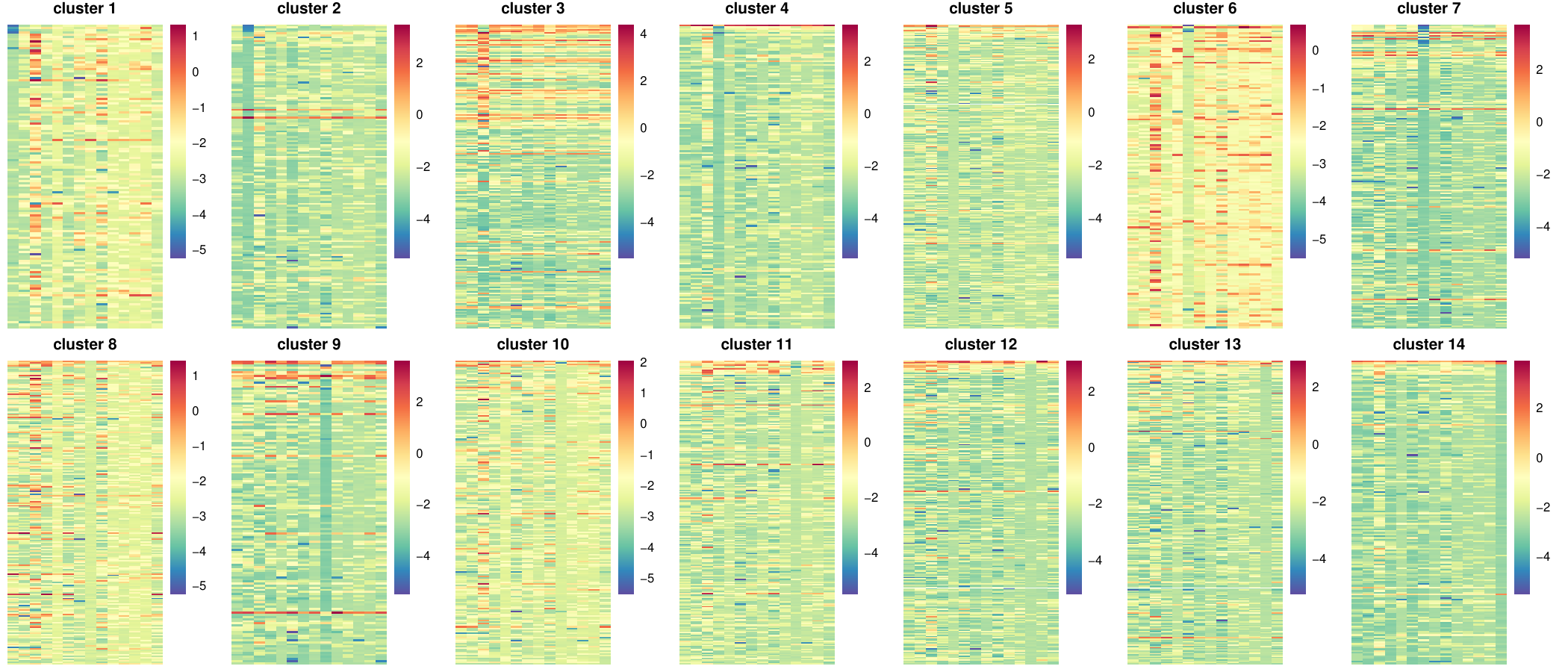}
\caption{Local marker genes: Heatmaps of posterior mean of the logarithm of mean parameters $\mu_{j,g}^*$ (top 2 rows) and dispersion parameters $\phi_{j,g}^*$ (bottom 2 rows). Rows correspond to genes and columns correspond to clusters. Each panel corresponds to a specific cluster. Only the local DE (or DD) marker genes for this specific cluster are plotted and they are ordered by decreasing minimum posterior tail probabilities $P_{g,j}^*$ and $L_{g,j}^*$ from top to bottom.}
\label{fig:local_genes_heatmap_by_cluster}
\end{figure}

Figure \ref{fig:local_genes_heatmap_by_cluster} displays the estimated mean expressions and dispersions for local marker genes in each cluster. Similar to global DE genes, the local DE genes in cluster 3 mainly exhibit higher mean expression levels. Moreover, for local DE genes in the other clusters, their mean expression levels are also higher in cluster 3. This trend is also observed for local DD genes in cluster 3.

\subsection{Latent Counts} \label{appendix:pax6-latent-count}

\cite{Tang2020} and \cite{Liu2024} provide posterior mean of the latent counts given the allocation variables, capture efficiencies and unique parameters 
\begin{align*}
\E(y_{c,g,d}^0 | y_{c,g,d},  z_{c,d} = j,\beta_{c,d}, \mu^*_{j,g}, \phi^*_{j,g}) = y_{c,g,d} \frac{ \mu^*_{j,g} + \phi^*_{j,g}}{ \mu^*_{j,g}\beta_{c,d}+ \phi^*_{j,g} } +\mu^*_{j,g}\frac{ \phi^*_{j,g}(1- \beta_{c,d})}{ \mu^*_{j,g}\beta_{c,d}+ \phi^*_{j,g} },
\end{align*}
which can be used to approximate the posterior mean of latent counts as
\begin{align} \label{eq:latent count}
\E(y_{c,g,d}^0 | \bY) \approx  \frac{1}{L} \sum_{l=1}^L \E(y_{c,g,d}^0 | y_{c,g,d},  z_{c,d}^{(l)} = j,\beta_{c,d}^{(l)}, \mu^{*\,(l)}_{j,g}, \phi^{*\,(l)}_{j,g}).
\end{align}

\begin{figure}[!h]
\centering
\begin{minipage}[h]{0.45\textwidth}
	\includegraphics[width=0.95\textwidth]{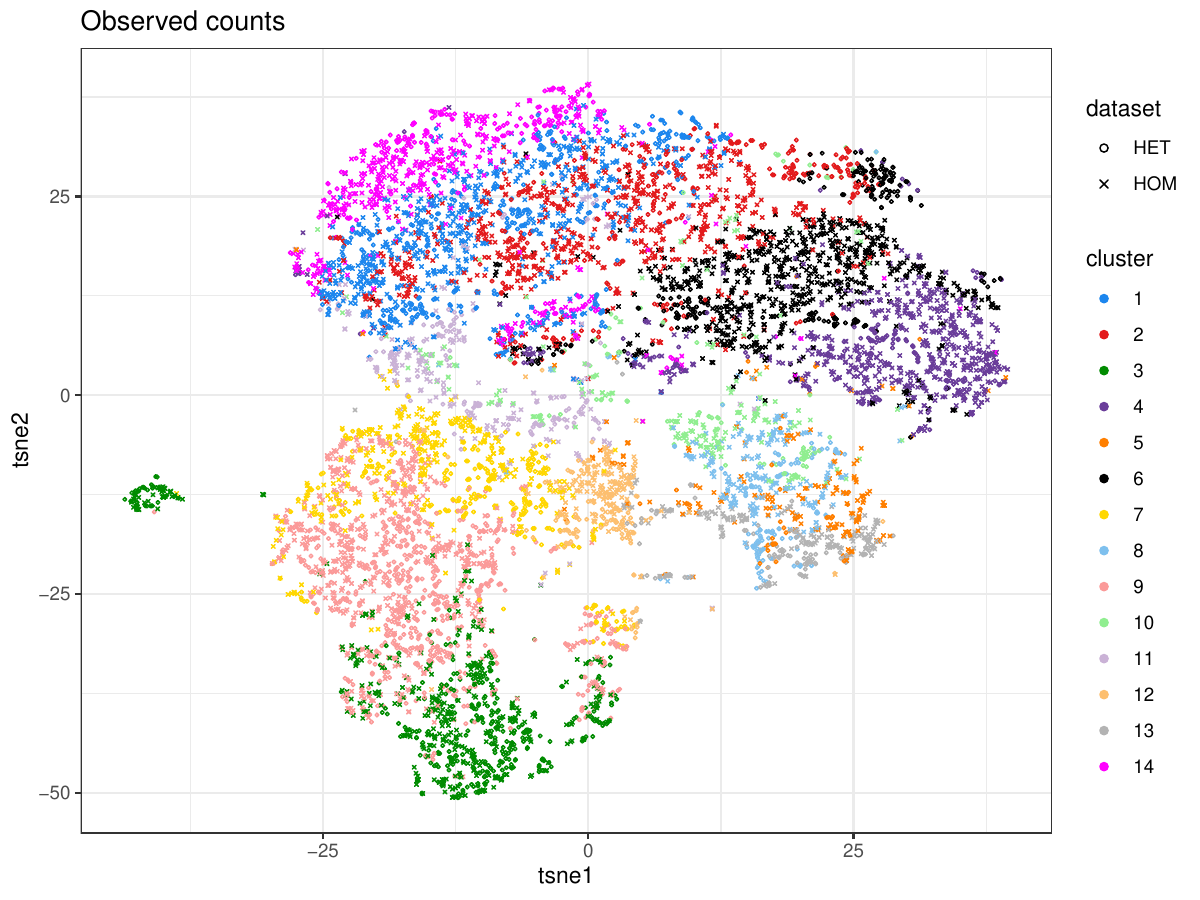}
\end{minipage}
\begin{minipage}[h]{0.45\textwidth}
	\includegraphics[width=0.95\textwidth]{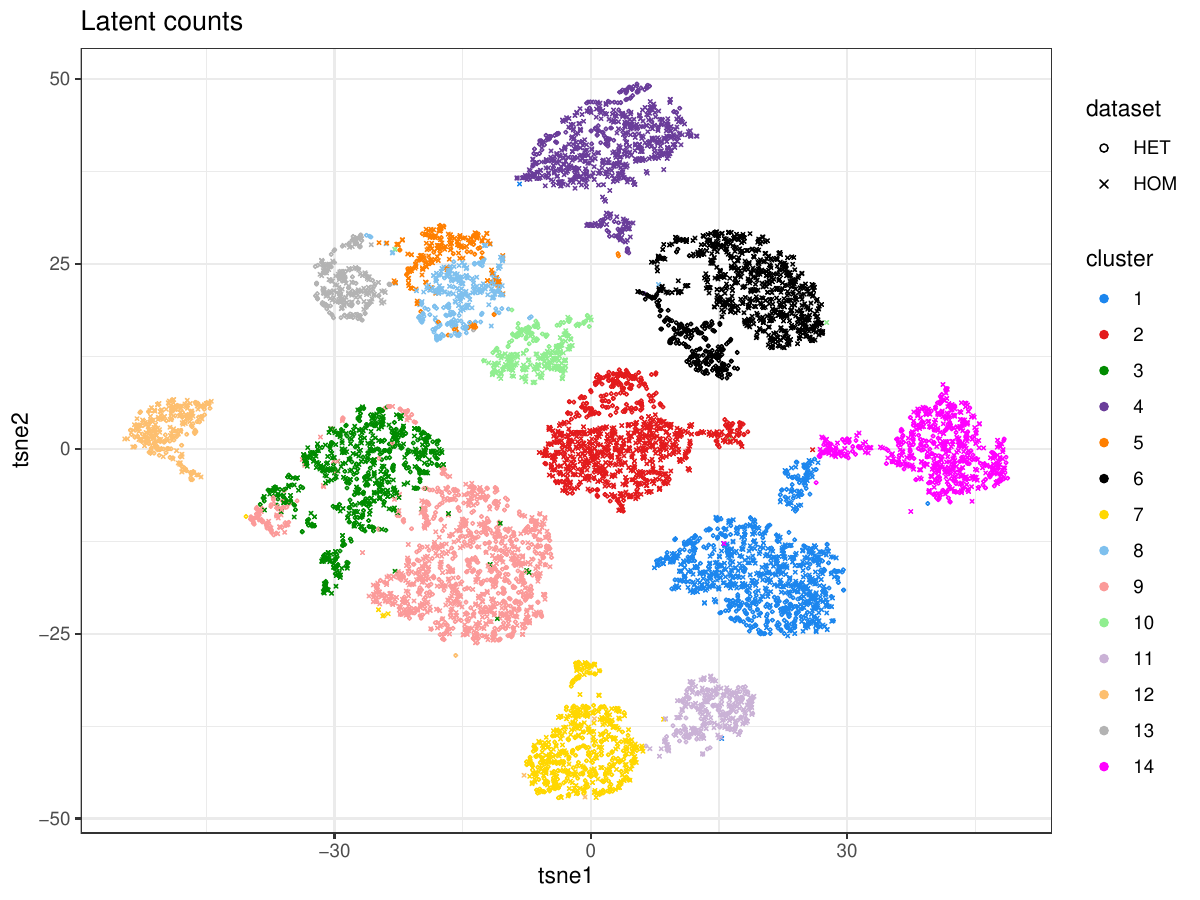}	
\end{minipage}
\caption{t-SNE plot for observed counts (left) and posterior mean of latent counts (right). Cells belonging to different clusters are shown in different colors. Different symbols indicate different experimental conditions. }
\label{fig:t-sne-plot-pax6}
\end{figure}

Figure \ref{fig:t-sne-plot-pax6} shows the t-SNE \citep{Van2008visualizing} plot for the observed and estimated latent counts from Eq. \eqref{eq:latent count} using all 1000 samples from the post-processing step. From the observed counts, some clusters are already quite separated, such as the purple and dark green clusters. The separation is much more apparent in the latent counts. 

\begin{figure}[!b]
\centering
\includegraphics[width=0.8\textwidth]{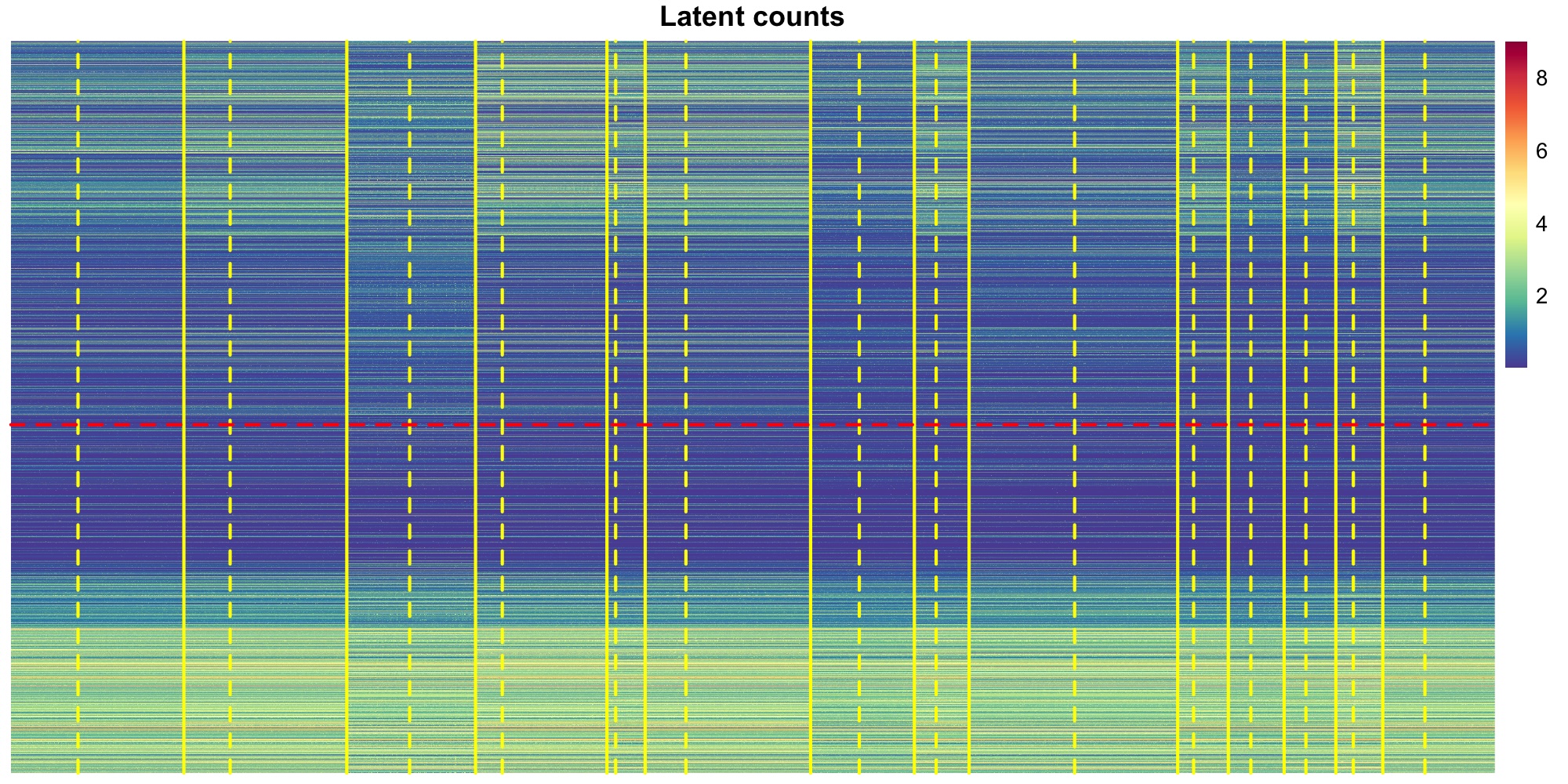}
\caption{Heatmap for the latent counts on the log scale after adding a pseudo-count of 1. Each row represents a gene and each column represents a cell. Genes are ordered by decreasing maximum posterior tail probabilities $P_g^*$ from top to bottom and those above the red dashed lines are global DE genes. Yellow solid lines separate clusters. Yellow dashed lines separate cells from different datasets within each cluster.}
\label{fig:heatmap_gene_latent_count_by_cluster}
\end{figure}

\begin{figure}[!h]
\centering
\includegraphics[width=0.8\textwidth]{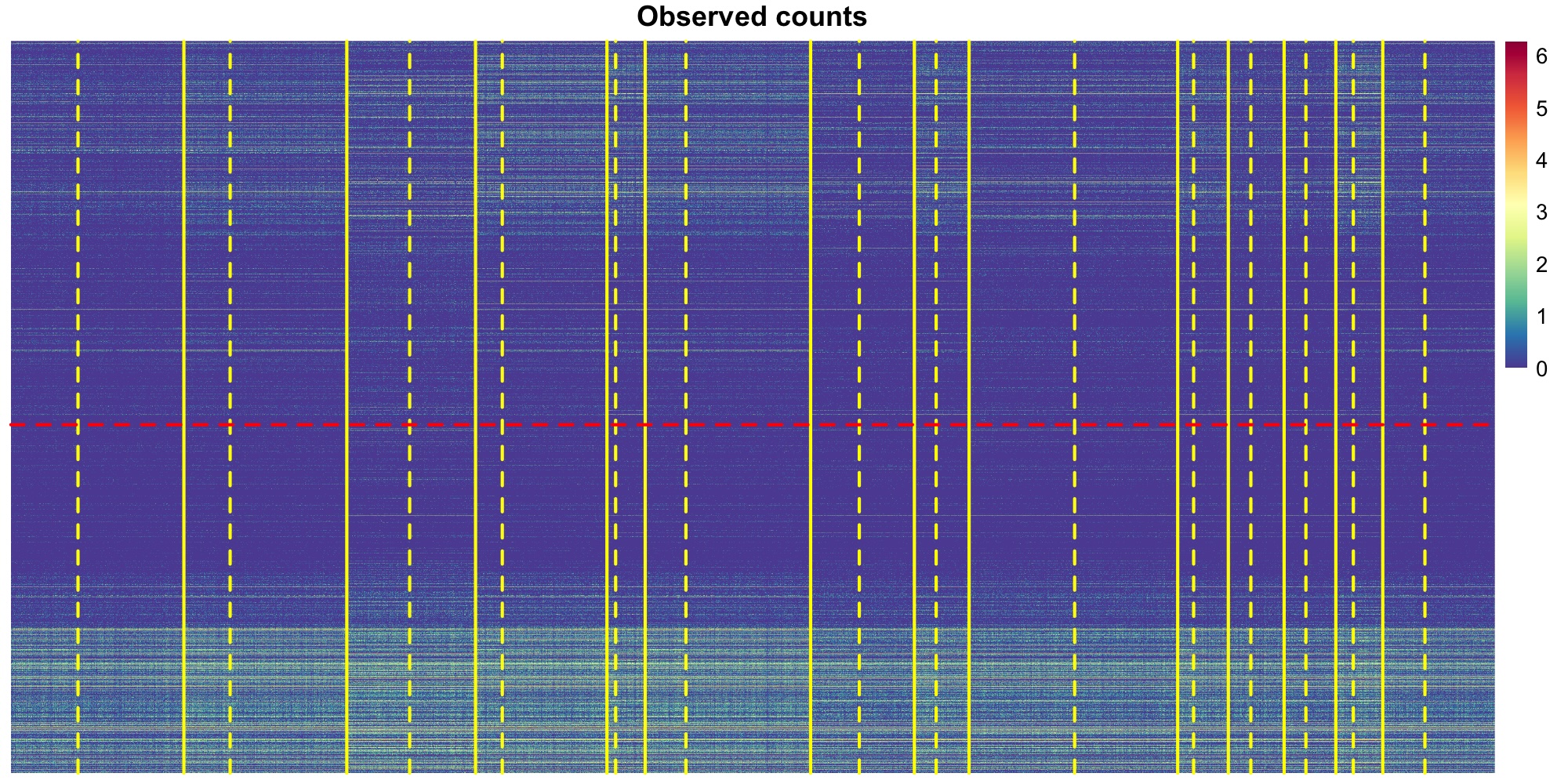}
\caption{Heatmap for the observed counts on the log scale after adding a pseudo-count of 1. Each row represents a gene and each column represents a cell. Genes are ordered by decreasing maximum posterior tail probabilities $P_g^*$ from top to bottom and those above the red dashed lines are global DE genes. Yellow solid lines separate clusters. Yellow dashed lines separate cells from different datasets within each cluster.}
\label{fig:heatmap_gene_observed_count_by_cluster}
\end{figure}

Figure \ref{fig:heatmap_gene_latent_count_by_cluster} and Figure \ref{fig:heatmap_gene_observed_count_by_cluster} show the estimated latent counts and observed counts for each cell on the log scale after adding a pseudo-count of 1. Global DE genes between different clusters show different patterns of latent counts, whilst within each cluster, the pattern is similar across datasets.

\subsection{Posterior Predictive Checks} \label{appendix:ppc-pax6}

\begin{figure}[!b]
\centering
\includegraphics[width=0.9\textwidth]{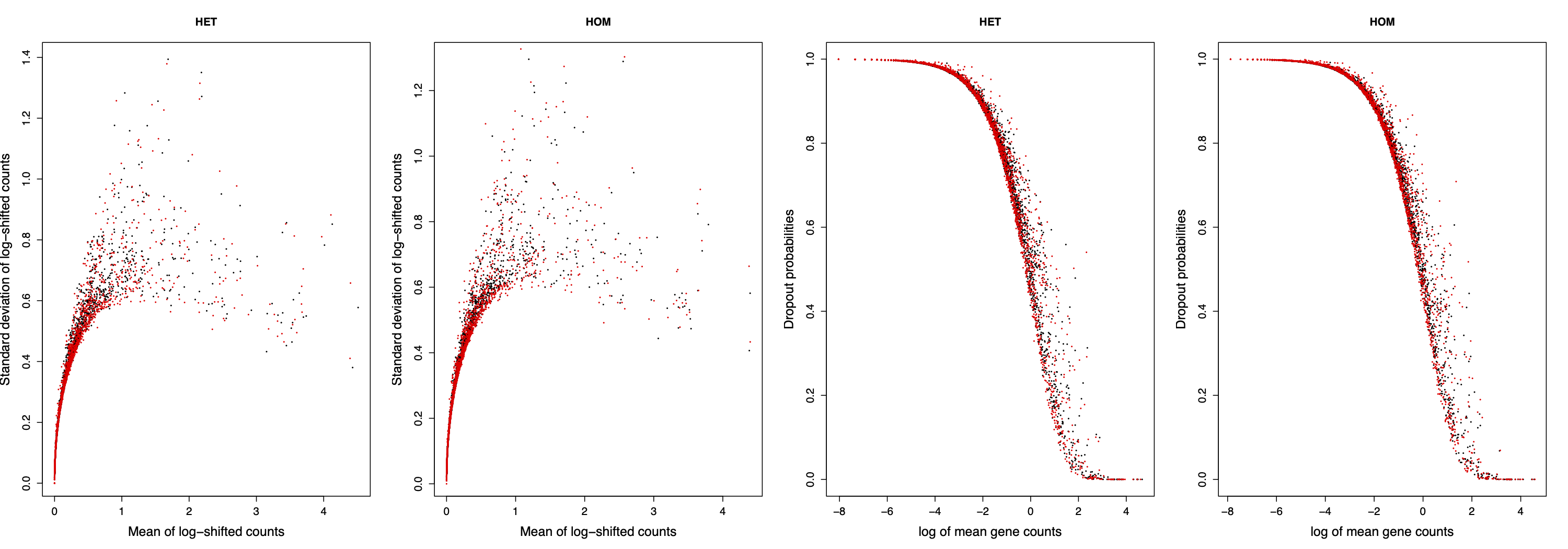}
\caption{Posterior predictive checks with one single replicated dataset. Left two plots show the relationship between mean and standard deviation of log-shifted counts in true (red) and replicated data (black) for HET and HOM. Right two plots show the relationship between log of mean counts and dropout probabilities. }
\label{fig:ppc_single1}
\end{figure}

\begin{figure}[!p]
\centering
\includegraphics[width=0.9\textwidth]{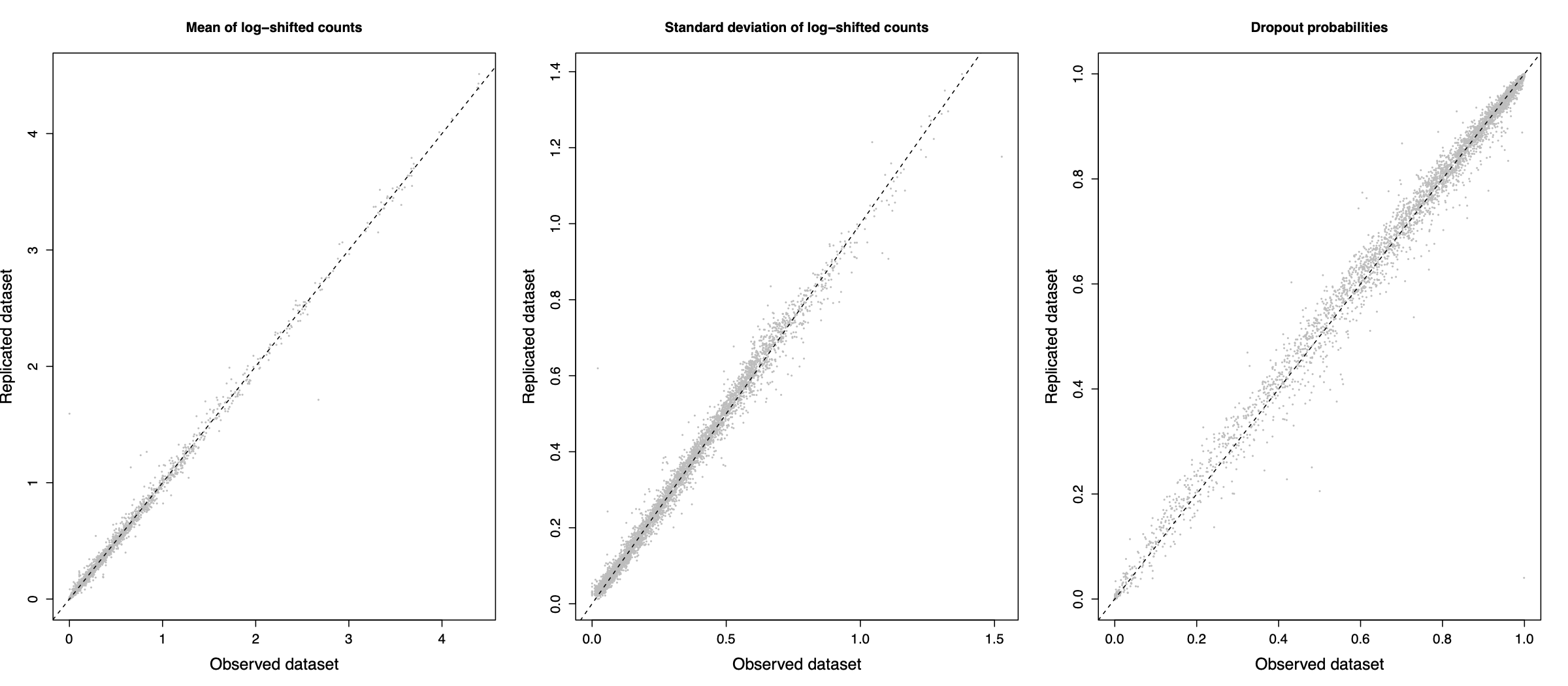}
\caption{Posterior predictive checks with one single replicated dataset. Each panel shows pointwise differences in a statistic between true and replicated data. The black dashed line corresponds to $y=x$.}
\label{fig:ppc_single2}
\end{figure}

The posterior predictive checks are conducted following from Section \ref{sec:ppc-pax6-method}. For a single replicate, Figure \ref{fig:ppc_single1} and Figure \ref{fig:ppc_single2} demonstrate that the replicated data exhibits similar relationships between pairwise statistics as observed in the true data. Additionally, the pointwise differences in statistics are nearly negligible, implying that the replicated data is consistent with the observed data.

\begin{figure}[!p]
\centering
\includegraphics[width=0.9\textwidth]{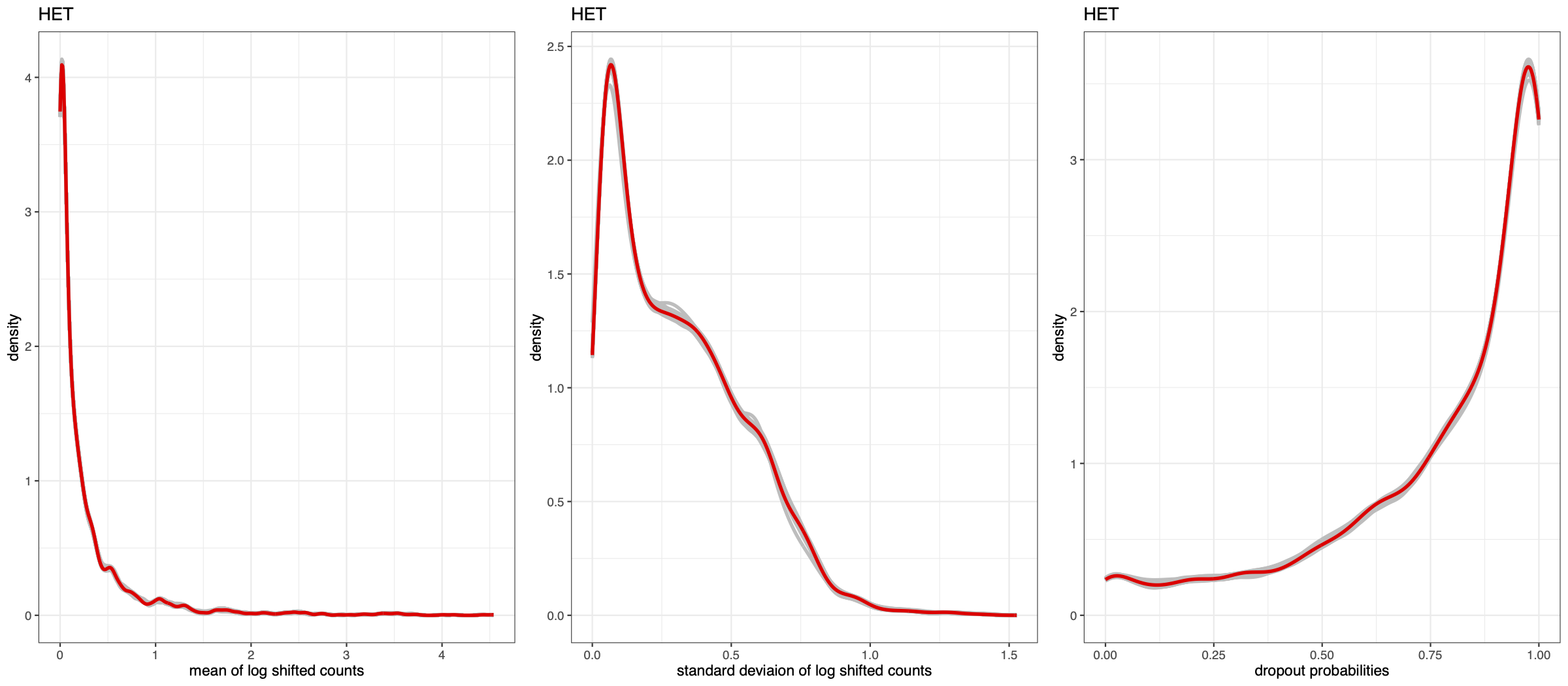}
\includegraphics[width=0.9\textwidth]{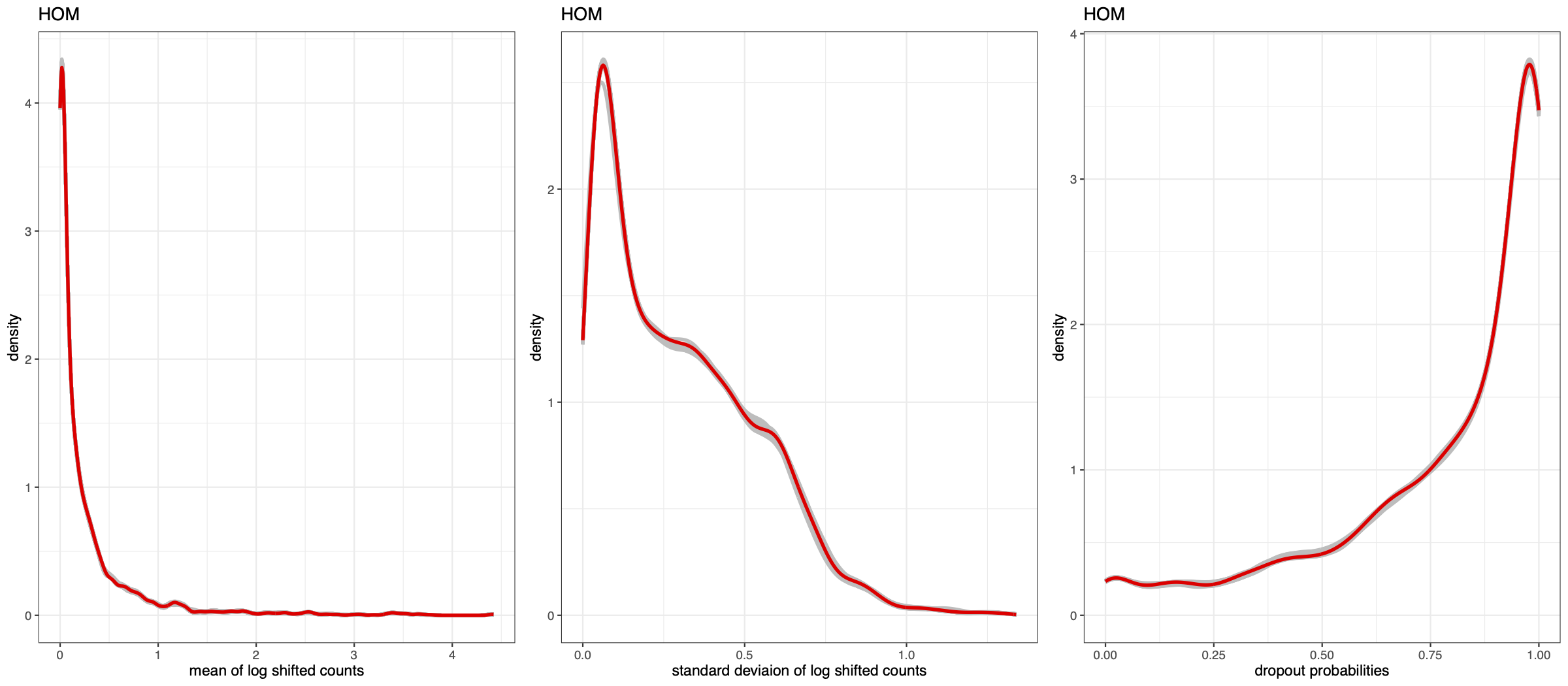}
\caption{Posterior predictive checks with multiple replicates for HET (top) and HOM (bottom). 
	Each panel shows the kernel density estimation of one statistic, with replicated and true datasets in grey and red, respectively. Left to right: mean of log shifted counts, standard deviation of log shifted counts and dropout probabilities.}
\label{fig:ppc_multiple}
\end{figure}

For multiple replicates, Figure \ref{fig:ppc_multiple} shows that estimated kernel of key statistics is similar between the simulated 200 datasets and the true observed data. Therefore, there is no strong disagreement between the model and the data.

\clearpage
\section{Additional Results for Calcium Imaging Data}
\label{appendix:additional-results-cidata}

\begin{figure}[htbp]
\centering
\includegraphics[width=0.8\textwidth]{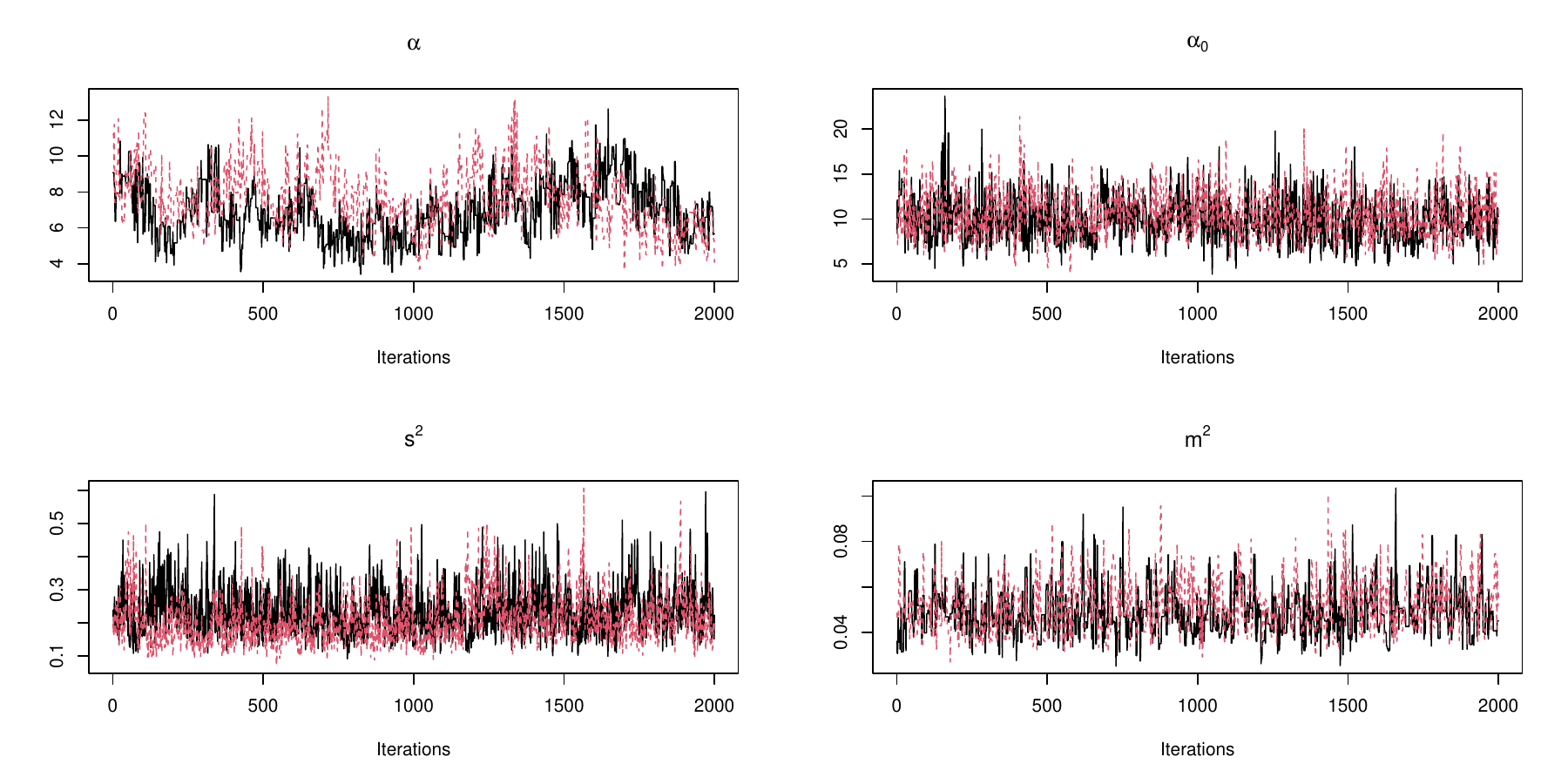}
\caption{Traceplots for concentration parameters and hyperparameters from two chains for estimating the data partition.}
\label{fig:cidata-trace}
\end{figure}

\begin{figure}[htbp]
\centering
\includegraphics[width=0.8\textwidth]{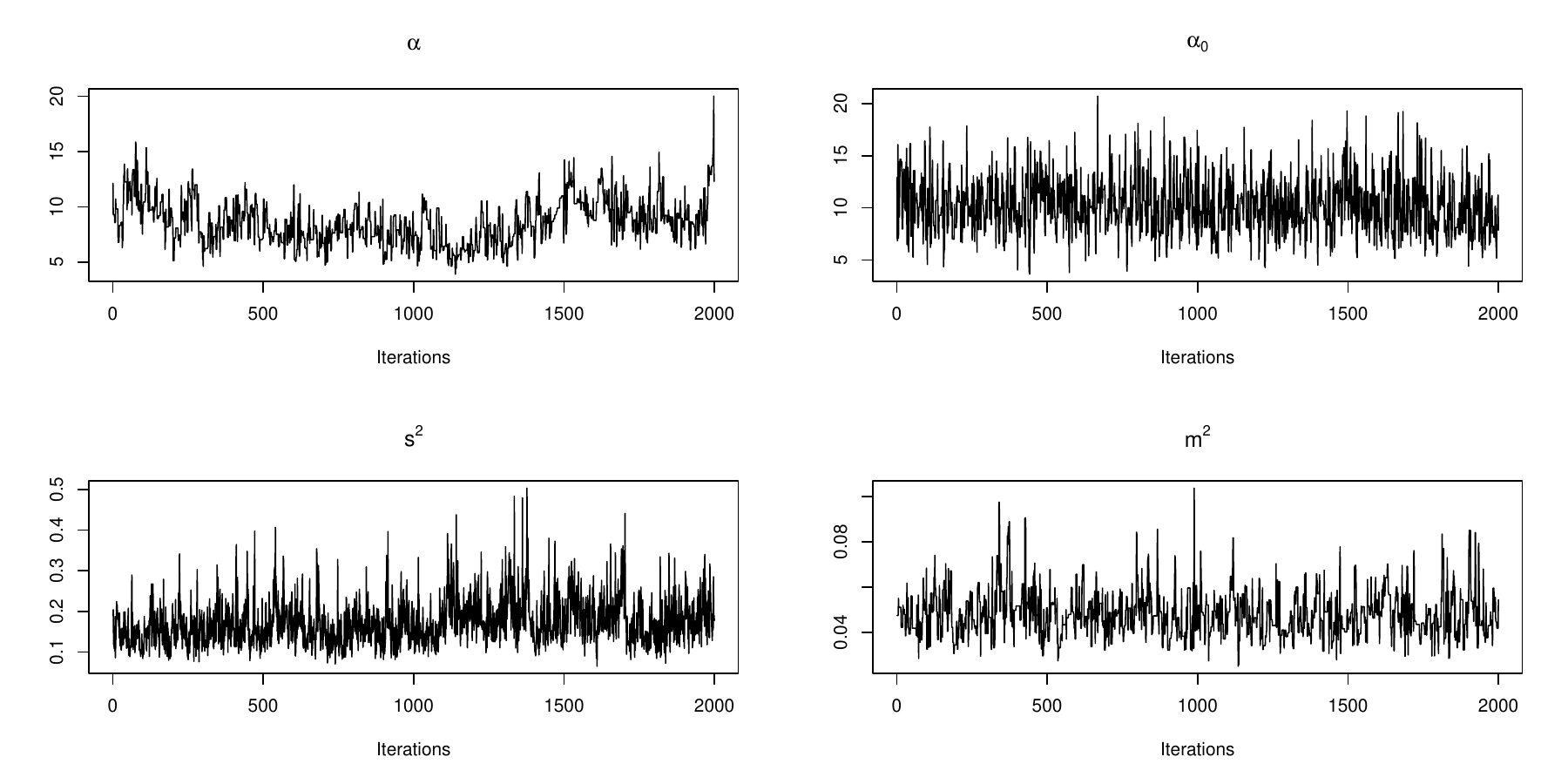}
\caption{Traceplots for concentration parameters and hyperparameters from the post-processing step.}
\label{fig:cidata-trace-post}
\end{figure}

In this section we provide additional results for clustering the calcium imaging data. To infer the clustering, two chains of length $10000$ have been run with a truncation level of $J=25$, followed by a burnin of $6000$ and a thinning of 2, yielding $2000$ posterior samples to estimate an optimal clustering based on VI. For the post-processing step, we run one chain of length 8000 and then apply a burnin of 4000 and thinning of 2, leading to 2000 samples. Traceplots for two chains for estimating the clustering are shown in Figure \ref{fig:cidata-trace}, and for the post-processing step they are shown in Figure \ref{fig:cidata-trace-post}, which suggest convergence. 

\subsection{General Results}

\begin{figure}[!htbp]
\centering
\includegraphics[width=0.95\textwidth]{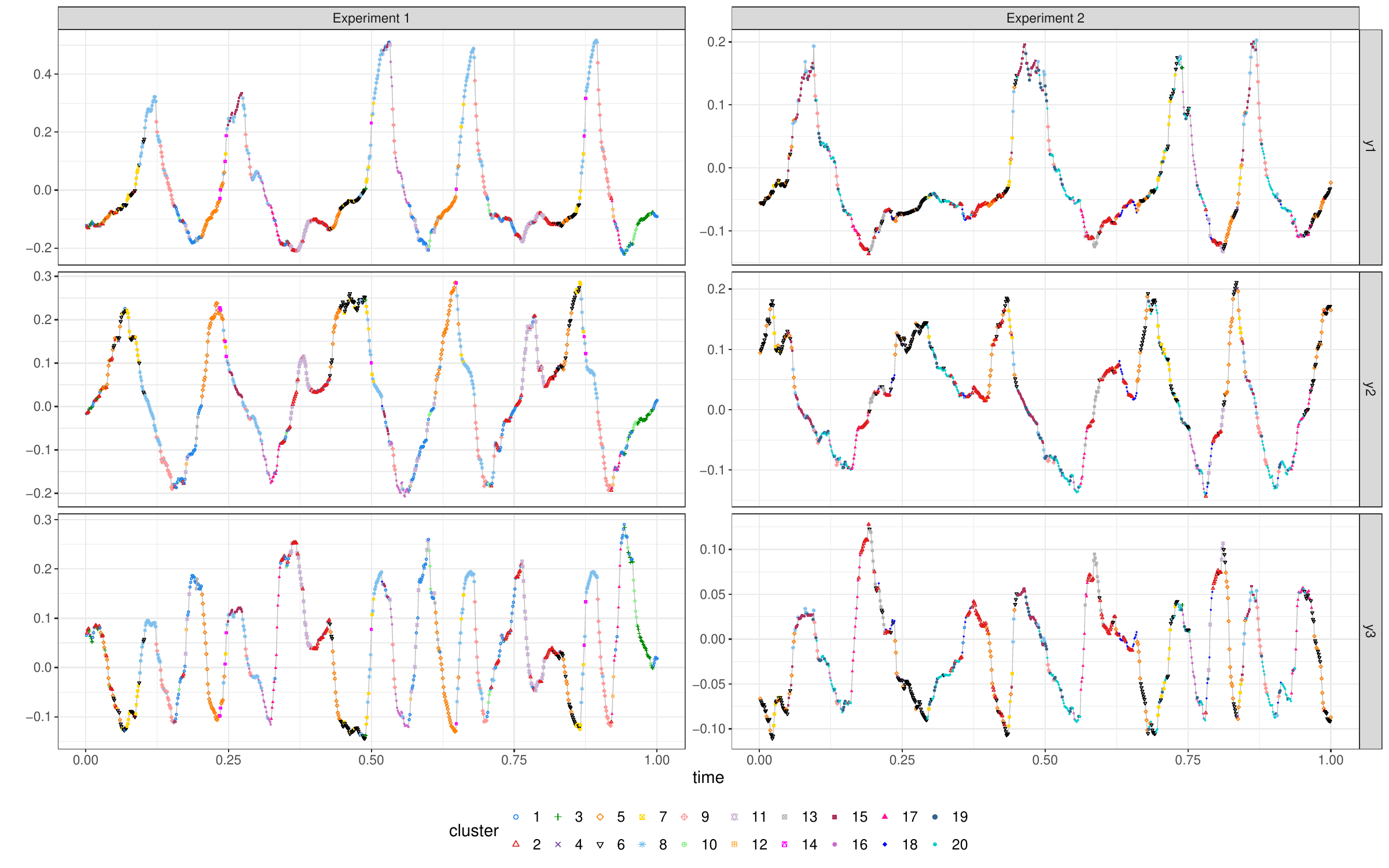}
\caption{Plot of each dimension against time $t$, with points colored by cluster labels. Left: Experiment 1. Right: Experiment 2.}
\label{fig:plot_y_vs_t_all}
\end{figure}

Figure \ref{fig:plot_y_vs_t_all} displays the time-series plot for all clusters. It is noticed that observations from the same cluster do share similar dependence on the past observed time point.

\begin{figure}[p]
\centering
\subfigure[C-HDP]{\includegraphics[width=0.95\textwidth]{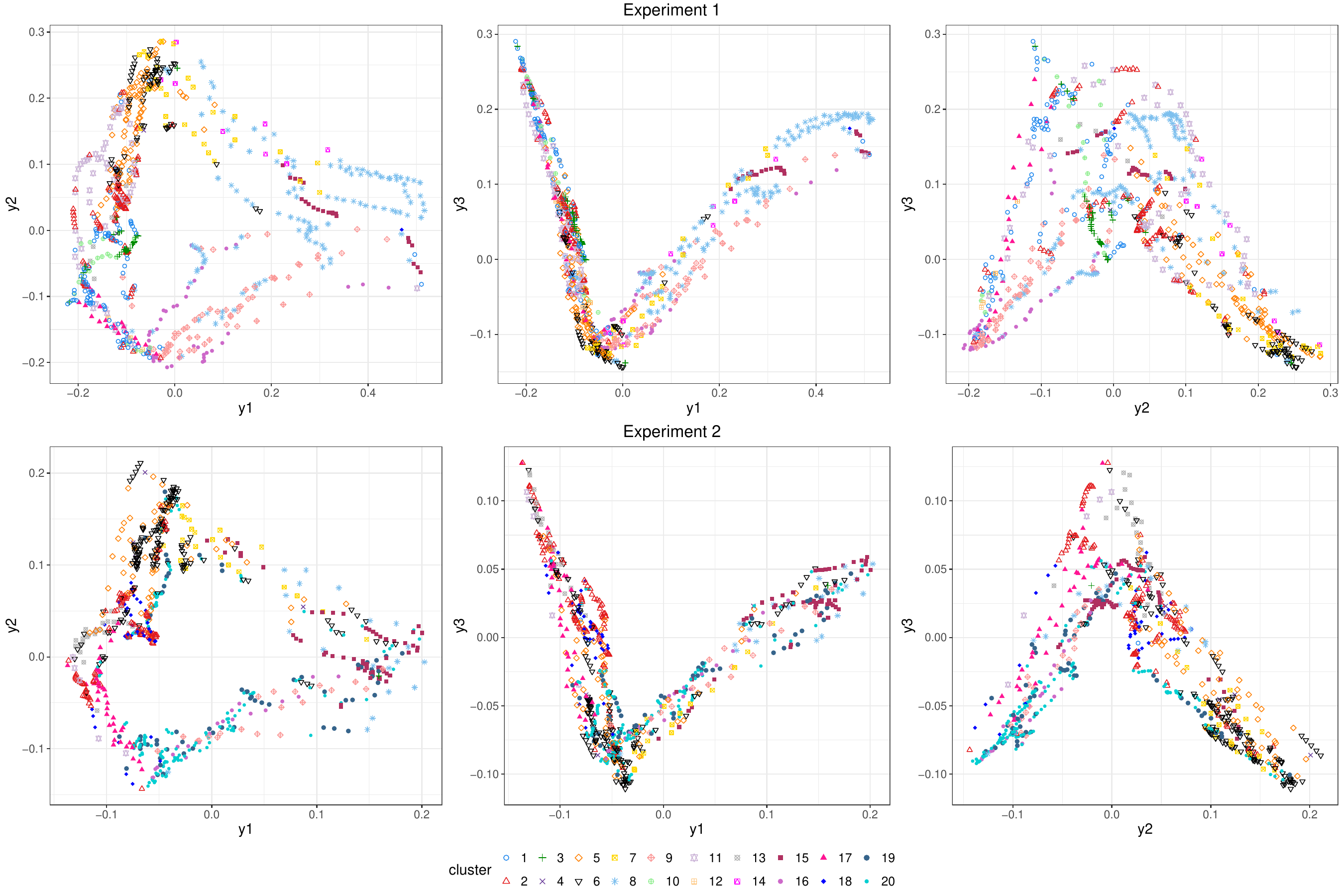}}
\subfigure[GMM]{\includegraphics[width=0.95\textwidth]{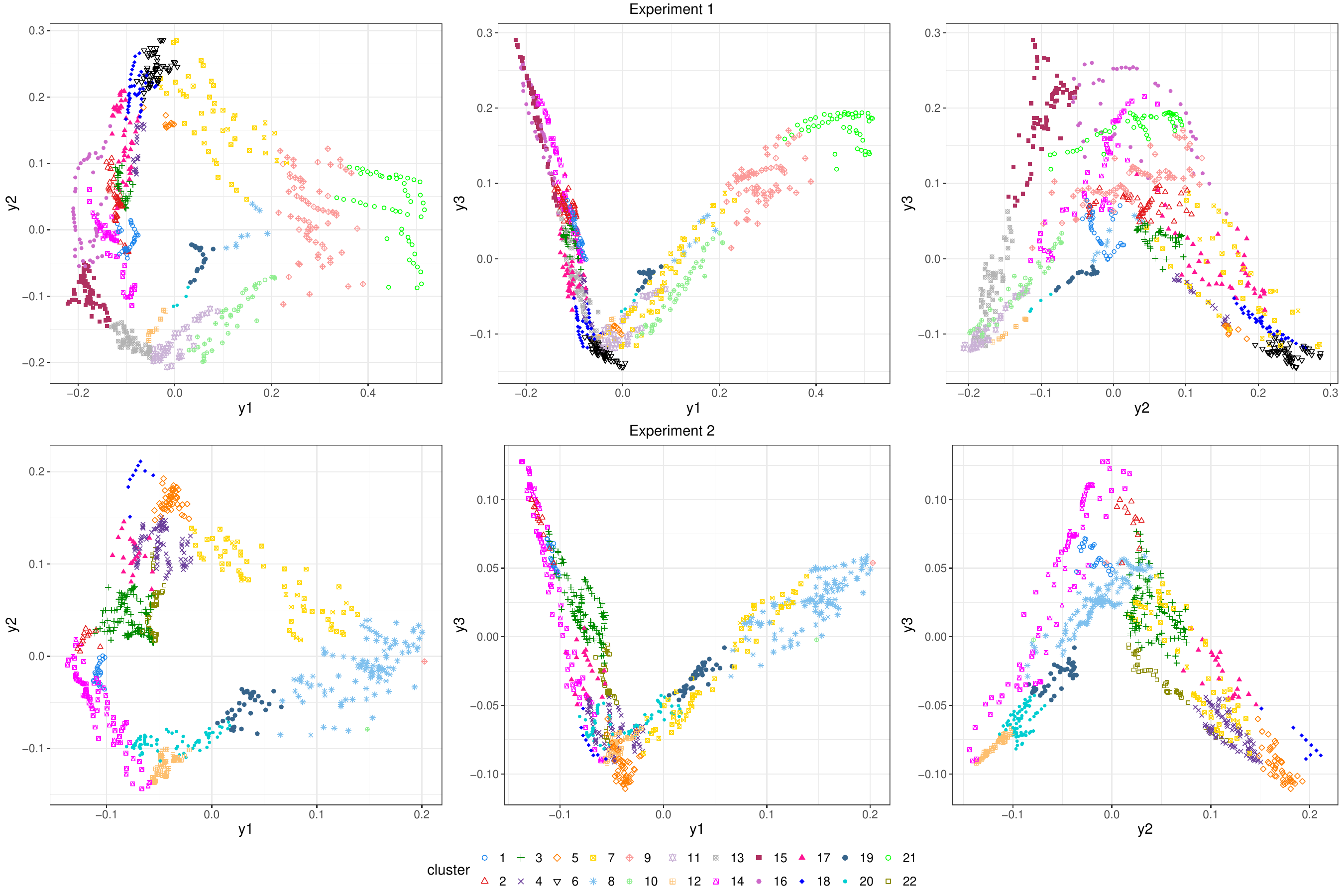}}
\caption{Pairwise scatterplots for two experiments, with observations colored by cluster membership from C-HDP (a) and GMM (b).}
\label{fig:pairplot-cidata}
\end{figure}

Figure \ref{fig:pairplot-cidata} shows the complete pairwise scatterplots from C-HDP and simple Gaussian mixture model. In C-HDP, the identified clusters shared between experiments tend to accumulate at similar positions in the lower-dimensional embeddings, whereas 
GMM identifies fewer shared clusters and just cuts time frames into groups of similar values.

\begin{figure}[tbp]
\centering
\includegraphics[width=0.95\textwidth]{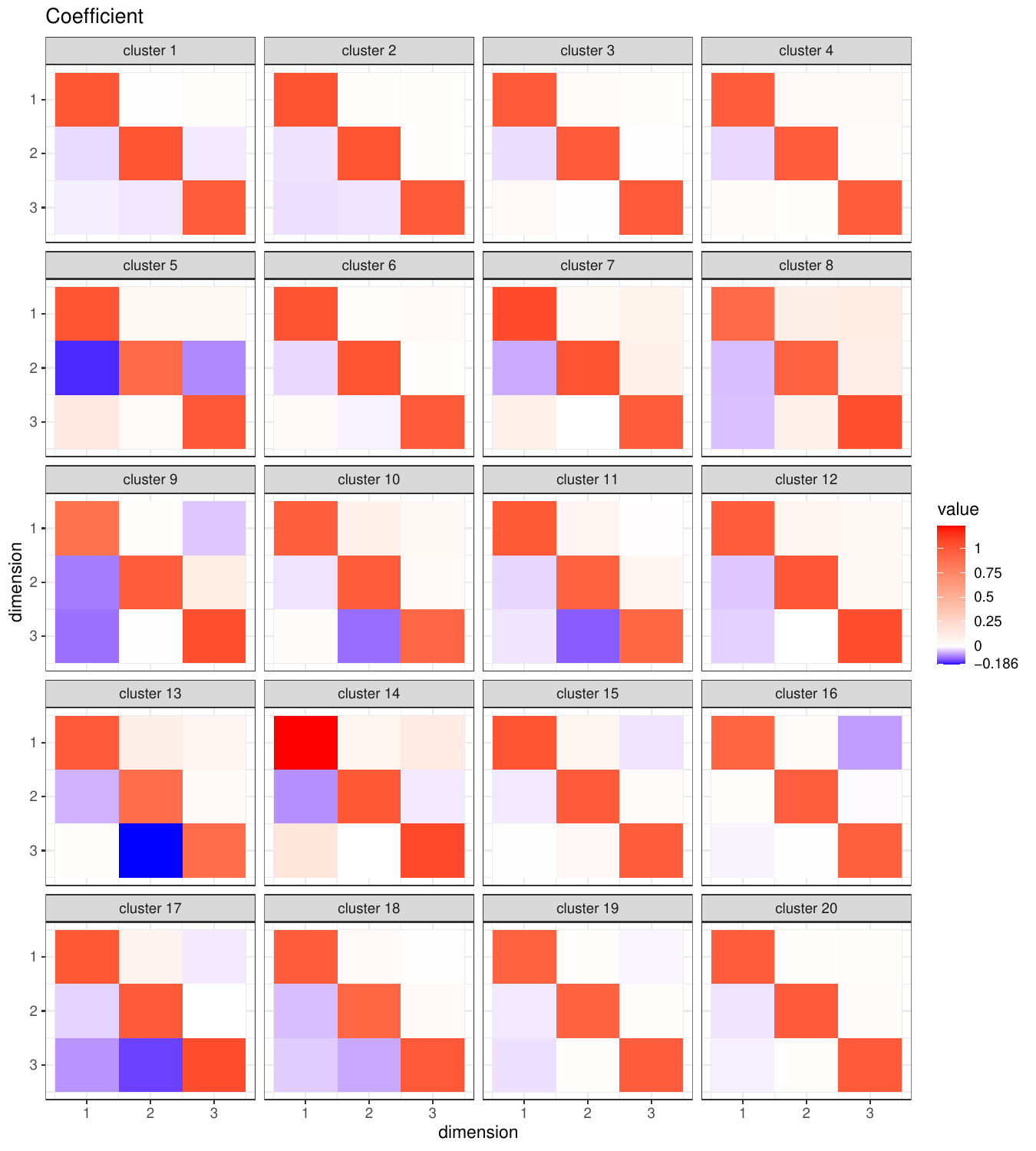}
\caption{Posterior mean of the coefficient matrix $\bB_j^*$ for each cluster (excluding the intercept). Values close to 0 are shown in white, with red for positive coefficients and blue for negative coefficients.}
\label{fig:heatmap_B}
\end{figure}

To investigate the differences between neural activity patterns, we visualize the posterior mean of the coefficient matrix $\bB_j^*$ for each cluster (Figure \ref{fig:heatmap_B}). Within each cluster, the contribution to each dimension is dominated by the corresponding dimension from the past time frame (red diagonal grids), with the strongest effect observed in cluster 14 (first dimension). In contrast, interactions between different lower-dimensional embeddings are generally weaker and tend to show more negative contributions. Specifically, cluster 13 exhibits the strongest negative between-dimension influence, followed by cluster 5.

\begin{figure}[p]
\centering
\subfigure[Experiment 1]{\includegraphics[width=0.95\textwidth]{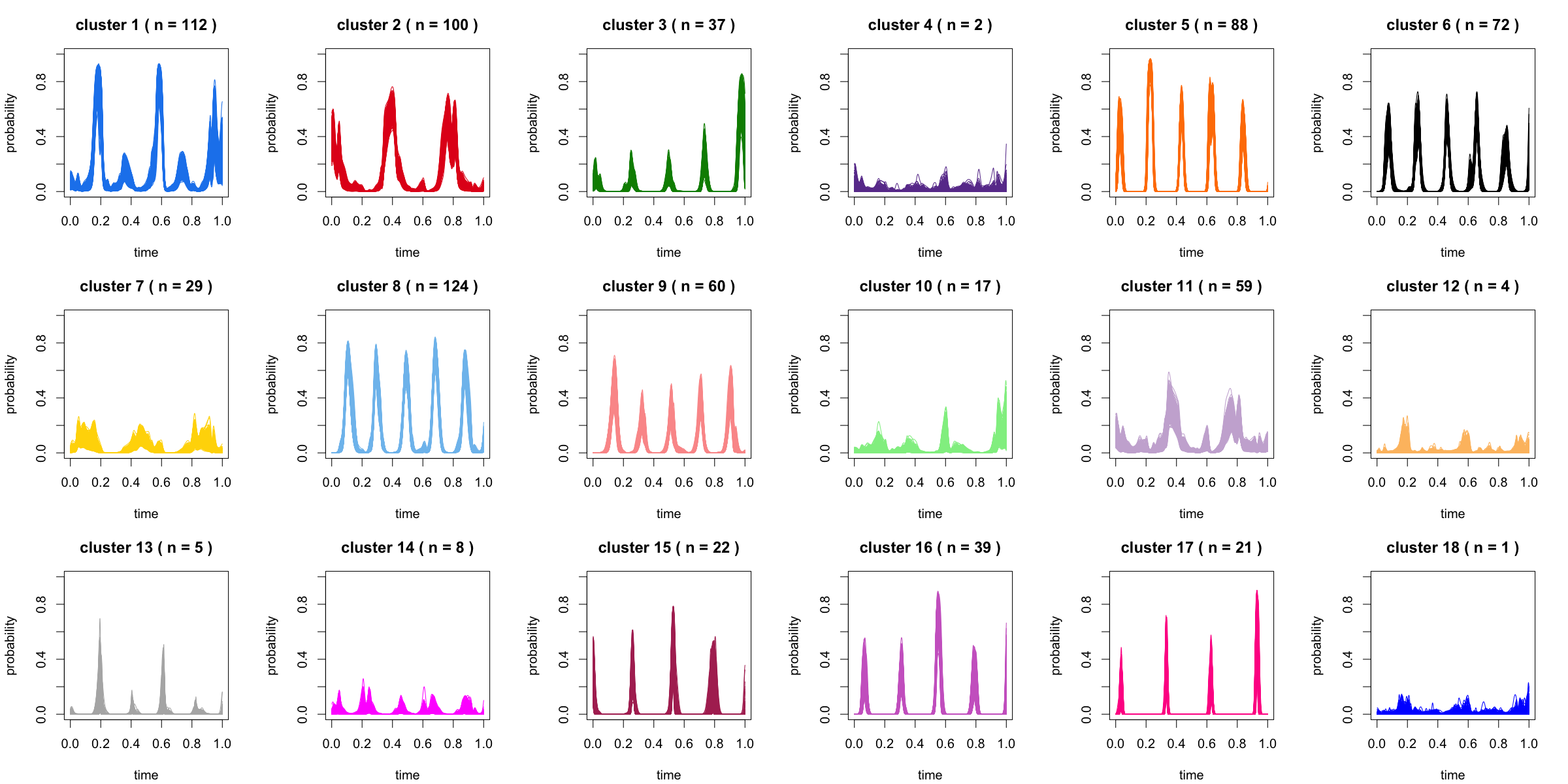}}
\subfigure[Experiment 2]{\includegraphics[width=0.95\textwidth]{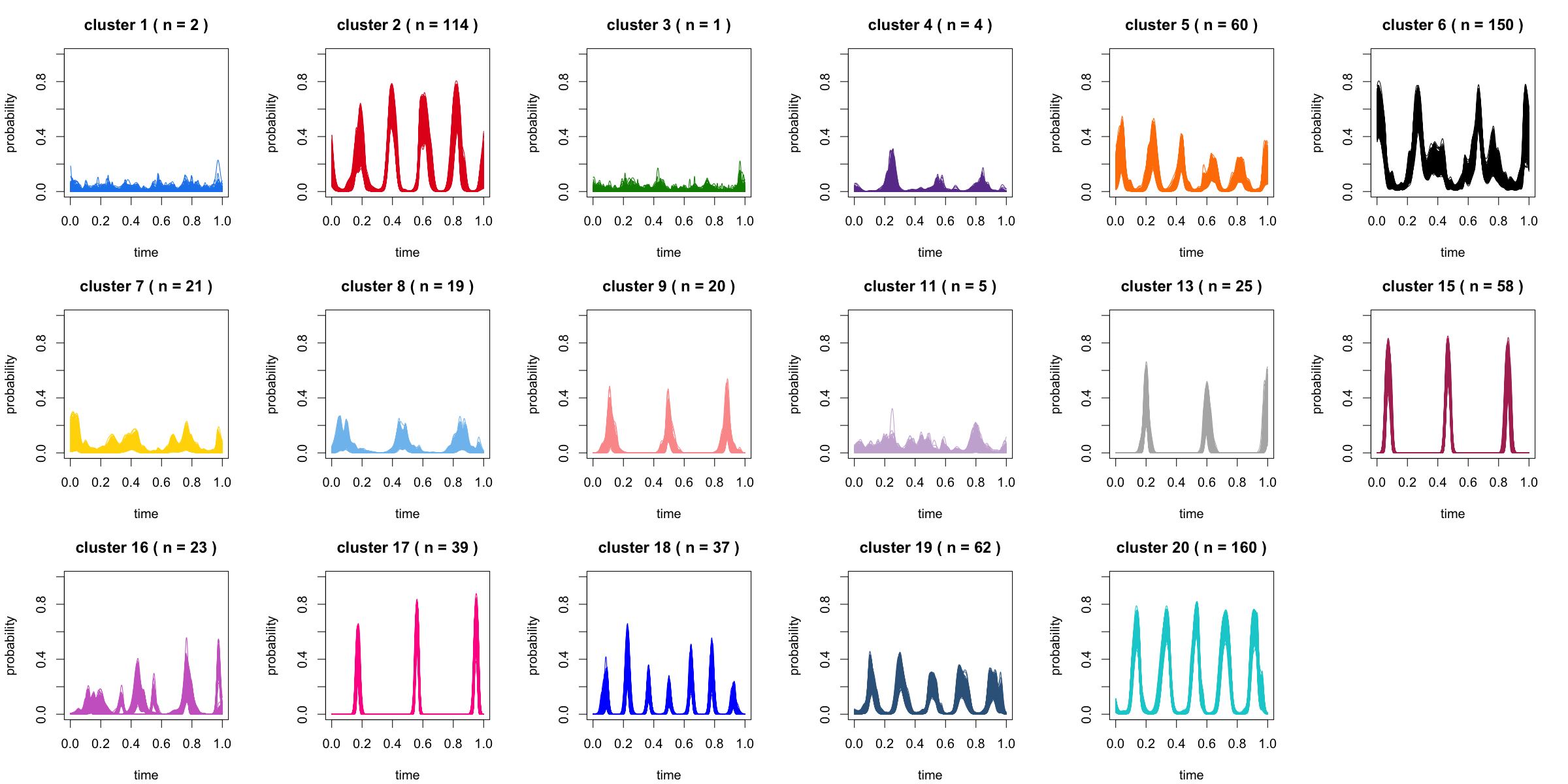}}
\caption{Time-dependent probabilities for each cluster in each experiment in the calcium imaging data. Only non-empty components are plotted. Cluster size is indicated in the title.}
\label{fig:p_vs_t_cidata}
\end{figure}

Figure \ref{fig:p_vs_t_cidata} shows posterior samples for time-dependent probabilities across all clusters, with probabilities close to zero for small-size clusters. In addition, clusters may have different periodicities, e.g. clusters 2, 6 and 9, suggesting varying frequencies associated with pattern across experiments. Although cluster 5 seems to have a similar periodicity, the probabilities are higher in the first experiment.

\subsection{Predictions} \label{appendix:cidata-prediction}

\begin{figure}[htb]
\begin{minipage}[h]{0.49\textwidth}
	\includegraphics[width=1\textwidth]{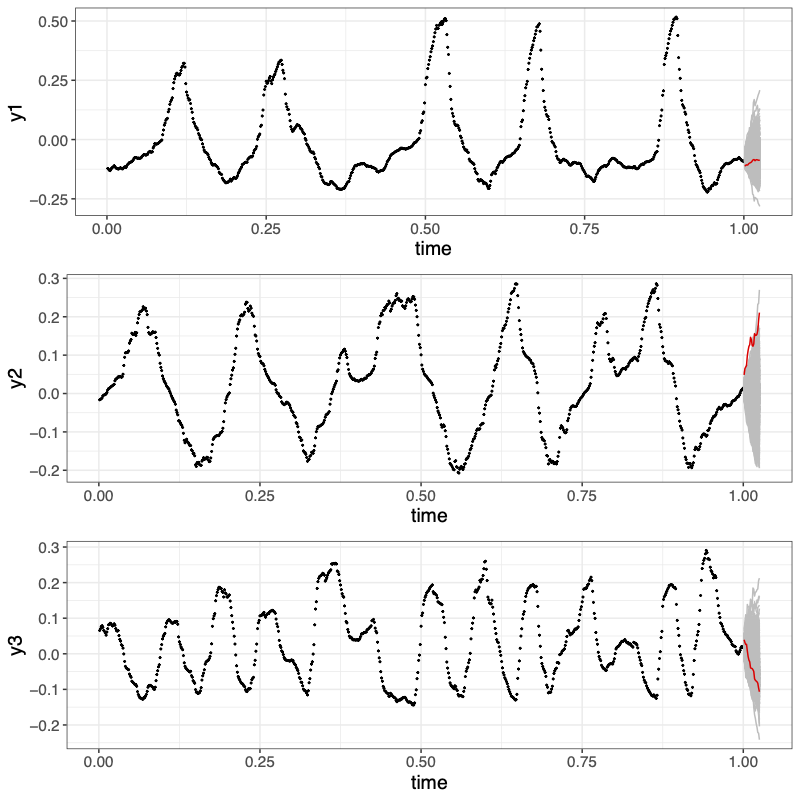}
\end{minipage}
\begin{minipage}[h]{0.49\textwidth}
	\includegraphics[width=1\textwidth]{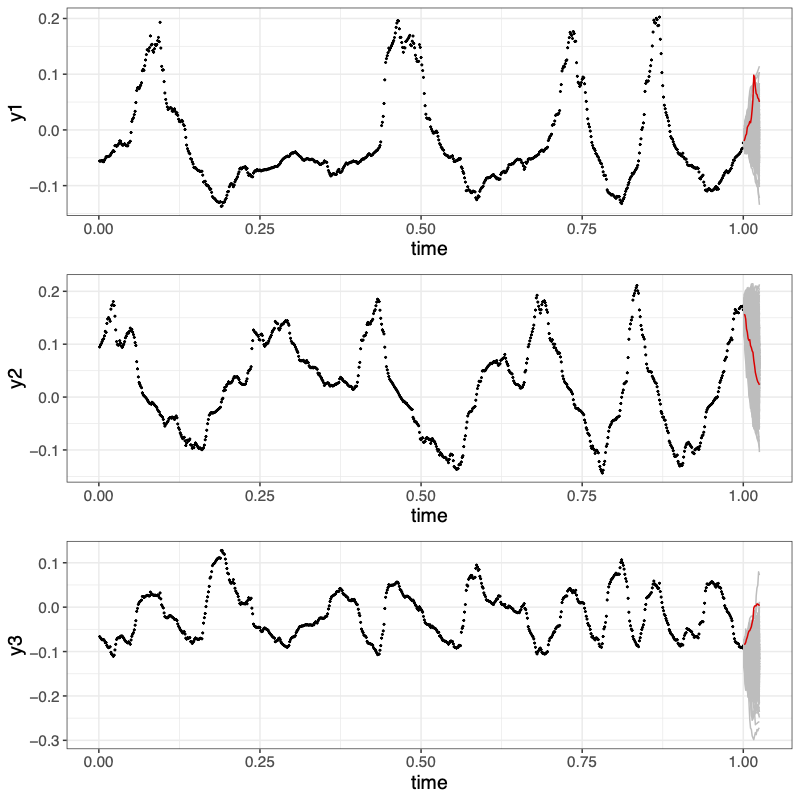}	
\end{minipage}
\caption{Predicting future trends. Red solid line denotes true future observations, with posterior predictive samples shown in grey. Black points denote the observed data used for model fitting. Left: Experiment 1. Right: Experiment 2.}
\label{fig:post_mean_of_ypred}
\end{figure}

Forecasting is one of the primary goals in time-series modelling, which allows us to understand the encoded neural activity in the near future. We predict the neural activity for 20 future times points, using the same time increment as the observed data. In particular, time-dependent probabilities are first used to generate allocation variables $z_{i,d}$, based on which future observed values are simulated. The true future neural activity are found to be generally covered by the samples of the predictions (Figure \ref{fig:post_mean_of_ypred}), and the uncertainty increases fast as predictions are made further away.

\begin{figure}[htb]
\centering
\includegraphics[width=0.8\textwidth]{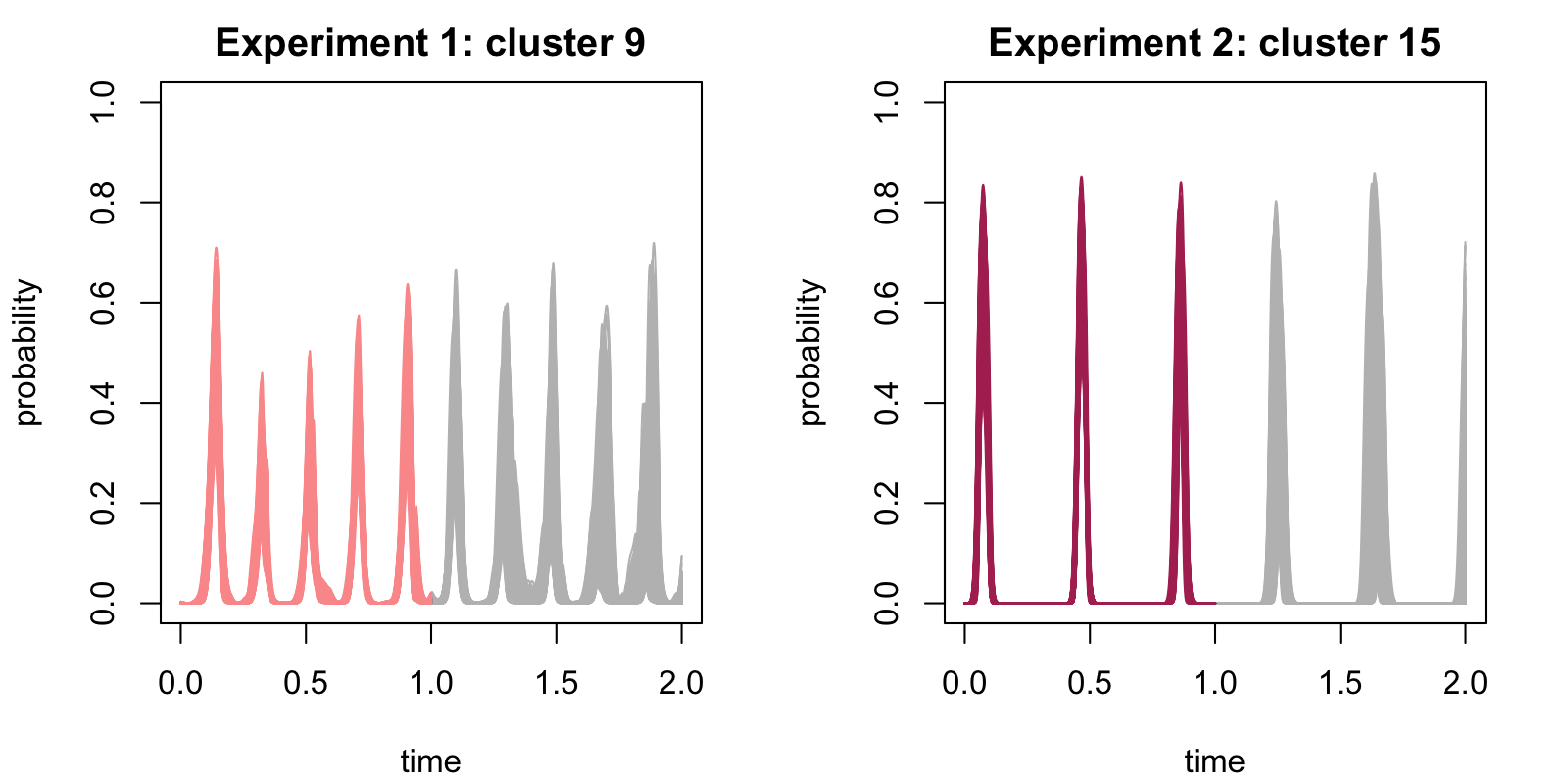}
\caption{Time-dependent probabilities for future time. Grey area shows predictions.}
\label{fig:p_vs_t_future_time}
\end{figure}

In addition to neural activity, we can also estimate the covariate-dependent probability for future time points. Unlike Gaussian kernels where probabilities at future time points are almost around zero due to absence of data, for periodic kernel, the repetitive pattern is similar for either interpolation and extrapolation, with similar uncertainty (Figure \ref{fig:p_vs_t_future_time}).

\subsection{Posterior Predictive Checks} \label{sec:ppc-cidata}

Conditional on the optimal clustering and observed data, we generate 200 replicated datasets using MCMC samples from the post-processing step. In particular, one replicate for observation $i$ in experiment $d$ is given by 
\begin{equation*}
\bmy_{i,d}^{rep,(l)} \sim \Norm\left(\left({\bL_{j}^*}^{(l)}\right)^T \bmx_{i,d},{\bSigma_{j}^{*}}^{(l)}\right),
\end{equation*}
where $\bmx_{i,d}=(1, y_{i-1,1,d},y_{i-1,2,d},y_{i-1,3,d})^T$, ${\bL_{j}^*}^{(l)}$ and ${\bSigma_{j}^{*}}^{(l)}$ denote the $l$-th posterior draw.

\begin{figure}[htb]
\begin{minipage}[h]{0.5\textwidth}
	\includegraphics[width=0.95\textwidth]{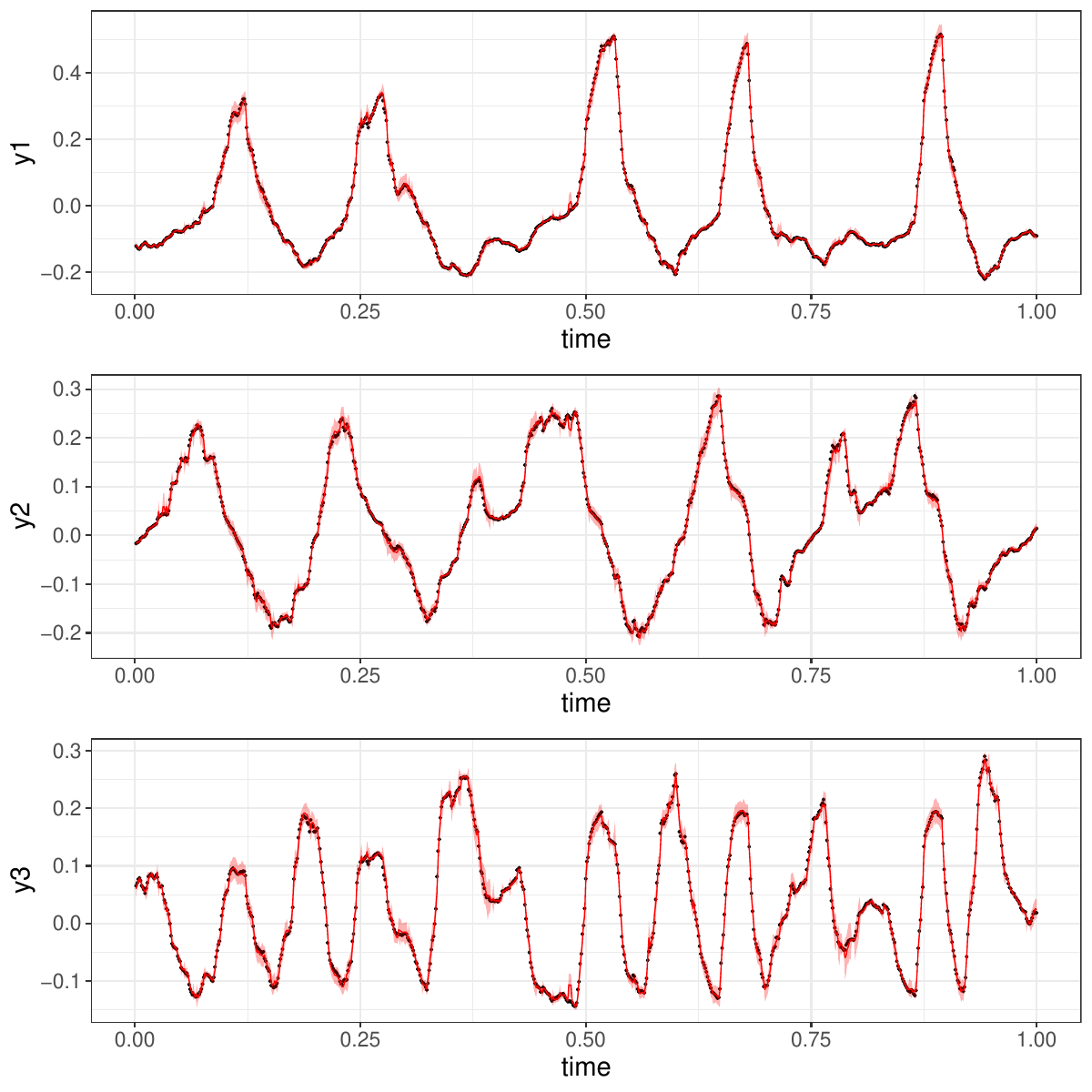}
\end{minipage}
\begin{minipage}[h]{0.5\textwidth}
	\includegraphics[width=0.95\textwidth]{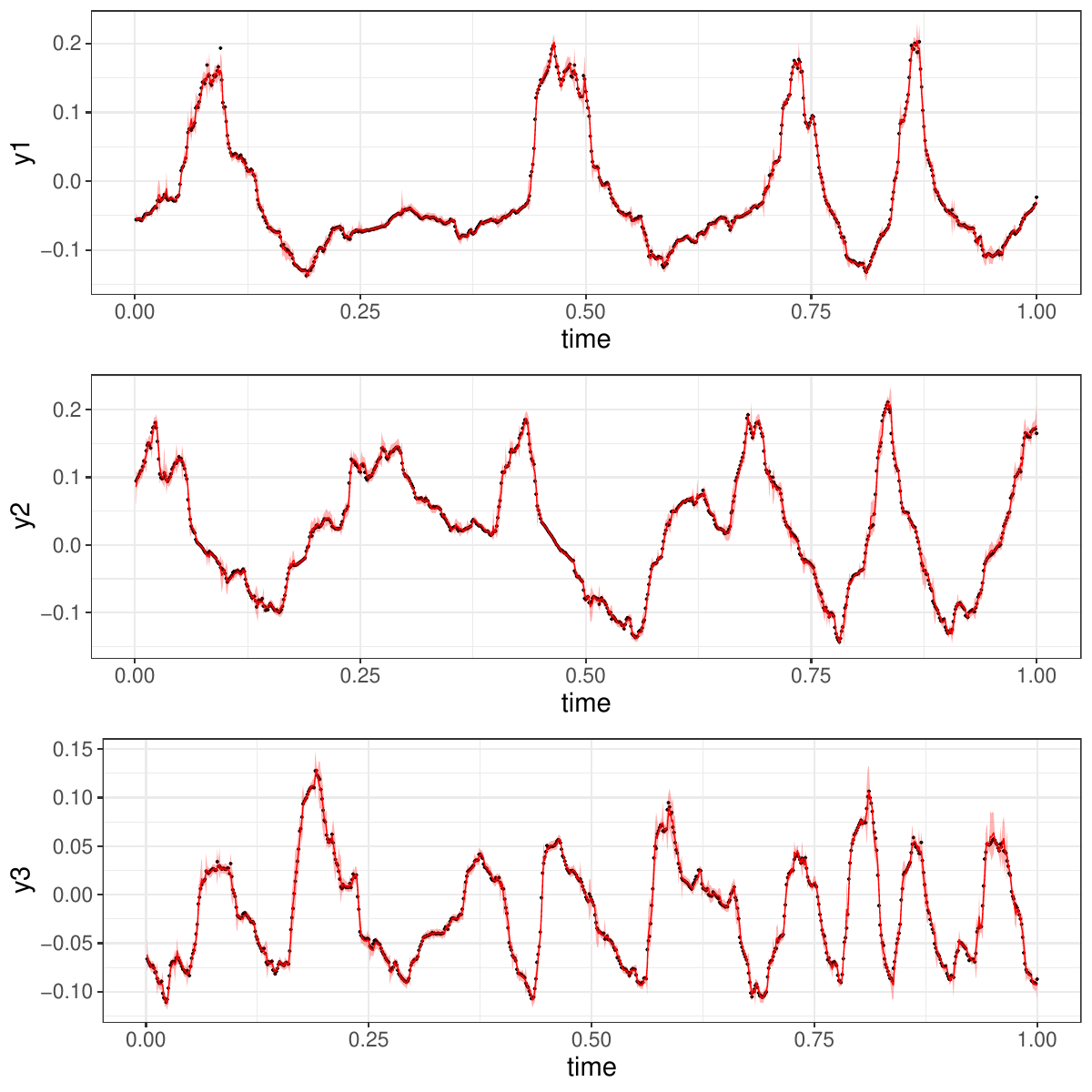}		
\end{minipage}
\caption{Posterior predictive checks based on 200 replicates. The red line denotes the posterior mean of the replicates, with 99\% HPD CIs shown in the red area. Black points denote the observed data. Left: Experiment 1. Right: Experiment 2.}
\label{fig:post_mean_of_yrep}
\end{figure}

From Figure \ref{fig:post_mean_of_yrep}, the replicated data closely resembles the true datasets. The observed activity is similar to the posterior mean of the replicates with short 99\% CIs, indicating no strong disagreement between the data and the model.

\end{document}